\documentclass[preprint,flushrt]{aastex}

\usepackage{natbib}
\usepackage{lineno}
\usepackage{graphics}
\usepackage{url}
\newcommand{\bl}{\begin{list}{}}

\newcommand{\el}{\end{list}}

\shorttitle{{\it Fermi}-LAT  $E>10 GeV$ source list}
\shortauthors{Abdo et al.}

\newcommand{\gray}{\mbox{$\gamma$-ray }}
\newcommand{\grays}{\mbox{$\gamma$-rays }}

\newcommand{\Fermic}{\emph{Fermi}}
\newcommand{\Fermi}{\Fermic\ }
\newcommand{\FermiLATc}{\Fermic~LAT}
\newcommand{\FermiLAT}{\FermiLATc\ }

\newcommand{\lapp}{\ensuremath{\stackrel{\scriptstyle <}{{}_{\sim}}}}
\newcommand{\gapp}{\ensuremath{\stackrel{\scriptstyle >}{{}_{\sim}}}}

\title{The First Fermi-LAT Catalog of Sources Above 10 GeV }
\author{
M.~Ackermann\altaffilmark{2{\dagger}}, 
M.~Ajello\altaffilmark{3}, 
A.~Allafort\altaffilmark{4}, 
W.~B.~Atwood\altaffilmark{5}, 
L.~Baldini\altaffilmark{6}, 
J.~Ballet\altaffilmark{7}, 
G.~Barbiellini\altaffilmark{8,9}, 
D.~Bastieri\altaffilmark{10,11}, 
K.~Bechtol\altaffilmark{4}, 
A.~Belfiore\altaffilmark{5,12,13}
% ...
}
% ...
\altaffiltext{3}{Space Sciences Laboratory, 7 Gauss Way, University of California, Berkeley, CA 94720-7450, USA}
\altaffiltext{4}{W. W. Hansen Experimental Physics Laboratory, Kavli Institute for Particle Astrophysics and Cosmology, Department of Physics and SLAC National Accelerator Laboratory, Stanford University, Stanford, CA 94305, USA}
\altaffiltext{5}{Santa Cruz Institute for Particle Physics, Department of Physics and Department of Astronomy and Astrophysics, University of California at Santa Cruz, Santa Cruz, CA 95064, USA}
\altaffiltext{6}{Universit\`a  di Pisa and Istituto Nazionale di Fisica Nucleare, Sezione di Pisa I-56127 Pisa, Italy}
\altaffiltext{7}{Laboratoire AIM, CEA-IRFU/CNRS/Universit\'e Paris Diderot, Service d'Astrophysique, CEA Saclay, 91191 Gif sur Yvette, France}
\altaffiltext{8}{Istituto Nazionale di Fisica Nucleare, Sezione di Trieste, I-34127 Trieste, Italy}
\altaffiltext{9}{Dipartimento di Fisica, Universit\`a di Trieste, I-34127 Trieste, Italy}
\altaffiltext{10}{Istituto Nazionale di Fisica Nucleare, Sezione di Padova, I-35131 Padova, Italy}
\altaffiltext{11}{Dipartimento di Fisica e Astronomia ``G. Galilei'', Universit\`a di Padova, I-35131 Padova, Italy}
\altaffiltext{12}{Universit\`a degli Studi di Pavia, 27100 Pavia, Italy}
\altaffiltext{13}{INAF-Istituto di Astrofisica Spaziale e Fisica Cosmica, I-20133 Milano, Italy}
% ...

\begin{abstract}

We present a catalog of $\gamma$-ray sources at energies above 10 GeV based on
data from the Large Area Telescope (LAT) accumulated during the first
three years of the {\it Fermi Gamma-ray Space Telescope} mission.  
The first \Fermic-LAT catalog of $>$10~GeV sources (1FHL) has
514 sources.
For each source we present location, spectrum,
a measure of variability, and associations
with cataloged sources at other wavelengths. We found that 449 (87\%)
could be associated with known sources,  of which 393 (76\% of the 1FHL sources) are
active galactic nuclei. Of the 27 sources associated with known pulsars, we find 20~(12) to 
have significant pulsations in the range $>$10~GeV~($>$25~GeV). In this work we also report that, at energies above 10 GeV, unresolved sources 
account for $27\pm8$\,\% of the isotropic $\gamma$-ray background, while the unresolved Galactic population contributes only at the few percent level to the Galactic diffuse background. We also highlight the subset of the 1FHL sources that are best  candidates 
for detection at energies above 50--100 GeV with current 
and future ground-based $\gamma$-ray observatories. 

\end{abstract}


\begin{document}

\section{\label{sec:intro}Introduction}


The primary catalog of $\gamma$-ray sources detected by the \Fermi Large Area Telescope (LAT), the second LAT source catalog \cite[hereafter 2FGL,][]{LAT_2FGL}, presents sources detected at energies above 100~MeV in the first two years of science operations.  Motivations for studying the $\gamma$-ray sky at even higher energies in LAT data are numerous, including finding the hardest-spectrum sources and characterizing them separately from their generally much brighter emission at lower energies.  Here we present a catalog of sources detected above 10~GeV in the LAT data.

This work is not the first systematic study of $\gamma$-ray sources in the GeV range.  \citet{Lamb1997} presented a catalog of 57 sources detected above 1~GeV in 4.5 years of data from the Energetic Gamma-Ray Experiment Telescope (EGRET) on the {\it Compton} Gamma-Ray Observatory.  Relative to the third EGRET catalog of sources detected above 100~MeV \citep[hereafter 3EG,][]{3EGCatalog} the localization regions are smaller and the fraction of sources for which no counterpart at other wavelengths could be confidently assigned is also smaller (53\% vs. 63\% of the 271 3EG sources).  Individual sources could not be detected at higher energies with EGRET but  \citet{Thompson2005} studied the distribution of the 1506 EGRET $\gamma$-rays above 10~GeV and found 187 to be within 1$\degr$ of a 3EG source.  \citet{Neronov2010} searched for sources at energies above 100~GeV in $\sim$2 years of LAT data for Galactic latitudes $|b| < 10\degr$, reporting 19 sources.  \citet{Neronov2011} reported strong correlations between $>$100~GeV LAT $\gamma$-rays and cataloged  $\gamma$-ray sources.

The current LAT data allow a much deeper exploration of the sky above 10~GeV than has been possible before, with an energy range that approaches the
$>$100~GeV (hereafter VHE) domain studied by  imaging atmospheric Cherenkov telescopes (IACTs).  Broadband studies of $\gamma$-ray sources
provide insights into the acceleration and radiation mechanisms operating at the highest energies. 
The relatively small fields of view and limited duty cycles of IACTs, and the low fluxes of VHE sources, makes target selection very important for source searches with IACTs. 
According to the TeVCat catalog\footnote{\url{http://tevcat.uchicago.edu/}} version 3.400, 105 sources have been detected at VHE\footnote{Including recently announced VHE detections the number is 143.}, which is approximately 20 times fewer than in the 2FGL catalog.
A catalog of $>$10~GeV  \FermiLAT detections may increase the efficiency of these searches with 
current generation of IACTs, namely H.E.S.S., MAGIC and VERITAS.

In our catalog of LAT sources above 10~GeV we report the locations, spectra, and
variability properties of the
514 sources significantly detected in this range during the first three years of the {\it Fermi} mission.
Many of these sources are
already included in the 2FGL catalog, although in that catalog their
characterization 
is dominated by the much larger numbers of
$\gamma$ rays detected in the
energy range 100~MeV--10~GeV. Consequently, the characteristics of the sources at the highest
\FermiLAT energies might be overlooked.
In addition, several of the sources reported here were
not listed in the 2FGL, possibly due to the 33\% less exposure.
We also develop a set of criteria to select the
sources that are the best candidates for detection at VHE with the
current generation of IACTs.

In \S~\ref{sec:FermiLAT} we describe the capabilities of the
\FermiLAT to perform astronomy at energies above 10 GeV.  Section~\ref{sec:sample}
describes the overall \Fermi sky above 10 GeV, the analysis procedures, 
the sources detected and the corresponding associations to known objects.  In \S~\ref{Characterization} 
we report on the overall characteristics of these sources, with special focus on active galactic nuclei (AGNs), which constitute the majority of the catalog.  Section~\ref{VHECandidates} presents the criteria for selecting sources that may be detectable with the current generation of IACTs operating above 100 GeV.  In \S~\ref{PopStudies} we report on the properties of the source populations above 10 GeV, and in \S~\ref{Conclusion} we summarize and conclude this work.

\cleardoublepage


\section{\label{sec:FermiLAT} Instrument \& Background}

The \FermiLAT is a $\gamma$-ray telescope operating from $20$\,MeV to
$>300$\,GeV. The instrument is a $4 \times 4$ array of identical
towers, each one consisting of a tracker (where the photons have a high probability of converting to
pairs, which are tracked to allow reconstruction of the $\gamma$-ray direction) and a segmented calorimeter (where the electromagnetic shower produces scintillation light, from which the $\gamma$-ray
energy can be estimated).
The tracker is covered
with an anti-coincidence detector to reject the charged-particle
background.  Further details on the LAT, its performance, and calibration 
are given by \citet{LAT09_instrument} and \citet{LAT12_calib}.
In  the following subsections we report on the event classification, the corresponding instrument response functions, the data selection, the exposure, and the resulting point-source sensitivity.  The sensitivity is derived using the approach presented by \citet{LAT_1FGL} for the first  \FermiLAT  source catalog, which is based on a standard likelihood function formalism.  
The likelihood combines the data with a model of the sky that includes
localized $\gamma$-ray sources and diffuse backgrounds and accounts for the instrument response functions and the exposure.


\subsection{LAT Event Class Selection}
\label{LatAnalysisIntro}

The $\gamma$-ray event selection used for this study benefitted from the experience
acquired by the \FermiLAT\ collaboration during the first years of
operation, which led to the development of the Pass 7 event classifications \citep{LAT12_calib}.
The {\tt P7CLEAN} event class was used in constructing this catalog as it provides a substantial reduction in residual cosmic-ray background (cosmic rays misclassified as $\gamma$-rays) above 10~GeV relative to the {  \tt P7SOURCE} event class used for 2FGL.  The isotropic background, which comprises both the diffuse $\gamma$-ray and residual cosmic-ray backgrounds, is a factor of approximately five
less than for the {  \tt P7SOURCE} event class, which was used for the 2FGL catalog analysis, for which the larger effective area at lower energies was the overriding consideration. The decrease in the isotropic background is dominated by the large reduction in residual
charged cosmic rays in the {  \tt P7CLEAN} class, approximately a factor of four at 10 GeV
and more than an order of magnitude at 100 GeV, as reported in \cite{LAT12_calib}. 
For the analyses we used the corresponding  {  \tt P7CLEAN\_V6} instrument
response functions.
The systematic uncertainty in the effective area above 10 GeV 
is estimated to be
10\% \citep{LAT12_calib}\footnote{See also \url{http://fermi.gsfc.nasa.gov/ssc/data/analysis/LAT_caveats.html}}.

\subsection{Performance of the LAT}
\label{LatInstrumentIntro}

The \FermiLAT  has a field of view of $\sim$2.4~sr, and is most sensitive 
(in $E^2 dN/dE$) for photon energies
of about 3~GeV.  Above this energy, up to
$\sim$300~GeV, the on-axis effective area for  {  \tt P7CLEAN\_V6} is
at least 0.7~m$^2$.  It rolls off to $\sim$0.65~m$^2$ by 500~GeV.  
At \gray\ energies below 10 GeV, the point-spread
function (PSF) is dominated by multiple Coulomb
scattering in the tracker (which varies inversely with the
electron energy).  Above 10~GeV the geometry of the tracker itself is
the dominant factor,
and so the PSF is not as strongly energy dependent as at lower energies.
The 68\% containment radius of the PSF ($Front$ and $Back$ averaged) is $\sim$$0\fdg$3 at 10~GeV,
narrowing to $\sim$$0\fdg2$ above 100~GeV.  
The energy resolution ranges from 8\% (68\% containment) at
10~GeV to approximately 15\% at 500~GeV due to the lack of containment of the
electromagnetic shower inside the calorimeter.
This does not appreciably affect the sensitivity, because the angular resolution and effective area depend only weakly on energy in this range.

\subsection{Data Selection and the Sky Above 10 GeV}
\label{FermiSky10GeV}


In this work we analyze $\gamma$ rays with energies in the range 10--500 GeV.
To limit the contamination from $\gamma$ rays produced by cosmic-ray interactions in the upper atmosphere, $\gamma$ rays with zenith angles greater than $105\degr$ were excluded.  {  To further reduce the residual $\gamma$ rays from the upper atmosphere} only data for time periods when the spacecraft rocking angle was less
than $52\degr$ were considered.  {  Time intervals with larger rocking angles are typically no more than tens of minutes long, occuring during orbits when the spacecraft was executing pointed observations instead of the standard sky-scanning survey mode.  The longest continguous time interval with rocking angle greater than $52\degr$ during the 3 years considered here (\S~\ref{LATExposure}) was 5 hours, during a pointed observation near the orbital pole.}

\begin{figure}[th]
\includegraphics[width=16cm]{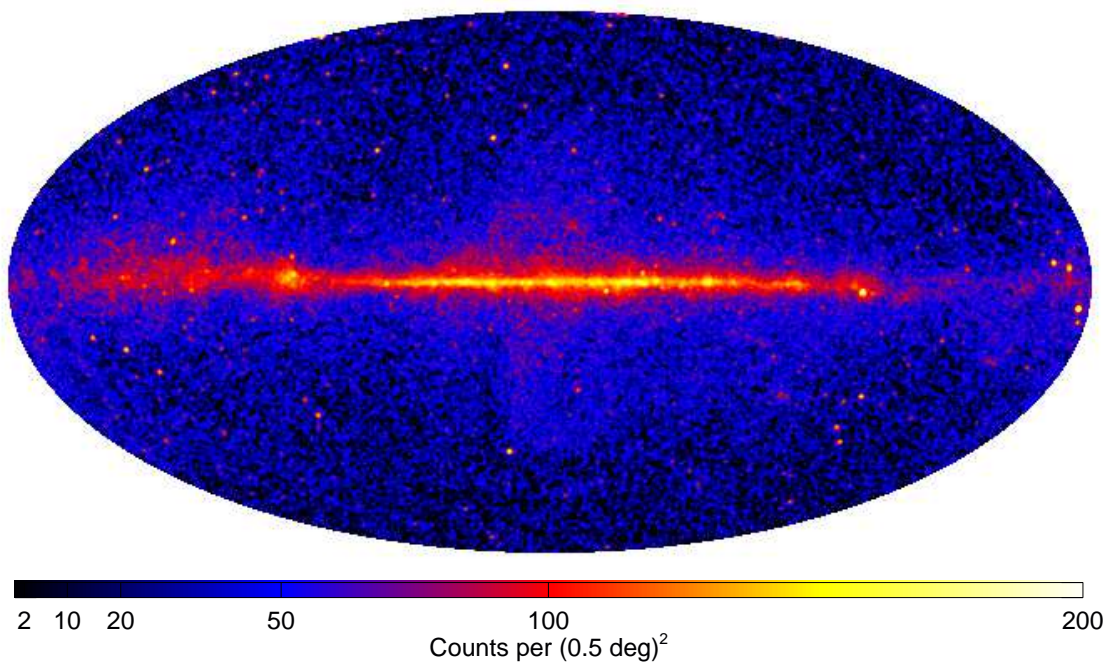}
\caption{Sky map of $\gamma$-ray counts above 10 GeV in Galactic coordinates in
Hammer-Aitoff projection.  {  The Galactic center (0,0) is at the center of the map and Galactic longitude increases to the left.} The binning is $0\fdg5$ and the image has
been smoothed with a 2-dimensional Gaussian of full width at half
maximum $0\fdg75$.}
\label{SkyMaps10GeV} 
\end{figure}

Figure~\ref{SkyMaps10GeV} shows the distribution of $\gamma$ rays above 10 GeV.  Since the
exposure is quite uniform (\S~\ref{LATExposure}), this
distribution reflects the spatial variations in the brightness of the sky.
The bright band along the Galactic equator is primarily due to diffuse $\gamma$-ray emission from cosmic-ray interactions with interstellar gas and radiation.  The isotropic background (extragalactic diffuse $\gamma$ rays and residual local contamination) becomes relatively more important at high latitudes, although structure in the Galactic diffuse emission is still evident, notably in the so-called {\it Fermi} bubbles, lobes of hard-spectrum emission above and below the Galactic center \citep{Su2010}.  Point sources of $\gamma$ rays are evident throughout the sky, with some concentration toward the Galactic equator.  


At energies above 10~GeV
the improved source-background contrast (with
respect to the 100 MeV--10 GeV range) provides
two benefits: {\it a)} the overall intensity of the diffuse
background (Galactic diffuse plus isotropic extragalactic and residual cosmic rays) falls approximately according to a power law of index $\sim$2.4
while the majority of the sources detectable
above 10 GeV have harder spectra (many of them with an index smaller than 2.0);
{\it b)} the PSF is narrowest at energies above 10~GeV (\S~\ref{LatInstrumentIntro}), and hence the photon signal from a \gray source is concentrated in a smaller region.
Therefore, above 10~GeV sources can be detected with only 4--5 $\gamma$ rays (\S~\ref{SpectralAnalysis}) and the analysis is less affected by
the uncertainties and/or inaccuracies in the model for the diffuse backgrounds.

\subsection{Exposure, Diffuse Gamma-Ray Backgrounds, and Point-source Sensitivity}
\label{LATExposure}
\begin{figure}[th]
\includegraphics[width=16cm]{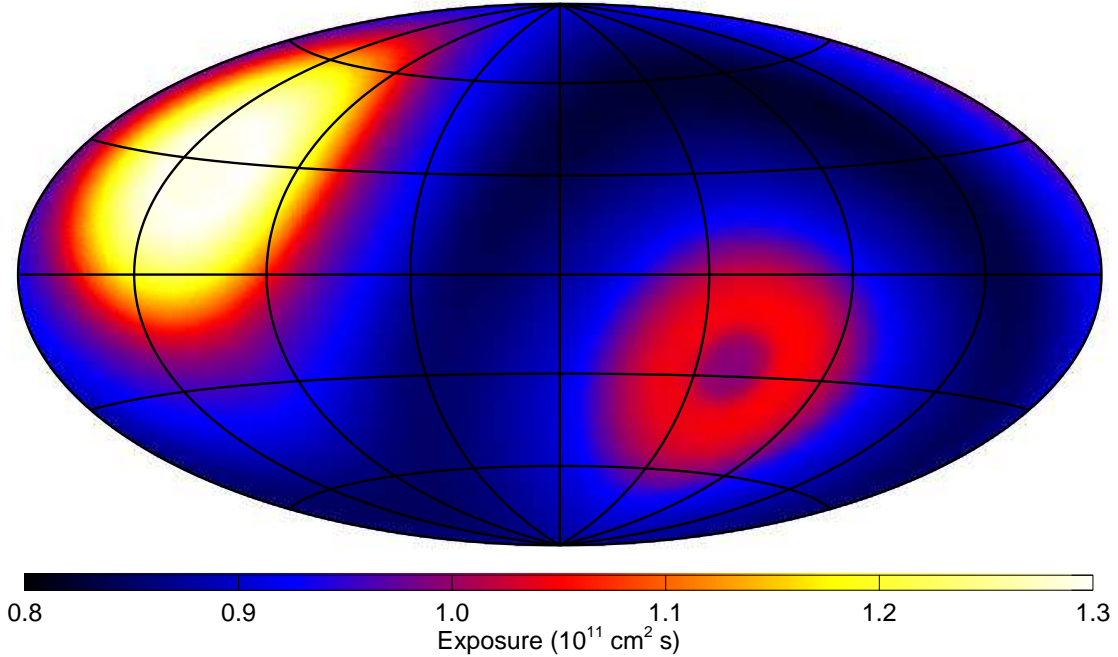}
\caption{\label{fig:LATExposure} 
Overall exposure at 10 GeV for the 3-year time period considered here, in Galactic coordinates in Hammer-Aitoff projection.  The same cuts on rocking angle and zenith angle as described in \S~\ref{FermiSky10GeV} have been applied.  The overall average is $9.5 \times 10^{10}$ cm$^2$s. }
\end{figure}

The time interval analyzed here is from the beginning of science operations, 2008 August 4 (MET 239557447)
to 2011 August 1 (MET 333849586), covering very nearly 3 years\footnote{Mission Elapsed Time (MET), the number of seconds since 00:00 UTC on 2001 January 1 (excluding leap seconds).}. 
The overall exposure for the 3-year interval is relatively uniform (Fig.~\ref{fig:LATExposure}), ranging from $-$15\% to +38\% of the average value of $9.5\times 10^{10}$ cm$^2$s, primarily as a function of declination.  The exposure at southern declinations is somewhat less because no observations are made during passages through the South Atlantic Anomaly.  In addition, the exposure near the northern celestial pole is enhanced because the majority of non-survey mode (pointed) observations have been made toward northern targets.  The exposure is slightly depressed in a $\sim$21$\degr$ diameter region near the southern celestial pole because of the 105$\degr$ limit on zenith angle for the $\gamma$ rays selected for analysis (\S~\ref{FermiSky10GeV}).

Proper quantification of the
diffuse backgrounds is necessary {  for} accurate source
detection and characterization. 
We used the publicly available models for the Galactic and isotropic
diffuse emissions for this analysis.
 These files, \texttt{gal\_2yearp7v6\_v0.fits} and \texttt{iso\_p7v6clean.txt},
can be retrieved from the \Fermi Science Support Center\footnote{See \url{http://fermi.gsfc.nasa.gov/ssc/data/access/lat/BackgroundModels.html}} (FSSC).
The same models were also used in producing the 2FGL catalog.

\begin{figure}[th]
\begin{center}
\includegraphics[width=8cm, height=4cm]{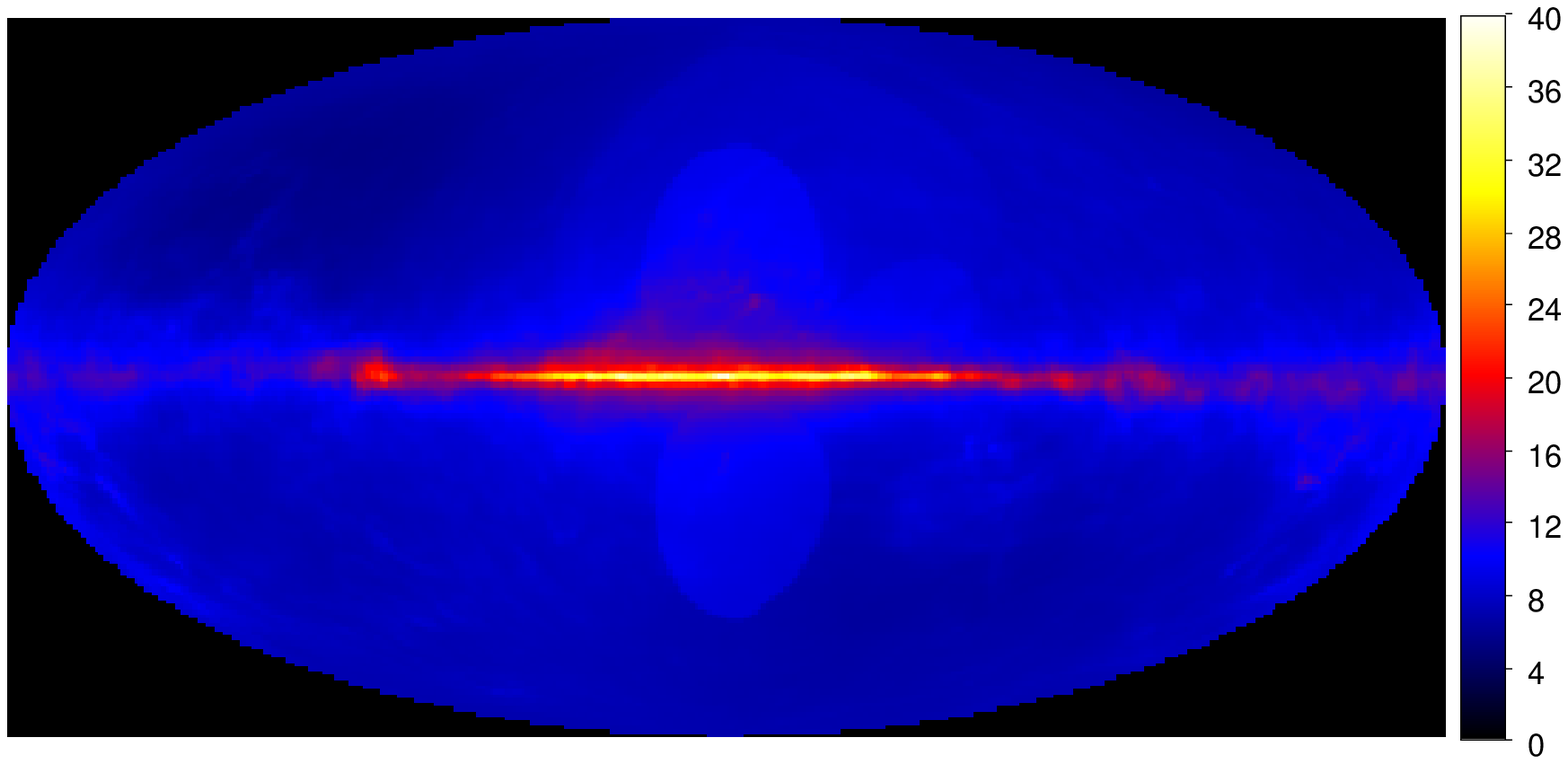}
\includegraphics[width=8cm, height=4cm]{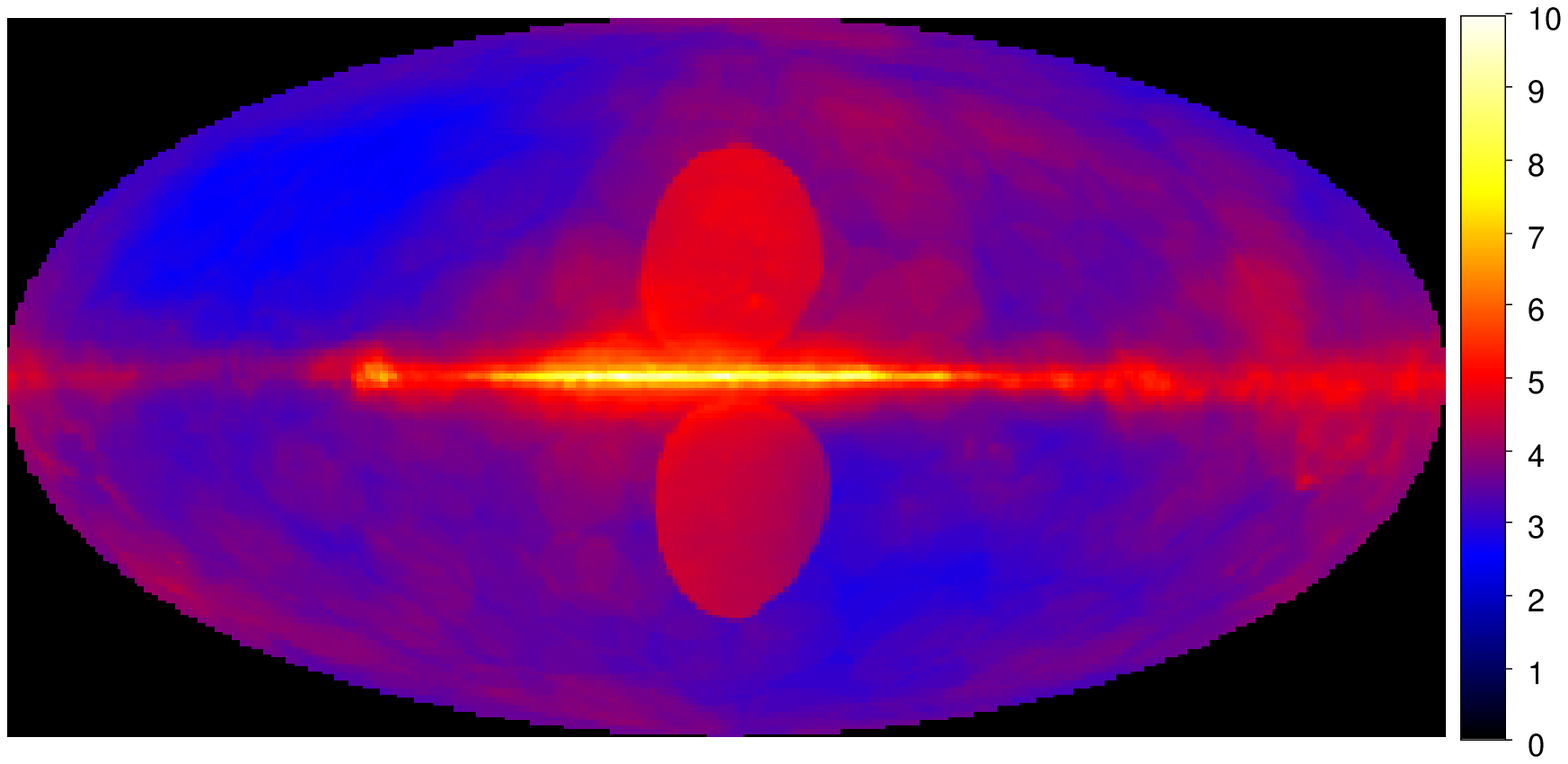}
\end{center}
\caption{\label{LATSensitivity} 
Minimum detectable photon flux (in 10$^{-11}$ ph cm$^{-2}$ s$^{-1}$)
for a $\gamma$-ray point source (with spectral index of 2.5) after 3
years for 10--500~GeV (left) and for 100--500~GeV (right).
The images are in Hammer-Aitoff projection in Galactic coordinates.
{  The images are available as FITS files from the FSSC.}  }
\end{figure}

{  The sensitivity of the LAT observations depends on the exposure, the diffuse backgrounds, and the PSF.}  The derived point-source flux sensitivity of the LAT for the 3-year interval is
depicted in Figure~\ref{LATSensitivity} for two energy ranges, 10--500 GeV and 100--500 GeV.
As for lower energies \citep[see, e.g.,][]{LAT12_calib}, 
these plots show that the sensitivity ranges by only a factor of two over
most of the sky, apart from the inner region of the
Galactic plane, where the intense diffuse \gray background greatly reduces the
point-source sensitivity.  The extended, lobe-like features of decreased sensitivity 
are due to the {\it Fermi} bubbles (\S~\ref{FermiSky10GeV}).
The specific shape in Figure~\ref{LATSensitivity} is determined by the template for the bubbles in the model for diffuse interstellar $\gamma$-ray emission used to evaluate the flux limits.\footnote{See \url{http://fermi.gsfc.nasa.gov/ssc/data/access/lat/Model_details/Pass7_galactic.html}}
The detection flux-threshold depends very little on the spectral shape
outside the Galactic plane (Fig.~\ref{DetThreshold}).
We note that for energies above 100 GeV, the 3-year
point-source sensitivity of the LAT, which is in the range
(2--4)$\times10^{-11}$ ph cm$^{-2}$ s$^{-1}$ for most of the sky, 
{  corresponds to about 6 hours of effective observing time for modern IACTs.}


\cleardoublepage


\section{\label{sec:sample} Analysis and Association Methodology}

The analysis follows broadly the same steps as the 2FGL catalog \citep{LAT_2FGL}. The significance of sources is measured by the test statistic $TS=2 \Delta \log \mathcal{L}$, comparing the likelihood with and without the source in the model.
Source detection and characterization began with the assembly of a list of `seeds' (\S~\ref{SeedsAndLocalization}), candidate sources that were selected for input to the likelihood analysis chain. The seeds were supplied to the standard maximum likelihood analysis that was used to jointly optimize the spectral parameters of the candidate sources and to judge their overall significances (\S~\ref{SpectralAnalysis}). The search for source variability differs from the 2FGL analysis owing to the limited statistics of the data (\S~\ref{Variability}). In the final step of the analysis we searched for candidate counterparts of these 1FHL sources with sources in previous LAT catalogs and sources in known $\gamma$-ray-emitting classes at other wavelengths (\S~\ref{SrcAssociations}).



\subsection{Seed Selection and Localization}
\label{SeedsAndLocalization}

The list of seeds and their locations were obtained in the same way as
for the 2FGL catalog analysis, i.e., through an iterative
3-step process: (1) identification of
potential $\gamma$-ray point sources, the `seeds'; (2) optimization of
the model of the $\gamma$-ray sky describing both the diffuse emission
and the potential sources; and (3) the creation of a residual TS map.
This iterative process was performed using the {\it pointlike} \citep{Kerr10} analysis
pipeline.  We briefly summarize the steps below.

The starting model was the collection of sources in the 2FGL catalog, to which we
added the new seeds obtained with the source-search algorithms {\it mr\_filter}
\citep{sp98}, {\it PGWave} \citep{dmm97,PGWave} and the {\it minimal spanning
tree} \citep{cmg08}. {  Each of the algorithms was applied to $\gamma$ rays in the 10--500~GeV range in the 3-year data set, and} all seeds found with at least one of these
were considered. The initial model was refined by an iterative process in which new seeds were identified in residual TS maps that covered the full sky, and seeds that were no longer significant in the model were removed.
The source-search algorithms were not used for the successive iterations.  As for the TS maps in 2FGL, the value of $TS$ at any given position was evaluated as the sum of test statistics for separate energy bands, $TS_i$, spanning the overall energy range.

In each iteration, the locations of the potential sources were
optimized during the third step, the creation of the residual TS map. 
In this step, the log likelihood was maximized with respect to position of
each seed, keeping the rest of the model (diffuse emission and other seeds) unchanged.  

The uncertainty in the localization of a seed was
determined by evaluating the variation of the likelihood function with
respect to the best-fit position.  To define the 95\% source location uncertainty region we fit an ellipse to the likelihood surface about the maximum, with offset $2 \Delta \log \mathcal{L} = -5.99$.  The eccentricities of the source-location regions are moderate, averaging 0.47, corresponding to a semi-minor-to-semi-major axis ratio of 0.89.  The ellipses have no preferred orientation on the sky.  The average solid angle of the 95\% confidence regions correspond to an effective position uncertainty of $0\fdg09$: the range is $0\fdg01$--$0\fdg22$.  

For 1FHL we have not applied corrections for systematic uncertainties for the source location region sizes.  As we show in \S~\ref{SrcAssociations}, for the 416 sources with firmly established associations and no spatial extension in LAT or IACT measurements, 19 (4.5\%) of the associations lie outside their calculated 95\% source location regions.  This is consistent with the nominal expectation, especially in consideration of the potential for slight bias from the role of angular offsets in assigning associations. 
For the 2FGL catalog analysis the systematics on source locations were somewhat larger, and a scale factor of 1.1 was applied.  For 2FGL the formal source location regions of the brightest pulsars were quite small and $0\fdg005$ was added in quadrature to account for potential residual misalignment of the LAT and spacecraft.  For the 1FHL catalog, this factor would have, at most, a minor contribution to all source location region sizes so we have not included it.


\subsection{Spectral Analysis of the Candidate Sources}
\label{SpectralAnalysis}

Starting from the list of seeds (\S~\ref{SeedsAndLocalization}),
we divided the sky into a number of regions of interest (RoI) covering
all source seeds; 561 RoIs were used for 1FHL.
Each RoI extends $2\degr$ beyond the seeds that are to be optimized within it in order to
cover the entire PSF as well as allow the background diffuse emission to be 
well fit.
Because the spatial resolution is good above 10 GeV, there is little
cross-talk between sources or between RoIs, so global convergence
was relatively easy to achieve. 

We explicitly model the known spatially extended sources as extended,
using the spatial extension from energies below 10 GeV, as reported in
previous works. 
In addition to the 12 extended sources included in the 2FGL analysis, we also included
10 that have been detected as extended sources since then.
Table~\ref{tbl:extended} lists the source names, spatial template
descriptions, and references for the dedicated analyses of these
sources. The 18 of these sources that are detected significantly
($TS>25$) above 10 GeV are tabulated with the point sources, with the
only distinction being that no position uncertainties are reported
(see \S ~\ref{Characterization}).

\begin{deluxetable}{lllcl}
\tabletypesize{\scriptsize}
\tablecaption{Extended Sources Modeled in the 1FHL Analysis
\label{tbl:extended}}
\tablewidth{0pt}
\tablehead{

\colhead{1FHL Name}&
\colhead{Extended Source}&
\colhead{Spatial Form}&
\colhead{Extent} &
\colhead{Reference}
}

\startdata
\nodata & SMC & 2D Gaussian & $0\fdg9$ & \citet{LAT10_SMC} \\
J0526.6$-$6825e & LMC & 2D Gaussian\tablenotemark{a} & $1.2$, $0.2$ & \citet{LAT10_LMC} \\
\nodata & S 147 & Map & \nodata & \citet{LAT12_S147} \\
J0617.2+2234e & IC 443 & 2D Gaussian & 0.26 & \citet{LAT10_IC443} \\
J0822.6$-$4250e & Puppis A & Disk & 0.37 & \citet{LAT12_extended} \\
J0833.1$-$4511e & Vela X & Disk & 0.88 & \citet{LAT10_VelaX} \\
J0852.7$-$4631e & Vela Junior & Disk & 1.12 & \citet{LAT11_VelaJr} \\
\nodata & Centaurus A (lobes) & Map & \nodata & \citet{LAT10_CenAlobes} \\
J1514.0$-$5915e & MSH 15$-$52 & Disk & 0.25 & \citet{LAT10_PSR1509} \\
J1615.3$-$5146e & HESS J1614-518 & Disk & 0.42 & \citet{LAT12_extended} \\
J1616.2$-$5054e & HESS J1616-508 & Disk & 0.32 & \citet{LAT12_extended} \\
J1633.0$-$4746e & HESS J1632-478 & Disk & 0.35 & \citet{LAT12_extended} \\
J1713.5$-$3951e & RX J1713.7$-$3946 & Map & \nodata & \citet{LAT11_RXJ1713} \\
J1801.3$-$2326e & W28 & Disk & 0.39 & \citet{LAT10_W28} \\
J1805.6$-$2136e & W30 & Disk & 0.37 & \citet{LAT12_W30} \\
J1824.5$-$1351e & HESS J1825$-$137 & 2D Gaussian & 0.56 & \citet{LAT11_J1825} \\
J1836.5$-$0655e & HESS J1837$-$069 & Disk & 0.33 & \citet{LAT12_extended} \\
J1855.9+0121e & W44 & Ring\tablenotemark{b} & (0.22, 0.14), (0.30, 0.19) & \citet{LAT10_W44} \\
J1923.2+1408e & W51C & Disk\tablenotemark{b} & (0.40, 0.25) & \citet{LAT09_W51C} \\
J2021.0+4031e & $\gamma$-Cygni & Disk & 0.63 & \citet{LAT12_extended} \\
J2028.6+4110e & Cygnus Cocoon & 2D Gaussian & 2.0 & \citet{LAT11_CygCocoon} \\
\nodata & Cygnus Loop & Ring & 0.7, 1.6 & \citet{LAT11_CygLoop} \\
\enddata

\tablenotetext{a}{Combination of two 2D Gaussian spatial templates.}
\tablenotetext{b}{{  The shape is elliptical; each pair of parameters $(a, b)$ represents the semi-major $(a)$ and semi-minor $(b)$ axes.}} 

\tablecomments{~List of all sources that have been modeled as extended sources. {  The Extent column indicates the radius for Disk sources, the dispersion for Gaussian sources, and the inner and outer radii for Ring sources.}   All spectra were modeled as power laws (as for point sources). Four were not detected above 10 GeV and do not have an 1FHL entry.}

\end{deluxetable}

Over the relatively narrow range 10 to 500 GeV, no source was found to
have significant spectral
curvature, so each spectrum was described by a power-law model.
Each RoI is too small to allow both the Galactic and
isotropic diffuse components to be properly characterized, so the isotropic level was fixed to the
best-fit value over the entire sky and we left free the Galactic normalization only.

This analysis was performed with the ScienceTools software package version v9r26p02.
We used binned likelihood functions, as in 2FGL, handling $Front$ and $Back$ events
separately, with $0.05\degr$ and $0.1\degr$ spatial binning respectively,
and 10 energy bins per decade.
The detection threshold was set to $TS > 25$,
corresponding to a significance just over $4 \sigma$ for 4 degrees of freedom
(two for the localization, and two for the spectrum).
Sources below that threshold were discarded from the model, except for the extended
sources, which we retained to model the background even when they were not
clearly detected above 10 GeV.
No constraint was enforced on the minimum number of $\gamma$ rays from detected sources,
because above 10 GeV
and outside the Galactic plane the detection is not background limited.
In practice the faintest sources were detected with only 4 $\gamma$ rays.

\begin{figure}[]
\begin{center}
\includegraphics[width=12cm]{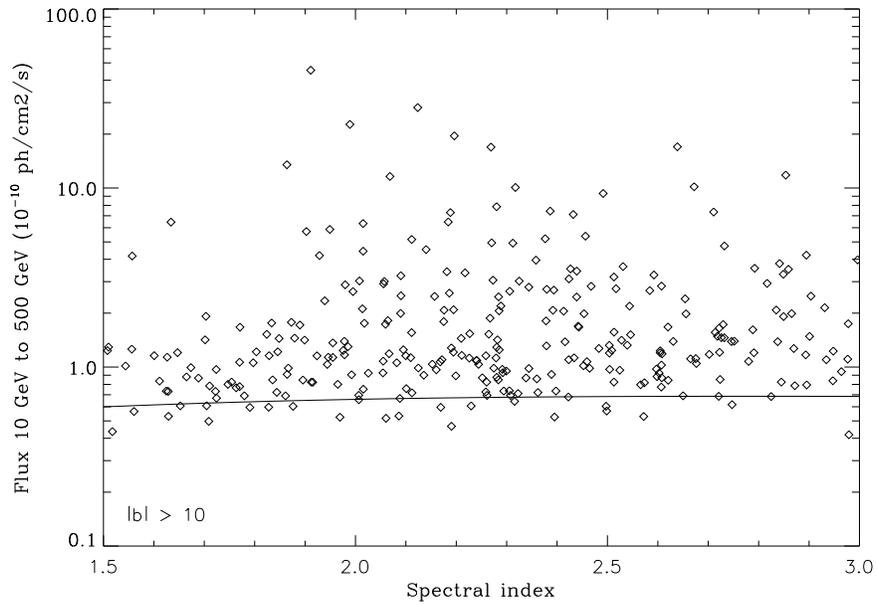}
\end{center}
\caption{\label{DetThreshold} 
Photon fluxes of all detected sources outside the Galactic plane ($|b| > 10\degr$) versus their photon spectral indices. The theoretical detection threshold for the average background is overlaid as the full line. As a result of the low intensity of the diffuse background and nearly constant PSF width over the entire range the detectability depends only very weakly on the spectral index.}
\end{figure}

At the end of the process 514 sources (including 18 of the extended sources
that we introduced manually) remained at $TS > 25$ among the 1705 input seeds.
The photon and energy fluxes over the full energy range were obtained by integrating
the power-law fits and propagating the errors.
The fluxes and spectral indices of the high-latitude sources ($|b| > 10\degr$) are shown in Figure~\ref{DetThreshold}.

\begin{figure}[th]
\begin{center}
\includegraphics[width=12cm]{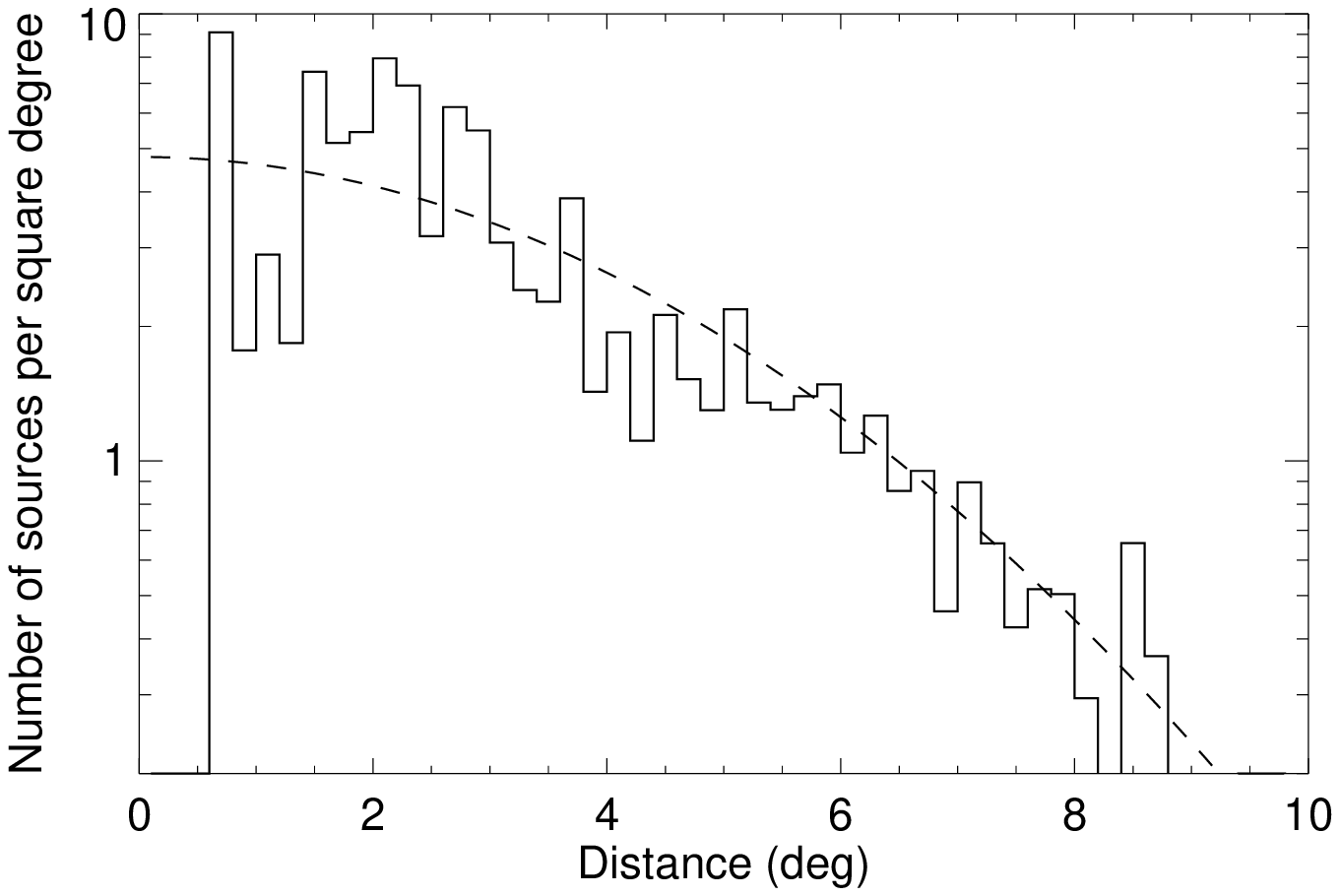}
\end{center}
\caption{\label{SourceConfusion} 
Distribution of nearest neighbor angular distances $D_n$ for all detected sources with $|b| > 10\degr$.  Each entry is scaled by $2\pi D_n \Delta D_n$, with $\Delta D_n = 0\fdg2$ the width of the bin in angular separation, in order to scale out solid-angle effects.  The dashed curve indicates the expected Gaussian distribution that would result for a random distribution of sources with no confusion.}
\end{figure}

Owing to the good angular resolution above 10 GeV (see \S~\ref{LatInstrumentIntro}), and the relatively low  density of sources (in
comparison with 2FGL), the detection of these sources is less affected by source confusion than was the case in the 2FGL catalog analysis.  Figure~\ref{SourceConfusion} shows that the distribution of nearest-neighbor source separations for $|b| > 10\degr$ is consistent with isotropic down to separations of $\sim$$0\fdg5$.  For the 2FGL analysis, source confusion became noticeable at $\sim$1$\degr$.  From fitting the observed distribution of nearest neighbor separations for separations greater than 1$\degr$, for which source confusion is not a consideration, we estimate that 5 sources were missed owing to source confusion, corresponding to a fraction of missed sources of 1.2\%.

After that global fitting over the full energy range we extracted photon fluxes
in three energy bands: 10--30~GeV, 30--100~GeV and 100--500~GeV.
These were obtained in the same way as fluxes in the 2FGL catalog, by holding fixed
all spectral indices and adjusting the normalizations only, including the Galactic diffuse.
We checked that the sum of photon fluxes is very well correlated with
the overall flux from the power-law fit. There is more scatter on the energy
flux, which depends more on the highest energy band where the statistical
uncertainties are largest.

Many sources, particularly above 100~GeV, are deep in the Poisson regime
(just a few events). As a result the likelihood profile is very asymmetric,
falling steeply from the maximum toward low fluxes but more gently toward large fluxes.
In order to reflect that situation in the catalog data products we report separate
$1 \sigma$ error bars toward low and high fluxes for individual bands,
obtained via MINOS in the Minuit\footnote{\url{http://lcgapp.cern.ch/project/cls/work-packages/mathlibs/minuit/home.html}} package.
When the test statistic in the band $TS_i < 1$ the $1 \sigma$ interval
contains 0, and in that case the negative error is set to Null.
For these non-significant sources we extract 95\% upper limits obtained using a Bayesian method \citep[following][]{helene83},
 by integrating $\mathcal{L}(F_i)$ from 0 up to
the flux that encompasses 95\% of the posterior probability.
With the probability chosen in
this way the 95\% upper limits $F_{95}$ are similar to $F_i + 2 \Delta F_i$ for
a hypothetical source with $TS_i=1$, where $F_i$ and $\Delta F_i$ are the best fit
and the $1 \sigma$ upper error bar obtained from MINOS.
Therefore in those cases we report $(F_{95}-F_i)/2$ in the upper error bar,
so that this column has approximately the same meaning for all sources.

\begin{figure}[t]
\begin{center}
\includegraphics[width=10cm]{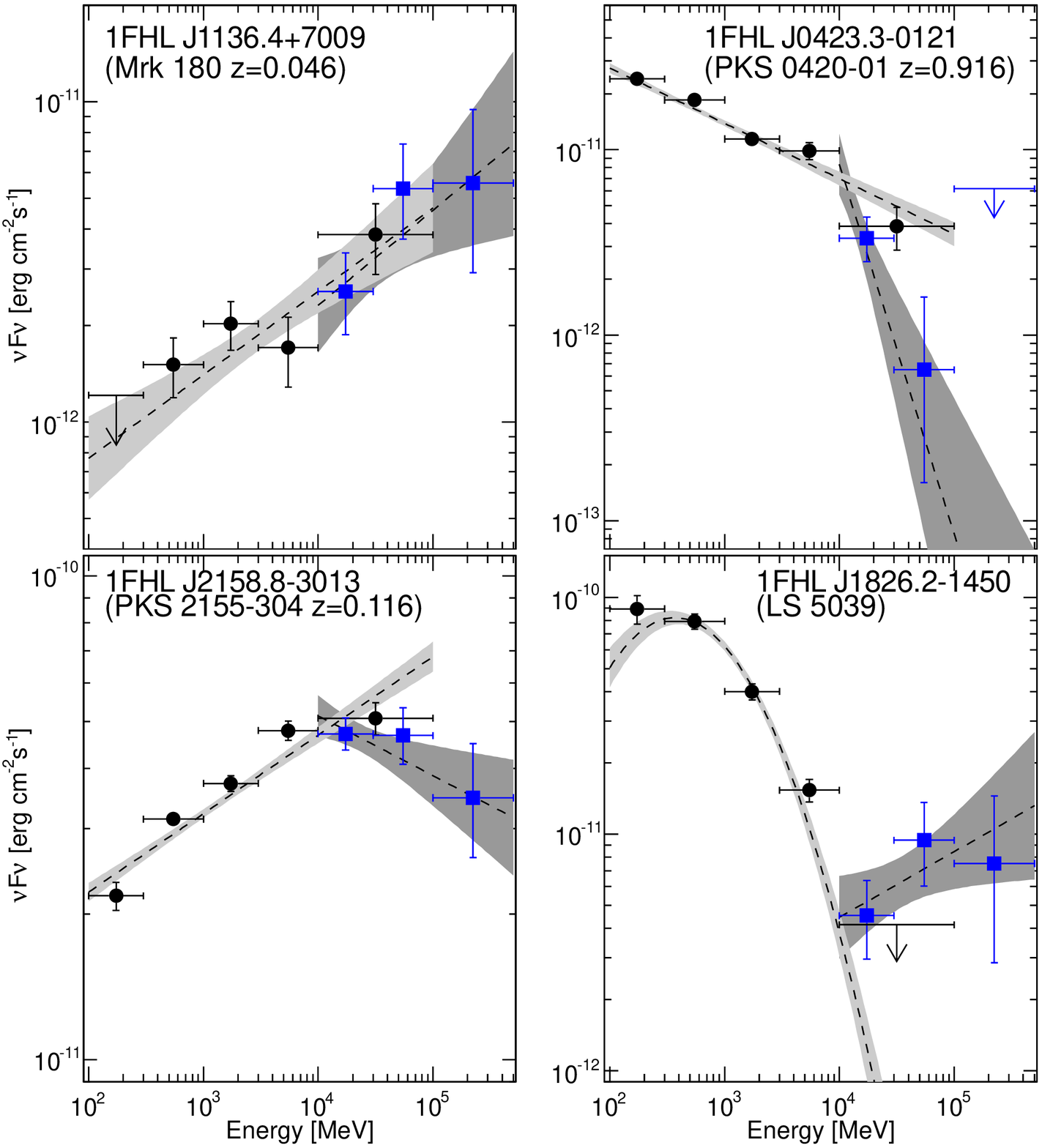}
\end{center}
\caption{\label{ExampleSpectra1FHL} 
Spectral energy distribution of four representative 1FHL sources with different spectral shapes
above 10 GeV:  the blazars Mrk 180 (z=0.046), PKS~2155$-$304
(z=0.116), and PKS~0420$-$01 (z=0.916), and the high-mass binary system LS~5039.
The black circles and light-gray bands depict the results
reported in the 2FGL catalog, while the blue squares and the
dark-gray bands depict the spectral results reported in this
work.  The panels are labeled with the 1FHL names and
the names of the corresponding associated sources (in parentheses). 
See text for further details.
}
\end{figure}

Figure~\ref{ExampleSpectra1FHL} compares the spectral measurements reported in
the 2FGL paper (in the 100~MeV to 100~GeV energy range) with the results reported here
in the 10--500~GeV energy range, for four representative sources.
95\% upper limits are plotted when $TS_i < 1$, as explained above.
In order to convert the photon fluxes in each band to $\nu F_\nu$ we proceeded as follows:
\begin{enumerate}
\item We converted the photon fluxes into energy fluxes in the same band, on the basis of the same power-law approximation used in the fit (photon index = Spectral\_Index of Table \ref{tbl:1fhlcolumns}).
\item We converted the energy fluxes into $\nu F_\nu$ by dividing by the logarithmic width of the band ln($E_{i+1}/E_i$) where $E_i$ and $E_{i+1}$ are the start and end points of the energy interval.
\end{enumerate}
We applied the same method to the 2FGL points, using the local spectral index at the bin center (in log) for the curved spectra.

The blazar Mrk 180 (z=0.046) has a 1FHL spectrum that is a continuation of its 2FGL
spectrum, while the classical TeV blazar PKS~2155$-$304 (z=0.116), which is a few times
brighter than Mrk 180, shows a clear turnover (from hard to
soft spectrum) at about 10~GeV. Given that PKS~2155$-$304 is a
relatively nearby source, this turnover is due to an internal
break in the emission mechanism of this source.
On the other hand, the
spectrum of the distant blazar PKS~0420$-$01 (z=0.916) shows a clear cutoff (strong turnover) around 10 GeV which, given the very high redshift of this
source, is likely dominated by the absorption of $\gamma$-rays in the
extragalactic background light (EBL).
The fourth panel of
Figure~\ref{ExampleSpectra1FHL} shows the 1FHL spectrum of the
high-mass binary system LS~5039, which has a completely different shape with respect to
the 2FGL spectrum, hence indicating the presence of a new spectral component \citep[see][]{LATHMB2012}.
Such deviations from the simple
spectral extrapolation from lower energies indicate the increasing dominance of
other physical processes occurring at the source, 
or in the environment crossed by the $\gamma$ rays, and hence they are very relevant for the proper understanding
of these sources. This is one of the important motivations
for producing the 1FHL catalog.

  
\subsection{Quantification of Variability with the Bayesian Block
  Algorithm}
\label{Variability}


The Bayesian Block algorithm for detecting and characterizing
variability in time series data
\citep{BayesianBlocks1998,BayesianBlocks} is particularly well-suited
for analyzing low count data, an important consideration for the 1FHL
catalog, for which more than half of the sources have fewer than 20
associated counts.  The algorithm partitions the time series data into
piecewise constant segments (blocks), each 
characterized by a rate (or flux) and duration. The locations of the
transitions between blocks are determined by optimizing a fitness
function for the partitions. The algorithm for finding the optimal
partitioning  is described by \citet{2005ISPL...12..105J}. For the
analysis of the 1FHL data, the fitness function used is the logarithm
of the maximum likelihood for each individual block under the
constant local rate hypothesis, as described by \citet{BayesianBlocks}.
Using the simulation results presented in that paper, an acceptable
fraction of false positives for detecting variability can be easily
specified. In the analysis presented here, a false positive threshold
of 1\% was used for all sources. This method also takes into account
the effective exposure associated to each event, thus correcting for
the exposure variations due to the motion of the field of view of the
LAT.

For each source, we used an RoI of $0\fdg5$~radius centered on the
best-fit coordinates to extract the events.  For sources with
neighboring 1FHL sources closer than $1^{\circ}$, we set the radius of
the RoI to the greater value of ($angular\:separation/2$) or
$0\fdg25$. Only 5 pairs of sources had their RoIs fixed at $0\fdg25$,
all of which are located in the Galactic plane. In addition to the
Bayesian Block analysis, for each source we also performed an aperture
photometry analysis using 50 equal time bins spanning the 3-year
interval. We did not do any background subtraction in either
analysis. {  Results of the Bayesian Block analysis are presented in
\S~\ref{BasicProperties}}.


\subsection{Associations}
\label{SrcAssociations}

The 1FHL sources were associated with (known) sources at other
wavelengths using similar procedures as for the 2FGL and 2LAC \citep{LAT_2LAC} 
catalogs.   As for these catalogs,
we keep the distinction between an \emph{association} and an \emph{identification}, the latter being more conservative.  Promoting an
association to an identification requires that correlated variability or source extension be found with observations at other wavelengths.

The associations were derived with two different procedures:  the
Bayesian and the likelihood-ratio association methods \citep{deRuiter1977,Sutherland1992}.
The Bayesian method and its implementation for associating LAT
sources with potential counterparts at other wavelengths is described in an appendix of the 1FGL paper \citep{LAT_1FGL}, and some refinements are
reported in the 2FGL paper. The likelihood-ratio method and its implementation are described in
the 2LAC paper. In the application of these two methods, potential
counterparts were retained as associations if they were found to have
{\it a posteriori} probabilities of at least 80\%.

For the Bayesian method, we used the 13$^{\rm th}$
edition of the Veron catalog \citep{VeronCetty2010}, version 20
of BZCAT \citep{Massaro2009}, the 2010 December 5 version of the
VLBA Calibrator Source List\footnote{
\url{http://astrogeo.org/vcs/}}, and version 3.400 of the TeVCat catalog. 
We also added new counterpart catalogs:  the Australia
Telescope 20-GHz Survey (AT20G) \citep{Murphy2010,Massardi2011} and
the Planck Early Release Compact Source Catalog \citep{Ade2011}.

For the likelihood-ratio method, the catalogs of potential counterparts were the NRAO
VLA Sky Survey \citep{Condon1998}, the second version of the 
wide-field radio imaging survey of the southern sky \citep{Mauch2003}, 
the PMN-CA catalog of southern radio sources \citep{Wright1996}, and the ROSAT all-sky survey bright
source catalogue \citep{Voges1999}. 
Note that these catalogs contain mostly extragalactic
sources and so the likelihood-ratio method was not very efficient in associating 1FHL
sources with Galactic sources.





In addition, we also evaluated correspondences with the 2FGL and
1FGL catalogs of LAT sources, and gave them higher priorities with respect to the other
(non-{\textit{Fermi}) catalogs. 
Therefore, whenever possible, we associated the 1FHL sources to previously-cataloged LAT sources, and for these cases we also adopted the source associations given in the
previously published \Fermi catalogs. 

The sources that could be associated with known or previously reported
sources (including unassociated 2FGL and 1FGL objects) total 484, of which 451 could be associated
with 2FGL sources, and 11 with 1FGL sources that are
not in the 2FGL catalog. We note that the number of
1FHL sources associated by the Bayesian method is 484, while the number that were associated
using the Likelihood-Ratio method is 441 (all of which were
also associated with the Bayesian method). This difference in
performance is attributable to the  likelihood-ratio method
being used only to find associations with extragalactic sources, while
the Bayesian method is more general and used specific catalogs of Galactic
sources. 
Three 1FHL sources each have associations with two
distinct sources with posterior probabilities
greater than 80\%:
1FHL J0217.4+0836 (associated with a BL Lac and an FSRQ), 1FHL
J0323.5$-$0107 (associated with two distinct BL Lacs), and
1FHL J0442.9$-$0017 (associated with a FSRQ and a BL Lac). 
We also note that the 1FHL catalog contains 52 (=$514-451-11$)
sources that could not be associated to objects reported in previous LAT catalogs
(with 11~months and 2~years of accumulated data for 1FGL and 2FGL respectively). We describe these
in \S~\ref{Characterization}.


\cleardoublepage

\section{The 1FHL Catalog}
\label{Characterization}

This section describes the contents of the 1FHL catalog and reports
the basic properties of the 1FHL sources.  The collective properties of the sources that do not have counterparts in the 2FGL catalog, the sources that are associated with AGNs, and the pulsars emitting above 10~GeV are also discussed.


\subsection{Description of the 1FHL Catalog}

Table~\ref{tbl:1fhlcolumns} describes the full contents of the 1FHL
catalog data product, which is available in FITS format from the FSSC.
Table \ref{MainCatalogTable} presents the catalog itself. Column names are identical (when the meaning is the same) or similar
to 2FGL columns \citep{LAT_2FGL}. The main exception is the Variability\_BayesBlocks entry
which is computed from the Bayesian Blocks analysis
(\S~\ref{Variability}). The $\gamma$-ray association column lists the corresponding source, if any, in the 2FGL, 1FGL, 3EG, or EGR \citep{EGRCatalog} catalogs.  Of the 46 high-confidence sources in the \citet{Lamb1997} GEV catalog of EGRET sources detected above 1 GeV, 35 have associations with 1FHL sources.  For the 1FHL catalog the source designations are 1FHL~JHHMM.m$\pm$DDMM, where FHL stands for
{\em Fermi High-energy (source) LAT}, where high energy means above 10
GeV. 


\begin{deluxetable}{llll}
\setlength{\tabcolsep}{0.04in}
\tablewidth{0pt}
\tabletypesize{\scriptsize}
\tablecaption{LAT 1FHL FITS Format\tablenotemark{*}:  LAT\_Point\_Source\_Catalog Extension\label{tbl:1fhlcolumns}}
\tablehead{
\colhead{Column} &
\colhead{Format} &
\colhead{Unit} &
\colhead{Description}
}
\startdata
Source\_Name & 18A & \nodata & \nodata  \\
RAJ2000 & E & deg & Right Ascension \\
DEJ2000 & E & deg & Declination \\
GLON & E & deg & Galactic Longitude \\
GLAT & E & deg & Galactic Latitude \\
Conf\_95\_SemiMajor & E & deg & Long radius of error ellipse at 95\% confidence level \\
Conf\_95\_SemiMinor & E & deg & Short radius of error ellipse at 95\% confidence level \\
Conf\_95\_PosAng & E & deg & Position angle of the 95\% long axis from celestial North, \\
 & & & positive toward increasing RA (eastward) \\
Signif\_Avg & E & \nodata & Source significance in $\sigma$ units (derived from TS) \\
Pivot\_Energy & E & GeV & Energy at which error on differential flux is minimal \\
Flux\_Density & E & cm$^{-2}$ GeV$^{-1}$ s$^{-1}$ & Differential flux at Pivot\_Energy \\
Unc\_Flux\_Density & E & cm$^{-2}$ GeV$^{-1}$ s$^{-1}$ & 1 $\sigma$  error on differential flux at Pivot\_Energy \\
Spectral\_Index & E & \nodata & Best fit photon number power-law index \\
Unc\_Spectral\_Index & E & \nodata &  1 $\sigma$ error on Spectral\_Index \\
Flux & E & cm$^{-2}$ s$^{-1}$ & Integral photon flux from 10 to 500~GeV \\
Unc\_Flux & E & cm$^{-2}$ s$^{-1}$ & 1 $\sigma$ error on integral photon flux from 10 to 500~GeV \\
Energy\_Flux & E & erg cm$^{-2}$ s$^{-1}$ & Energy flux from 10 to 500~GeV obtained by spectral fitting \\
Unc\_Energy\_Flux & E & erg cm$^{-2}$ s$^{-1}$ & 1 $\sigma$ error on energy flux from 10 to 500~GeV \\
Flux10\_30GeV & E & cm$^{-2}$ s$^{-1}$ & Integral flux from 10 to 30~GeV \\
Unc\_Flux10\_30GeV & 2E & cm$^{-2}$ s$^{-1}$ & 1 $\sigma$ errors on integral flux from 10 to 30~GeV\tablenotemark{a} \\
Sqrt\_TS10\_30GeV & E & \nodata & Square root of the Test Statistic between 10 and 30~GeV \\
Flux30\_100GeV & E & cm$^{-2}$ s$^{-1}$ & Integral flux from 30 to 100~GeV \\
Unc\_Flux30\_100GeV & 2E & cm$^{-2}$ s$^{-1}$ & 1 $\sigma$ errors on integral flux from 30 to 100~GeV\tablenotemark{a} \\
Sqrt\_TS30\_100GeV & E & \nodata & Square root of the Test Statistic between 10 and 30~GeV \\
Flux100\_500GeV & E & cm$^{-2}$ s$^{-1}$ & Integral flux from 100 to 500~GeV \\
Unc\_Flux100\_500GeV & 2E & cm$^{-2}$ s$^{-1}$ & 1 $\sigma$ errors on integral flux from 100 to 500~GeV\tablenotemark{a}\\
Sqrt\_TS100\_500GeV & E & \nodata & Square root of the Test Statistic between 100 and 500~GeV \\
Variability\_BayesBlocks & I & \nodata & Number of Bayesian Blocks\tablenotemark{b} found (1 for non-variable) \\
Extended\_Source\_Name & 18A & \nodata & Cross-reference to the ExtendedSources extension for extended sources, if any \\
ASSOC\_GAM & 18A & \nodata & Name of corresponding source in gamma-ray catalog, if any \\
TEVCAT\_FLAG & 2A & \nodata & P if positional association with non-extended source in TeVCat \\
   & & \nodata & E if associated with an extended source in TeVCat, N if no TeV association \\
   & & \nodata & C if the source survives the TeV candidate selection criteria specified in \S~\ref{VHECandidates}. \\ 
ASSOC\_TEV & 21A & \nodata & Name of TeV association, if any \\
CLASS1 & 4A & \nodata & Class designation for most likely association; see Table~\ref{TableStatisticsSrcClasses} \\
CLASS2 & 4A & \nodata & Class designation for alternate association, if any \\
ASSOC1 & 26A & \nodata & Name of identified or most likely associated source \\
ASSOC2 & 26A & \nodata & Name of alternate association, if any \\
\enddata
\tablenotetext{a} {Separate 1 $\sigma$ errors are computed from the likelihood profile toward lower and larger fluxes. The lower error is set equal to Null if the 1 $\sigma$ interval contains 0.}
\tablenotetext{b} {The probability threshold for the Bayesian Blocks
  analysis is given by the \texttt{VARPROBA} keyword.}
\tablenotetext{*} {The FITS file is available in the electronic edition of the Astrophysical Journal Supplements, and also from the FSSC.}
\end{deluxetable}

\begin{deluxetable}{lrrrrrrrrrrrcccclccl}
\setlength{\tabcolsep}{0.02in}
\tabletypesize{\scriptsize}
\rotate
\tablewidth{0pt}
\tablecaption{LAT Catalog of Sources Above 10 GeV\label{MainCatalogTable}}
\tablehead{
\colhead{Name 1FHL} &
\colhead{R.A.} &
\colhead{Decl.} &
\colhead{$l$} &
\colhead{$b$} &
\colhead{$\theta_{\rm 1}$} &
\colhead{$\theta_{\rm 2}$} &
\colhead{$\phi$} &
\colhead{$\sigma$} &
\colhead{$F_{10}$} &
\colhead{$\Delta F_{10}$} &
\colhead{$S_{10}$} &
\colhead{$\Delta S_{10}$} &
\colhead{$\Gamma_{10}$} &
\colhead{$\Delta \Gamma_{10}$} &
\colhead{Var} &
\colhead{$\gamma$-ray Assoc.} &
\colhead{TeV} &
\colhead{Class} &
\colhead{ID or Assoc.}
}
\startdata
 J0007.3+7303 &   1.827 & 73.060 & 119.682 & 10.467 & 0.024 & 0.023 & $-$9 &     31.8 &   125.1 &  10.3 &   31.6 &    3.1 &  3.73 & 0.24 & 1 & 2FGL J0007.0+7303  & E & HPSR & LAT PSR J0007+7303         \\
 J0007.7+4709 &   1.947 & 47.155 & 115.271 & $-$15.067 & 0.073 & 0.058 & 43 &      7.1 &    14.5 &   4.1 &    3.8 &    1.3 &  3.57 & 0.74 & 1 & 2FGL J0007.8+4713  & \nodata & bzb & MG4 J000800+4712           \\
 J0008.7$-$2340 &   2.194 & $-$23.674 &  50.306 & $-$79.770 & 0.120 & 0.114 & $-$65 &      4.5 &     8.2 &   3.4 &    3.2 &    2.0 &  2.57 & 0.69 & 1 & 2FGL J0008.7$-$2344  & \nodata & bzb & RBS 0016                   \\
 J0009.2+5032 &   2.316 & 50.541 & 116.110 & $-$11.772 & 0.075 & 0.066 & $-$88 &     10.6 &    27.2 &   5.4 &   12.3 &    3.8 &  2.38 & 0.30 & 1 & 2FGL J0009.1+5030  & C & bzb & NVSS J000922+503028        \\
 J0018.6+2946 &   4.673 & 29.776 & 114.500 & $-$32.559 & 0.144 & 0.121 & $-$60 &      4.6 &     7.5 &   3.1 &    4.7 &    3.1 &  2.02 & 0.49 & 1 & 2FGL J0018.5+2945  & C & bzb & RBS 0042                   \\
 J0022.2$-$1853 &   5.555 & $-$18.899 &  82.190 & $-$79.380 & 0.083 & 0.068 & 39 &      7.0 &    12.2 &   4.1 &    9.2 &    4.9 &  1.85 & 0.37 & 1 & 2FGL J0022.2$-$1853  & C & bzb & 1RXS J002209.2$-$185333      \\
 J0022.5+0607 &   5.643 & 6.124 & 110.019 & $-$56.023 & 0.119 & 0.108 & $-$22 &      6.3 &    14.1 &   4.5 &    5.7 &    2.7 &  2.53 & 0.51 & 1 & 2FGL J0022.5+0607  & C & bzb & PKS 0019+058               \\
 J0030.1$-$1647 &   7.525 & $-$16.797 &  96.297 & $-$78.550 & 0.118 & 0.092 & 74 &      4.3 &     5.6 &   2.8 &    5.9 &    4.7 &  1.56 & 0.50 & 1 & \nodata & C & \nodata & \nodata \\
 J0033.6$-$1921 &   8.407 & $-$19.361 &  94.245 & $-$81.223 & 0.047 & 0.044 & $-$55 &     15.4 &    42.0 &   7.3 &   28.9 &    8.2 &  1.93 & 0.21 & 1 & 2FGL J0033.5$-$1921  & P & bzb & KUV 00311$-$1938             \\
 J0035.2+1514 &   8.806 & 15.234 & 117.143 & $-$47.455 & 0.079 & 0.071 & $-$77 &      6.9 &    14.6 &   4.4 &    5.2 &    2.2 &  2.73 & 0.54 & 1 & 2FGL J0035.2+1515  & \nodata & bzb & RX J0035.2+1515            \\
 J0035.9+5950 &   8.990 & 59.838 & 120.987 & $-$2.975 & 0.043 & 0.039 & $-$19 &     13.3 &    34.9 &   6.0 &   29.8 &    8.0 &  1.74 & 0.19 & 1 & 2FGL J0035.8+5951  & P & bzb & 1ES 0033+595               \\
 J0037.8+1238 &   9.473 & 12.645 & 117.778 & $-$50.091 & 0.113 & 0.098 & $-$18 &      4.3 &     7.1 &   3.1 &    2.1 &    1.2 &  3.22 & 0.96 & 1 & 2FGL J0037.8+1238  & \nodata & bzb & NVSS J003750+123818        \\
\enddata
\tablecomments{R.A. and Decl. are celestial coordinates in J2000 epoch, $l$ and $b$ are Galactic coordinates, in degrees; $\theta_1$ and $\theta_2$
are the semimajor and semiminor axes of the 95\% confidence source
location region; $\phi$ is the position angle in degrees east of north;
$F_{10}$ and $\Delta F_{10}$ are photon flux (10~GeV -- 500~GeV) in
units of $10^{-11}$ cm$^{-2}$ s$^{-1}$; 
$S_{10}$ and $\Delta S_{10}$ are the energy flux (10~GeV --500~GeV) 
in units of $10^{-12}$ erg cm$^{-2}$ s$^{-1}$; $\Gamma_{10}$ and
$\Delta \Gamma_{10}$ are the photon power-law index and uncertainty
for a power-law fit;
Var is the
number of change points in the Bayesian Blocks light curve (see the text); 
$\gamma$-ray Assoc. lists associations with
other catalogs of GeV $\gamma$-ray sources; TeV indicates an association
with a point-like or small angular size TeV source (P) or extended
TeV source (E); this column also indicates good candidates for TeV detections (C), as defined in \S~\ref{VHECandidates}; Class  designates the astrophysical class of the associated source (see the text);
ID or Assoc. lists the primary name of the associated source or
identified counterpart.
Three 1FHL sources have two associations listed here; the two distinct associated source
  names and class types are reported separated by the symbol ``\&''.
This table is published in its entirety in a machine-readable form in the electronic edition of the Astrophysical Journal Supplements.
A portion is shown here for guidance regarding its form and content.}
\end{deluxetable}

The designators for the source associations and identifications are listed in 
Table~\ref{TableStatisticsSrcClasses} along with the source counts. Because of the limited capability for
variability and morphological studies (due to the low photon counts
above 10 GeV), for 1FHL sources with counterparts in the 2FGL catalog we adopted the same associations and identifications as for 2FGL.
Similarly we also used the designator ``spp'' to denote the class of the six sources that
have positional associations with {  supernova remnants} (SNRs) of angular diameters $>$20$\arcmin$ and/or {  pulsar wind nebulae} (PWNe).  Owing to the increased chance of coincidental associations with the SNRs and the ambiguity of SNR vs. PWN associations for some of the sources, the potential associations are reported separately, in Table~\ref{tbl:spp}.
Only two new class designators were included in the 1FHL catalog.  For 20 pulsars, pulsed emission was detectable above
10 GeV (see
\S~\ref{Pulsars10GeV}), and we use ``HPSR'' as the class designator.  Also, we use the designator
``SFR'', for star-forming region, and apply it to the Cygnus Cocoon (1FHL~J2028.6+4110e).

\begin{deluxetable}{lcccccc}
\setlength{\tabcolsep}{0.04in}
\tabletypesize{\scriptsize} 
\tablecolumns{5} 
\tablewidth{0pc}
\tablecaption{LAT 1FHL Sources by Class\label{TableStatisticsSrcClasses}}
\tablehead{ 
  &   Identified &  & Associated &  & Total & Fraction of \\
\colhead{Class Description}                    
& \colhead{Designator}      & \colhead{Number}
& \colhead{Designator}      & \colhead{Number} 
& \colhead{Number} & \colhead{full catalog [\%]}
}
\startdata 
Blazar of the BL Lac type  & BZB & 7 & bzb & 252  & 259 & 50.4 \\ 
Blazar of the FSRQ type  &  BZQ & 13 &  bzq\tablenotemark{a} & 58 &  71  &  13.8\\
Active galaxy of uncertain type & AGU & 1 & agu & 57 & 58 & 11.3 \\
Pulsar, identified by pulsations above 10 GeV   & HPSR & 20 & \nodata
& \nodata & 20  & 3.9\\
Pulsar, identified by pulsations in LAT (excluding HPSR)   & PSR & 6 &
\nodata & \nodata  & 6 & 1.2 \\
Pulsar, no pulsations seen in LAT yet & \nodata & \nodata & psr & 1 
& 1 & 0.2\\
Supernova remnant &  SNR& 6 & snr & 5 & 11 &  2.1 \\
Pulsar wind nebula &  PWN & 3 & pwn & 3 & 6 & 1.2\\
Unclear whether SNR or PWN & \nodata & \nodata & spp & 6 & 6 & 1.2\\
Radio galaxy & RDG & 1 & rdg & 4  &  5  & 1.0 \\
High-mass binary & HMB & 3 & hmb & 0 & 3 & 0.6\\
Normal galaxy & GAL & 1 & gal & 0 & 1 & 0.2\\
Star forming region & SFR & 1 & sfr & 0 & 1 & 0.2\\
LBV star & LVB & 0 & lvb & 1 & 1 & 0.2\\
Unassociated source &  \nodata & \nodata & \nodata & 65 & 65 & 12.6\\
\enddata

\tablenotetext{a}{1FHL J1312.8+4827, classified here as bzq, may in
  fact be a narrow-line Seyfert 1 galaxy (Sokolovsky et al., {\it
    submitted}).}
\tablecomments{For the three 1FHL sources with two associations (see
  \S~\ref{SrcAssociations} and Table \ref{MainCatalogTable}), we 
  consider only the first associated source (which is the one with the highest
  probability of association).}
\end{deluxetable}

\begin{deluxetable}{llllll}
\tabletypesize{\scriptsize}
\tablecaption{Potential Associations for Sources Near SNRs
\label{tbl:spp}}
\tablewidth{0pt}
\tablehead{

\colhead{1FHL Name}&
\colhead{2FGL Name }&
\colhead{SNR Name}&
\colhead{PWN Name} & 
\colhead{TeV Name} & 
\colhead{Common Name} 
}

\startdata
J1111.5$-$6038		&	J1112.1$-$6040		&  G291.0$-$00.1  &  G291.0$-$0.1 & \nodata & \nodata \\
J1552.6$-$5610 		 &    J1552.8$-$5609 	 	&  G326.3$-$01.8 & \nodata &   \nodata &   Kes 25 \\	     
J1640.5$-$4634		&	J1640.5$-$4633		 & G338.3$-$00.0  & G338.3$-$0.0	&  HESS J1640$-$465 & \nodata \\
J1717.9$-$3725		&	J1718.1$-$3725		 & G350.1$-$00.3    & \nodata & \nodata & \nodata \\
J1745.6$-$2900		& 	J1745.6$-$2858		 & G000.0+00.0  &	G359.98$-$0.05  & \nodata &	Sgr A East \\
J1834.6$-$0703		&	J1834.7$-$0705c 	&  G024.7+00.6   & 		\nodata & \nodata & \nodata \\
\enddata

\end{deluxetable}

A remarkable
characteristic of this catalog is that the blazars and blazar
candidates\footnote{The fraction of non-beamed AGNs is expected to be
only few percent, and so most of the AGNs of unknown type are expected to
be blazars of either FSRQ or BL Lac type.} amount to $\sim$75\% of the entire catalog ($\sim$86\% of the
associated sources), indicating that this source class largely dominates
the highest-energy LAT sky.
It is worth mentioning that the four 1FHL sources associated with
radio galaxies have also shown characteristics that are typical for
blazars, either in radio morphology 
(prominent flat-spectrum core with one-sided jet),
in optical spectrum, or in $\gamma$-ray variability (sporadic 
short-term flux variability with timescales of \lapp 1 day).
This is the case for PKS~0625$-$35
\citep[e.g. see][]{Wills2004},
M~87 \citep[e.g. see][]{Abramowski2012_M87MW2010}, 
NGC~1275
\citep[e.g. see][]{Kataoka2010_NGC1275Fermi,Aleksic2012_NGC1275}, 
and IC~310 \citep[e.g. see][]{Kadler2012,Shaw2013,Aleksic2013_IC310RapidVariability}.
The fifth radio galaxy, Cen A, is exceptional because of its proximity
and also a presence of \gray\ emitting giant lobes clearly resolved
with the LAT \citep{LAT10_CenAlobes}. Blazar-like properties of the
active nucleus in the source, which has been also detected in the VHE
band \citep{HESSCenADiscovery}, are subject to ongoing debate.
The only non-AGN extragalactic source is the nearby Large Magellanic Cloud
 (LMC) galaxy, which, given its proximity, has an extension of
 $2\degr$.  

The second largest source class is pulsars, with 5.2\% of the
catalog total.  SNRs and PWNe together are only 4.5\% of
the catalog. 


We note that, of the  65 1FHL sources that could not be associated with sources of known natures,
 five are associated with extended (Galactic) unidentified H.E.S.S. sources, 26 are
associated with unidentified 2FGL sources (including 1 associated with
one of the five previously-mentioned Galactic H.E.S.S. unidentified sources), 5 are associated with unidentified 1FGL sources, and
2 are associated with unidentified sources from the 3EG catalog.
The remaining 28 sources could not be associated with any $\gamma$-ray source reported previously. 
We note that the fraction of unassociated 1FHL sources is only
$\sim$13\% (65 out of 514), while that of the unassociated 2FGL sources was $\sim$31\% (575
out of 1873). 
The smaller fraction of unassociated 1FHL sources might be related to
the lower source density and good source localization, which facilitates the association of the
sources, as well as the brightness of the 1FHL sources at lower frequencies (particularly
optical and X-ray) in comparison to that of the 2FGL sources.


Figure~\ref{AngularSeparationWithMWSrc} shows the distribution of angular separations
between the associated 1FHL sources and their
counterparts. The total number of sources shown in this
distribution is 416. Of the 449 1FHL sources with associations, we
removed the 16 1FHL extended sources (see
Table~\ref{tbl:extended})\footnote{Two of the 18 1FHL extended sources are
  unassociated, and so are not included in the initial sample of 449 sources.}, the 6 sources classified as
``spp'' (see Table~\ref{tbl:spp}), and 11 1FHL sources that are
positionally coincident with extended TeV sources (all of which are
\gray\ pulsars: 5 PSR and 6 HPSR).  These 33 sources were
removed because the emission centroid in one energy range does not
necessarily coincide with the centroid (or location for point
sources) in the other energy range.  The angular separation for each source was
normalized with the quantity 
$r_{95}/\sqrt{-2\ln 0.05}$, where $r_{95}=\sqrt{\theta_1
  \theta_2}$  is the geometric mean of $\theta_1$ and $\theta_2$, the semi-major and
semi-minor radii of the location ellipse at 95\% confidence level.
The expected distribution of the angular difference with respect to
the real associations, when the distances are normalized as described
above, is described by a Rayleigh function
with $\sigma$=1. This function is also depicted in
Figure~\ref{AngularSeparationWithMWSrc}. The agreement between this
model curve and the observed distribution is quantified by a $\chi^2 /NDF$=27/19 (p-value=0.10), implying a successful association of
the 1FHL sources.

\begin{figure}
\begin{center}
\includegraphics[width=10.0cm]{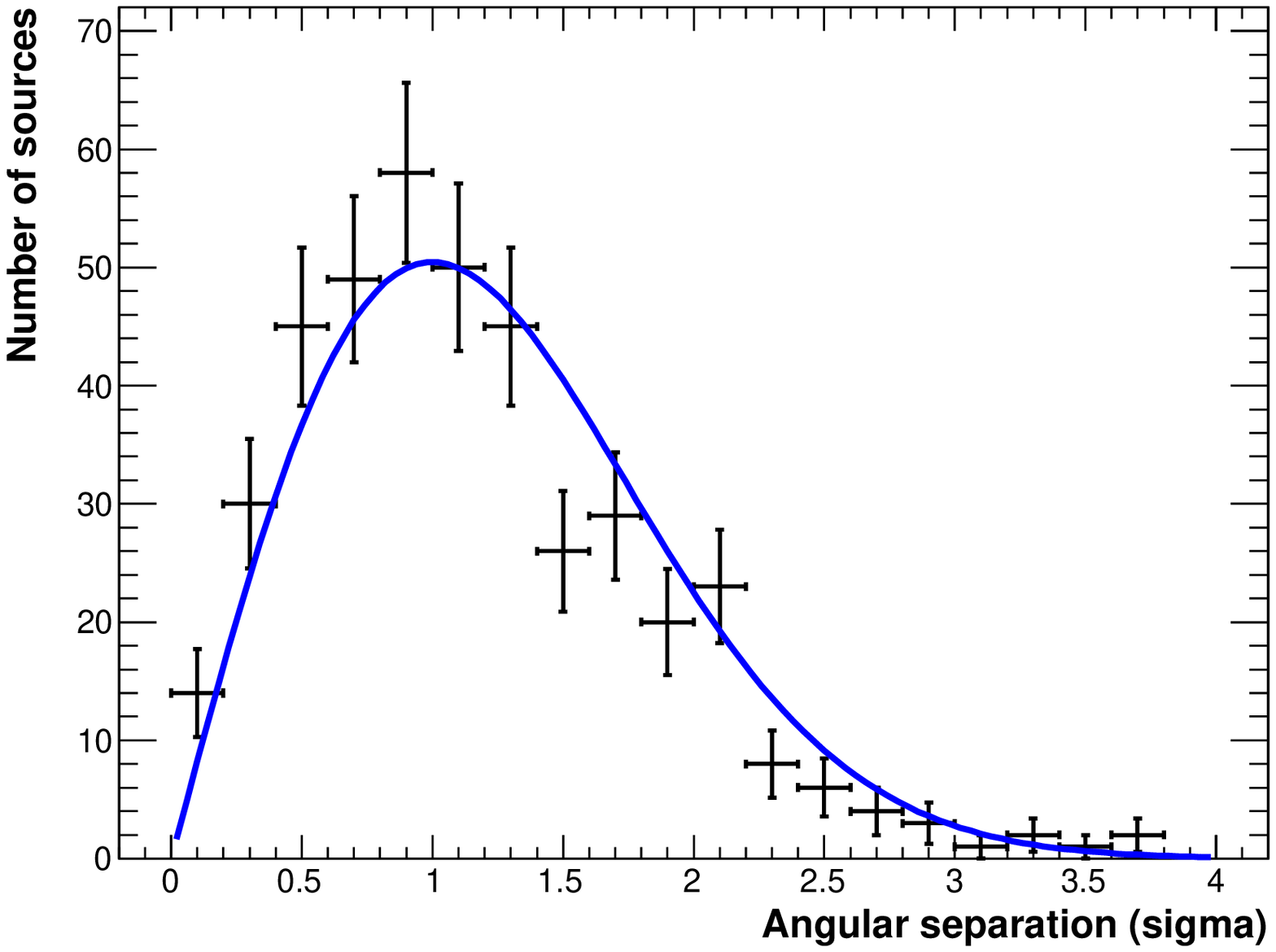}
\end{center}
\caption{\label{AngularSeparationWithMWSrc} 
Distribution of the angular separation between the 1FHL sources and 
the objects with which they are associated.
Only point sources were included in this distribution. 
The angular separation is normalized with the quantity
$r_{95}/\sqrt{-2\ln 0.05}$, where $r_{95}$ is the location uncertainty at
the 95\% confidence level. The blue curve is the
expected distribution of real associations. See text for details.}
\end{figure}

\begin{figure}  
\begin{center}
\includegraphics[width=18.0cm]{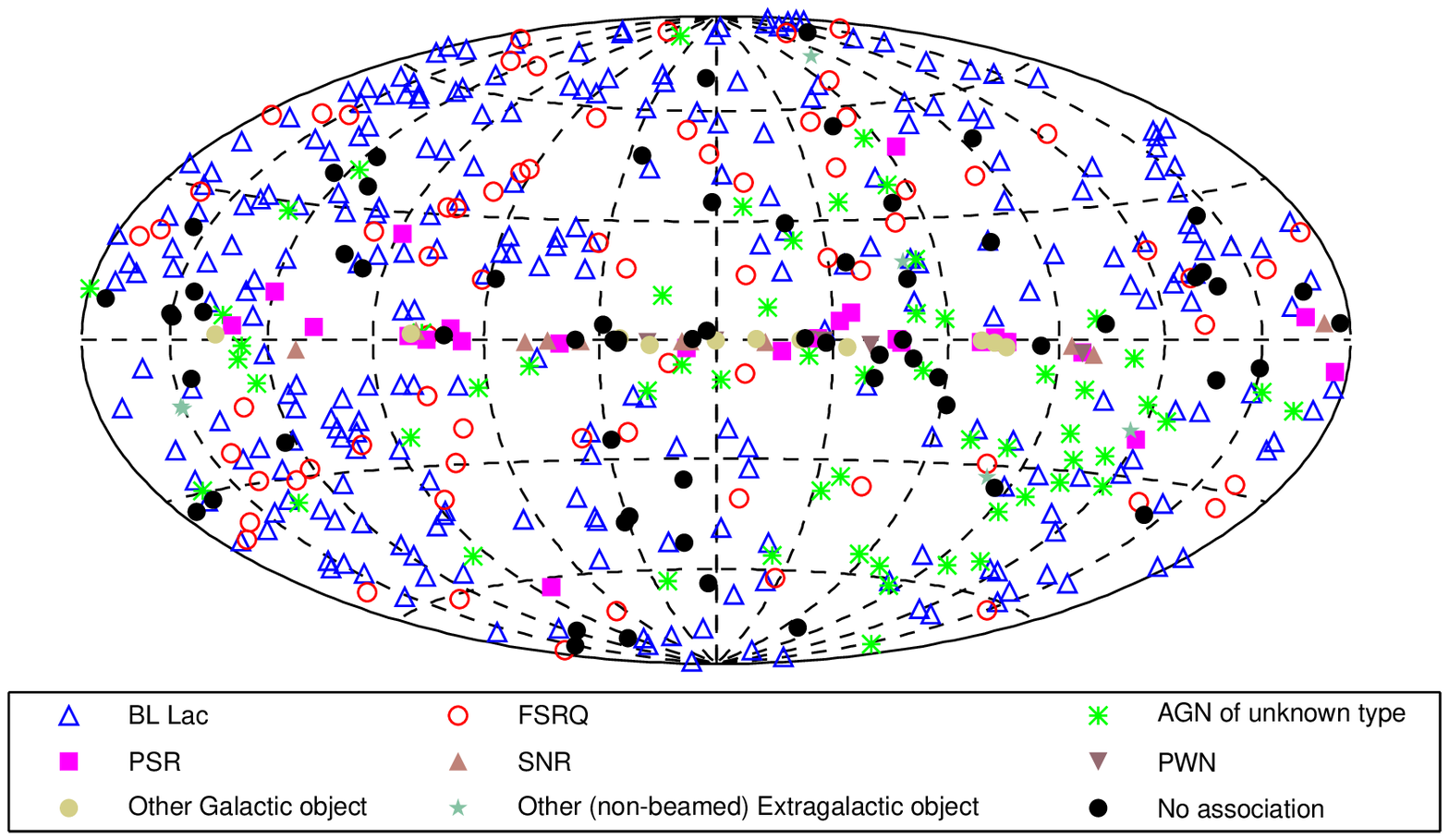}
\end{center}
\caption{\label{SkymapSummaryClasses} 
Sky map showing the sources by their source class, as reported in Table \ref{TableStatisticsSrcClasses}.  {  The projection is Hammer-Aitoff in Galactic coordinates.}}
\end{figure}

The locations on the sky of the sources in the above-mentioned  classes are
depicted in Figure~\ref{SkymapSummaryClasses}.  To a good approximation,
the Galactic sources are located essentially in the Galactic plane
(apart from some pulsars), while
the blazars are distributed roughly uniformly outside
the Galactic plane.
The source statistics are relatively
low, which precludes strong statements on the source
distributions. However, when considering the blazars, which constitute the majority
of 1FHL sources, an asymmetry between the northern and
southern Galactic hemispheres seems evident: the number
of BL Lacs and FSRQs is larger in the northern hemisphere, while the
number of AGNs of unknown types seems to be larger in the southern
hemisphere. The Galactic latitude distributions for these source
classes are depicted in Figure~\ref{SinBSummaryClasses}, showing that
the fraction of BL Lacs and FSRQs in the southern hemisphere is 42\% (108
sources out of 259) and
39\% (28 sources out of 71) respectively.
The fraction of AGNs of unknown type in the southern hemisphere is
71\% (41 out of 58), suggesting that many of these
sources must be BL Lacs and/or FSRQs. 

A similar north/south asymmetry with a larger number of sources was previously observed and reported in  2LAC
 and attributed to the slightly different exposure and the known non-uniformities of the counterpart catalogs.
In this work, we also consider AGNs with $|b|<10\degr$ (which were
excluded from the 2LAC paper), and they show another asymmetry: the fraction of known BL Lacs and FSRQs
is smaller at low latitudes, while the number of AGNs of unknown type
is slightly higher (at the level of 2 standard deviations). 
The lower fraction of BL Lacs and FSRQs at low Galactic latitudes is
certainly affected by the lower sensitivity
of LAT to detect sources in this region due to the higher diffuse
background (see Fig.~\ref{LATSensitivity}). Yet in this work we find that the asymmetry in the counterpart catalogs must also
play a role in the lower fraction of blazars at
low Galactic latitudes, as indicated by the higher fraction of AGNs of unknown
type for these latitudes.

The unassociated sources are fairly uniformly
distributed outside the Galactic plane, with a substantial increase in density for
$|b| < 11\fdg5$
($|\sin b| <0.2$).
It is to be expected that a large fraction of the low-latitude
unassociated sources are pulsars, SNRs and PWNe; but given the distributions shown in
Figure~\ref{SinBSummaryClasses}, unassociated
blazars are undoubtedly also present at low Galactic latitudes.


\begin{figure}  
\begin{center}
\includegraphics[width=10cm]{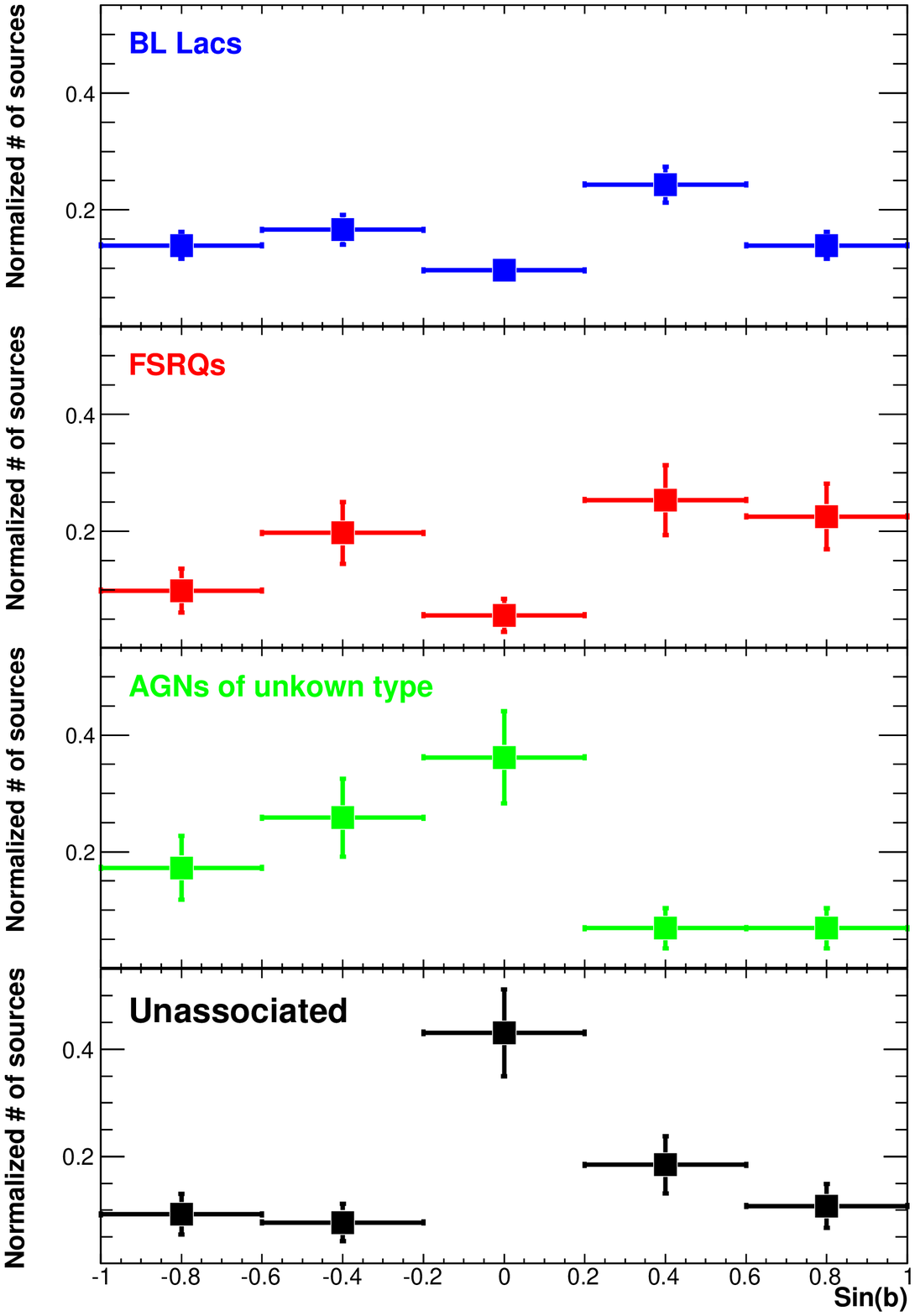}
\end{center}
\caption{\label{SinBSummaryClasses} 
Galactic latitude distributions of BL Lacs, FSRQs, AGNs of unknown type,
and unassociated 1FHL sources. The distributions were normalized to
the total numbers of source associations in each of these source classes, namely
259, 71, 58, and 65, respectively.}
\end{figure}


\subsection{Basic Properties of the 1FHL Sources}
\label{BasicProperties}

Figure~\ref{DistributionOfSignificanceMainGroups} shows the distribution
of significances ($\sigma$, derived from the $TS$ values on the assumption of 
four degrees of freedom) for the 1FHL sources grouped as extragalactic,
Galactic, and unassociated sources. 
There are no big differences between extragalactic and Galactic. In
contrast, the sources without associations differ from those
with associations; they are clustered at the lowest
significances, with most of them showing a significance smaller than 8$\sigma$.

\begin{figure}[th]
\begin{center}
\includegraphics[width=10cm]{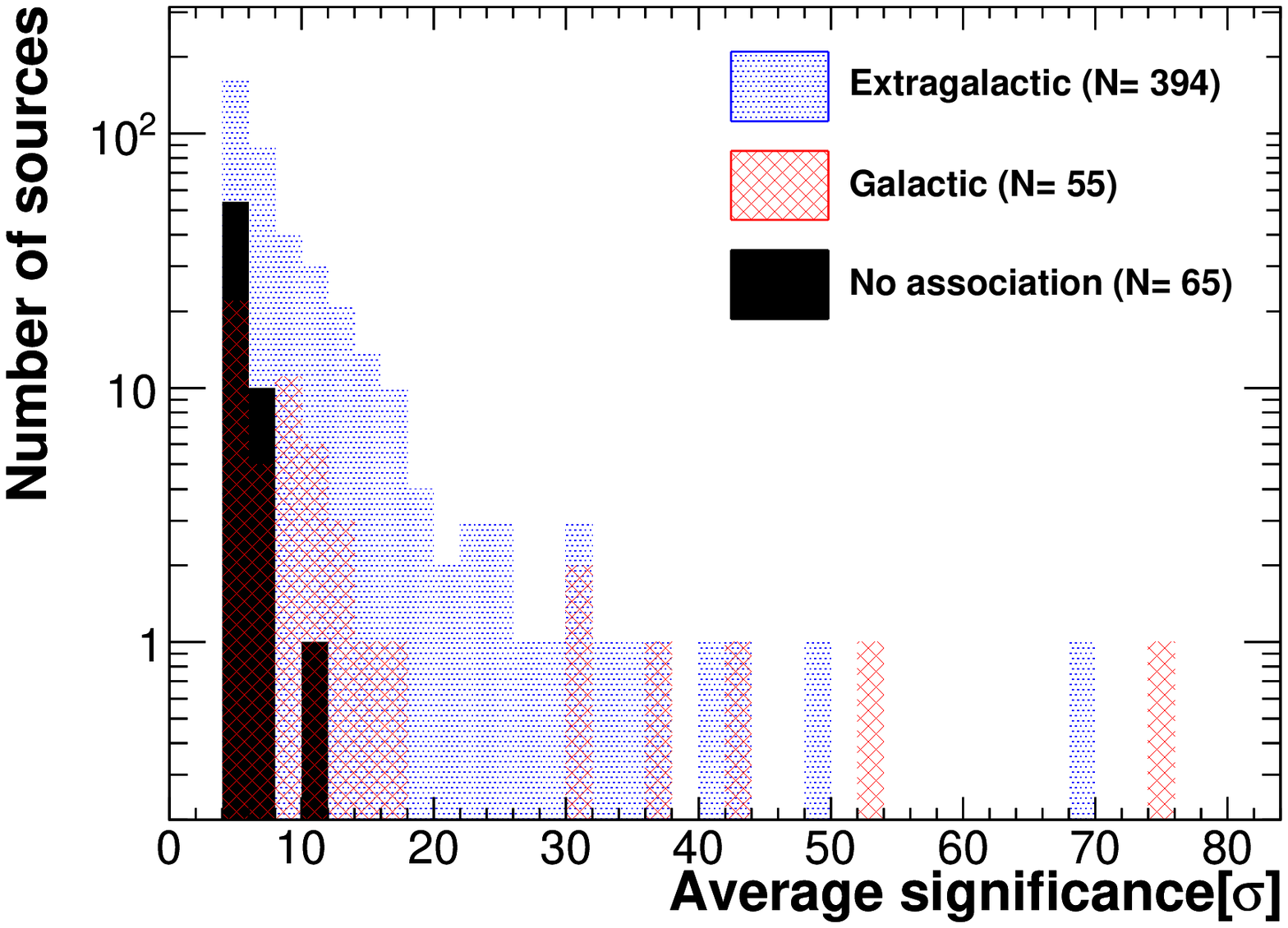}
\end{center}
\caption{\label{DistributionOfSignificanceMainGroups} 
Distribution of the significances of the 1FHL sources. The three
histograms report the significances for three groups of
sources: Extragalactic, Galactic, and unassociated sources.}
\end{figure}

Figure~\ref{DistributionOfFluxAndIndexForTSAbove25} shows the
distribution of the measured photon fluxes and photon indices for the
various source classes from the 1FHL list, grouped as in Figure~\ref{DistributionOfSignificanceMainGroups}.
Three sources with very soft spectral indices stand out:
1FHL~J2311.0+3425 (index 11$\pm$5), 1FHL~J1907.7+0600 (index 7$\pm$2), and
1FHL~J1635.0+3808 (index 6$\pm$2). The first and third are
associated with distant FSRQs 
(B2~2308+34 and 4C~+38.41, both with $z \sim$1.8), while the second is
associated with a \gray\ pulsar (LAT~PSR~J1907+0602). 
These three sources are significantly detected in the 10--30 GeV range, but
not detected in the ranges 30--100 GeV and 100--500 GeV. 
Consequently, the spectra resulting from our analysis are extremely
soft, and have large statistical uncertainties due to the lack of
high-energy photons. 
  
The distribution of spectral indices for 1FHL sources with associations
in the Milky Way have no obvious differences from those with blazar
associations, while the
measured fluxes for the Galactic sources clearly tend to be greater.
(The lowest fluxes are found only for
sources with extragalactic associations, or no associations.)
This is not an intrinsic property of the Galactic sources, but rather
due to the worse photon flux sensitivity in the
Galactic plane (due to the brighter diffuse backgrounds), as reported in
\S~\ref{LATExposure}. 

\begin{figure}[th]
\begin{center}
\includegraphics[width=7cm]{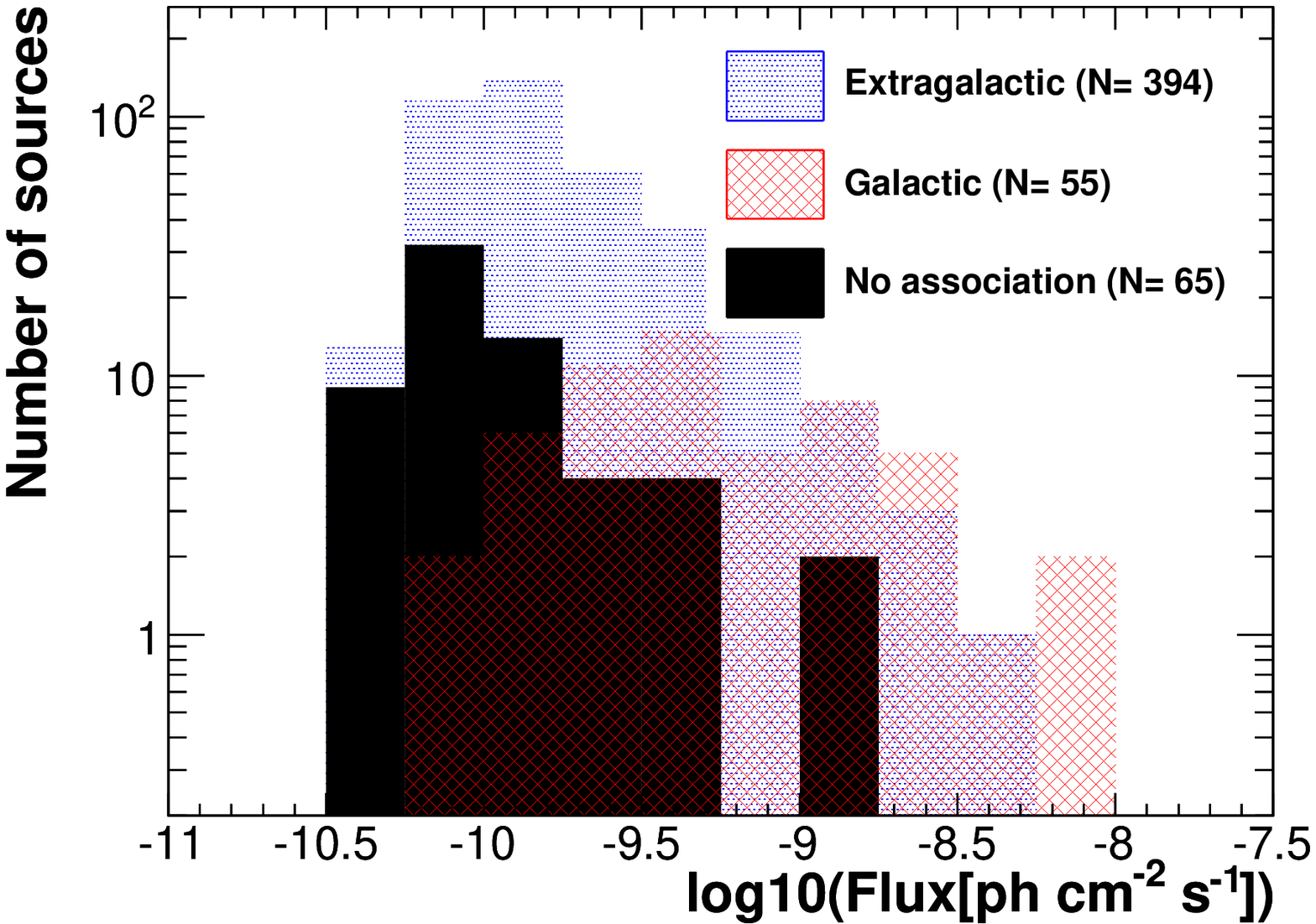}
\includegraphics[width=7cm]{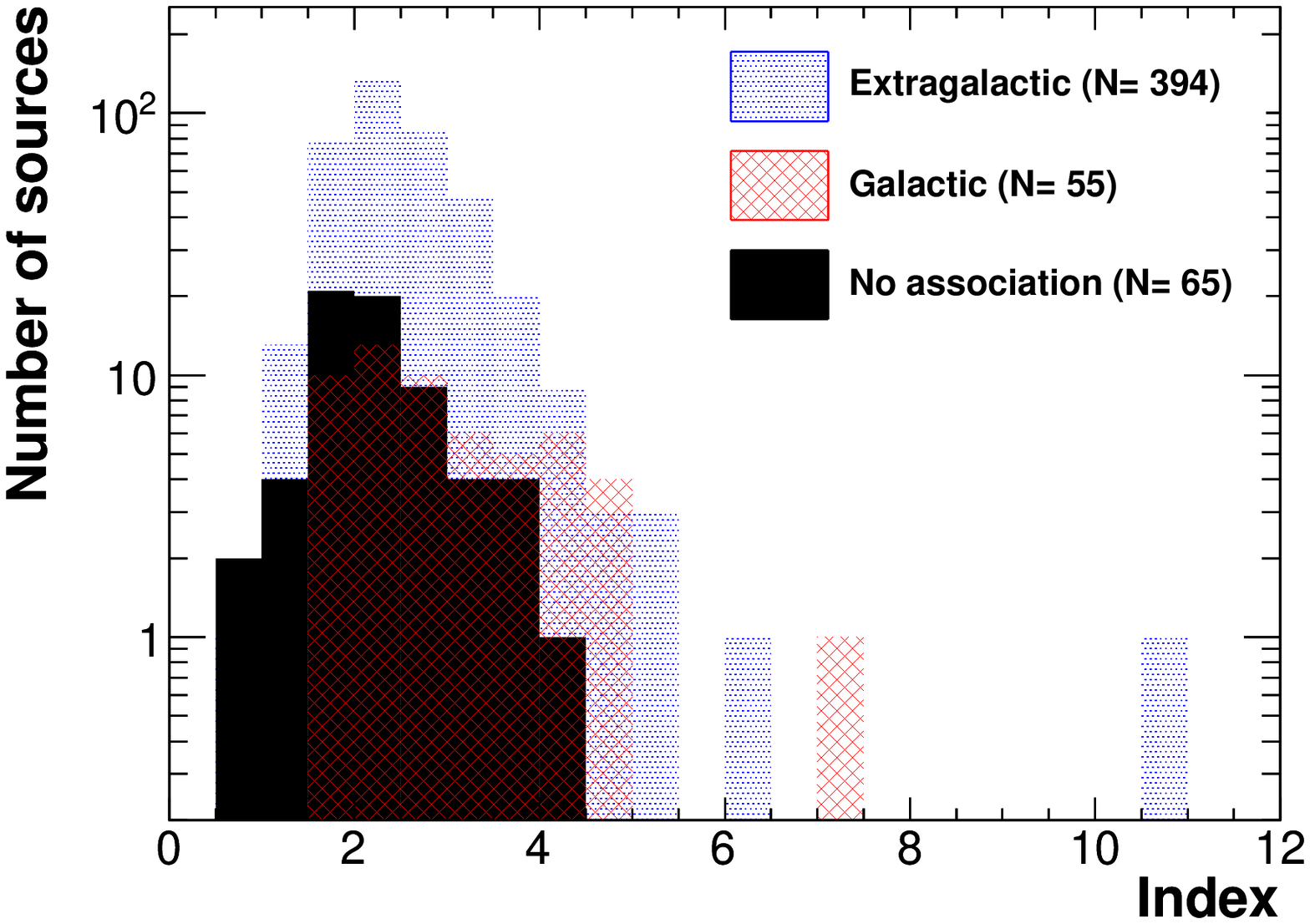}
\end{center}
\caption{\label{DistributionOfFluxAndIndexForTSAbove25} 
Distribution of the measured photon fluxes (left) and photon
index (right) for the 1FHL sources. The three
histograms report the significances for three different groups of
sources: Extragalactic, Galactic and sources without associations.}
\end{figure}

{  A search for variability was performed using the Bayesian Block
algorithm as described in \S~\ref{Variability}.} A total of 43 sources
show evidence for variability, i.e., have two or more blocks, and they
all belong to the blazar class.  For these sources, the numbers of
events within the RoIs range from 10 (1FHL~J0210.9$-$5100 and
1FHL~J1635.0+3808) to 178 (1FHL~J0222.6+4302), with a median value of
30.  Most of the light curves for the variable sources (39/43) contain
two or three blocks, while the light curves for the remaining 4 (4/43)
contain four, five, six, and ten blocks each. The number of Bayesian
Blocks measured for each of the 1FHL sources is reported in
\S~\ref{1FHLAGNs}.  With our chosen false-positive rate, a total of
5--6 sources would be expected to have more than 1 block by chance.

{  Figure \ref{VariabilityBBLC} shows the light curves of 9 sources with
different variability characteristics representative of the larger
sample.  Five of these sources are particularly interesting:}

\begin{itemize}

\item {  The light curve of 1FHL~J0222.6+4302 (3C 66A) displays two
  prominent flares.  The first flare occurred in 2008 October and was
  detected in the VHE band by VERITAS
  \citep{2008ATel.1753....1S,2011ApJ...726...43A}. The second flare
  occurred in 2009 May, but the source was too close to the Sun for
  VHE observations.}

\item {  The source 1FHL~J0238.7+1639 (AO 0235+164) was detected in a
  high state during the first three months of the {\it Fermi} mission
  before transitioning to a lower state and eventually fading below
  the threshold for detection after 2009 September.}

\item {  The most frequently variable source in the catalog is
  1FHL~J1224.8+2122 (4C~$+$21.35). The Bayesian Block algorithm
  detected ten blocks, indicating four short and strong flares over
  the course of a few months.  No events were detected from this
  source before 2009 March 1. The second flare was detected above 100
  MeV by the {\FermiLATc} in 2010 April
  \citep{2010ATel.2584....1D}. The third flare is the brightest
  detected by the Bayesian Block analysis and occurred on 2010 May 25
  when three 
  $\gamma$ rays
  (above 10~GeV) were detected
  within a ten hour span. This flare was reported by AGILE above
  100~MeV \citep{2010ATel.2641....1B}. The last flare occurred between
  2010 June 17 and 2010 June 19, when seven $\gamma$-ray-like events
  arrived within a ~29 hour interval. It was detected by AGILE
  \citep{2010ATel.2686....1S} and
  \FermiLATc~\citep{2010ATel.2687....1I} above 100 MeV. The high
  activity from this flare was also observed at VHE on 2010 June 17 by
  MAGIC \citep{2010ATel.2684....1M,2011ApJ...730L...8A}.  MAGIC
  detected significant variability with a flux-doubling time of only
  10 minutes.}

\item {  The source 1FHL J2253.9+1608 (3C 454.3) is among the brightest
  detected above 10~GeV. A higher-flux state starting in 2010 November
  and lasting 3 months was detected in both the Bayesian Block and
  aperture photometry analyses. A short and bright flare occurred
  during this period starting on 2010 November 19 and lasting only two
  days. This short/bright flare above 10 GeV is very similar to those
  observed from 4C~$+$21.35, indicating that the FSRQ 3C~454.3 might
  also have been detected at VHE had it been observed during this
  period. However, detection of 3C~454.3 would have been more
  difficult due to its strong spectral break at GeV energies, even
  during large flares
  \citep[see][]{2010ApJ...721.1383A,2011ApJ...733L..26A} and the
  greater redshift ($z$=0.859) of this source.  The flare above 10~GeV
  was also detected above 100~MeV by
  \FermiLATc~\citep{2010ATel.3041....1S} and AGILE
  \citep{2010ATel.3043....1S}.}

\item {  One of the brightest sources in the 1FHL catalog is the
  high-frequency-peaked blazar 1FHL J1104.4+3812 (Mrk~421).  Despite
  having 383 events within the RoI, the source is not detected as
  variable by the Bayesian Block analysis above 10 GeV (see
  Fig. \ref{VariabilityBBLC}). The aperture photometry indicates a
  period of higher activity centered around late 2009 to early 2010. A
  dedicated analysis with a false positive threshold of 5\% confirms
  this higher flux state, which matches well the period of enhanced
  VHE activity observed by MAGIC and VERITAS in November 2009, and
  January, February and March 2010
  \citep{2011ICRC....8...62G,SunGamma2012}.  However, our variability
  analysis above 10 GeV fails to detect the extremely bright, day-long
  VHE flare detected by VERITAS on 2010 February 17, when Mrk~421
  increased its flux by about a factor of 20 with respect to its
  typical value \citep{2010ATel.2443....1O}.}
\end{itemize}

The results from the Bayesian Block analysis cannot be directly
compared with the likelihood analysis performed to derive monthly
light curves for the 2FGL catalog. Despite the differences in the
methods and the time intervals (two years versus three years), we
highlight some comparisons.  Of the 43 variable sources detected above
10 GeV, only 2 sources do not have counterparts in 2FGL
(1FHL~J0318.8+2134 and 1FHL~J1532.6$-$1317). Both sources show higher
fluxes in the third year, i.e., after the time interval of the 2FGL
analysis.  Of the remaining 41 sources, only 5 did not show evidence
for variability in 2FGL (1FHL~J0203.6+3042, 1FHL~J0316.1+0904,
1FHL~J0809.8+5217, 1FHL~J1603.7$-$4903, and
1FHL~J1748.5+7006). Therefore, it appears that the population of
sources variable above 10 GeV is also variable in the 2FGL energy band
(100 MeV - 100 GeV).  Although the most frequently variable source
above 10 GeV (1FHL~J1224.8+2122) has the second largest
$TS_\textrm{\small var}$ (13030) in 2FGL, the number of Bayesian
Blocks and $TS_\textrm{\small var}$ are not strongly correlated. For
example, several sources with two or three blocks have much larger
$TS_\textrm{\small var}$ values than sources with four, five, or six
blocks.

\begin{figure}
\begin{center}
\includegraphics[width=\textwidth]{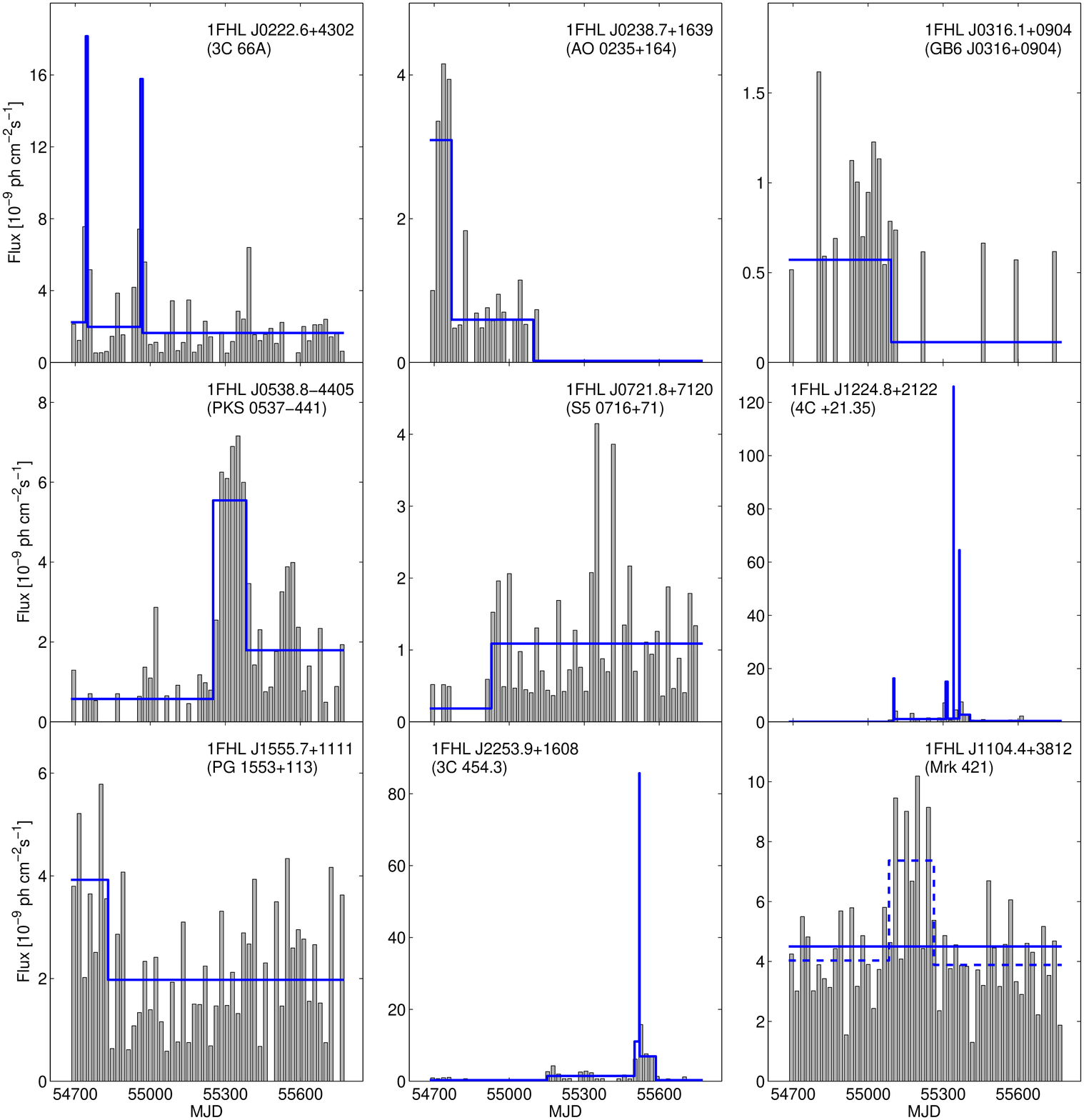}
\end{center}
\caption{Light curves for a subset of the variable sources. The histograms
  correspond to the aperture photometry analysis, and the solid lines
  correspond to the Bayesian Block analysis using a $1\%$ false
  positive threshold.  The panels are labeled with the 1FHL names and
  the names of the corresponding associated sources (in parentheses). The dashed line 
  in the panel for Mrk~421 corresponds to a 5\% false positive threshold
  (see text for details).
}
\label{VariabilityBBLC}
\end{figure}

\subsection{1FHL Sources Not in the 2FGL Catalog}
\label{non2FGL}

The 1FHL catalog Table~\ref{MainCatalogTable} contains 63 sources
not associated with 2FGL sources\footnote{Out of the 63
  sources, 11 are
  associated with 1FGL sources that did not reach a $TS$ value of 25
  in the 2FGL analysis, which used 2 years of LAT data.}.
Among these sources, 
spatial extension at MeV/GeV/TeV energies has been recently reported for nine.
For eight of these, extension had previously been resolved by the LAT
(see Table \ref{tbl:extended}). 
The nine sources are listed in Table \ref{SpecialNon2FGLSrc}.  
In the 2FGL catalog, each of these sources is modeled as a point source,
and as a result our association pipeline failed to link these 1FHL sources
with the 2FGL counterparts despite angular separations of less than $0\fdg3$
(typically less than $0\fdg2$).  For this reason, we split the 
list of 63 non-2FGL sources into
two groups: non-2FGL\_a, the 9 sources reported in 
Table~\ref{SpecialNon2FGLSrc}; and non-2FGL\_b, the remaining 54
(point-like) sources without 2FGL counterparts.

\begin{deluxetable}{lrrrllclrrrr}
\rotate
\setlength{\tabcolsep}{0.04in}
\tablewidth{0pt}
\tabletypesize{\tiny}
\tablecaption{1FHL {\it Extended} Sources Without 2FGL Counterparts\label{SpecialNon2FGLSrc}}
\tablehead{
\colhead{1FHL Name} & \colhead{R.A.} & \colhead{Decl.} & 
\colhead{$\sigma$} & \colhead{Extended Source} &
\colhead{ASSOC\_TEV} & \colhead{CLASS1} & 
 \colhead{2FGL Name} &  \colhead{R.A. (2FGL)} & \colhead{Decl. (2FGL)} & 
\colhead{$\sigma$} & \colhead{Ang. Sep.} \\
 & \colhead{[deg]} & \colhead{[deg]} &  
 & & &  &  &  \colhead{[deg]} & \colhead{[deg]} & 
\colhead{(2FGL) } & \colhead{[deg]}
}
\startdata
J2021.0+4031e & 305.270 & 40.520 & 15.7 & gamma Cygni & VER J2019+407
& snr &  J2021.5+4026 & 305.392  & 40.441  &  129.7 & 0.12 \\ 
J0852.7$-$4631e & 133.200 & $-$46.520 & 11.1 & Vela Junior & RX J0852.0$-$4622 & snr & J0851.7$-$4635 & 132.941  & $-$46.592  &  5.5 & 0.19 \\ 
J1633.0$-$4746e & 248.250 & $-$47.770 & 10.9 & HESS J1632$-$478 & HESS J1632$-$478 & pwn & J1632.4$-$4753c & 248.114  & $-$47.891  &  8.8 & 0.15 \\ 
J1615.3$-$5146e & 243.830 & $-$51.780 & 10.8 & HESS J1614$-$518 & HESS J1614$-$518 &  & J1615.2$-$5138 & 243.801  & $-$51.635  &  14.7 & 0.15 \\ 
J1713.5$-$3951e & 258.390 & $-$39.850 & 8.3 & RX J1713.7$-$3946 & RX J1713.7$-$3946 & SNR &  J1712.4$-$3941 & 258.111  & $-$39.687  &  5.1 & 0.27 \\ 
J1616.2$-$5054e & 244.060 & $-$50.910 & 7.8 & HESS J1616$-$508 & HESS J1616$-$508 & pwn &  J1615.0$-$5051 & 243.758  & $-$50.852  &  15.2 & 0.20 \\ 
J1836.5$-$0655e & 279.140 & $-$6.920 & 7.6 & HESS J1837$-$069 & HESS J1837$-$069 &  &  J1837.3$-$0700c & 279.347  & $-$7.011  &  8.2 & 0.22 \\ 
J0822.6$-$4250e & 125.660 & $-$42.840 & 6.9 & Puppis A &   & snr &  J0823.0$-$4246 & 125.766  & $-$42.770  &  10.2 & 0.10 \\ 
J1634.7$-$4705 & 248.690 & $-$47.089 & 4.2 &  & HESS J1634$-$472 &  &  J1635.4$-$4717c & 248.850  & $-$47.297  &  7.7 & 0.24 \\ 
\enddata
\tablecomments{The entries are sorted in reverse order of detection significance reported in the main 1FHL catalog table \ref{MainCatalogTable}.  
Each of the sources is associated with an extended (Galactic) VHE source.
All were classified as point sources in
  the 2FGL catalog, while 8 (out of 9) were recently found to have a
  significance extension at MeV/GeV energies, as we noted in table \ref{tbl:extended}.
  For these 8 sources,
  the table reports the Extended\_Source\_Name used in table \ref{tbl:extended}
  }
\end{deluxetable}

Figure~\ref{DistributionOfSignificanceNon2FGL} shows the distribution of
the detection significances for all of the 1FHL sources, grouped in
several classes: all sources, sources with 2FGL counterparts, and the
 two groups of non-2FGL sources described in the text, 
non-2FGL\_a and non-2FGL\_b. The distribution peaks at the threshold
of $\sim$4 $\sigma$ ($TS=25$), and extends to about 40 $\sigma$ with three sources
having formal significances greater than 50 $\sigma$. This plot shows that the $\gamma$-ray
sources that were not reported in the 2FGL catalog cluster 
at the significance threshold.

\begin{figure}[th]
\begin{center}
\includegraphics[width=10cm]{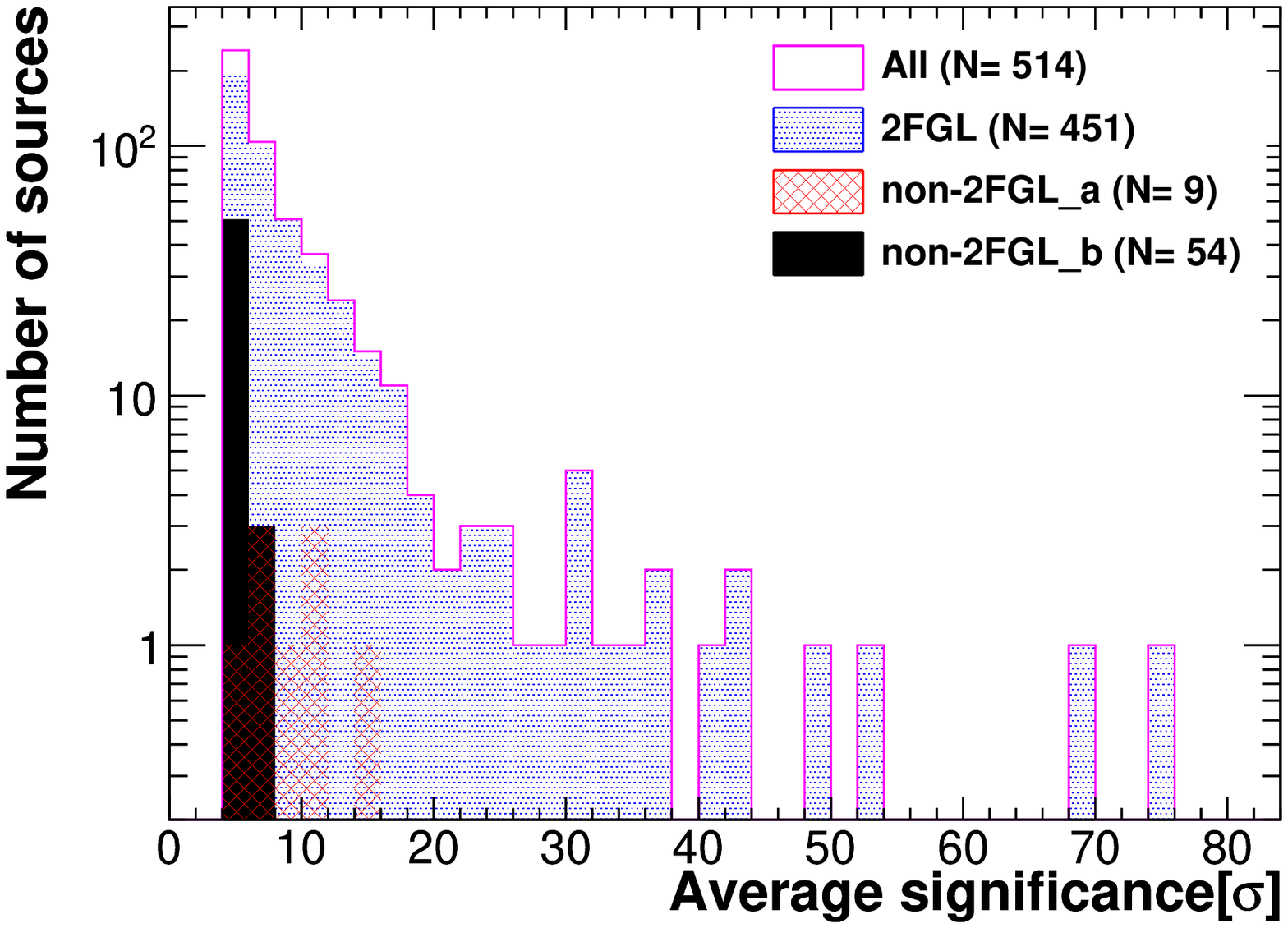}
\end{center}
\caption{\label{DistributionOfSignificanceNon2FGL} 
Distribution of the significances of the 1FHL sources. The four
histograms report the significances for all 1FHL sources (All); 1FHL
sources whose locations match 2FGL sources (2FGL); 1FHL
extended sources that do not match 2FGL sources, but are less than $0\fdg3$ from 2FGL sources 
(non-2FGL\_a); 1FHL point sources which do not match 2FGL sources (non-2FGL\_b).}
\end{figure}

Figure~\ref{DistributionOfFluxAndIndexNon2FGL} reports the distributions
of flux and index for the 1FHL sources that are not in 2FGL, grouped as in
Figure~\ref{DistributionOfSignificanceNon2FGL}.
The group non-2FGL\_b has the lowest fluxes and smallest indices. In
particular, this group hosts the four sources with the smallest indices
(\lapp 1): 1FHL J1314.9$-$4241 (associated with the blazar 
MS 13121$-$4221),  1FHL J1856.9+0252 (associated with the presumed PWN
HESS J1857+026), and 1FHL J2159.1$-$3344 and 
1FHL J0432.2+5555 (not associated with any known sources).
All these sources are very weak and have hard spectra in the $>$10 GeV energy
range.

\begin{figure}[th]
\begin{center}
\includegraphics[width=7cm]{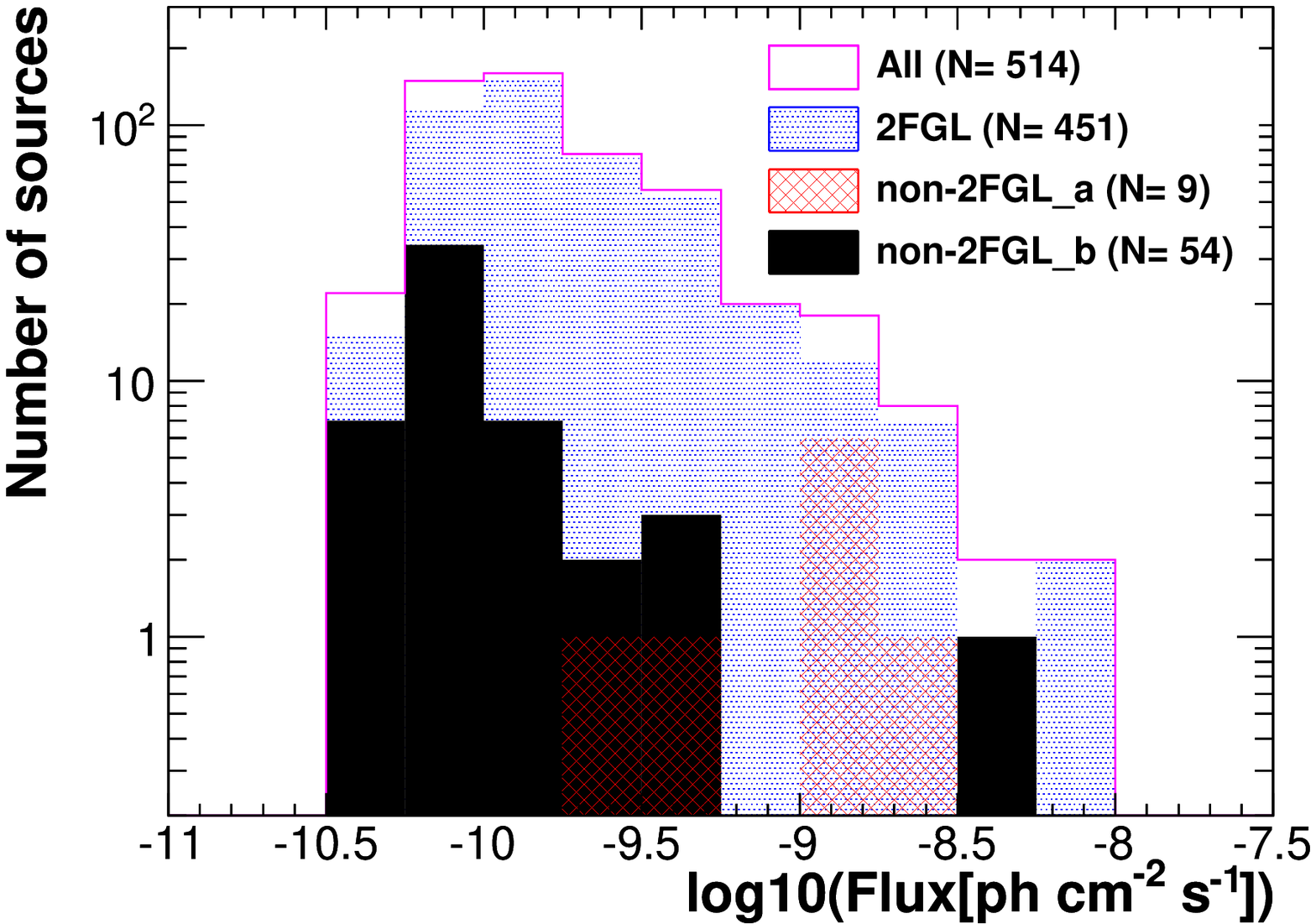}
\includegraphics[width=7cm]{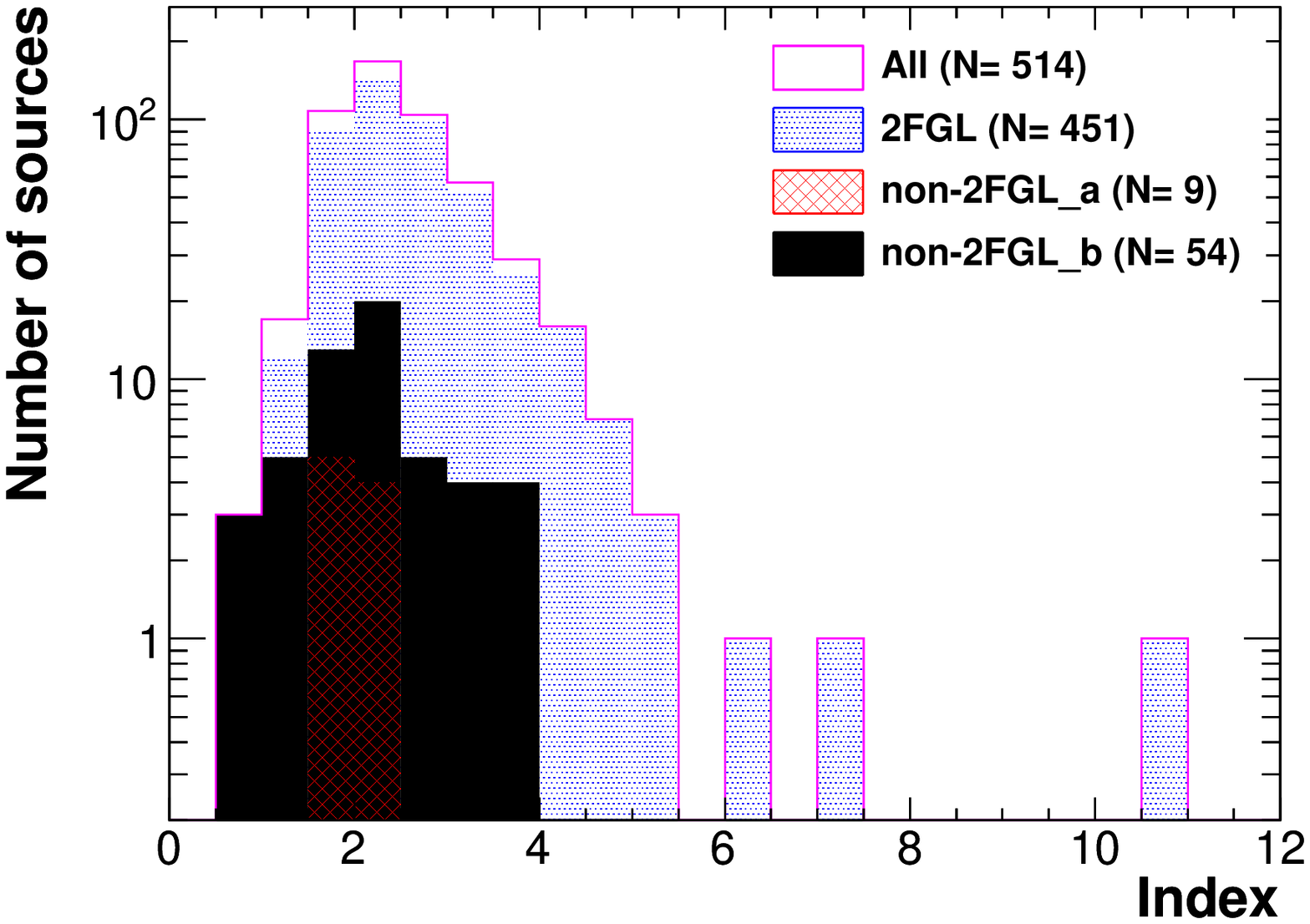}
\end{center}
\caption{\label{DistributionOfFluxAndIndexNon2FGL} 
Distribution of the measured photon fluxes (left) and photon
index (right) for the 1FHL sources. 
The four
histograms report the significances for all 1FHL sources (All); 1FHL
sources whose locations match any in the 2FGL catalog (2FGL); 1FHL
extended sources which do not match any in the 2FGL, but are less than $0\fdg3$ from 2FGL sources
(non-2FGL\_a); 1FHL point sources that do not have corresponding 2FGL
sources (non-2FGL\_b).}
\end{figure}

Figure~\ref{SkymapNon2FGL} shows the distribution on the sky of the 63 1FHL sources
without 2FGL counterparts. Apart from the 9 extended sources from
Table~\ref{SpecialNon2FGLSrc} (4 SNRs, 2 PWNe,
and 3 sources without associations),
most sources are located outside the Galactic plane: 9 blazars, 8 blazar candidates,
and a large fraction of the 36 unassociated sources. 

We conclude that most of the new $\gamma$-ray sources reported in the
1FHL catalog (not reported previously in the 2FGL catalog) are likely to be blazars
with weak, hard-spectrum emission that might have been more active in
the third year. As reported in \S~\ref{Variability}, only two non-2FGL sources
(1FHL J0318.8+2134 and 1FHL J1532.6$-$1317) have significantly greater
average fluxes in the third year of
LAT observations than they did during the first
two years (the time interval for the 2FGL catalog). 
The limited counting statistics above 10~GeV, however, make variability hard to confirm. 

\begin{figure}[th]
\begin{center}
\includegraphics[width=18.0cm]{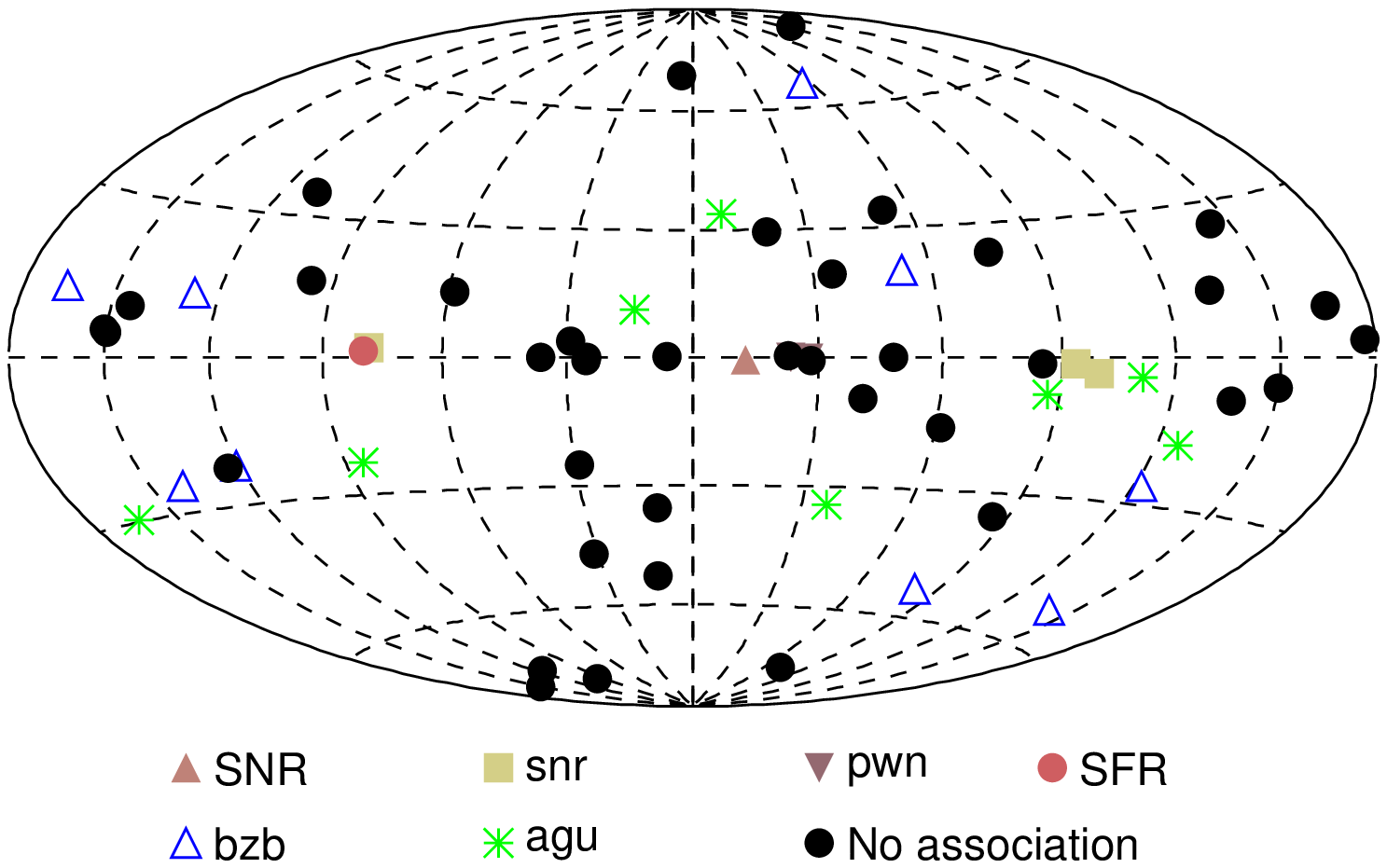}
\end{center}
\caption{\label{SkymapNon2FGL} 
Sky map showing the locations of the 1FHL sources that do not have counterparts in the 2FGL catalog
(groups non-2FGL\_a and non-2FGL\_b). The sources are depicted with
the source classes described in Table~\ref{TableStatisticsSrcClasses}.
{  The projection is Hammer-Aitoff in Galactic coordinates.}
}
\end{figure}

\subsection{The 1FHL AGNs}
\label{1FHLAGNs}

The 1FHL catalog is strongly dominated by AGNs,
with 393\footnote{The only non-AGN extragalactic source is the LMC.} sources
associated with AGNs.
Among them, blazar and blazar
candidates\footnote{Most of the ``agu'' sources are expected to be
  blazars.} represent 86\% of the sources that have associations.
In this section we report on the overall $\gamma$-ray properties of
these AGNs.

As for the 2FGL catalog, most of the extragalactic 1FHL sources
are \emph{non-thermal-dominated} or
\emph{jet-dominated} sources; that is, the broad-band emission is
produced by high-energy particles accelerated in the magnetized jet of material
ejected from the central engine.
Non-thermal AGNs are classified by the frequency of the peak of the
synchrotron emission, which is related to the maximum energy of the
accelerated electrons. Here we used the convention proposed
by \citet{LATSEDs_BrightBlazars2010} and
classify the AGNs  as {\it low-synchrotron-peaked}
 (LSP), {\it intermediate-synchrotron-peaked} (ISP), and
{\it high-synchrotron-peaked} (HSP) if the peak of the synchrotron
emission $\nu_{syn_{peak}}$ is located below 10$^{14}$~Hz, in the range
10$^{14}$--$10^{15}$~Hz, or above 10$^{15}$~Hz, respectively.
This 
is commonly designated as 
the {\it spectral energy distribution (SED) classification}, and it is complementary to the broadly used
{\it optical classification}, which uses the presence/absence of
emission lines to classify sources as FSRQ or BL Lac-type.

Table~\ref{TableCatalogAGNFirstNEntries} reports the optical and SED
classifications, as well as  the redshifts (if
available) and the measured variability for all of the 1FHL AGN sources. 
The variability is quantified as described in \S~\ref{Variability}, with the most
variable sources having the highest number of blocks, and the
sources with no significant variability having only one block.
The optical and SED classifications, and redshifts, were obtained
primarily from the 2LAC paper \citep{LAT_2LAC}, with the information for the
non-2FGL AGNs being obtained from the BZCAT \citep{Massaro2009}.
Moreover, we also used the recent work by \citet{Shaw2013} to obtain the optical classification and redshift information for some
sources. 
We noted that seven AGNs with 1FHL associations have redshifts reported by
\citet{LAT_2LAC} that are in conflict with the information reported by \citet{Shaw2013}.
For four sources (1FHL J0007.7+4709, 1FHL J0508.1+6737,
1FHL J1312.2$-$2158, and 1FHL J2116.2+3339) the newly-reported redshifts
by \citet{Shaw2013} do not match those of \citet{LAT_2LAC}.
For three others (1FHL J0909.3+2312, 1FHL
J2016.3$-$0907, and 1FHL J2323.8+4210), the values given by
\citet{LAT_2LAC} violate lower limits
reported by \citet{Shaw2013}. 
We 
report both values
(or value and lower limit) in Table
\ref{TableCatalogAGNFirstNEntries} as ``z1 \& z2'' (``z1 \&
LowLimit\_z2''), {  where the first entry is retrieved from
  \citet{LAT_2LAC} and the second one (value or lower limit) from \citet{Shaw2013}.}

\begin{deluxetable}{lrrlcccc}
\setlength{\tabcolsep}{0.04in}
\tablewidth{0pt}
\tabletypesize{\scriptsize}
\tablecaption{Characteristics of the 1FHL Sources With AGN Associations\label{TableCatalogAGNFirstNEntries}}
\tablehead{
\colhead{1FHL Name} &
\colhead{R.A.} &
\colhead{Decl.} &
\colhead{Assoc.} &
\colhead{Optical Class.} &
\colhead{SED Class.} &
\colhead{Redshift}  &
\colhead{Variability\_BayesBlocks}
}
\startdata
J0007.7+4709 & 1.947 & 47.155 & MG4 J000800+4712 & BL Lac & LSP & 0.28
\& 2.1 & 1 \\
J0008.7-2340 & 2.194 & $-$23.674 & RBS 0016 & BL Lac & \nodata & 0.147 & 1 \\
J0009.2+5032 & 2.316 & 50.541 & NVSS J000922+503028 & BL Lac & \nodata & \nodata & 1 \\
J0018.6+2946 & 4.673 & 29.776 & RBS 0042 & BL Lac & HSP & \nodata & 1 \\
J0022.2-1853 & 5.555 & $-$18.899 & 1RXS J002209.2-185333 & BL Lac & HSP & \nodata & 1 \\
J0022.5+0607 & 5.643 & 6.124 & PKS 0019+058 & BL Lac & LSP & \nodata & 1 \\
J0033.6-1921 & 8.407 & $-$19.361 & KUV 00311-1938 & BL Lac & HSP & 0.610 & 1 \\
J0035.2+1514 & 8.806 & 15.234 & RX J0035.2+1515 & BL Lac & HSP & \nodata & 1 \\
J0035.9+5950 & 8.990 & 59.838 & 1ES 0033+595 & BL Lac & HSP & \nodata & 1 \\
J0037.8+1238 & 9.473 & 12.645 & NVSS J003750+123818 & BL Lac & HSP & 0.089 & 1 \\
J0040.3+4049 & 10.096 & 40.827 & 1ES 0037+405 & BL Lac & HSP & \nodata & 1 \\
J0043.7+3425 & 10.936 & 34.429 & GB6 J0043+3426 & FSRQ & \nodata & 0.966 & 1 \\
\enddata
\tablecomments{R.A. and Decl. are celestial coordinates in J2000
  epoch, Assoc. is the name of the associated (or identified)
  source counterpart, Optical Class. is the optical
  classification of the AGN, SED class is the SED
  classification (whenever available), and Variability\_BayesBlocks is the number of
  Bayesian Blocks (see \S~\ref{Variability}).
  Four of the sources have two distinct redshifts reported in the
  literature and their redshifts are listed here separated by the symbol
  ``\&''. Three sources have redshifts in the literature that
  violate the spectroscopic lower limits reported in \cite{Shaw2013};
  these are listed as ``z1 \& LowLimit\_z2''. Three 1FHL sources have double associations; the two distinct associated source
  names and characteristics are reported separated by the symbol ``\&''.
{  This table is published in its entirety in the electronic edition
  of the Astrophysical Journal Supplements,  and available as a FITS
  file from the FSSC.
  A portion is shown here for guidance regarding its form and content.}}
\end{deluxetable}

Table \ref{TableSEDClassAndRedshift} summarizes the number of 1FHL AGN
sources belonging to the various SED classifications with and without
redshift determinations. Among all blazars, the dominant SED class is
HSP, which makes up
$\sim$41\% of the 1FHL AGNs. This is not a surprising result because
HSPs typically have a {\it hard} spectrum
(power-law index $\lapp$2) and hence they are expected to be the
AGN source class that emit the highest-energy photons. Table
\ref{TableSEDClassAndRedshift} 
also shows that the 1FHL catalog has 208 ($\sim$53\%) sources associated with AGNs of known
 redshifts\footnote{
This number does not include the seven sources with conflicting
redshift information reported above.}, from which the fractions of
sources with measured redshifts are 47\%, 46\% and 41\% for HSP, ISP, and sources without
SED classification, respectively, and 76\% for the LSP class. The fraction of
LSPs with available redshifts is larger because 58 of the 99 LSPs  are actually FSRQs which, {\it by definition}, have measured redshifts,
while no ISP or HSP are FSRQs and the FSRQ optical classification overlaps
exclusively with the LSP SED classification.

\begin{deluxetable}{lcc}
\setlength{\tabcolsep}{0.04in}
\tabletypesize{\scriptsize} 
\tablecolumns{3} 
\tablewidth{0pc}
\tablecaption{Summary of SED Classifications and Available Redshifts for 1FHL Sources With AGN Associations\label{TableSEDClassAndRedshift}}
\tablehead{ 
\colhead{SED Classification}                    & \colhead{Number of Sources}
& \colhead{Number with Measured Redshift (fraction)} 
}  
\startdata 
HSP  &  162 &  76 (47\%)\\ 
ISP  &    61 &   28 (46\%)\\
LSP  &    99 &   75 (76\%)\\
Not Classified & 71 & 29 (41\%) \\
\hline 
{\bf Total} & {\bf 393} & {\bf 208 (53\%)} \\
\enddata

\end{deluxetable}

Figure \ref{DistributionOfIndexAt100MeVAnd10GeV} shows the
distribution of the measured power-law indices of the 1FHL blazars in
the energy ranges 100 MeV to 100 GeV (extracted from the 2FGL
catalog table) and 10--500 GeV (from Table~\ref{MainCatalogTable}). 
The figure does not show the nine 1FHL sources that are associated with the five radio galaxies and the other four non-blazar AGNs. 
Note that the number of entries in the distributions
from the left panel is less than that in the distributions in
the right panel. This is because
the 1FHL catalog
contains 17 AGN associations (9 BL Lacs and 8 blazar candidates) that do not exist in the 2FGL catalog
(\S~\ref{non2FGL}). 
The figure shows a clear spectral softening 
for each source class when the minimum energy is increased from 100~MeV 
to 10~GeV. This is due both to intrinsic softening of the
spectra of many sources\footnote{The intrinsic softening can occur
  because of internal $\gamma$-$\gamma$ absorption, which is energy
  dependent, or because of a steep decrease with energy of the number of
  high-energy particles (presumedly electrons/positrons) that are
  responsible for the high-energy $\gamma$ rays.} 
  and to $\gamma$-ray
attenuation in the optical/UV EBL for distant ($z>0.5$)
sources. We also note that in both panels the photon indices of
the FSRQs cluster at the largest index
values, while BL Lacs have the smallest index values. 
So even when the spectra are characterized using photons above 10 GeV, we
find that about 30\% of the BL Lacs (77  out of 259) have indices harder than 2.
The index distribution of the blazar candidates (``agu'' sources)
is similar to that of BL Lacs, which suggests that a large
fraction of these blazar candidates are actually BL Lacs.

\begin{figure}[th]
\begin{center}
\includegraphics[width=7cm]{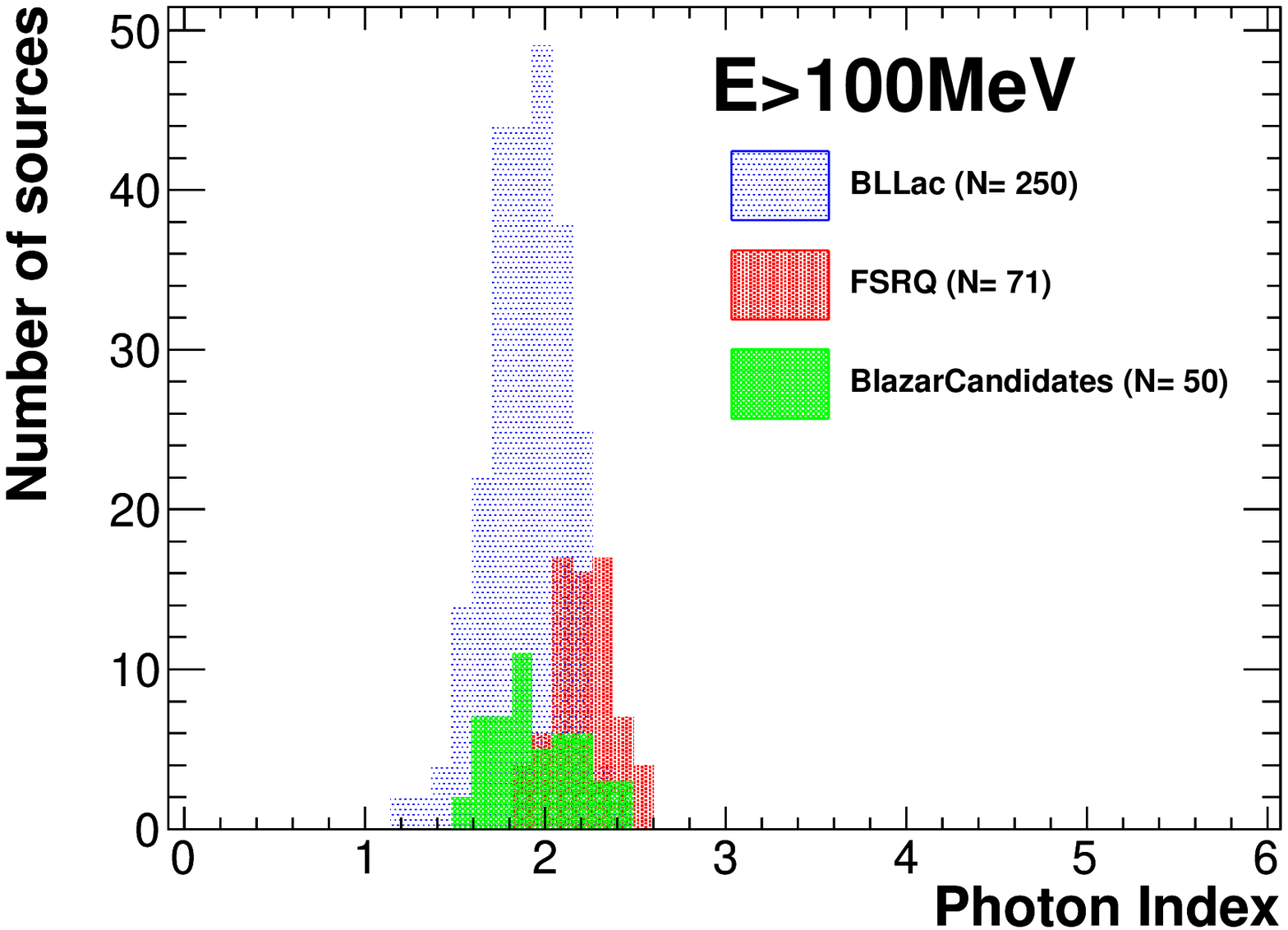}
\includegraphics[width=7cm]{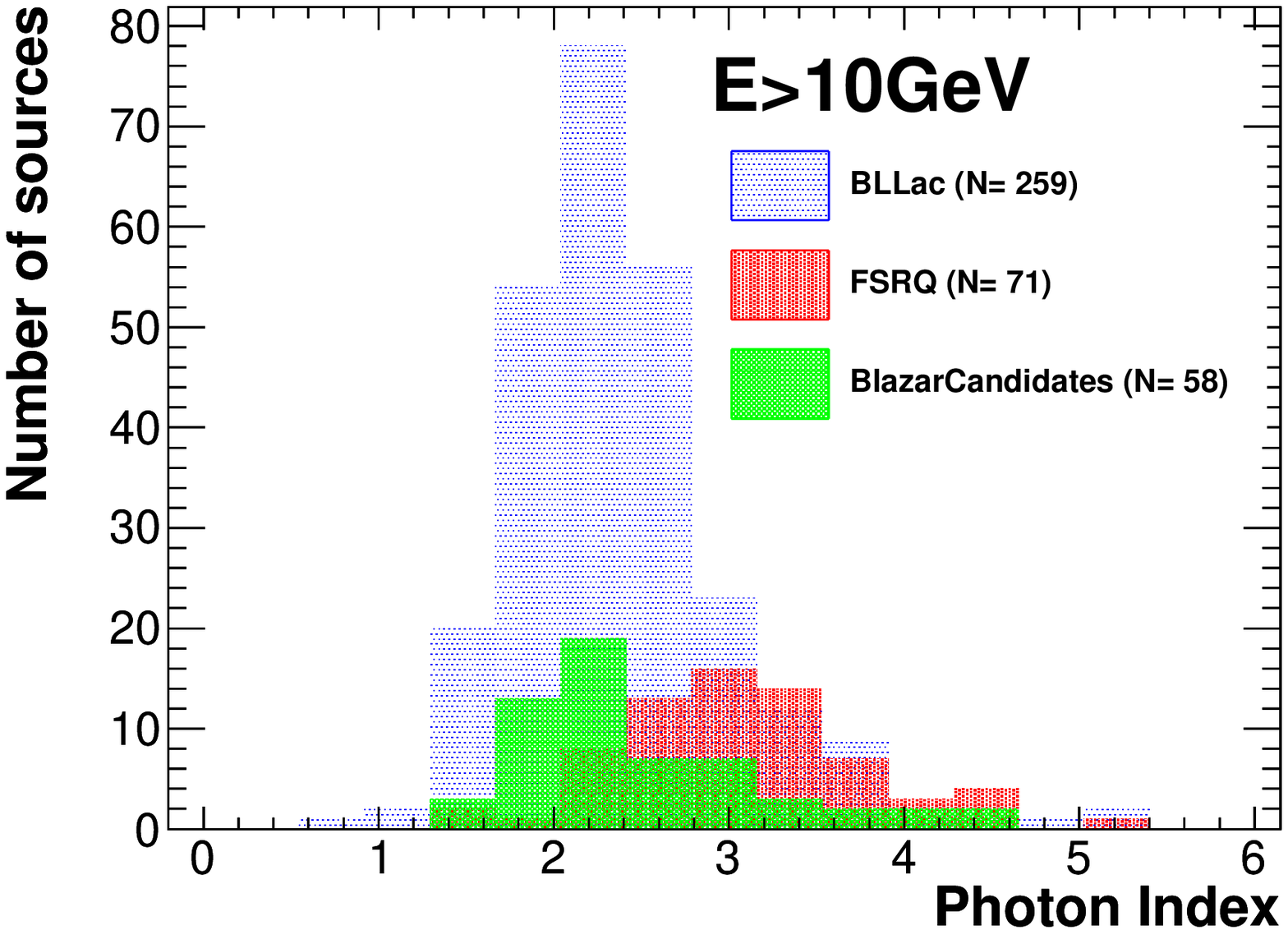}
\end{center}
\caption{\label{DistributionOfIndexAt100MeVAnd10GeV} 
Distribution of measured photon index for selected groups of 1FHL AGN sources above 100 MeV
({\it left}, extracted from the 2FGL catalog) and above 10 GeV 
({\it right}, extracted from Table~\ref{MainCatalogTable}). The three
histograms show the distributions for three different groups of
AGN associations: BL Lacs, FSRQs, and AGU or blazar candidates. See text for further details.}
\end{figure}

Figure \ref{DistributionOfRedshiftForAGN1FHLObjects} shows the
redshift distribution for the BL Lacs and FSRQs from the 1FHL catalog. 
For simplicity, we did not include in this plot the redshift distribution of
the five 
radio galaxies, which cluster at low redshifts. Neither did we include the
redshift distribution for the nine blazar candidates (``agu''), which
span z=0--1. 
We note that most of the BL Lacs have
redshifts less than 0.5, while most of the FSRQs have redshifts
greater than 0.5. The lack of BL Lacs at high redshift could be due to
the different characteristics of BL Lacs relative to FSRQs, which are known
to have a stronger intrinsic $\gamma$-ray brightness \citep{LATSEDs_BrightBlazars2010}. However, we also note
that the observed redshift distribution of BL
Lacs has an important bias due to the
difficulty of measuring their redshifts.
About half the BL Lacs associated with 1FHL sources do not have known redshifts, while all 
of the FSRQ associations have measured redshifts.

The right-hand panel in
Figure~\ref{DistributionOfRedshiftForAGN1FHLObjects} shows the redshift
distribution for BL Lacs split into the various SED
classifications, namely HSP, ISP, and LSP. The figure also depicts the
redshift distribution for the seven
sources without SED classifications. The figure shows clearly that
the distribution of HSPs (those with the highest
synchrotron peak frequency) peaks at the lowest redshifts. 
We note that the above-mentioned trends, as well as 
the overall shape of the redshift distributions for BL Lacs and FSRQs
and for the different subclasses of BL Lacs, are very similar to
those shown in Figure~12 of the 2LAC paper
\citep{LAT_2LAC}, hence indicating that selecting sources emitting
above 10 GeV does not introduce any bias/distortion
in the redshift properties of the sample of blazars detected by
\FermiLAT. 

\begin{figure}[th]
\begin{center}
\includegraphics[width=7cm]{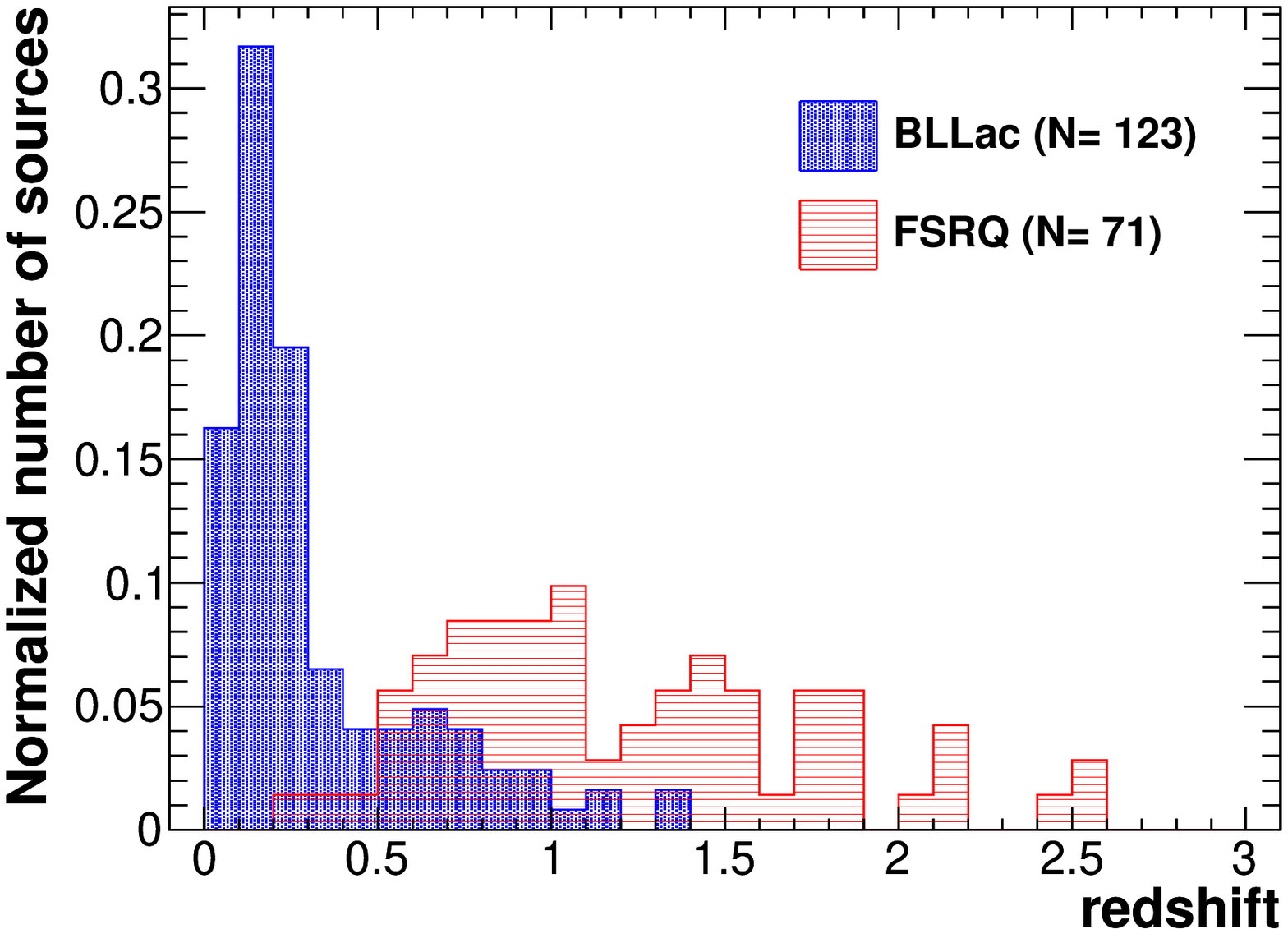}
\includegraphics[width=7cm]{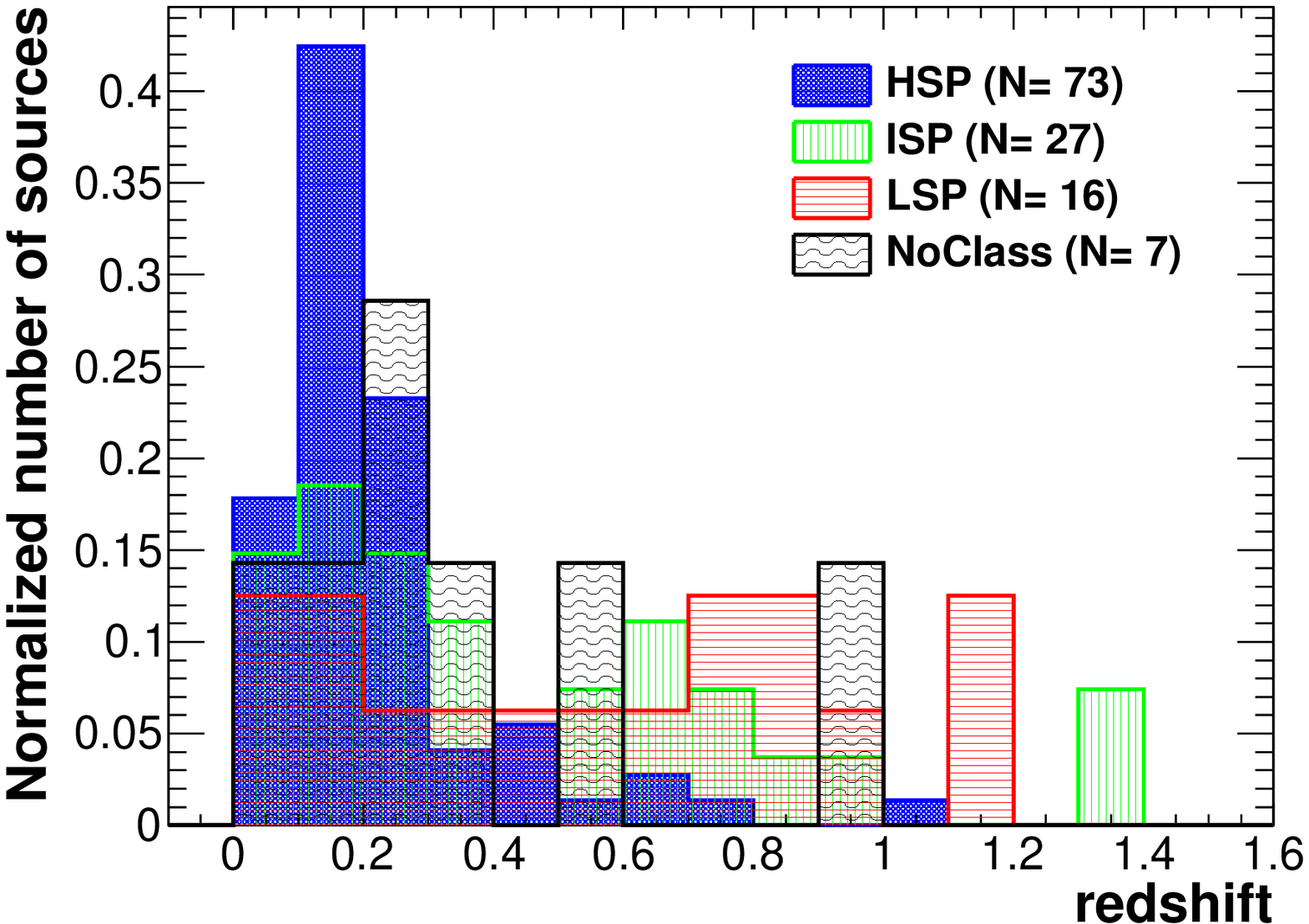}
\end{center}
\caption{\label{DistributionOfRedshiftForAGN1FHLObjects} 
Distribution of redshifts for selected groups of 1FHL AGN sources. 
The {\it left} panel shows the normalized redshift distributions for BL Lacs
(blue dotted-filled, 123 sources) and FSRQs (red horizontal-line-filled, 71 sources). The {\it right} panel
shows the normalized distributions for BL Lacs classified as
HSP (blue dotted-filled, 73 sources), ISP (green vertical-line-filled,
27 sources), LSP (red horizontal-line-filled, 16 sources) and sources
without SED classification (black wavy-line-filled, 7 sources). 
}
\end{figure}

Figure \ref{IndexVsRedshift} shows a scatter plot of the photon index
($E>100$~MeV and $E>10$~GeV) versus the redshift for the various blazar subclasses: FSRQs, HSP-BL Lacs, 
ISP-BL Lacs, LSP-BL Lacs and BL Lacs without SED classification. 
There is no redshift evolution in the spectral shape characterized
with photon energies above 100 MeV, which is in agreement with the results reported in
Figure 19 of the 2LAC paper\footnote{The data used to produced the left
  panel from Figure~\ref{IndexVsRedshift} are the same as used in
  the 2LAC, differing only in the selection of the blazar
sample: only 194 1FHL blazars (FSRQs+BL Lacs) are being used here.}. 
However, the photon index computed
with energies above 10 GeV has a redshift dependence: sources get softer with increasing
redshift. This trend is not apparent in the BL Lac sample,
which clusters at relatively low redshifts (mostly below 0.5);
but it is noticeable in the sample of FSRQs, which extends up to redshift 2.5.
A potential reason for this evolution  of the $>$10 GeV spectral shape
(but not for the $>$100 MeV spectral shape) is the attenuation of the $\gamma$ rays on optical/UV
photons of the EBL, which is energy dependent and 
affects photons only above a few tens of GeV. 
A cosmological evolution of
the FSRQ sample that introduces an intrinsic softening of the spectra
may also play a role. However, for consistency with the
experimental observations reported in Figure~\ref{IndexVsRedshift}, such
a  cosmological evolution of FSRQs should affect only
the emission above 10 GeV.

\begin{figure}[th]
\begin{center}
\includegraphics[width=7cm]{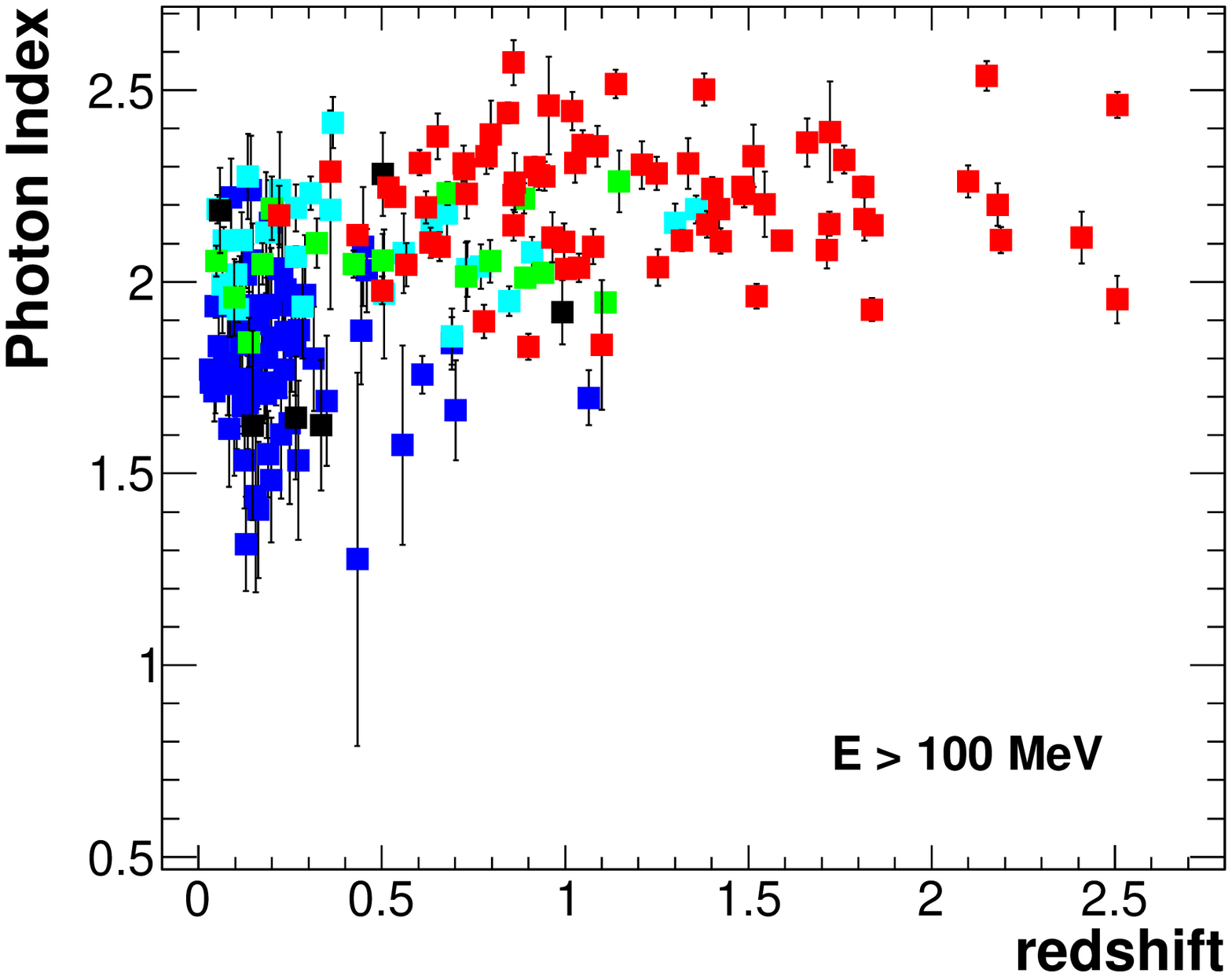}
\includegraphics[width=7cm]{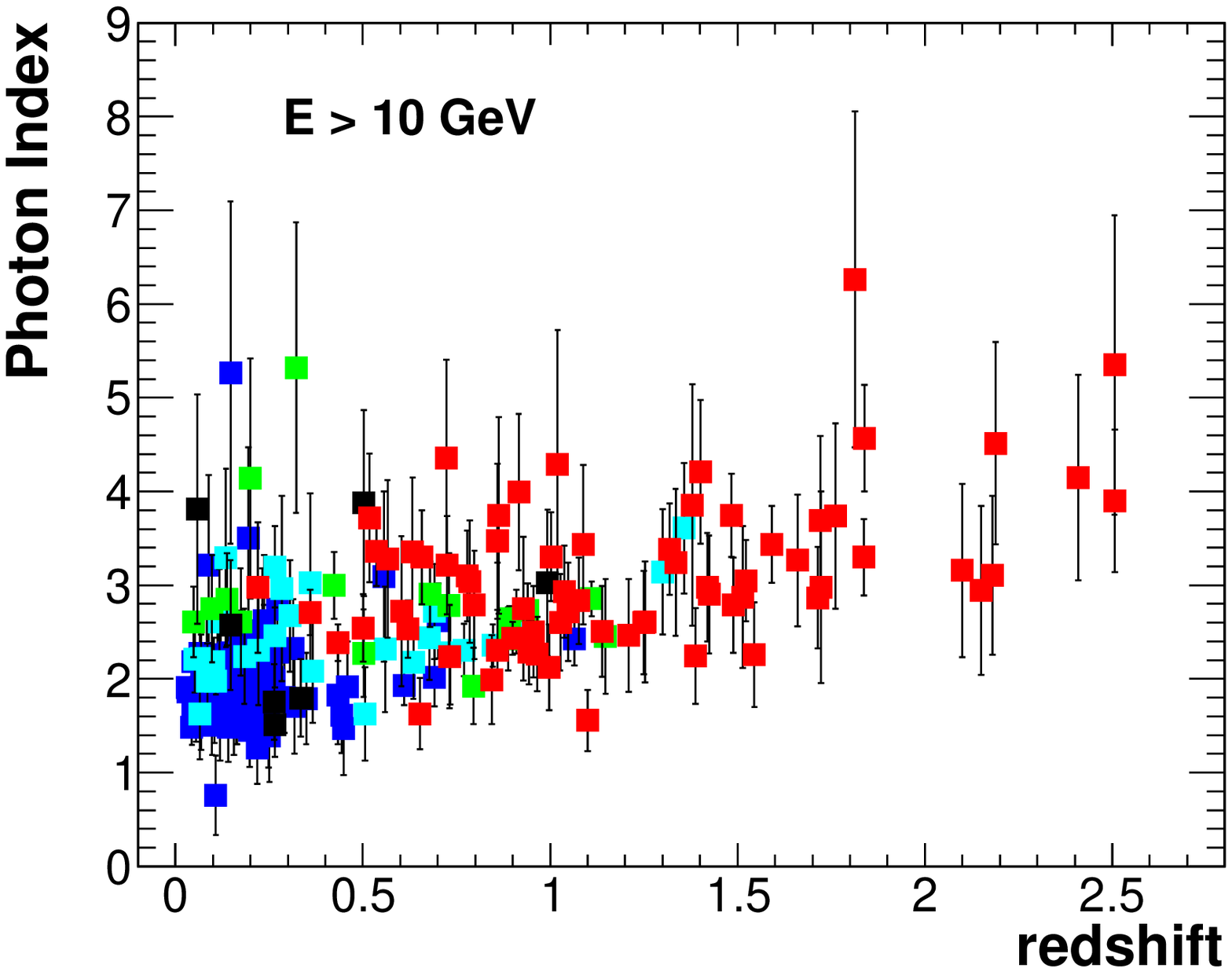}
\end{center}
\caption{\label{IndexVsRedshift} 
Power-law index {  from the observed source spectra} versus redshift for the 1FHL sources with available
redshifts. The {\it left} panel shows the power-law index describing the
spectral shape above 100 MeV (extracted from the 2FGL catalog) and the
{\it right} panel shows the power-law index describing the spectral
shape above 10 GeV (\emph{this work}). In both panels, red indicates
FSRQs (71 sources), dark-blue for HSP-BL Lacs (73), 
light-blue for ISP-BL Lacs (27),  green for LSP-BL Lacs (16), and black for BL Lacs with unclassified SEDs (7).}
\end{figure}

The last column of Table~\ref{TableCatalogAGNFirstNEntries} quantifies
the variability of the 1FHL sources, determined as described in
\S~\ref{Variability}. Among the 43 1FHL AGN sources identified as variable we find 22 LSPs
(22\% of the 99 1FHL LSP associations), 7 ISPs (13\% of the 61 1FHL ISP
associations), 6 HSPs (4\% of the 162 1FHL HSP sources), and 8 sources with
no SED classifications (11\% of the 71 1FHL sources with unclassified SEDs).
One of the outstanding characteristics is that 
most of the 1FHL sources identified as variable belong to the blazar subclass
LSP, not to the subclass HSP, which is the dominant blazar
subclass, and which has a larger number of high-energy photons. 
We stress that the three classic VHE blazars 
most variable
above a few hundred GeV, namely Mrk~421, Mrk~501 and
PKS~2155$-$304, are not found to be variable in the 1FHL catalog. This is
surprising, given that these three also have the largest numbers of detected photons
above 10 GeV:  432, 247, and 132, respectively (as evaluated from the likelihood analysis).
Moreover, the fraction of 1FHL LSPs identified as variable
($\sim$22\%) is substantially
higher than the fraction of 1FHL HSPs identified as variable ($\sim$4\%). 
This trend was already observed in the 2FGL blazars at energies
above 100 MeV, and reported in the 2LAC paper (e.g., see Figs. 26 and 27 of that work). 
Therefore, we can confirm that, across the entire energy range of the
LAT, the LSPs are more variable than the
HSPs.  These experimental observations show that the variability in the falling segment of the high-energy (inverse
Compton) SED bump is greater than that in the rising segment of the SED bump.

\subsection{Pulsars Above 10 GeV}
\label{Pulsars10GeV}

Pulsars are the second-largest class of associated sources in the 1FHL catalog. The detection with IACTs of pulsations from the
Crab, first at $>$25~GeV \citep{Crab_MAGIC}, and more recently at VHE \citep{Crab_VERITAS,Crab_MAGIC_2012} makes
the study of high-energy ($>$10~GeV) emission from $\gamma$-ray
pulsars with the LAT especially timely. A similar study conducted on EGRET data above 10 GeV revealed 37 events coincident with 5 $\gamma$-ray 
pulsars~\citep{Thompson2005}. 

The second \FermiLAT catalog of $\gamma$-ray pulsars \citep[][hereafter referred to as 2PC]{2PC},
includes results for 117 $\gamma$-ray pulsars detected in 3 years of LAT data. 
In this section we focus on pulsar emission above 10~GeV.

Pulsars are naturally associated with SNRs and PWNs, both of
which also can be bright VHE emitters. In addition to knowing how many pulsars are {\it associated} with 1FHL
sources, we would also like to determine which of these pulsars can be
{\it identified} with the 1FHL sources, by showing pulsations above
10~GeV (HPSR).

{  The pulsation analysis described here relies on the 2PC pulsar timing models\footnote{Available at
  \url{http://fermi.gsfc.nasa.gov/ssc/data/access/lat/ephems/}}. In
addition to studying 1FHL sources associated with pulsars, the analysis was extended to include a number of 2PC pulsars that are
candidate HPSRs, despite having no associated 1FHL source. Out of the 27 pulsars associated
with 1FHL sources (listed in Table~\ref{pulsar_table}), two (PSRs
J1536$-$4948 and J2339$-$0533) are not included in 2PC and are therefore left out of
this analysis. }

In order to test for high-energy pulsations we used a likelihood ratio
test, comparing the distribution in pulsar phase of the high-energy events
with the low-energy pulse profile. We considered the standard H-test~\citep{deJager89}
but found it to be less sensitive. This is not too surprising,
given that the H-test involves no assumptions about the pulse
profile\footnote{Its usage is generally recommended in cases such as
  the standard LAT searches for $\gamma$-ray pulsars, for which there is no {\it a priori} knowledge about the
shape of the $\gamma$-ray light curve.} while the likelihood ratio
test benefits from the available information on the low-energy pulse
profile, even if this may not necessarily be exactly the same as the
high-energy profile. We used high-energy ($>10$ GeV) photons within an RoI of $0\fdg6$ radius for
front-converting ({\it Front}) events and $1\fdg2$ for back-converting
({\it Back}) events, roughly corresponding to the 95\% containment
angles of the reconstructed incoming photon direction for normal
incidence above 10~GeV. For the low-energy profile, we assumed the
probability distribution function (PDF), with
phase $\phi$, obtained in the 2PC using the weighted LAT photons
above 100 MeV (where the {\it weight} of each photon corresponds to the
probability that it comes from the pulsar; see 2PC for details):
\begin{equation}
PDF_{LE}\left(\phi\right)=d+\sum_{i=1}^{n}c_{i}\cdot
f_{i}\left(\phi\right), 
\end{equation}
a combination of $n$ skewed Gaussian and Lorentzian distributions
$f_{i}$. The overall normalization of the PDF is defined such that
$d+\sum_{i=1}^{n}c_{i}=1$, where $d$ represents the unpulsed (or
``DC'') component of the pulsar. For the high-energy PDF, we considered the family of distributions given by:
\begin{equation}
PDF_{HE}\left(\phi\right)=\left(1-x\right)+x\cdot\frac{PDF_{LE}\left(\phi\right)-d}{1-d}
\end{equation}
with $0\leq x\leq1$. We maximized a likelihood function derived from
$PDF_{HE}$ with respect to $x$, to give $\mathcal{L}(\hat{x})$, and
compared it to the null hypothesis, for $x=0$, that there is no pulsation,
i.e. $PDF_{HE}\left(\phi\right)=1$. By construction, the likelihood
under the null hypothesis is $\mathcal{L}\left(0\right)=1$, so the
test statistic, defined as
$TS=-2\ln(\mathcal{L}(0)/\mathcal{L}(\hat{x}))$, can be simplified to
$TS=2\ln\mathcal{L}\left(\hat{x}\right)$. 

We converted the measured value of $TS$ into a tail probability (or
p-value, P) by assuming (by virtue of Wilks' theorem) that the TS
follows a $\chi^{2}$ distribution with 1 degree of freedom.
Since we are only testing for a positive correlation (one-sided test) between the
low and high energy pulse profiles (whereas a negative correlation is
equally likely in the null hypothesis), we divide the (two-sided) p-values by 2.
We set a threshold of P=0.05 to claim evidence
for pulsations, with P$_{10}$ representing the p-value obtained using $>$10 GeV events and P$_{25}$ corresponding to
the p-value obtained using $>$25 GeV events. Given that we are using
an asymptotic approximation to convert between the measured TS values
and the corresponding p-values, we report only p-values greater than
$2.0\times10^{-9}$ ($\sim6\sigma$); rather than provide unreliable
numbers in the tails of the distribution, we prefer to quote the rest
only as upper limits. 

We validated the procedure with Monte Carlo simulations. Given a high-energy profile with a certain number of events, {\it n}, we generated 
random sets of {\it n} phases uniformly distributed between 0 and 1. We then performed exactly the same test on these 
fake data sets and measured the rate of false positives. We repeated the simulations for every pulse profile and verified that the 
asymptotic distribution is valid in all cases with more than 2 events. 

In the case of J1836+5925, only two events are detected above 25 GeV. Although the asymptotic approximation
fails to reveal significant pulsations (P=5.5$\times10^{-2}$,
above {  the} significance threshold of 0.05), {  the} Monte Carlo simulations
demonstrate that the false positive rate is actually 1.0$\times10^{-2}$, so the
$>$25 GeV pulsations, in fact, pass {  the} threshold. 
Table~\ref{pulsar_table} summarizes the results of {  the pulsation} analysis.
Out of the 25 $\gamma$-ray pulsars associated with 1FHL sources for
which {  the pulsation analysis was performed}, 20 show evidence for pulsations
above 10 GeV (P$_{10}<$0.05) and 12 of these (listed in bold in
Table~\ref{pulsar_table}) show evidence for pulsations above 25
GeV (P$_{25}<$0.05).  Figure~\ref{pulsar_fig1} shows the
pulse profiles of these 20 pulsars, including the weighted low-energy ($>$100 MeV) pulse
profile, along with the folded $>$10 GeV and $>$25 GeV photons.

{  Five $\gamma$-ray pulsars associated with 1FHL sources show no
pulsations above 10 GeV: 
\begin{itemize}
\item PSR~J0205+6449,  associated with the SNR 3C~58, is thought to be one of the youngest pulsars in the Galaxy and is shown
in 2PC to have a GeV PWN. 
\item PSR~J1023$-$5746 is coincident with HESS J1023$-$575 and is identified as a promising GeV PWN candidate \citep[e.g.,][]{LAT10_blind8,LAT11_PWNcat}.
\item PSR~J1112$-$6103  is identified in 2PC as having significant extended off-peak emission.
\item PSR~J1418$-$6058 in the Kookaburra complex is coincident with the Rabbit PWN and thought to be powering the PWN candidate HESS~J1418$-$609.
\item PSR~J1420$-$6048, also in the Kookaburra complex, is in the vicinity of HESS J1420$-$607  and is a promising LAT PWN candidate (Acero et al., {\it submitted}).
\end{itemize}
In short, the $>$10~GeV emission from these five 1FHL sources is more likely to be
from PWNs than from the pulsars themselves.}

\begin{deluxetable}{lccrrrrrrrr}
\tabletypesize{\footnotesize}
\tablewidth{0pt}
\tablecaption{1FHL sources associated with $Fermi$-LAT pulsars \label{sources}}
\tablehead{
\colhead{1FHL} & \colhead{PSR} &
\colhead{P} & \colhead{$l$} &
\colhead{$b$} & \colhead{n$_{10}$}
& \colhead{P$_{10}$} &
\colhead{n$_{25}$} &
\colhead{P$_{25}$} & \colhead{Ref.} \\
 &  & [ms] & [deg] & [deg] &  &  &  &  & &
}
\startdata
J0007.3+7303 & {\bf J0007+7303}$^{\#}$ & 316 &119.7 & +10.5 & 179 & $<2\times10^{-9}$ & 20 & 1.7$\times10^{-3}$ & [1, 2, 3] \\
J0205.7+6448 & J0205+6449 &  65.7 & 130.7 & +3.1 &  38 & $>0.05$ & 12   & $>0.05$ & [4]\\
J0534.5+2201 & {\bf J0534+2200}$^{\dag\#}$ & 33.6 & 184.6 & --5.8 & 674 &6.3$\times10^{-8}$ &191 & 2.4$\times10^{-2}$ & Crab [5, 6, 7] \\
J0614.0--3325 & {\bf J0614--3329} & 3.15 & 240.5  & --21.8 & 26 &$<2\times10^{-9}$ & 3 & 2.0$\times10^{-2}$ & [8]\\
J0633.9+1746 & {\bf J0633+1746}$^{\#}$ & 237 & 195.1  & +4.3 & 260 & $<2\times10^{-9}$ & 11 & 1.4$\times10^{-5}$ & Geminga [9]\\
J0835.3--4510 & {\bf J0835--4510}$^{\dag\#}$ & 89.4 & 263.6  & --2.8 & 1005 & $<2\times10^{-9}$ & 56 & $<2\times10^{-9}$ & Vela [10, 11]\\
J1022.6--5745 & J1023--5746 &  112 & 284.2 & --0.4 & 152 &$>0.05$&46  & $>0.05$ & [12]\\
J1028.4--5819 & {\bf J1028--5819}$^{\#}$ & 91.4 & 285.1  & --0.5 & 164 & $<2\times10^{-9}$ & 41 & 4.0$\times10^{-2}$ & [13]\\
J1048.4--5832 & {\bf J1048--5832} & 124 &287.4  & +0.6 & 85 & 9.7$\times10^{-6}$ & 22 & 2.1$\times10^{-2}$ & [14]\\
J1112.5--6105 & J1112--6103 &  65.0 &  291.2 & --0.5& 112 &$>0.05$ &28 & $>0.05$ & \\
J1231.2--1414 & J1231--1411 & 3.68 & 295.5 & +48.4 & 15 & 5.3$\times10^{-7}$ & 4 & $>0.05$ & [8]\\
J1413.4--6205 & {\bf J1413--6205} & 110 & 312.4 & --0.7 & 278 & 4.4$\times10^{-3}$ & 64 & 1.5$\times10^{-2}$ & [12]\\
J1418.6--6059 & J1418--6058 & 111 &  313.3 & +0.1 & 324 & $>0.05$&72& $>0.05$& [2] \\
J1420.1--6047 & J1420--6048 & 68.2 &  313.5 & +0.2 & 278 & $>0.05$&65&$>0.05$ & [15]  \\
J1514.3--4945 & J1514--4946 & 3.58 & 325.2 & +6.8 & 24 & 1.7$\times10^{-4}$ & 3 & $>0.05$ & [16]\\
J1536.4--4951 & J1536--4948 & 3.08 & 328.2 & +4.8 & \nodata &\nodata& \nodata     & \nodata& Not in 2PC    \\
J1620.7--4928 & J1620--4927 & 172 & 333.9 & +0.4 & 297 & 9.4$\times10^{-3}$ & 77 & $>0.05$ & [17]\\
J1709.7--4429 & J1709--4429$^{\#}$ & 103 & 343.1 & --2.7 & 272 & $<2\times10^{-9}$ & 25& $>0.05$ & [18]\\
J1809.8--2329 & {\bf J1809--2332} & 147 & 7.4  & --2.0 & 119 & $<2\times10^{-9}$ & 18  & 4.3$\times10^{-2}$ & [2] \\
J1836.4+5925 & {\bf J1836+5925} & 173 & 88.9 & +25.0 & 36 & 1.0$\times10^{-4}$ & 2 &1.0$\times10^{-2}$\tablenotemark{*} & [2, 19]\\
J1907.7+0600 & J1907+0602$^{\#}$ & 107 & 40.2 & --0.9 & 158 & 2.3$\times10^{-4}$ & 36 & $>0.05$ & [2, 20, 21] \\
J1953.3+3251 & J1952+3252 & 39.5 & 68.8 & +2.8 & 48 & 1.2$\times10^{-5}$ & 7 & $>0.05$ & [18]\\
J1958.6+2845 & J1958+2846 & 290 & 65.9 & --0.4 & 64 & 1.0$\times10^{-2}$ & 11 & $>0.05$ & [2] \\
J2021.0+3651 & {\bf J2021+3651}$^{\#}$ & 104 & 75.2 & +0.1 & 107&$<2\times10^{-9}$ &20& 7.6$\times10^{-3}$ & [21, 22, 23]\\
J2032.1+4125 & J2032+4127$^{\#}$ & 143 & 80.2 & +1.0 & 210 & 5.6$\times10^{-8}$ & 54 & $>0.05$ & [2, 24] \\
J2229.0+6114 & {\bf J2229+6114}$^{\#}$ & 51.6 &106.7 & +3.0 & 86 & $<2\times10^{-9}$ &14 & 6.1$\times10^{-3}$ & [14, 25]\\
J2339.8--0530 &  J2339--0533 & 2.88 & 81.1 & --62.4 & \nodata &\nodata & \nodata & \nodata & Not in 2PC  \\

\enddata
\tablecomments{1FHL source; associated pulsar (in {\bf
    bold} if seen at $>$25 GeV); a $^{\dag}$ ($^{\#}$) implies a LAT-detected (TeV-detected) PWN; P is the pulsar period, in
  milliseconds; Galactic longitude ($l$) and latitude ($b$) in degrees; n$_{10}$ (n$_{25}$) is the number of $>$10 (25) GeV photons (within a 95\% containment radius) and P$_{10}$ (P$_{25}$) the
  corresponding tail probability. We quote only p-values $<0.05$ and
  $>2\times10^{-9}$ ($\sim6\sigma$). (*) For PSR~J1836+5925, the two
  $>$25 GeV events result in a p-value=5.52$\times10^{-2}$ according
  to the asymptotic approximation, but Monte Carlo simulations show
  that the true p-value is 1.0$\times10^{-2}$, so the pulsations can
  be considered significant. {  References }[1]\,\cite{LAT08_CTA1},
  [2]\,\cite{LAT09_blind}, [3]\,\cite{CTA1_VERITAS}, [4]\,\cite{J0205}, [5]\,\cite{LAT10_Crab},
  [6]\,\cite{Crab_MAGIC}, [7]\,\cite{Crab_VERITAS}, [8]\,\cite{Ransom11_3MSP}, [9]\,\cite{LAT10_Geminga}, [10]\,\cite{LAT09_Vela},
  [11]\,\cite{LAT10_Vela}, [12]\,\cite{LAT10_blind8},
  [13]\,\cite{LAT09_PSR1028}, [14]\,\cite{LAT09_PSR1048},
  [15]\,\cite{Weltevrede2010}, [16]\,\cite{Kerr_MSPs}, [17]\,\cite{Pletsch_blind9},
  [18]\,\cite{3EGRETPSRs}, [19]\,\cite{J1836}, [20]\,\cite {J1907}, [21]\,\cite{MGRO_BSL},
  [22]\,\cite{Dragonfly_AGILE}, [23]\,\cite{Dragonfly_LAT}, [24]\,\cite{Camilo09}, [25]\,\cite{J2229_VERITAS}.}
\label{pulsar_table}
\end{deluxetable}

Because a pulsation search is more sensitive than a simple source
detection search, we extended the analysis to include pulsars from 2PC
whose spectra show possible emission above 10~GeV, even if they have
no associated 1FHL source. There are 14 additional pulsars in 2PC with at
least one spectral bin above 10~GeV detected with TS$\geq$4, a
$\sim2\sigma$ detection\footnote{See ~\cite{2PC} for further details, including plots, regarding the spectral analysis of 
these and other LAT pulsars.}.  
The results of the pulsation analysis for these pulsars are listed in
Table~\ref{pulsar_table2}. 
Eight out of the 14 pulsars selected in this way show evidence for
pulsations above 10~GeV,
 and one of them (J1954+2836) shows evidence for pulsations even above
 25~GeV. The pulse profiles of these eight pulsars are shown in Figure~\ref{pulsar_fig2}. 

\begin{deluxetable}{lccrrrrrrrr}
\tabletypesize{\footnotesize}
\tablewidth{0pt}
\tablecaption{$Fermi$-LAT $\gamma$-ray pulsars with hints of $>10$ GeV
  emission but no 1FHL association\label{sources}}
\tablehead{
\colhead{PSR} &\colhead{P} & \colhead{$l$} &\colhead{$b$} & \colhead{n$_{10}$}& \colhead{P$_{10}$} &\colhead{n$_{25}$} &\colhead{P$_{25}$} & \colhead{Ref.} \\
 & [ms] & [deg] & [deg] &  &  &  &  &
 }
\startdata
J0218+4232 & 2.32 & 139.5 & --17.5 & 79 &$>0.05$ & 23 & $>0.05$ & [1] \\
J0633+0632 & 297 &205.1 & --0.9  & 24 &1.3$\times10^{-2}$ & 5 & $>0.05$ & [2] \\
J1509--5850 & 88.9 & 320.0 & --0.6 & 187 & $>0.05$ & 52 & $>0.05$ & [3] \\
J1747-2958 & 98.8 & 359.3 & --0.8 & 272 & 3.8$\times10^{-2}$ & 64 & $>0.05$ & \\ 
J1803--2149$^{\#}$ & 106 & 8.1 & +0.2  & 270 & 1.4$\times10^{-2}$ &76 & $>0.05$ & [4] \\
J1826--1256 & 110 & 18.6 &--0.4 & 304 & $>0.05$ & 80 &$>0.05$  & [2] \\
J1838--0537 & 146 & 26.5 & +0.2 & 321 & $>0.05$ & 96 & $>0.05$ &  [5]\\
{\bf J1954+2836} & 92.7 &65.2 & +0.4 & 66 & 4.8$\times10^{-6}$& 12 &6.5$\times10^{-3}$ & [6, 7, 8]\\
J2017+0603 & 2.90 & 48.6 & --16.0 & 16 & 1.4$\times10^{-2}$ & 2 & $>0.05$ & [9]\\
J2021+4026 & 265 & 78.2 & +2.1 & 289 & $>0.05$ & 77 & $>0.05$ & [2]\\
J2043+1711 & 2.38 & 61.9 & --15.3 & 0 & $>0.05$ & 0 & $>0.05$ & [10]\\
J2111+4606 & 158 & 88.3 & --1.5 & 33 & 3.4$\times10^{-3}$& 11 & $>0.05$ & [4]\\
J2238+5903 & 163 & 106.6 & +0.5 & 51 & 4.0$\times10^{-2}$ & 14 &$>0.05$ & [2]\\
J2302+4442 & 5.20 & 103.4 & -14.0 & 19 & 5.1$\times10^{-4}$ & 2 & $>0.05$ & [9]\\

\enddata
\tablecomments{PSR is the name of $\gamma$-ray pulsar (in {\bf bold} when detected above 25 GeV). A $^{\dag}$ next to
  the name means a GeV PWN is detected by the LAT, while a $^{\#}$ means a TeV ($>$100 GeV) PWN is detected by ground-based
instruments (see {\tt http://tevcat.uchicago.edu}); P is the pulsar period, in
milliseconds; The Galactic longitude ($l$) and latitude ($b$), are given in
degrees. n$_{10}$ (n$_{25}$) is the number of photons (within the 95\% containment radius of the PSF) above 10 (25) GeV and P$_{10}$ (P$_{25}$) gives the
  corresponding tail probabilities, against a null hypothesis of no
  pulsations. We quote only p-values $<0.05$ and $>2\times10^{-9}$
  ($\sim6\sigma$). {  References }[1]\,\cite{LAT09_MSP}, [2]\,\cite{LAT09_blind}, [3]\,\cite{Weltevrede2010},
  [4]\,\cite{Pletsch_blind9},[5]\,\cite{J1838}, [6]\,\cite{LAT10_blind8},[7]\,\cite{Dragonfly_LAT},[8]\,\cite{J1954_MAGIC},[9]\,\cite{Cognard11},[10]\,\cite{Guillemot2012}.}
\label{pulsar_table2}
\end{deluxetable}

The effect of the spectral cutoff in pulsars is manifested by the
dramatic drop in photon statistics  from 10 GeV to 25 GeV (cf. columns 6 and
8 of Table~\ref{pulsar_table} and columns 5 and 7 of Table~\ref{pulsar_table2}). A change in pulse profile at higher
energies ($>$10 GeV), compared to low energies ($>$100 MeV), is also apparent, with the
widths of the peaks typically narrowing and the height of the first peak
decreasing in significance. These features of the high-energy profiles have
been reported for the brightest $\gamma$-ray pulsars like Vela, the
Crab, and Geminga \citep{LAT09_Vela,LAT10_Crab,LAT10_Geminga},
but we show here that they are present in other pulsars too, including
MSPs, like J0614$-$3329. In the case of the Crab, the LAT pulse profile
shown in Figure~\ref{pulsar_fig1} is heavily contaminated by the
emission from the PWN. An analysis beyond the scope of this paper
 would be required to disentangle the two spectral components
and provide a more sensitive analysis of the Crab pulsar in the $>$10 GeV energy
range. In the
case of Vela, another feature that is apparent in the high-energy profile (see Fig.~\ref{pulsar_fig1})
is an energy-dependent shift of the position of the third (``middle'') peak, which moves
toward the second peak with increasing energy, as reported
by \cite{LAT09_Vela}. This change in profile at higher energies
highlights a shortcoming of the analysis described here. The choice of the low-energy ($>100$ MeV) pulse
profile as a template for the pulation search in the high-energy events was based in part on
the assumption that the difference between the two profiles would be relatively modest. 

Beyond 25 GeV, the drop in statistics for $\gamma$-ray pulsars becomes
even more dramatic. Nevertheless, a number of pulsars in this study
still have evidence of pulsations above 25 GeV. By scanning in energy (in steps of 1 GeV) we
determined, for each pulsar, the energy beyond which the tail probability
increases above 5\%. Given the very small statistics, we relied on
Monte Carlo simulations to obtain the p-values and corresponding energy
thresholds. For the same reason, we caution against considering these as significant detections. 
Table~\ref{pulsar_table3} summarizes the results of this scan.

\begin{deluxetable}{lccccc}
\tabletypesize{\footnotesize}
\tablewidth{0pt}
\tablecaption{$Fermi$-LAT $\gamma$-ray pulsars detected above 25 GeV \label{sources}}
\tablehead{
\colhead{PSR} &
\colhead{E$_{\mathrm{max}}$} & 
\colhead{$E^{detected}_{max}$} &
\colhead{$\Phi_{\gamma_{\mathrm{max}}}$} & 
\colhead{Notes}
}
\startdata

J0007+7303$^{\#}$ & 28 & 788 &  0.64 &  \\
J0534+2200$^{\dag\#}$ & 26 & 784 & 0.33 & Crab \\
J0614--3329 & 63 & 63.6  & 0.68 &  \\
J0633+1746$^{\#}$ & 33 & 52.7  & 0.05 & Geminga \\
J0835--4510$^{\dag\#}$ & 37 & 752  & 0.28 & Vela \\
J1028--5819 & 27 & 386 & 0.49 & \\
J1048--5832 & 35 & 201 & 0.28 & \\
J1413--6205 & 29 & 331 & 0.28 &  \\
J1809--2332 & 26 & 159 & 0.07 &  \\
J1836+5925 & 26 & 97.9 & 0.05 & \\
J1954+2836 & 62 & 95.7 & 0.57 &  \\ 
J2021+3651$^{\#}$ & 26 & 113 & 0.64 &  \\
J2229+6114$^{\#}$ & 31 & 169 & 0.17 &  \\

\enddata
\tablecomments{PSR is the name of $\gamma$-ray pulsar; a $^{\dag}$ implies a GeV PWN is detected by the LAT, while a
  $^{\#}$ implies an associated TeV PWN, detected by ground-based
  instruments above 100 GeV (see {\tt http://tevcat.uchicago.edu}); E$_{\mathrm{max}}$ is the maximum energy (in GeV) above which  $P<$0.05 is still
  obtained while $E^{detected}_{max}$ is the highest-energy event detected (in GeV) and $\Phi_{\gamma_{\mathrm{max}}}$ is the corresponding pulsar phase of this event.}
\label{pulsar_table3}
\end{deluxetable}

The presence of a PWN will complicate studies of pulsations at the highest energies. 
The high-energy $\gamma$-ray emission from PWNs can be particularly significant relative to the pulsars they are associated with,
especially for some young, energetic pulsars (e.g., the Crab). An associated PWN
can thus represent a significant background, limiting the
sensitivity of a pulsation search. In the case of the Crab, the PWN is
particularly bright, both at GeV and TeV energies. 
Thus, although the Crab pulsar has been detected (in fact, is the only
pulsar detected) by IACTs above 100 GeV, the LAT results, shown in Table~\ref{pulsar_table}, are not as significant
as those for a number of other $>$25~GeV $\gamma$-ray pulsars. Indeed,
with the current analysis, we are unable to detect pulsations beyond
26~GeV for the Crab.  In Tables~\ref{pulsar_table} and
\ref{pulsar_table2} 
we flag those with a LAT-detected GeV PWN or a TeV ($>$100 GeV) PWN detected by IACTs.  The maximum energy and phase columns of
Table~\ref{pulsar_table3} suggest that we may be detecting
events from a number of PWNs. For example, in the case of J0007+7303, the 
highest-energy event is 788 GeV, arriving at phase 0.64,
far from the pulsar peaks, suggesting that a PWN origin is more
likely than a PSR origin. VERITAS recently reported the detection of such a PWN
above 100 GeV~\citep{CTA1_VERITAS}.

\begin{figure}[th]
        \centering
                \includegraphics[width=1.55in]{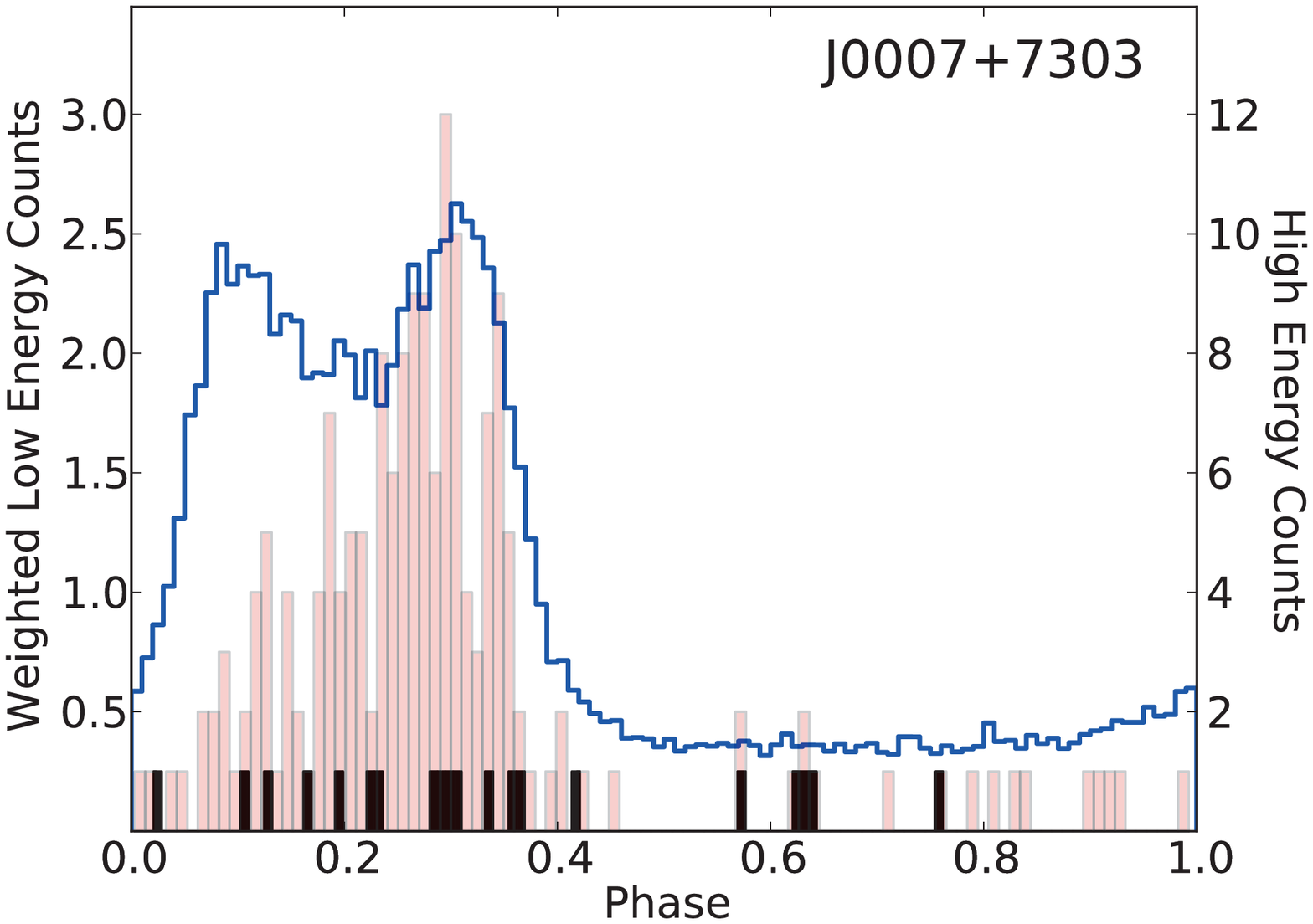}
                \includegraphics[width=1.55in]{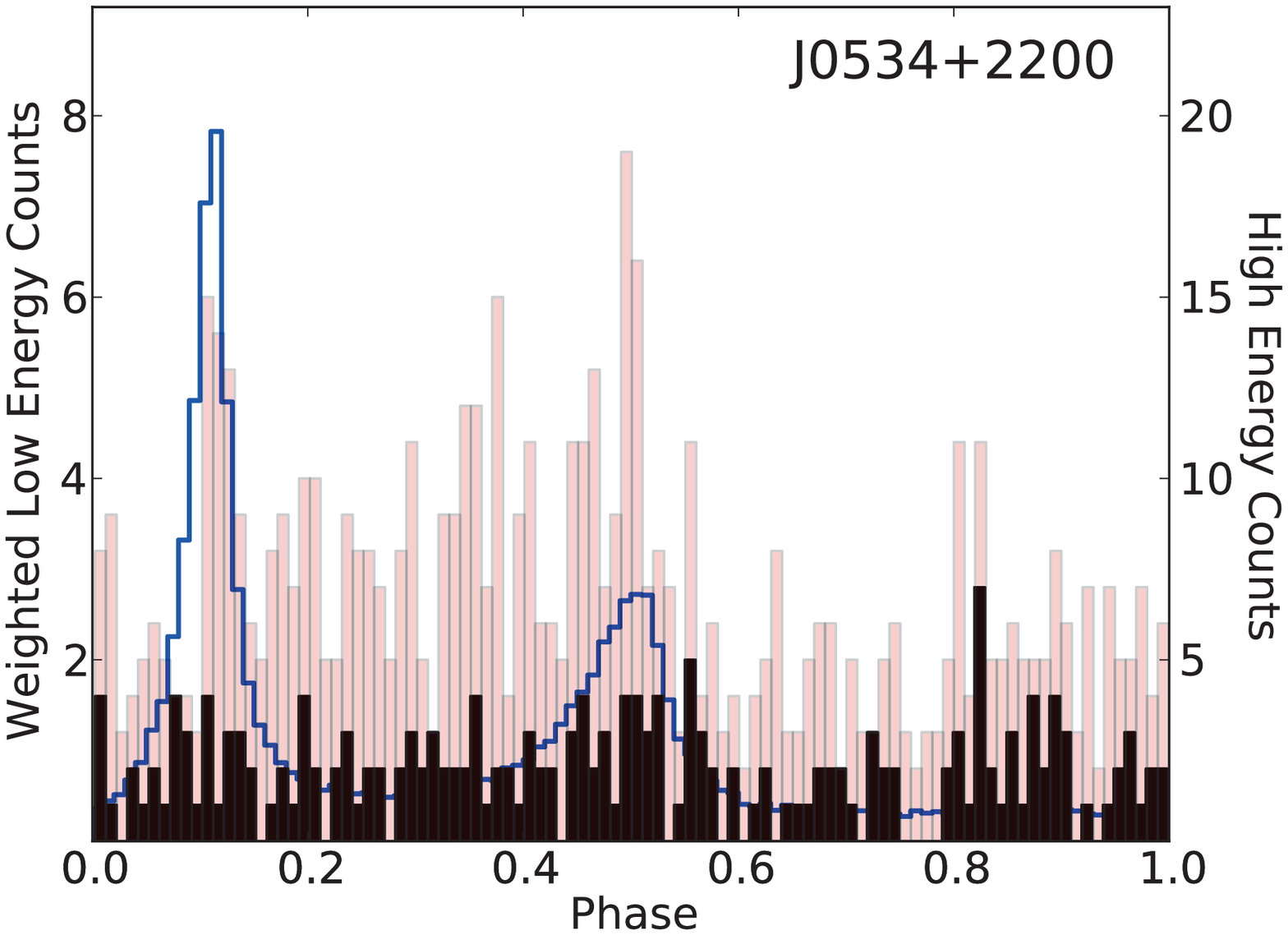}
                \includegraphics[width=1.55in]{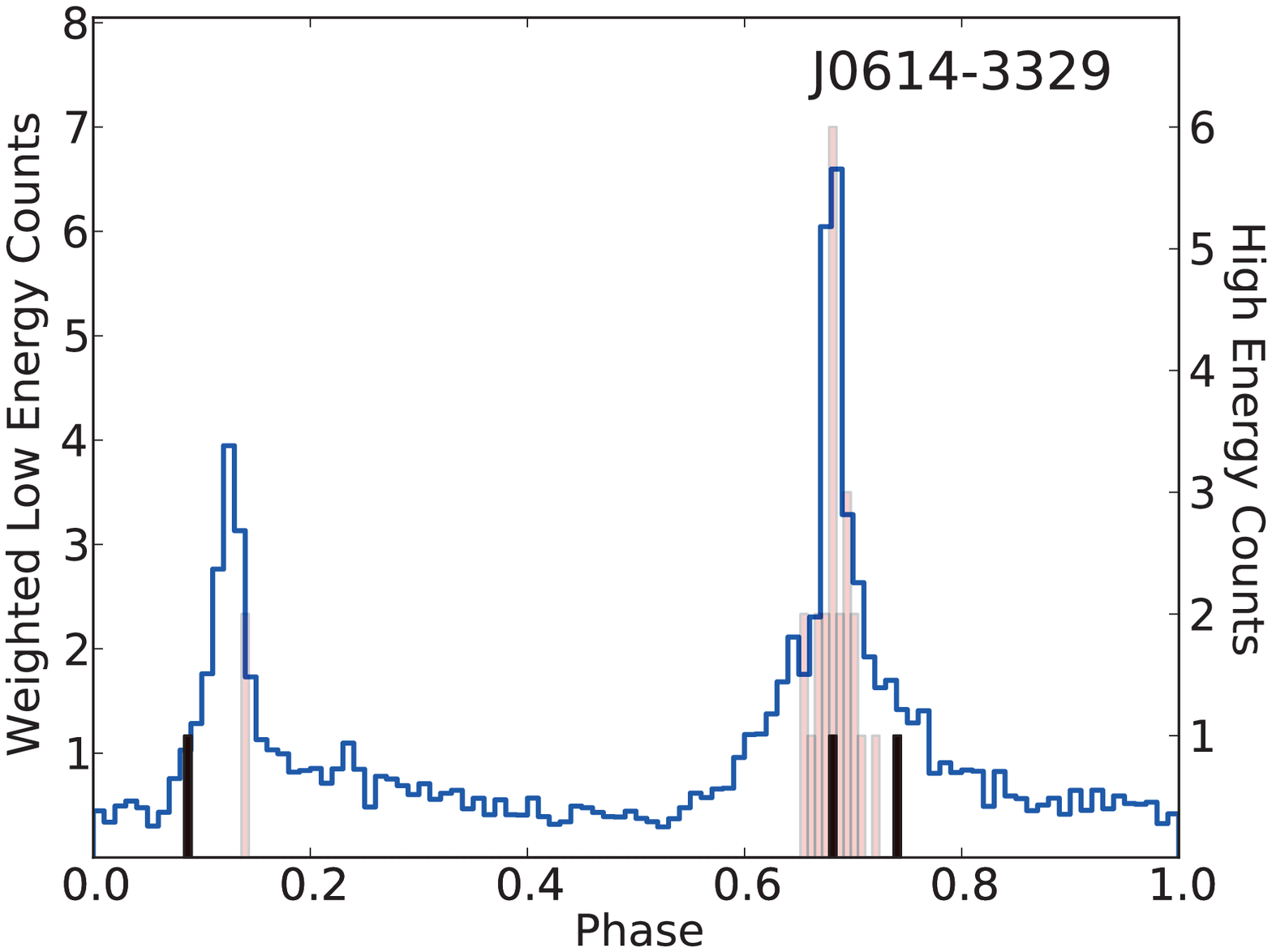}
                \includegraphics[width=1.55in]{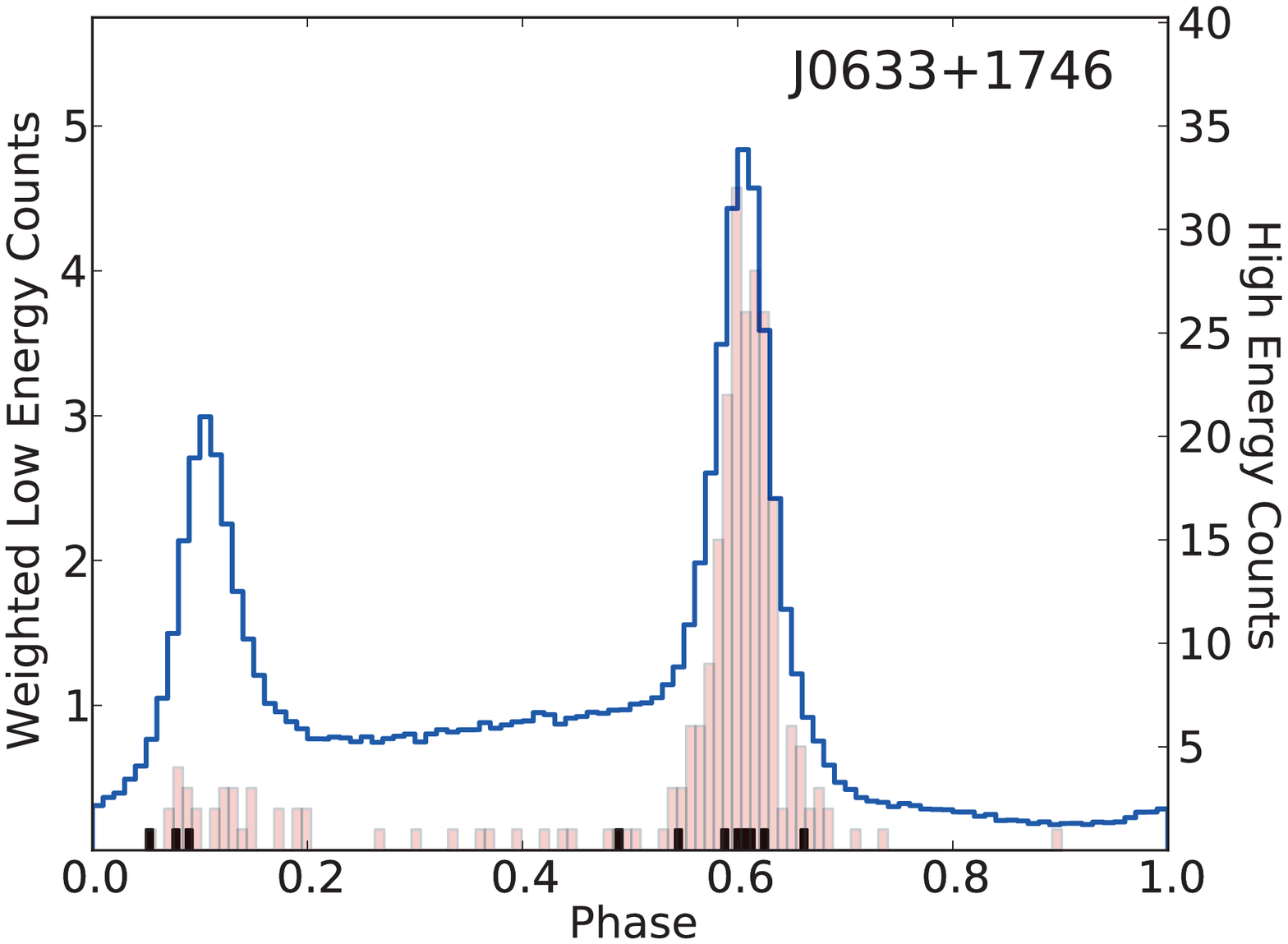}
                \includegraphics[width=1.55in]{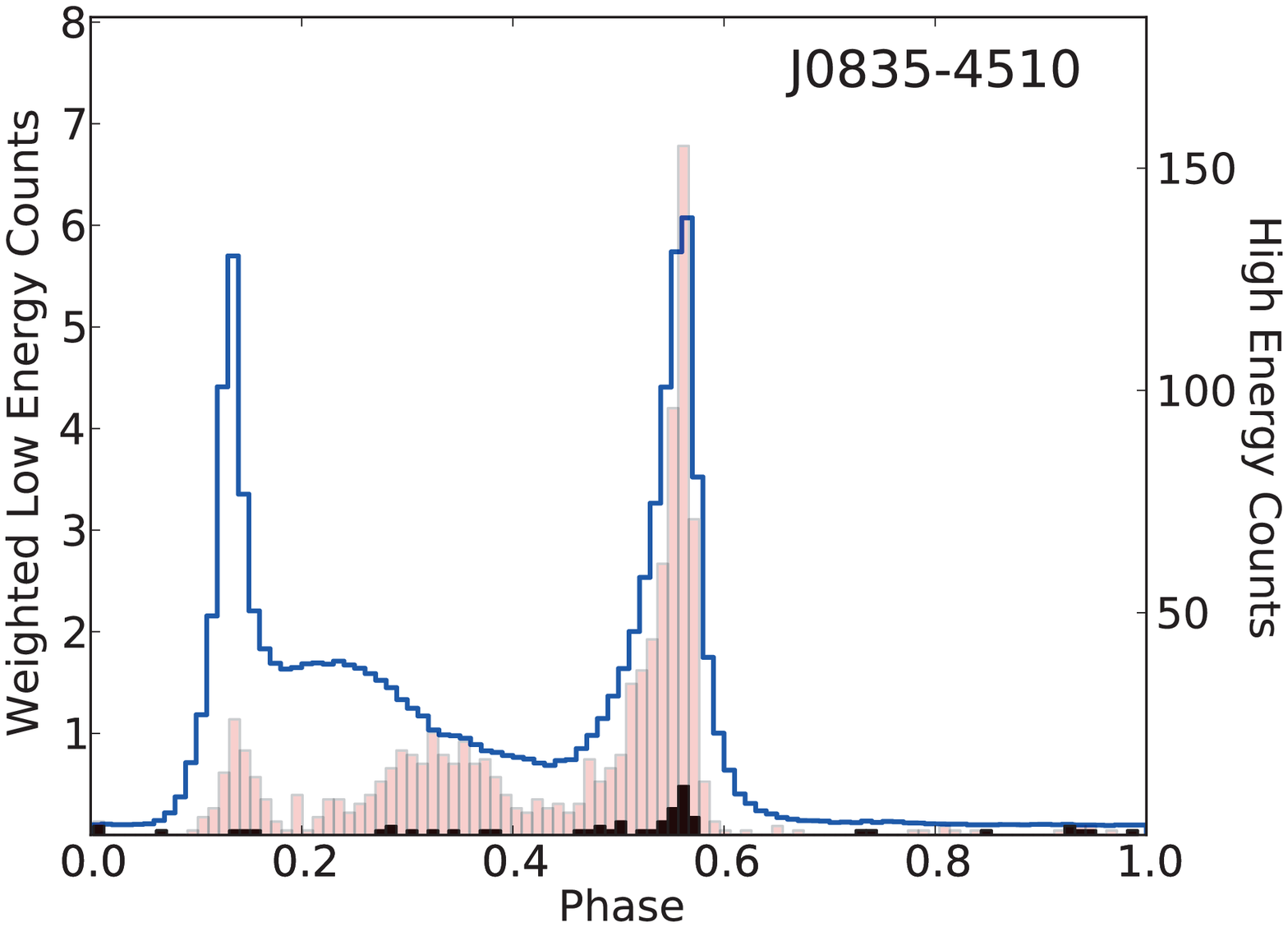}
                \includegraphics[width=1.55in]{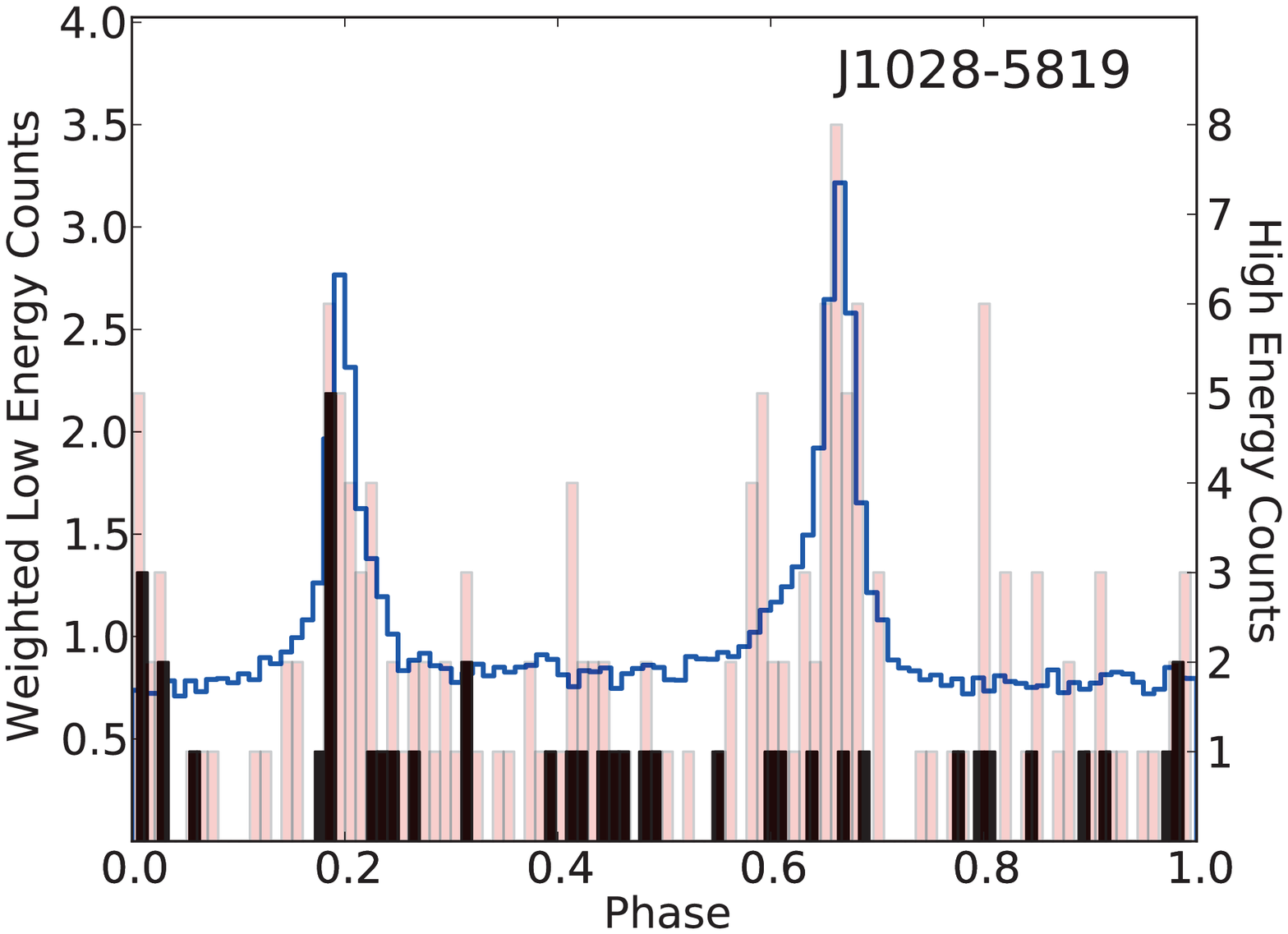}
                \includegraphics[width=1.55in]{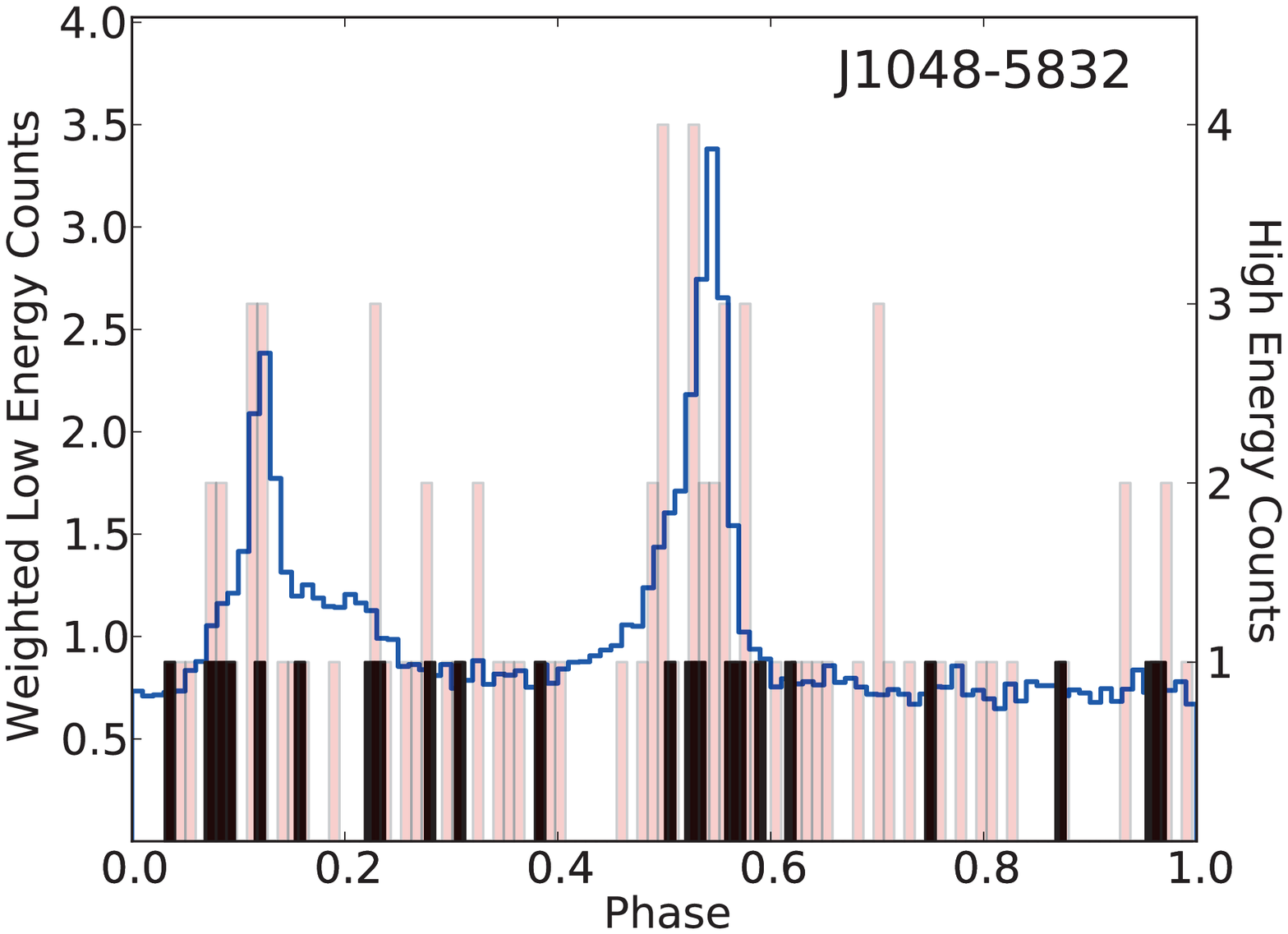}
                \includegraphics[width=1.55in]{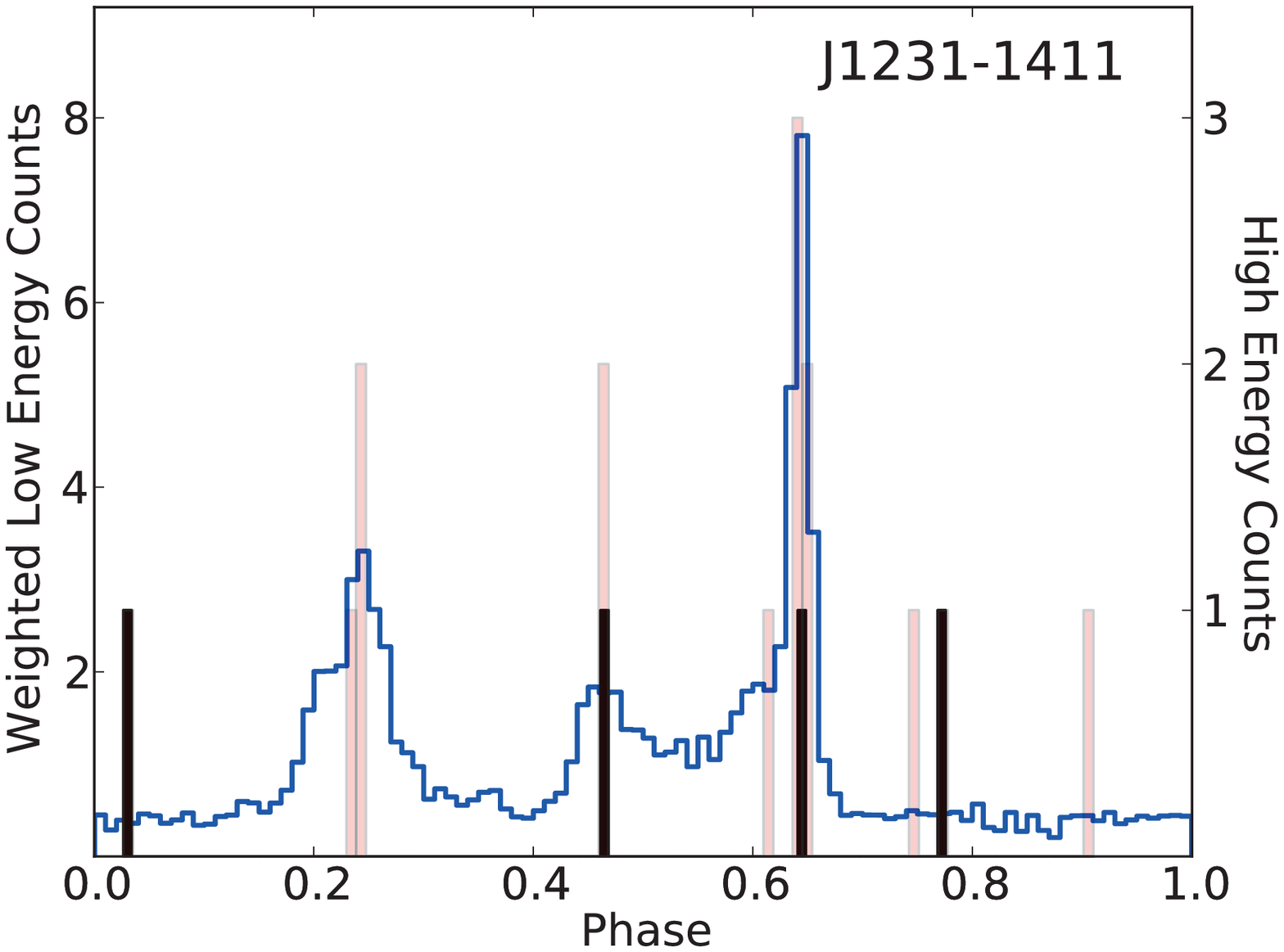}
                \includegraphics[width=1.55in]{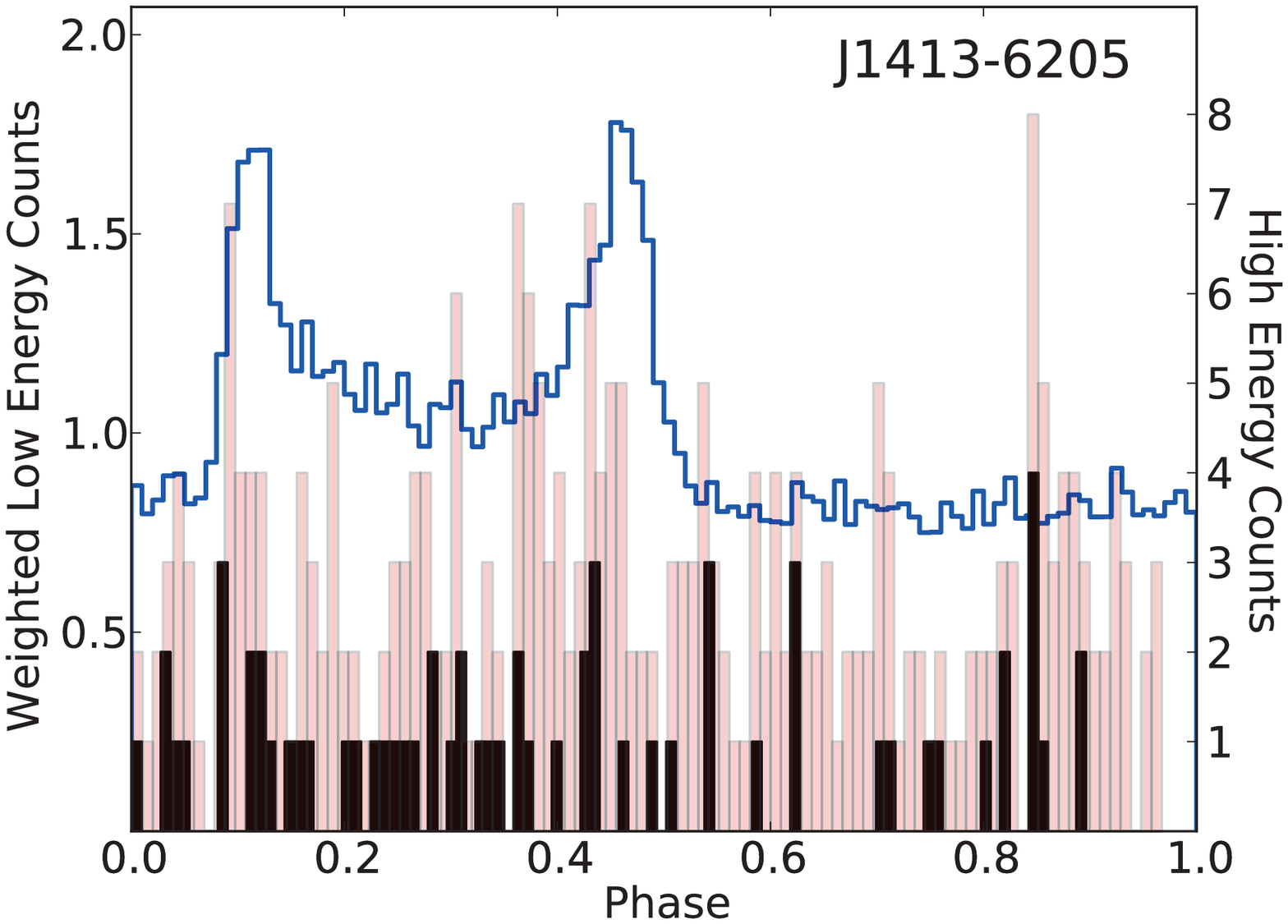}
                \includegraphics[width=1.55in]{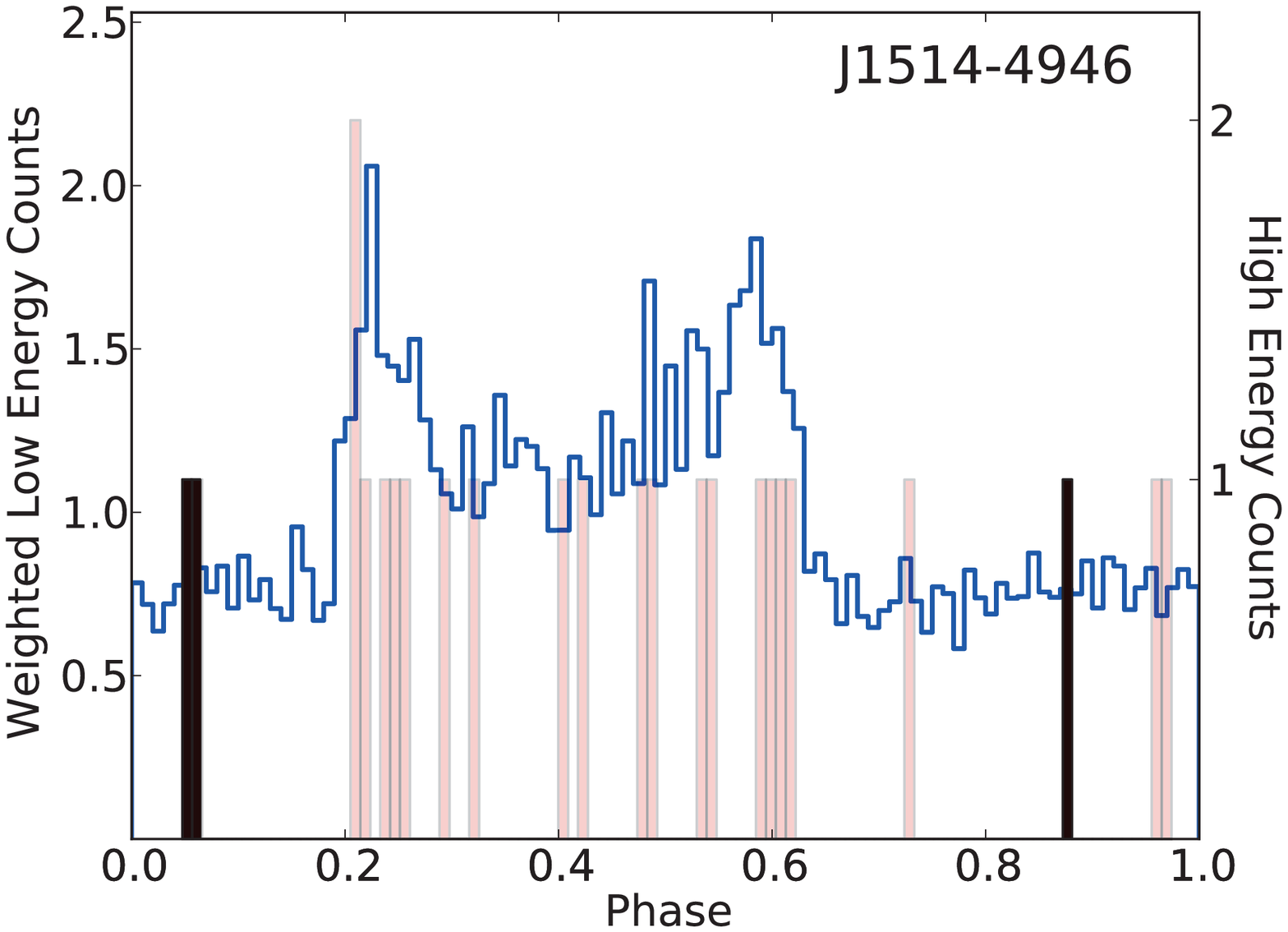}
                \includegraphics[width=1.55in]{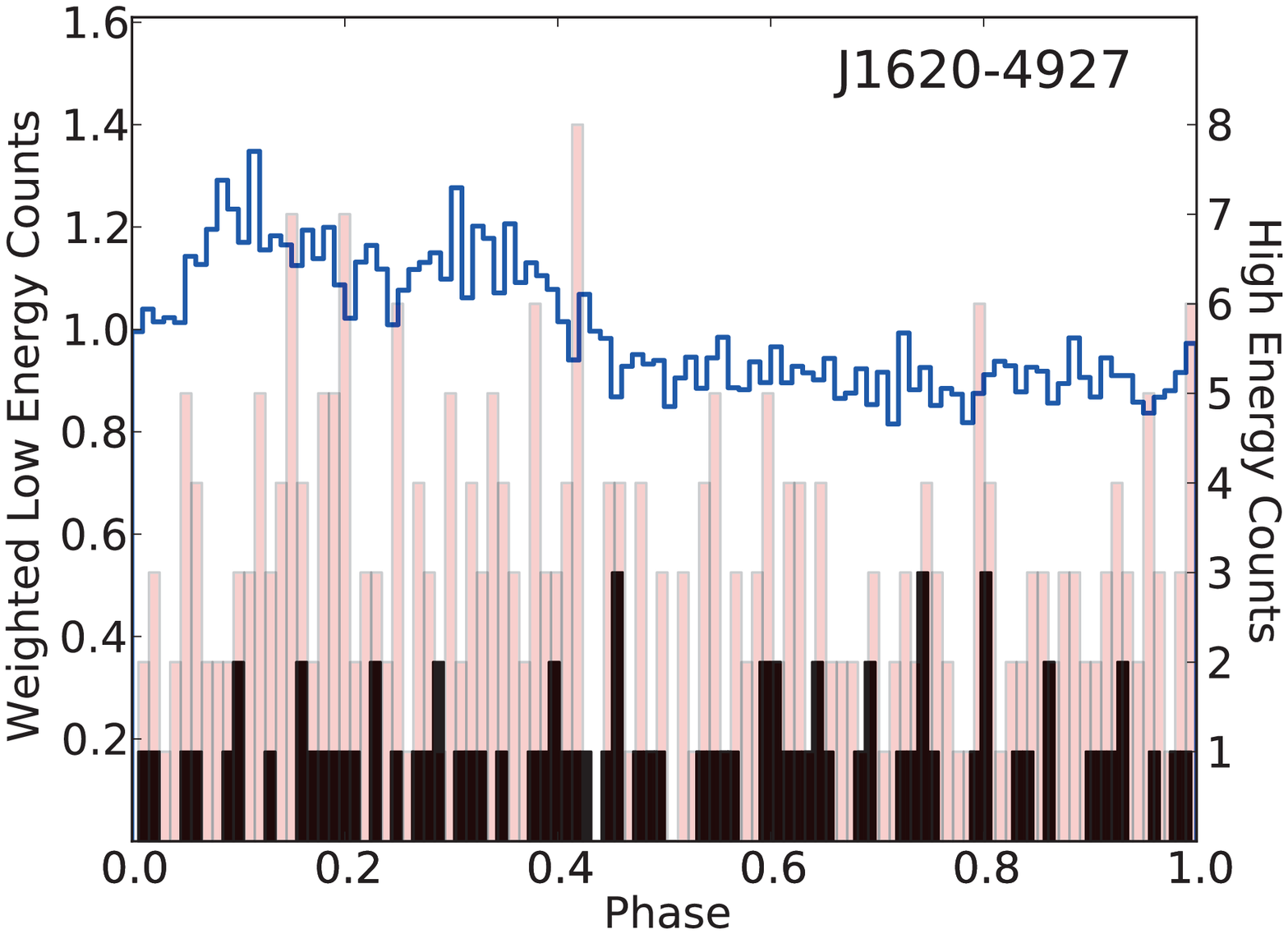}
                \includegraphics[width=1.55in]{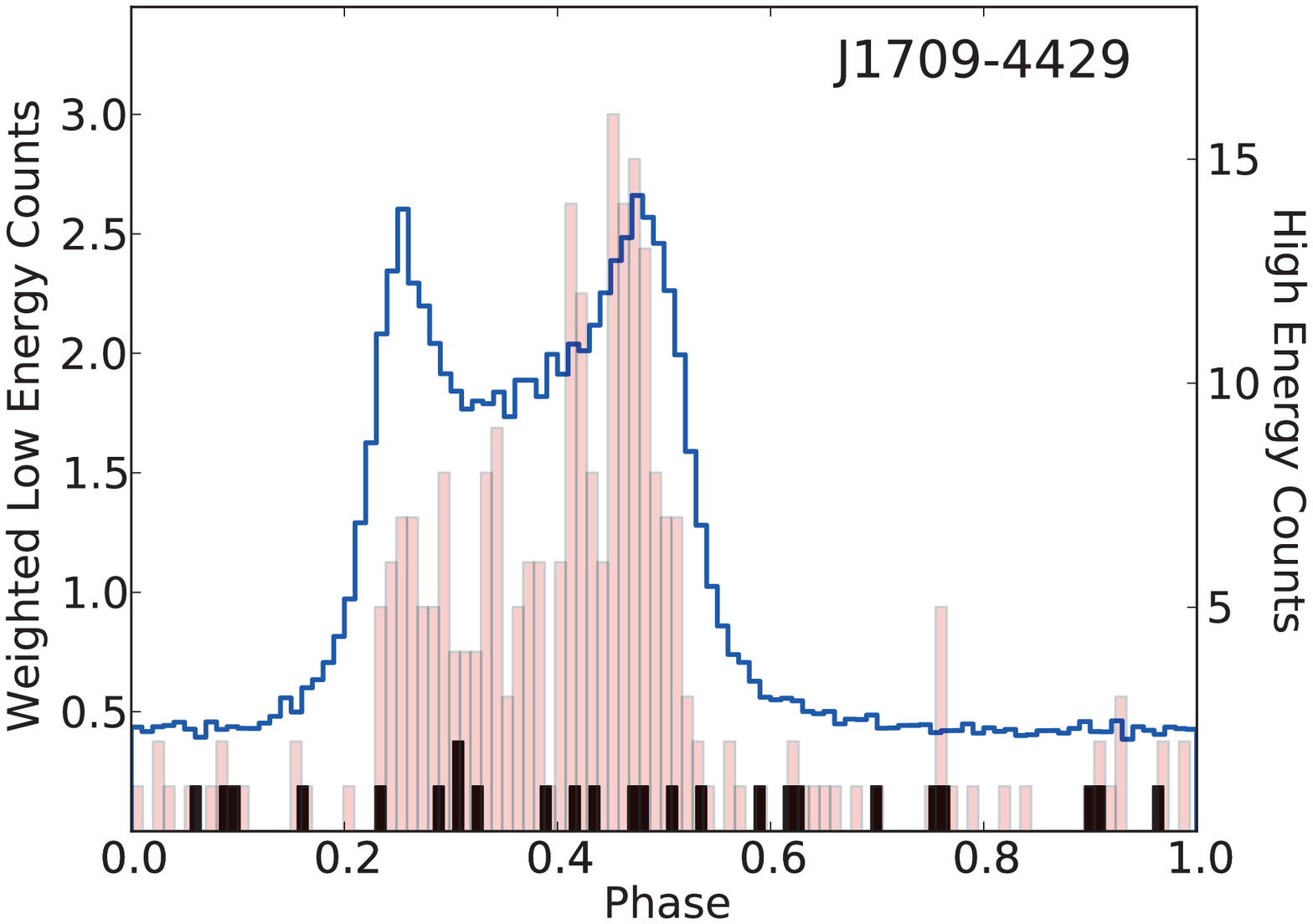}
                \includegraphics[width=1.55in]{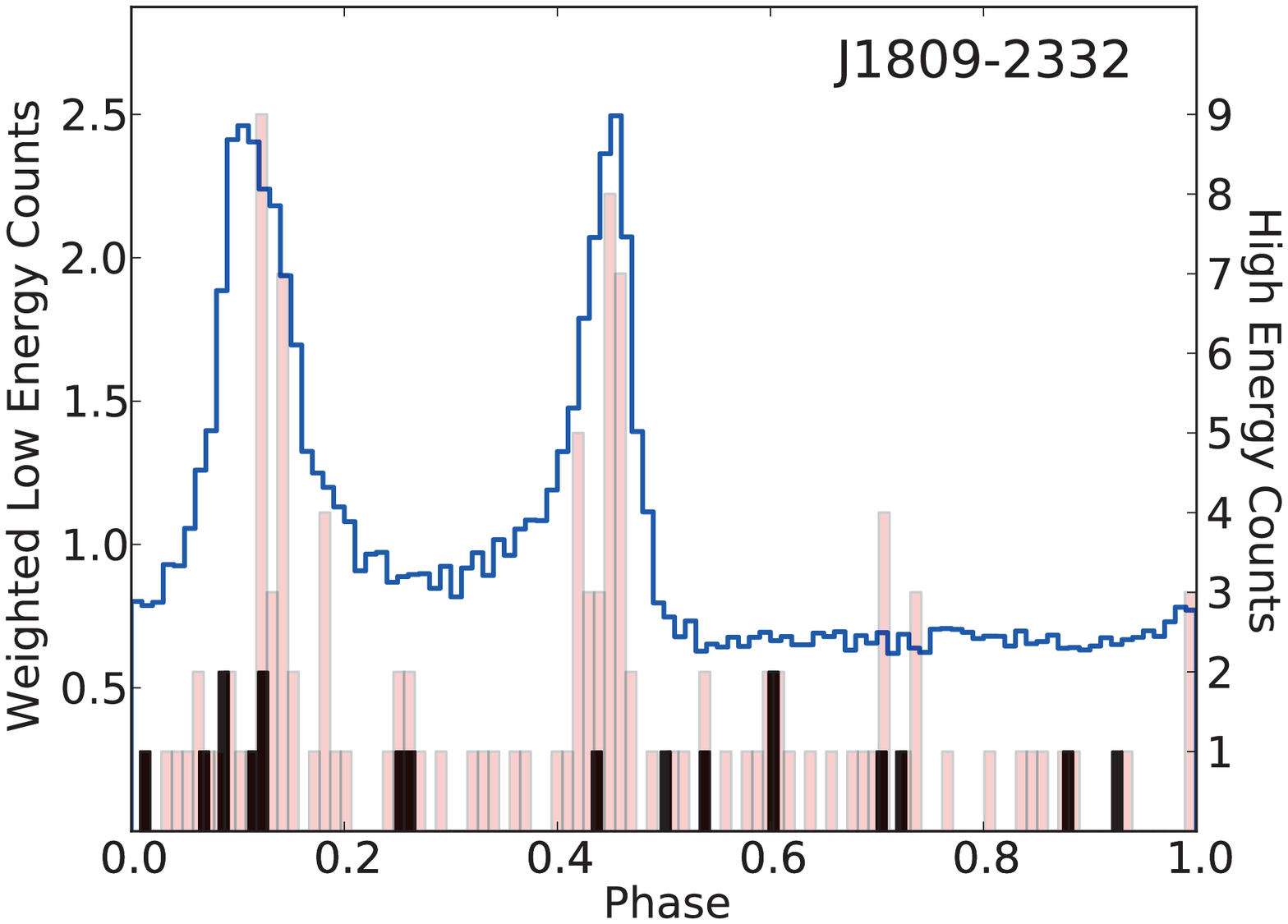}
                \includegraphics[width=1.55in]{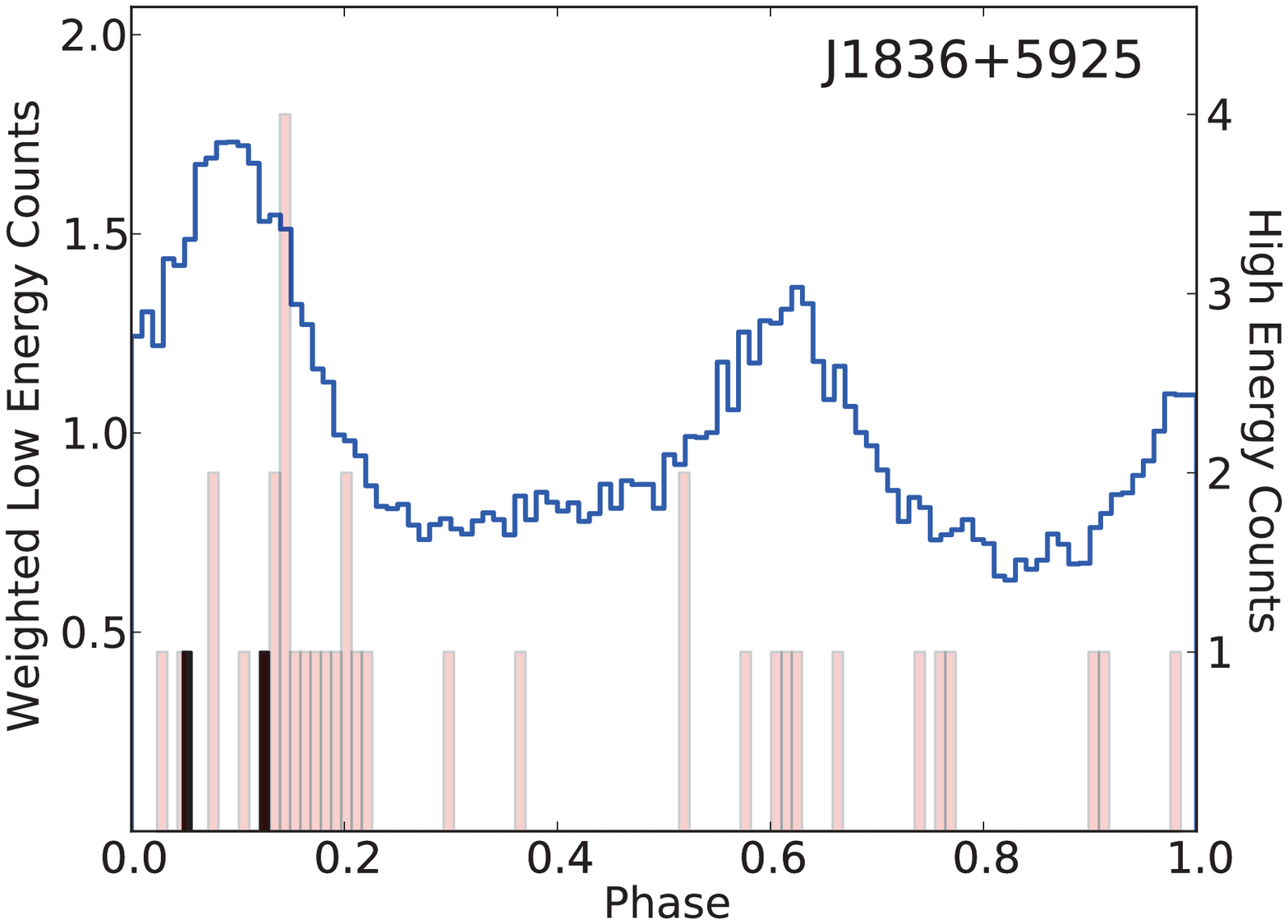}
                \includegraphics[width=1.55in]{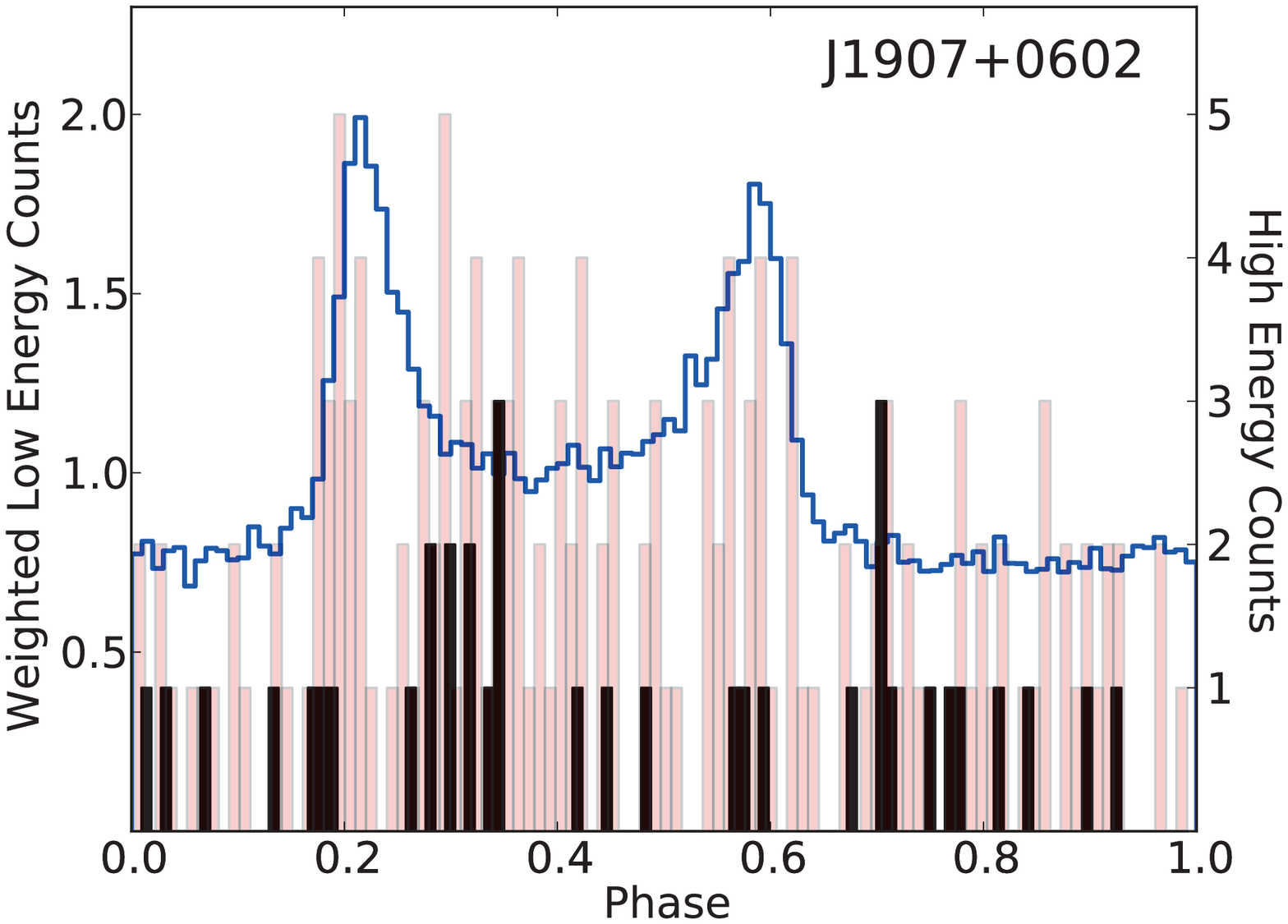}
                \includegraphics[width=1.55in]{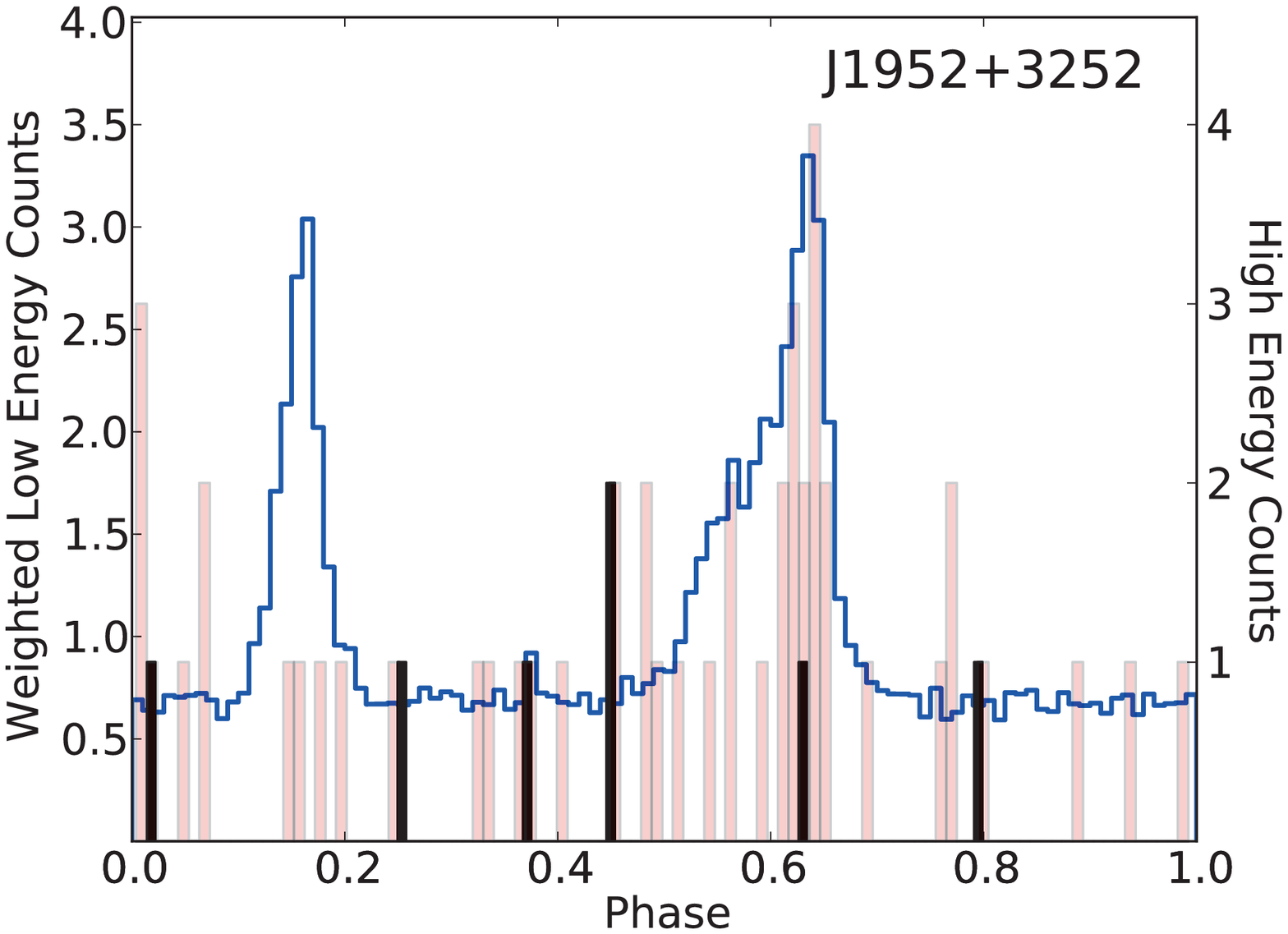}
                \includegraphics[width=1.55in]{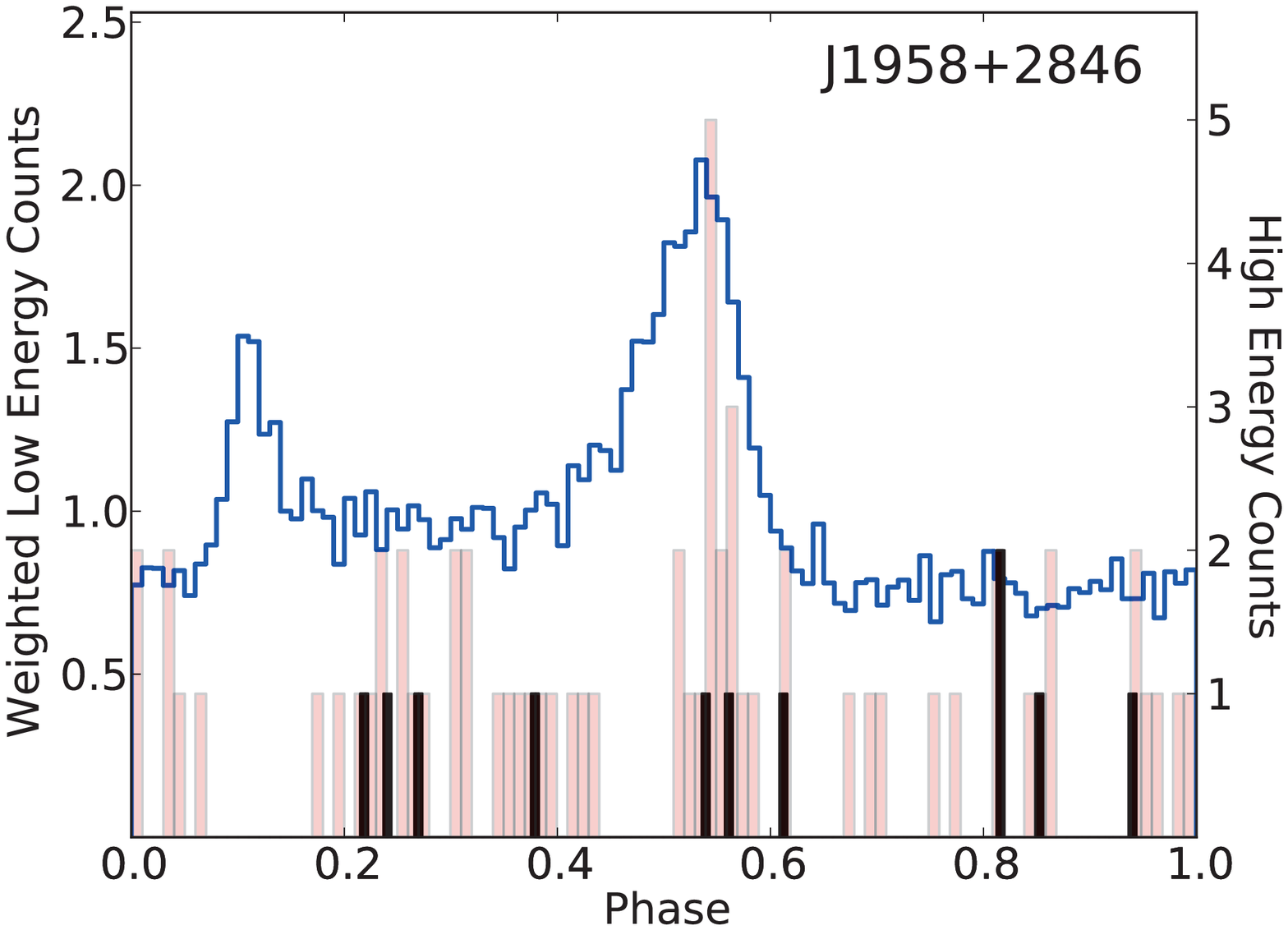}
                \includegraphics[width=1.55in]{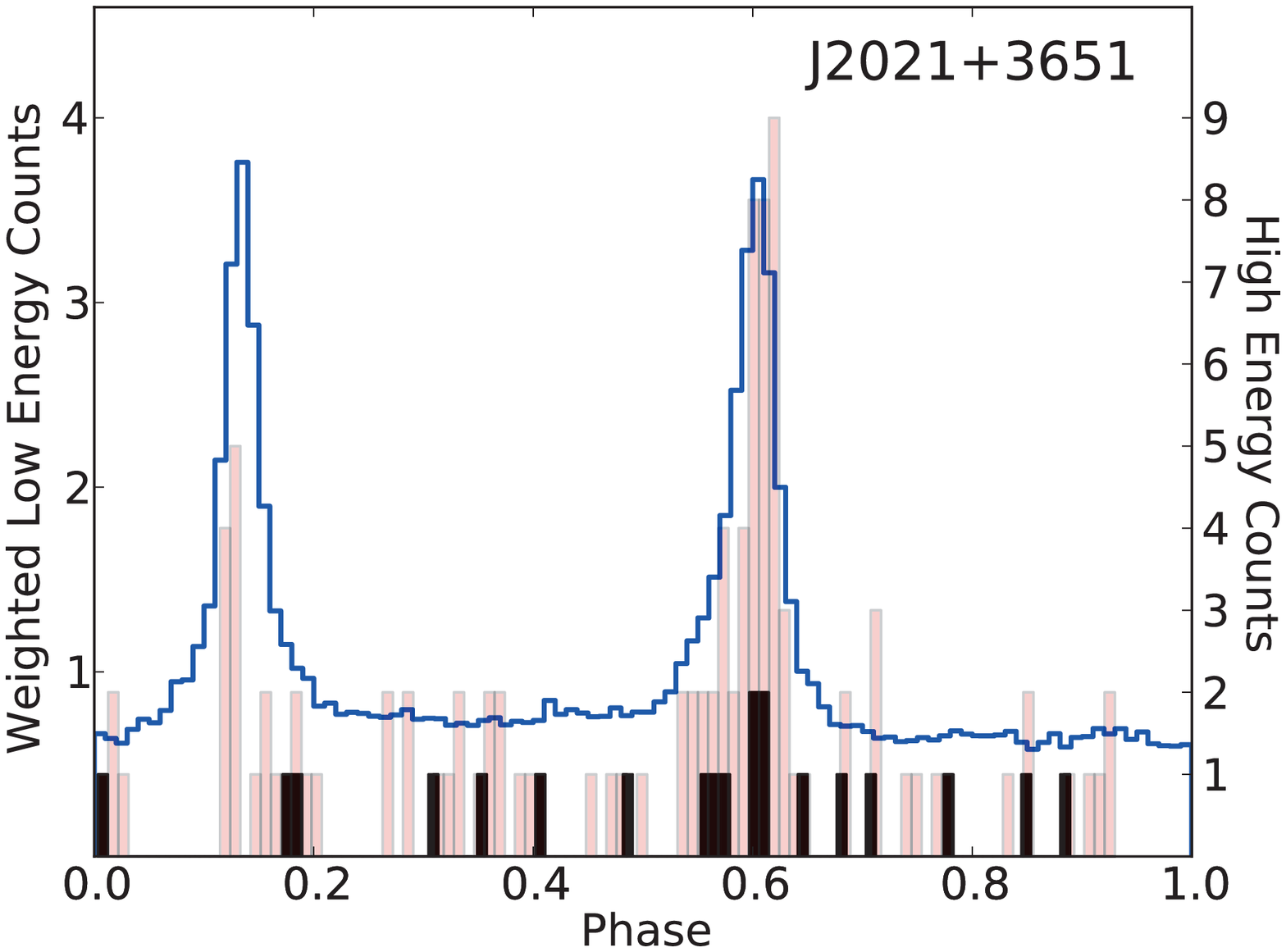}
                \includegraphics[width=1.55in]{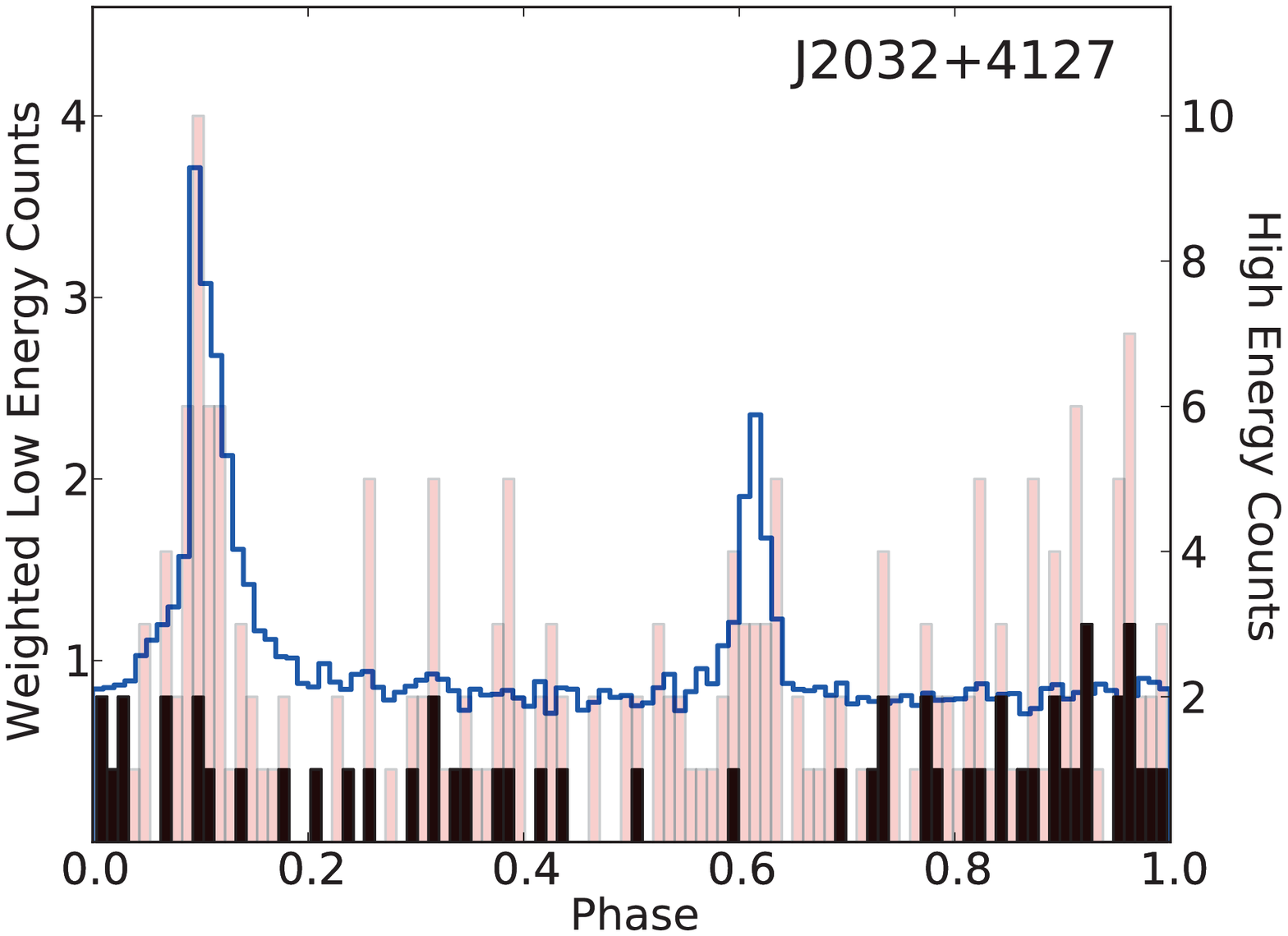}
                \includegraphics[width=1.55in]{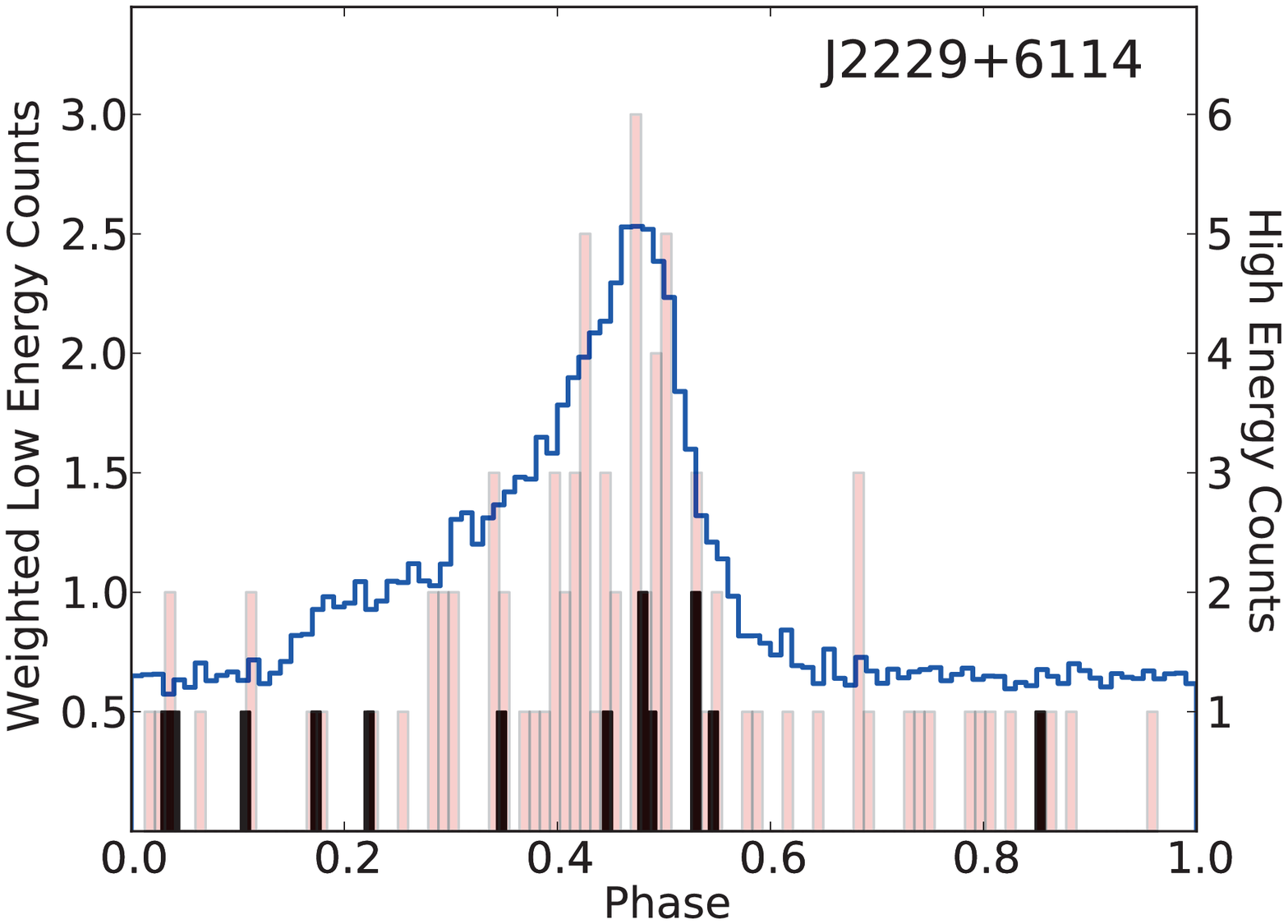}
\caption{Folded pulse profiles of $\gamma$-ray pulsars associated with
  1FHL sources, obtained with 3 years of {  \tt P7CLEAN} {\it Fermi}-LAT data. The blue histogram
  (y-axis scale on the left) represents the weighted ``low energy''
  ($>$100 MeV) light curve (using the 2PC spectral model). The filled histograms (y-axis scale on the
  right) show the events above 10 GeV (pink) and 25 GeV (black). }
\label{pulsar_fig1}
\end{figure}

\begin{figure}[th]
        \centering
                \includegraphics[width=1.55in]{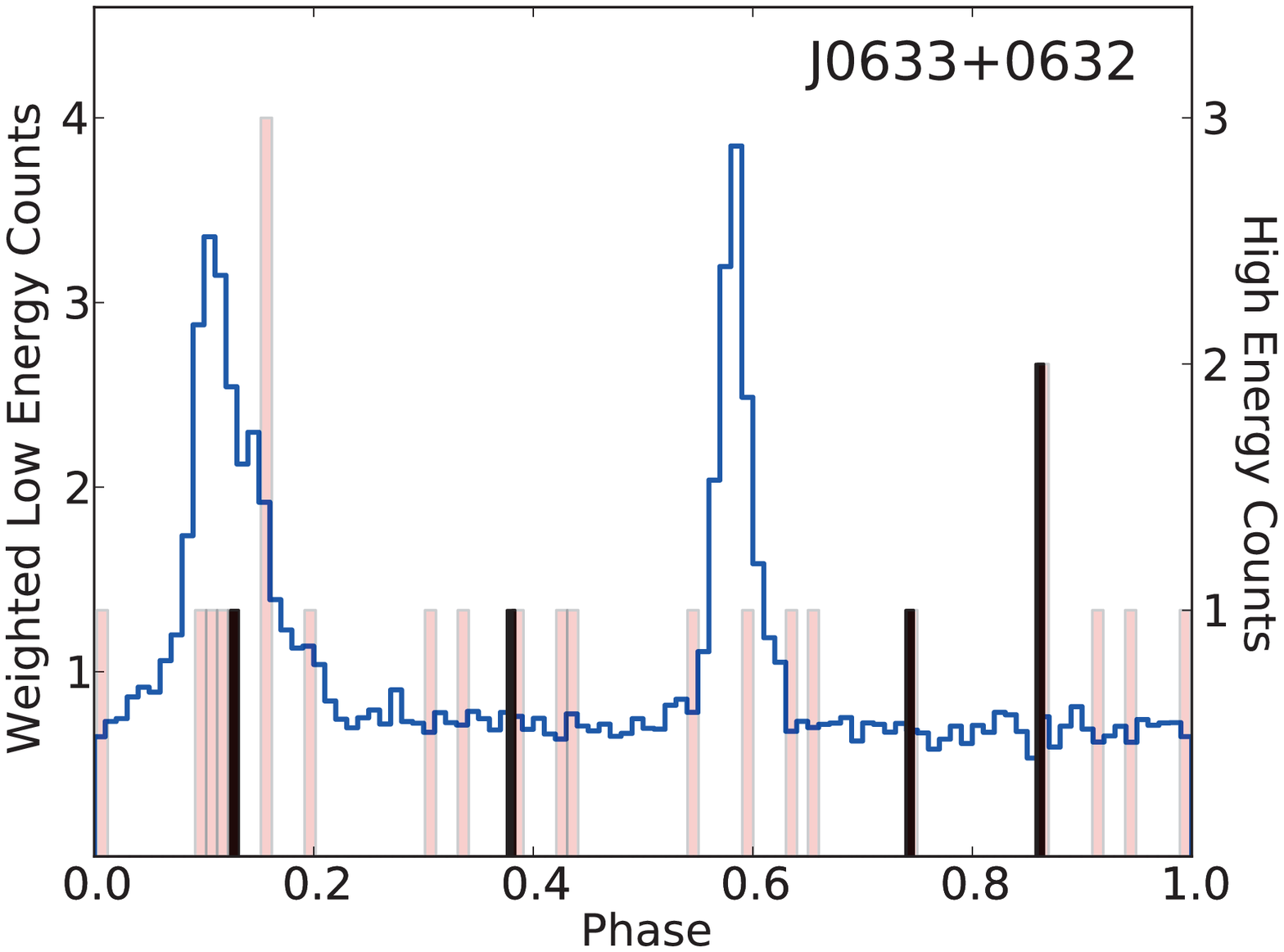}
                \includegraphics[width=1.55in]{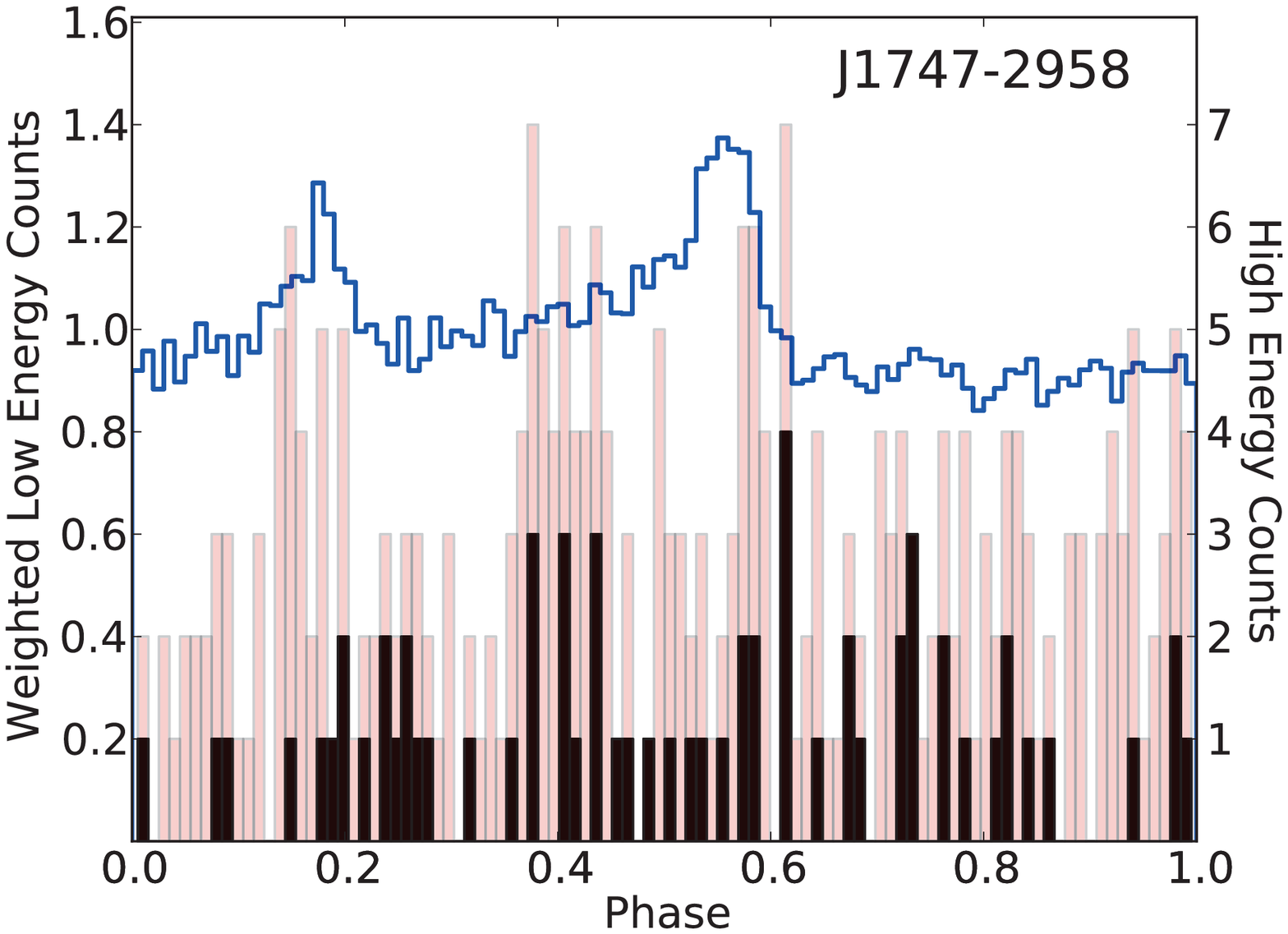}
                \includegraphics[width=1.55in]{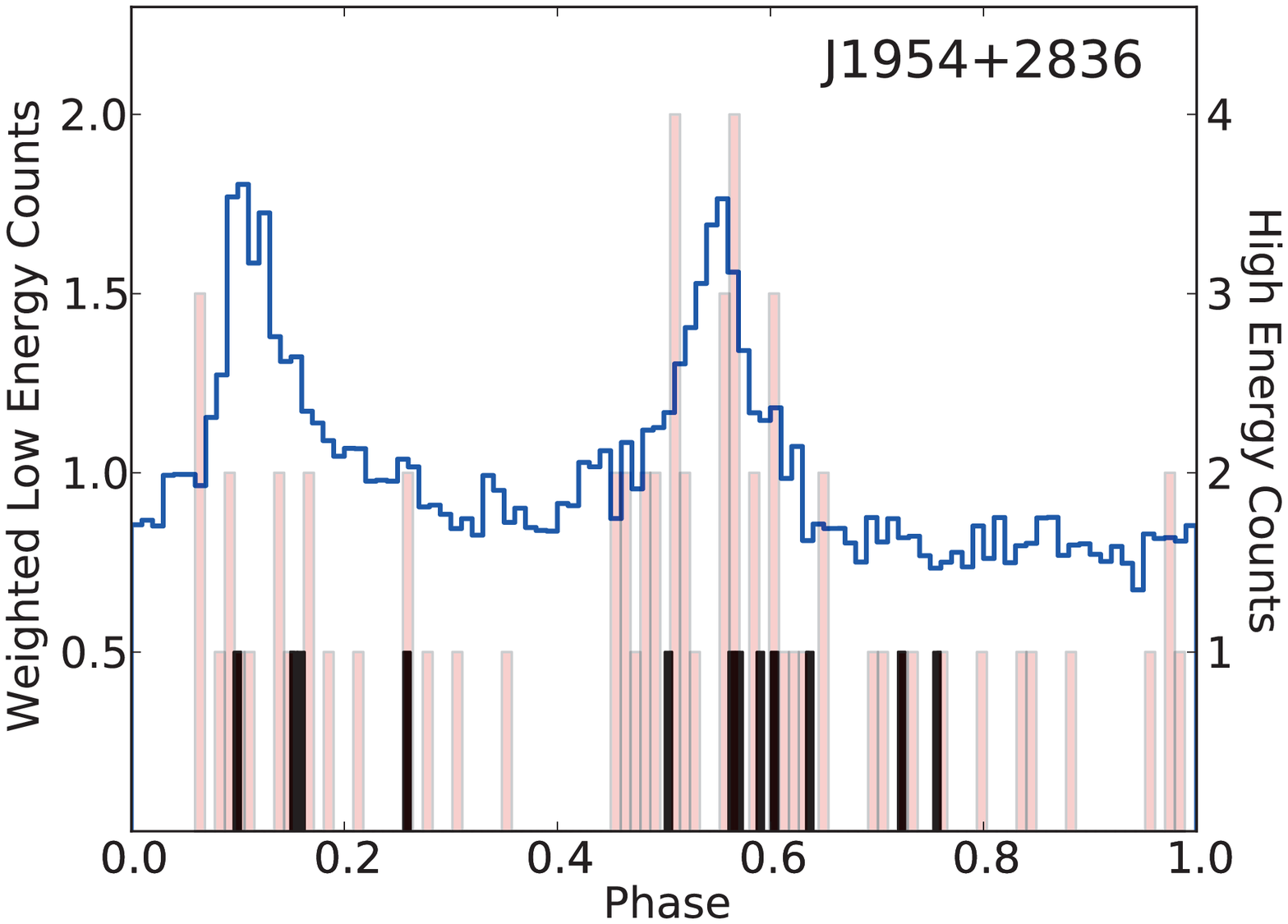}
                \includegraphics[width=1.55in]{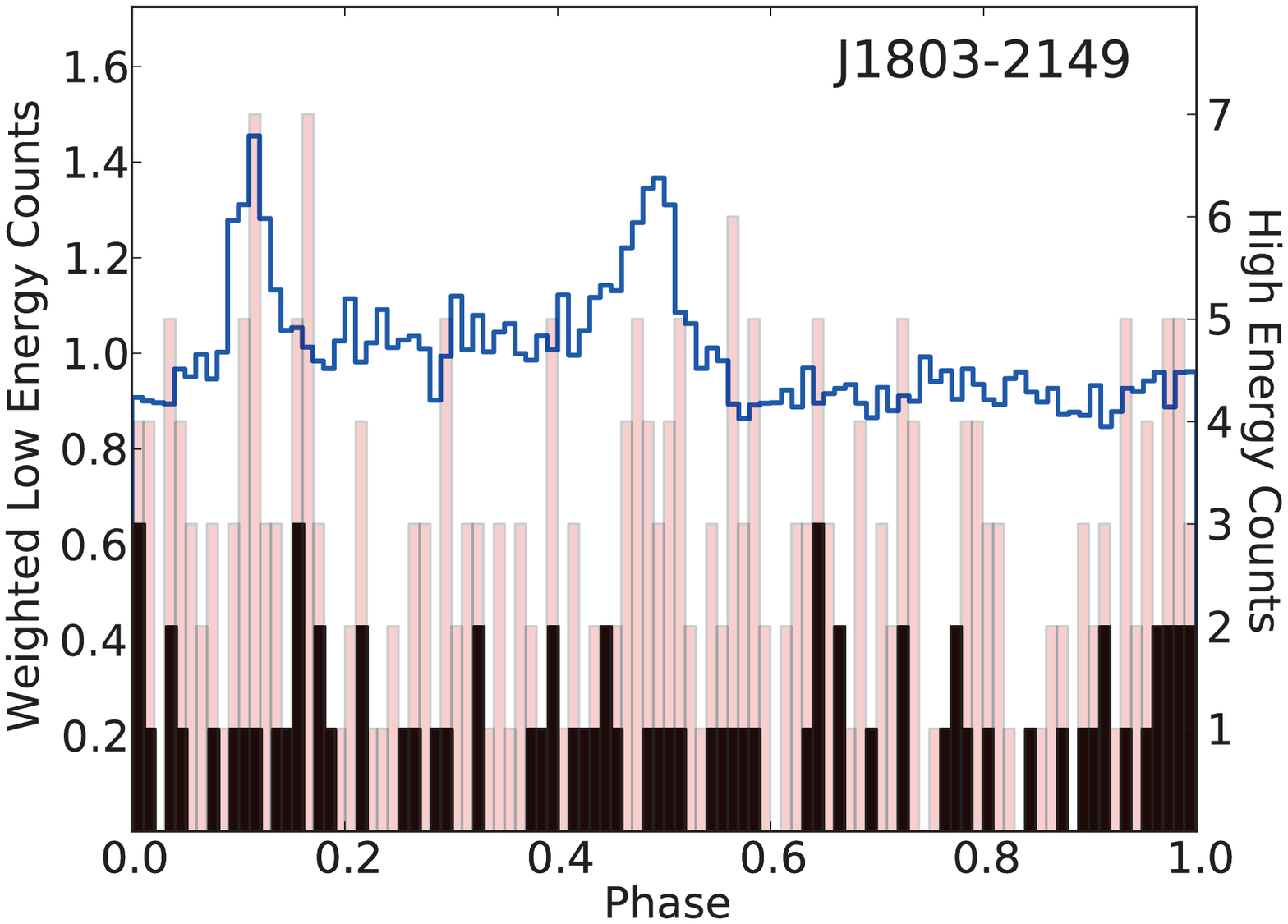}
                \includegraphics[width=1.55in]{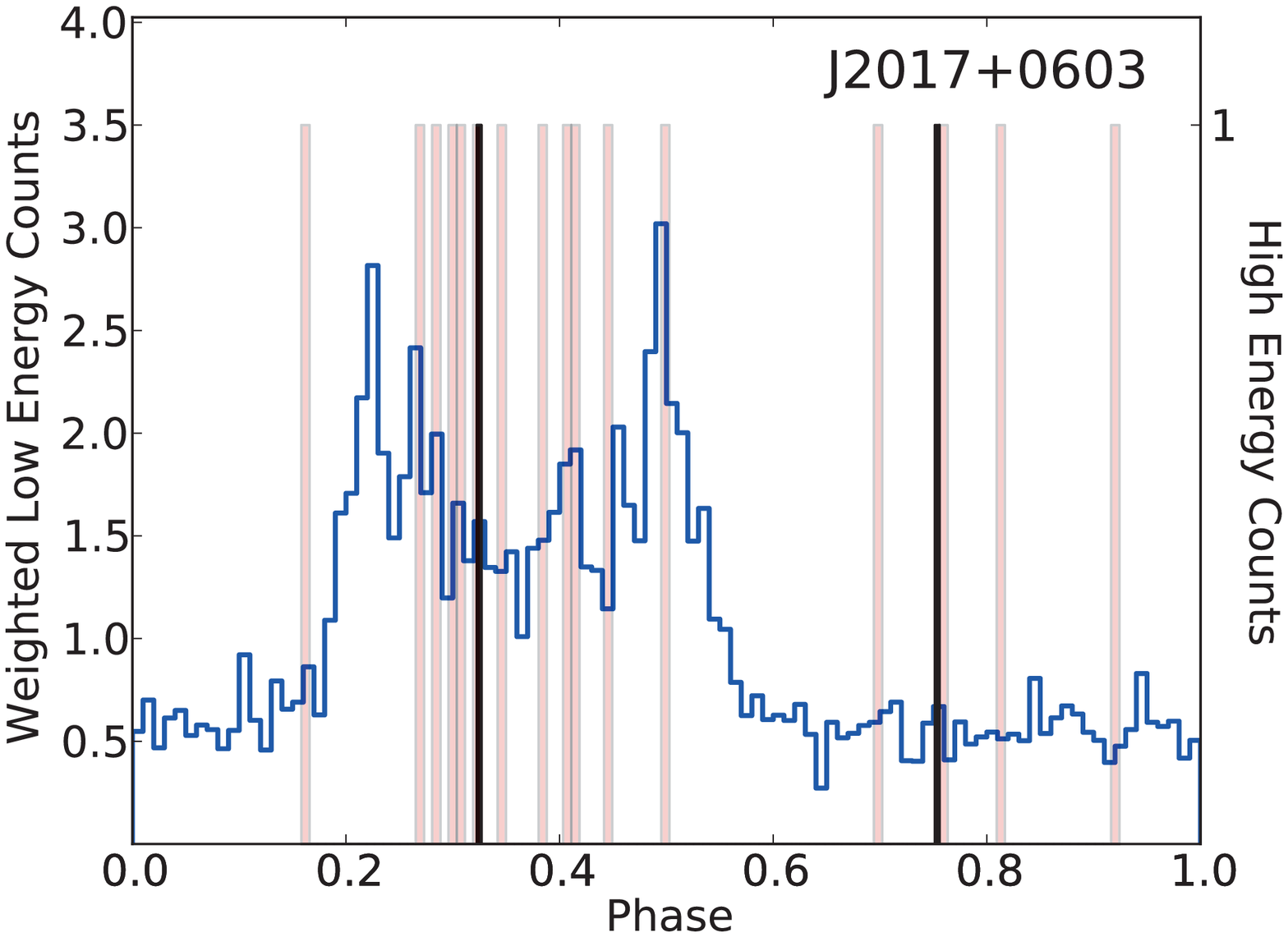}
                \includegraphics[width=1.55in]{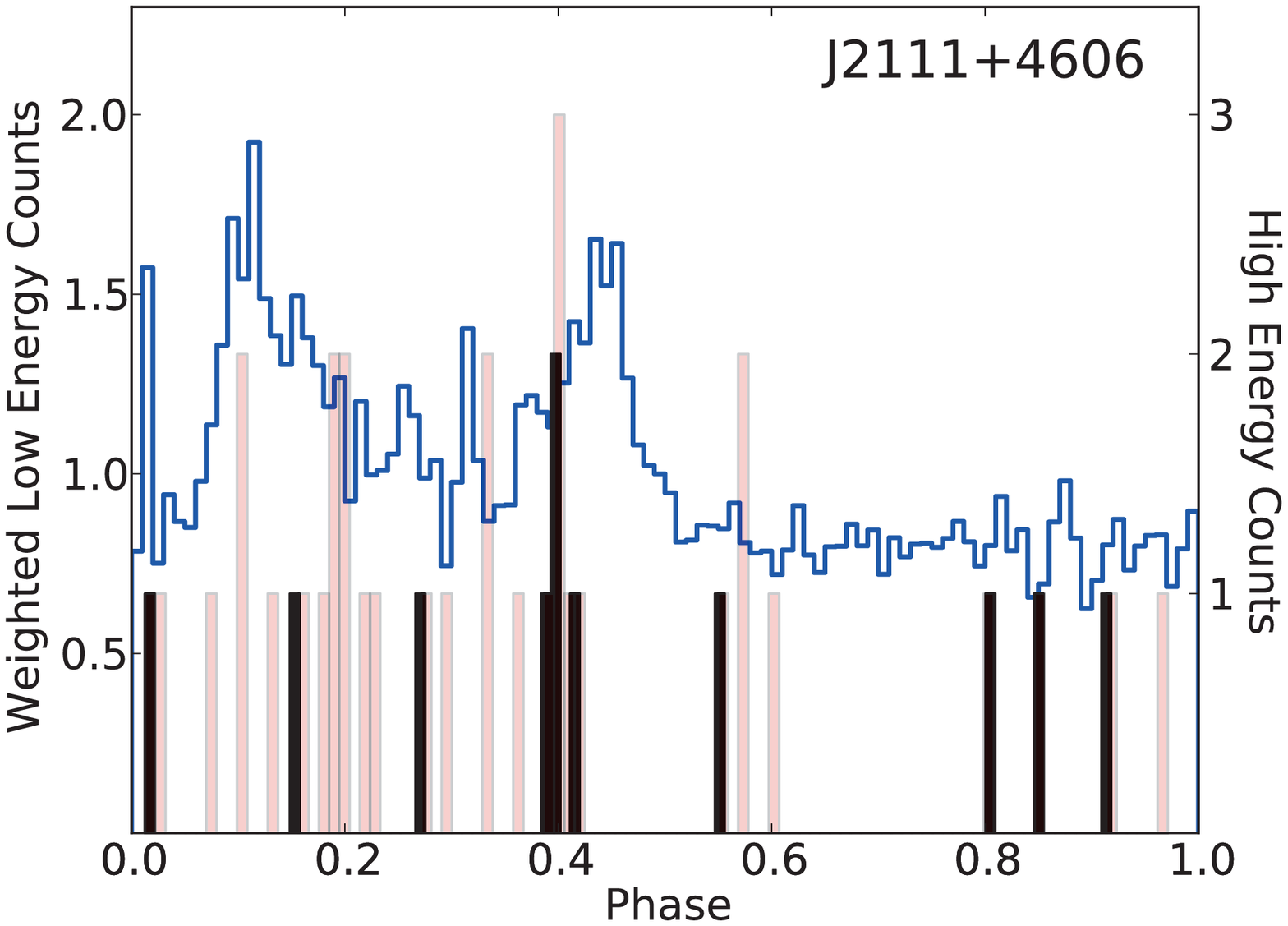}
                \includegraphics[width=1.55in]{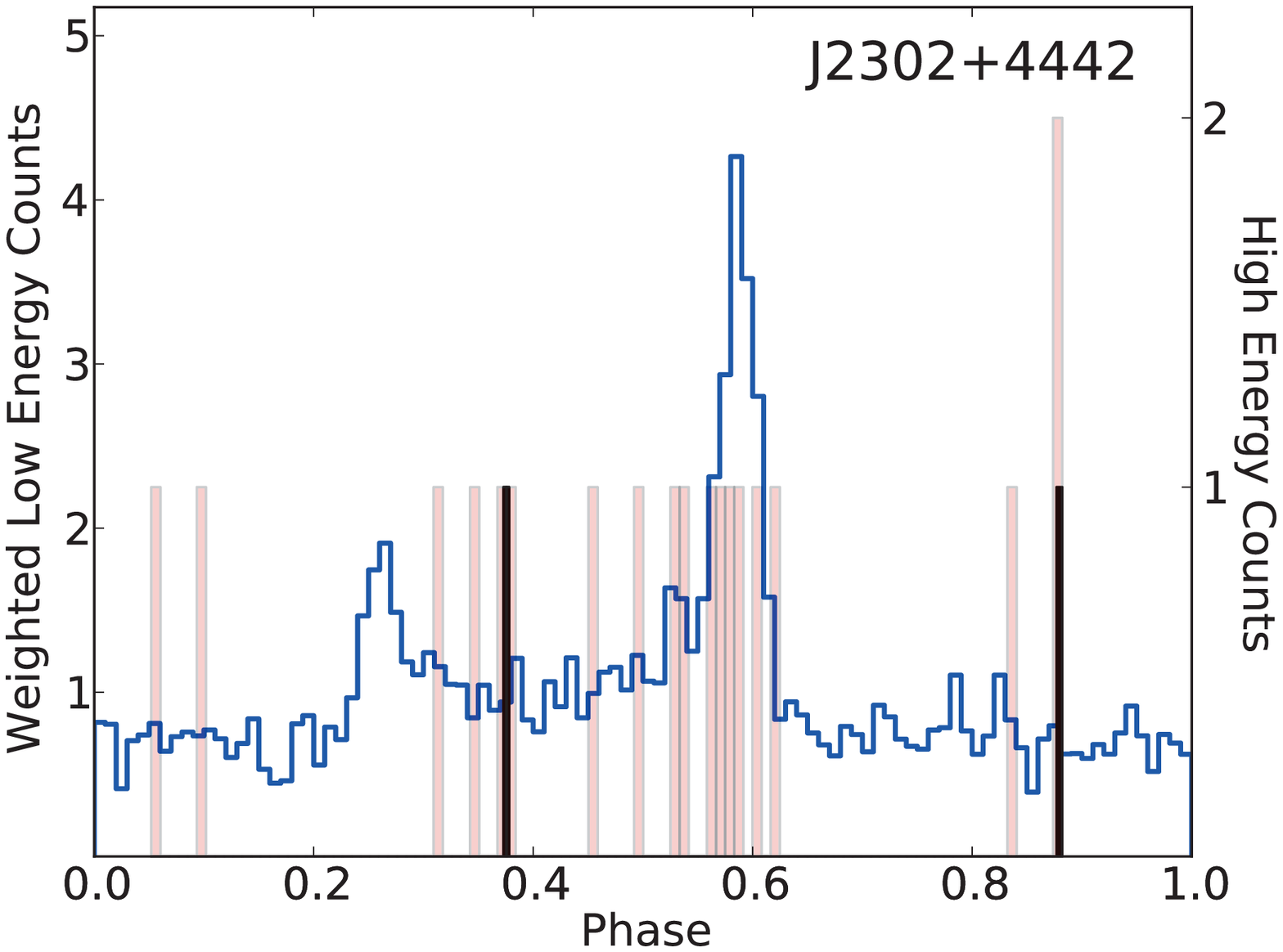}
                \includegraphics[width=1.55in]{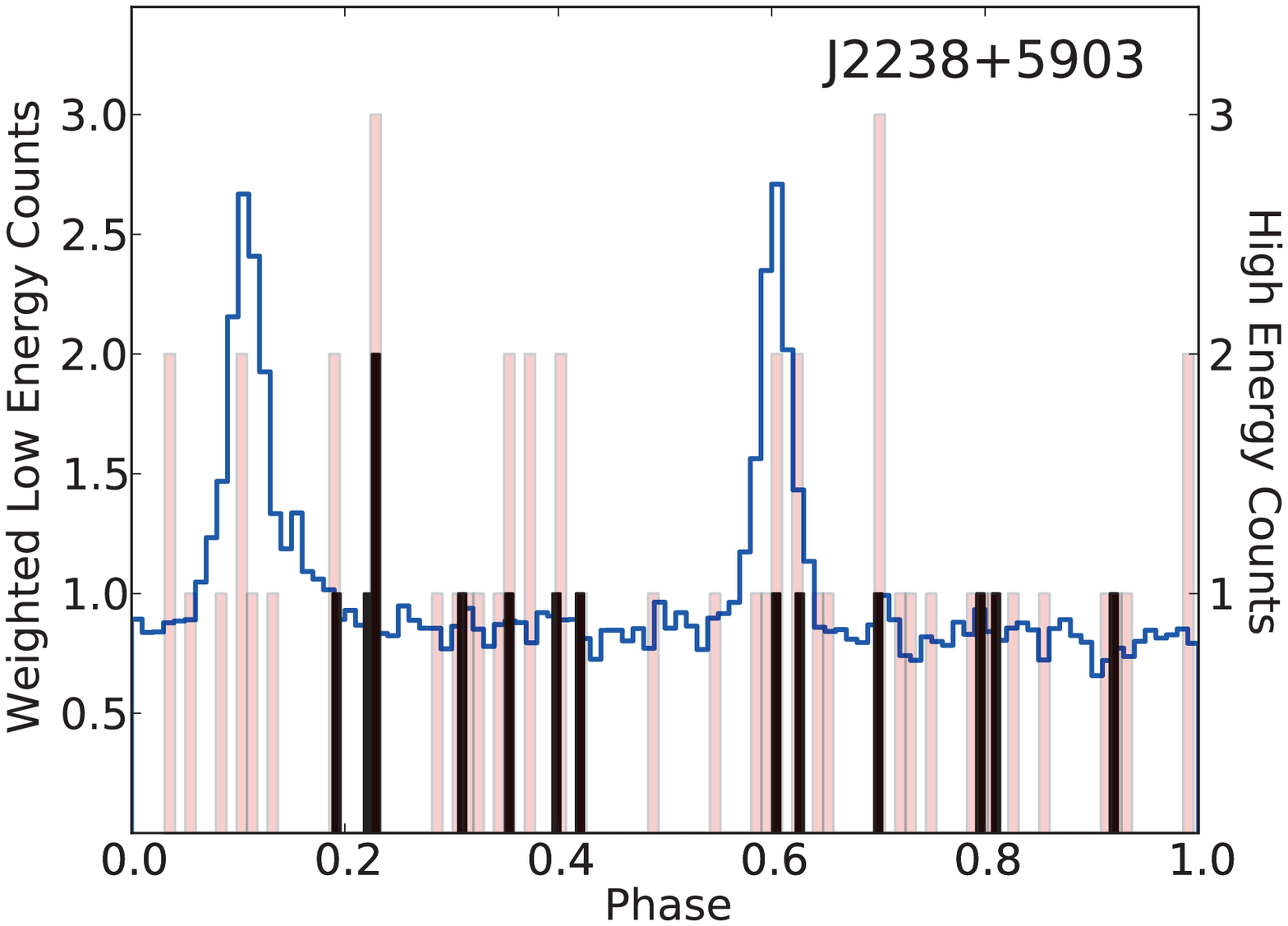}
\caption{Folded pulse profiles of $\gamma$-ray pulsars with no
  corresponding 1FHL sources, obtained with 3 years of {  \tt P7CLEAN} {\it Fermi}-LAT data. The blue histogram
  (y-axis scale on the left) represents the weighted ``low energy''
  ($>$100 MeV) light curve (using the 2PC spectral model). The filled histograms (y-axis scale on the
  right) show the events above 10 GeV (pink) and 25 GeV (black).}
\label{pulsar_fig2}
\end{figure}

The 28 HPSRs discussed in this section include members of every class of pulsar
detected so far by the LAT and include the 5 brightest EGRET-detected
pulsars\footnote{PSR~J1057$-$5226 (B1055$-$52) is the only EGRET pulsar
  not detected above 10 GeV.}: 5 young (non-recycled) radio-loud $\gamma$-ray pulsars, 13 young (non-recycled) radio-quiet $\gamma$-ray pulsars, and 5 $\gamma$-ray MSPs.
It is not obvious how to select the best candidates for the detection of pulsations at VHE with IACTs, since any such selection must depend on many assumptions, and spectral extrapolations from 10 GeV
upward are notoriously unreliable. 
Improving the analysis for pulsations to address the shortcomings discussed above
is left for future publications.  
A dedicated treatment of the separate PWN and PSR spectral components would likely improve the sensitivity of our search for
pulsations, especially for those pulsars affected by a high level of PWN emission.
Finally, a different choice of templates with which to
compare the high-energy pulse profile (e.g., the $>1$ GeV pulse
profile), taking into account the evolution with energy of the pulse
profile should also improve the sensitivity of the pulsation
search.  The sensitivity will, in any case, improve as the LAT data
continue to accumulate.

\cleardoublepage

   
\cleardoublepage


\section{Good Candidates for VHE Detection}
\label{VHECandidates}

Astrophysical interest in $\gamma$ rays extends beyond the energy range that is easily accessible to a space-based instrument like the \FermiLATc, which is limited by the size and mass of a satellite. Ground-based \gray\ telescopes that use the Earth's atmosphere as a detector have enormous collecting areas and can operate successfully at energies where the LAT simply runs out of photons. Present and future VHE telescopes include both particle detector arrays (e.g. Tibet AS, ARGO-YBJ, HAWC and LHASSO) and IACTs, which are presently the most sensitive VHE instruments. The survey capability of the \FermiLAT\ at high energies provides a valuable complement to these IACTs, which are pointed instruments. 
It is worth noting that the 2FGL catalog of sources detected above 100 MeV has 1873 entries while the number of sources detected above 100~GeV and reported in TeVCat (version 3.400) is only 143 (including announced but not published VHE detections). Therefore, the LAT catalogs, 
{  and particularly this one above 10 GeV, offer candidate} VHE targets. 
The 10~GeV minimum energy used for the 1FHL catalog analysis is a good compromise between having an adequate number of photons measured by LAT and being close to the energy range where IACTs operate. In this section we describe the best VHE candidates among the full set of the 1FHL catalog sources.

The most advanced IACTs are currently H.E.S.S. and VERITAS (arrays of four $\sim$12~m
telescopes, \citealp{HESS,VERITAS}), and MAGIC (two telescopes of $\sim$17~m diameter, \citealp{MAGIC}).
H.E.S.S. and VERITAS have energy thresholds\footnote{
The energy threshold is conventionally defined as the peak in the differential energy trigger rate for a ``Crab nebula-like'' spectrum. } of 
$\sim$100~GeV (and typically measure $\gamma$-ray\ spectra above 140~GeV) while MAGIC has an energy threshold of 60~GeV (and typically measures $\gamma$-ray\ spectra above 80~GeV).  {  H.E.S.S added a 28-m diameter telescope to the existing array in Summer 2011. The resulting H.E.S.S II array has been operational since September 2012 and should allow the system to reduce the energy threshold below 50 GeV.}
The planned Cherenkov Telescope Array (CTA) will be even more powerful in terms of sensitivity and operational \gray\ energy range \citep{CTAPerformance}. The currently operating IACTs have effective fields of view of less than 4$\degr$ degrees and so usually operate in a targeted observation mode. The photon fluxes in the VHE range are very low and hence relatively long observing times ($\gapp$ 5--10~hours) are typically required to make detections.
Because IACTs can operate only on clear, essentially moonless nights\footnote{MAGIC can operate during nights with moderate moonlight with a reduced PMT HV, and VERITAS can operate even during bright moon by using an optical filter in front of the PMT camera. Such operation increases the energy threshold and reduces the sensitivity of the observations.}, the duty cycles are typically only about 10-12\%, which corresponds to $\lapp$ 1000 hours observing time annually. 
The Galactic plane is the only extended region that has been systematically scanned with the latest generation of IACTs.

\begin{figure}[th]
\begin{center}
\includegraphics[width=18.0cm]{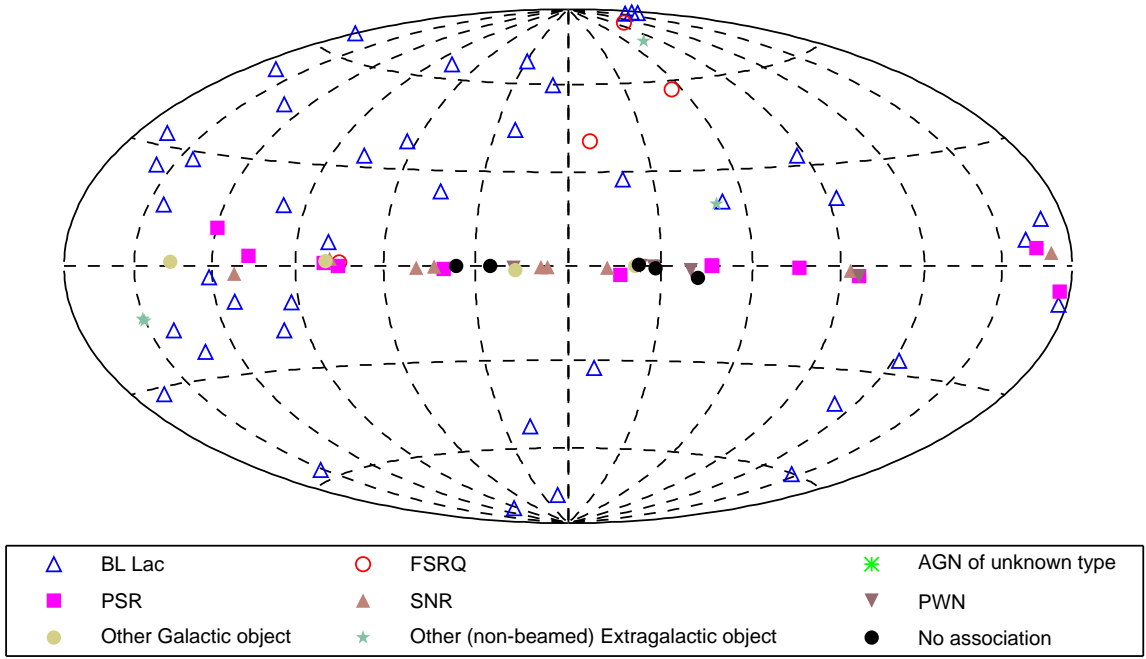}
\end{center}
\caption{\label{SkymapTeVSrc} 
Sky map showing the 1FHL sources that have been detected at VHE by IACTs. The markers represent the source classes reported in Table \ref{TableStatisticsSrcClasses}.
{  The projection is Hammer-Aitoff in Galactic coordinates.}
}
\end{figure}


The number of 1FHL sources that have associations with known VHE emitters is 84, which is about 2/3 of all the known VHE emitters. IACTs are responsible for the discovery of 81, while the other three were first detected in the VHE range by MILAGRO, a water Cherenkov detector \citep{MILAGRO}.
These sources are depicted in Figure~\ref{SkymapTeVSrc}.  We note that in the 2FGL catalog, coincidentally, 84 sources were associated with VHE emitters (not accounting for the association of 2FGL J2229.0+6114 with two VHE sources).  
In addition, 14 2FGL sources have been reported to be VHE emitters since the publication of the 2FGL catalog (see TeVCat). Therefore, of the 1873 sources in the 2FGL catalog, 98 now are associated with VHE sources, while of the 514 sources in the 1FHL, 84 have VHE counterparts.  The VHE sources in common total 80.  The 1FHL sources with VHE associations that are not in the 2FGL catalog are the blazars PKS~0548$-$322 and MS~13121$-$4221, the cocoon of freshly-accelerated cosmic rays in the Cygnus X star-forming region \citep{LAT11_CygCocoon}, and the unidentified source HESS ~J1857+026 \citep[presumed to be a PWN,][]{HESSJ1857}. On the other hand, most of the 2FGL sources with VHE associations that are not in the 1FHL catalog are GeV pulsars that were associated 
with spatially extended, Galactic TeV sources. Only 3 point-like TeV sources with associations in the 2FGL catalog do not also have associations in the 1FHL catalog:  the blazar 1ES~0414+009, and the starburst galaxies NGC~253 and M82.  Each of these required very long exposures for VHE detection: $\sim$70 hours with H.E.S.S. to detect the blazar and $\sim$130 hours each with H.E.S.S. and VERITAS to detect the starburst galaxies.

This comparison shows that by limiting the energy range to $>$10~GeV, the 1FHL sources do not miss many VHE sources. 
Naturally, among the 1FHL sources, some are more feasibly detectable at VHE. In the subsections below we describe the criteria that we used to select the most promising VHE source candidates among the 1FHL sources, and report  the results.

\subsection{Criteria for Selection of TeV candidates}

Figure~\ref{DistributionFlux10And50GeV} shows the distribution of fluxes above 10~GeV ($F_{10}$) and above 50~GeV ($F_{50}$) for the 1FHL sources.
The quantities $F_{10}$ are directly provided by the likelihood analysis (Table~\ref{MainCatalogTable}), while the values of $F_{50}$ are calculated from the power-law spectra. The figure shows that the known TeV sources cluster at the highest fluxes, this correlation being clearer for $F_{50}$.  Such a relation is quite natural since the energy 50~GeV is  close to the analysis energy threshold of the current generation of  IACTs. Therefore, $F_{50}$ is a very good indicator of the VHE flux. 

\begin{figure}[th]
\begin{center}
\includegraphics[width=7cm]{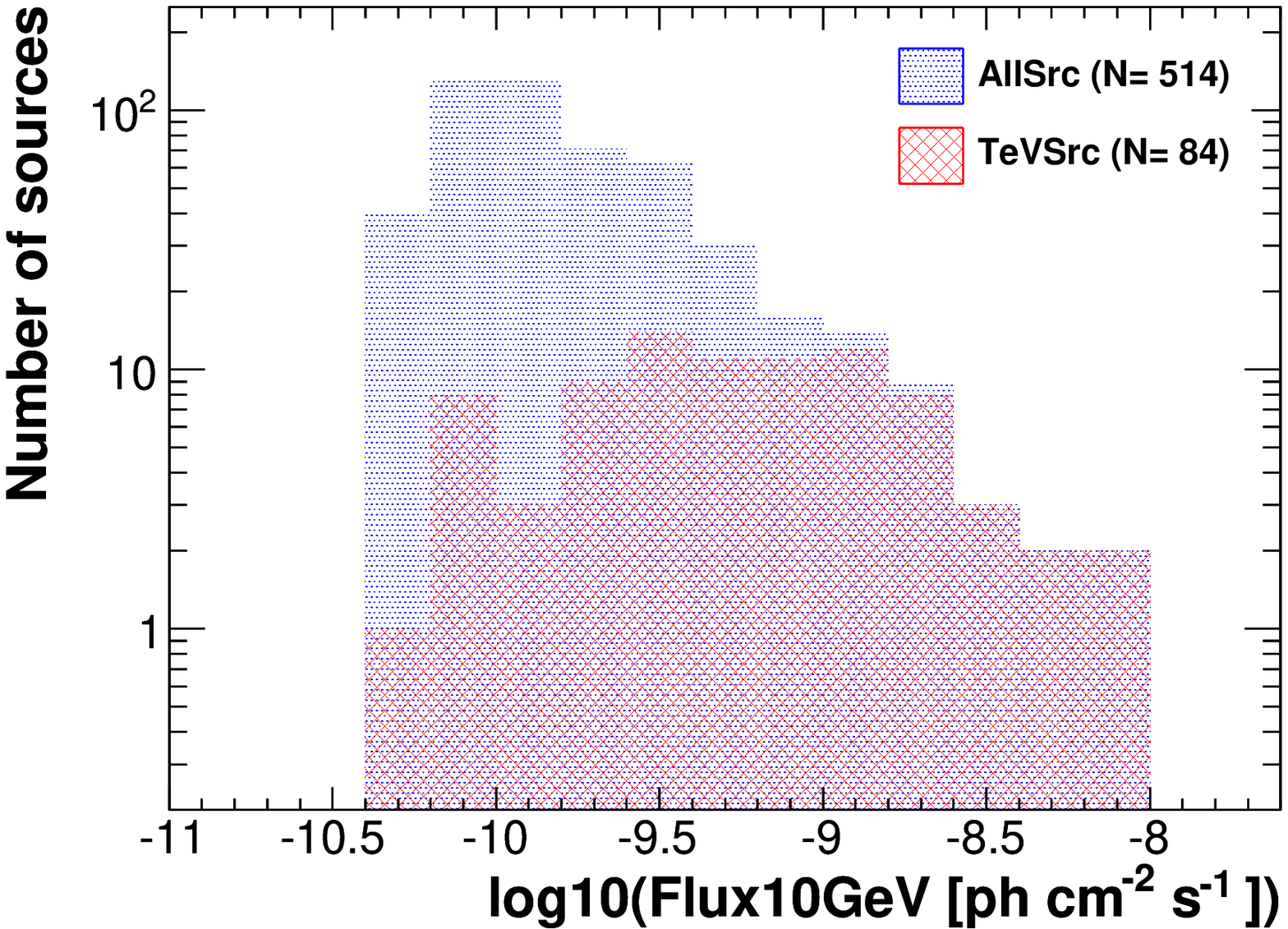}
\includegraphics[width=7cm]{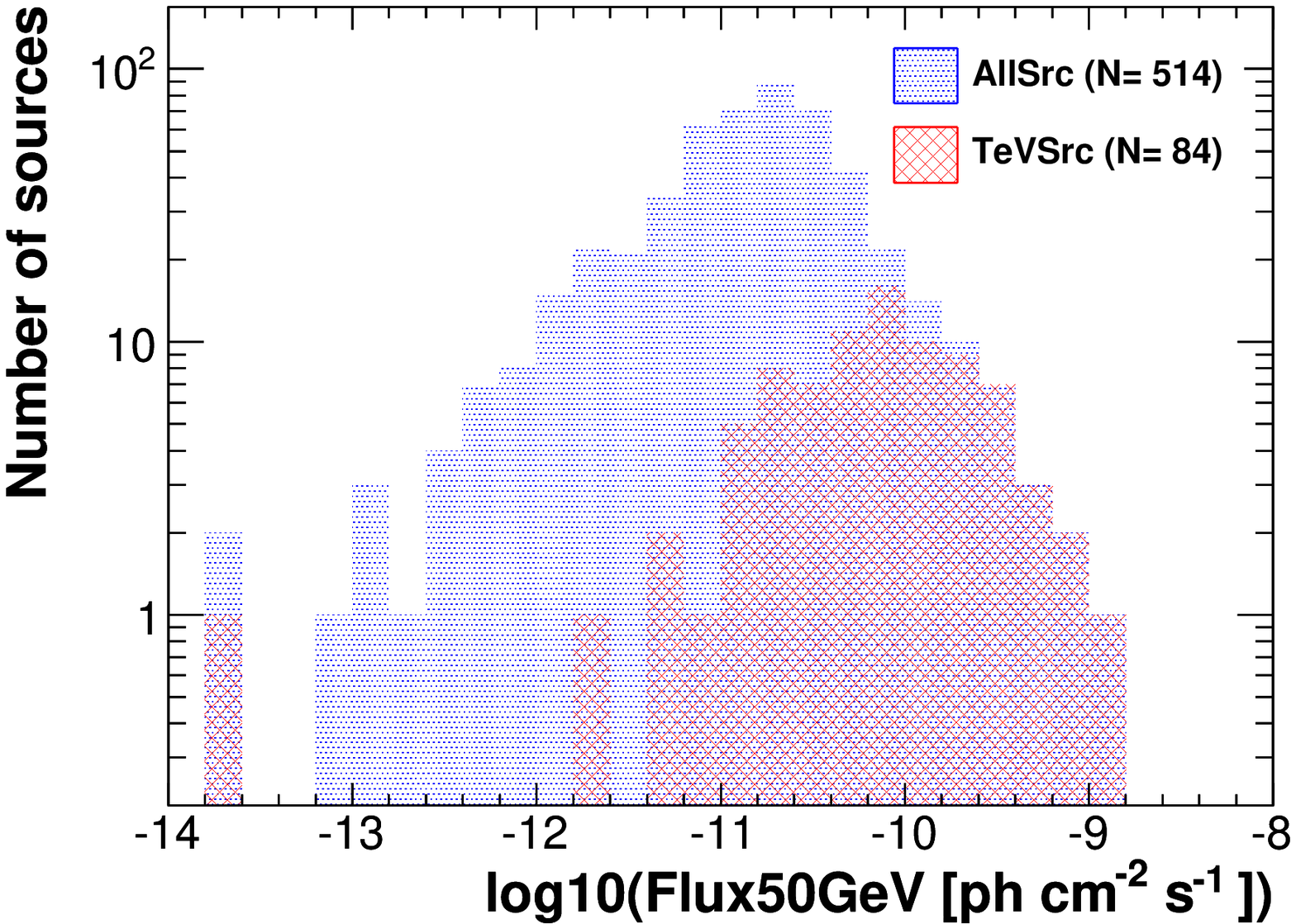}
\end{center}
\caption{\label{DistributionFlux10And50GeV} 
Distributions of measured flux above 10~GeV (left) and estimated flux above 50~GeV (right). 
The blue histograms depict all the 1FHL sources, while the red filled histograms show the 1FHL sources that have already been detected at VHE.}
\end{figure}

Two additional quantities can also be used to select good candidates for VHE detection. 
These are the spectral index above 10~GeV ($\Gamma_{10}$; see Table~\ref{MainCatalogTable}) and the {\it pseudo significance} of the signal above 30~GeV, $Sig_{30}$, which we define as $(TS_{30\_100}+TS_{100\_500})^{1/2}$, where $TS_{30\_100}$ and $TS_{100\_500}$ are  TS values for the 30--100~GeV and 100--500~GeV energy bands, respectively, reported in the catalog data product.  
The distributions of these quantities for all the 1FHL sources are shown in Figure~\ref{DistributionIndexAndSigma30GeV}.
The known VHE sources cluster at low $\Gamma_{10}$ values and at high $Sig_{30}$ values.
Even though these quantities are not as powerful discriminators as $F_{50}$, they can help to remove from consideration sources that are not likely VHE emitters.

\begin{figure}[th]
\begin{center}
\includegraphics[width=7cm]{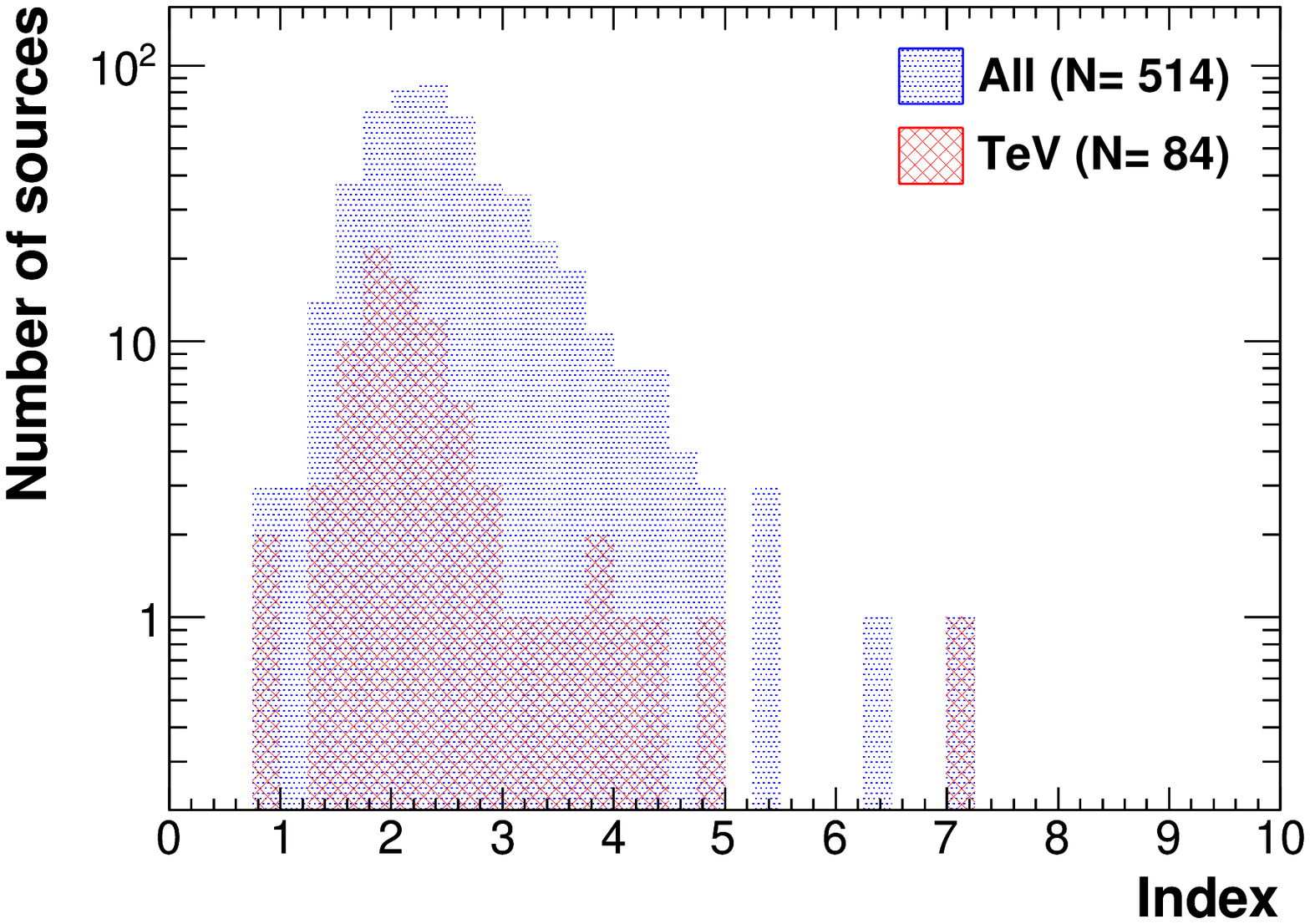}
\includegraphics[width=7cm]{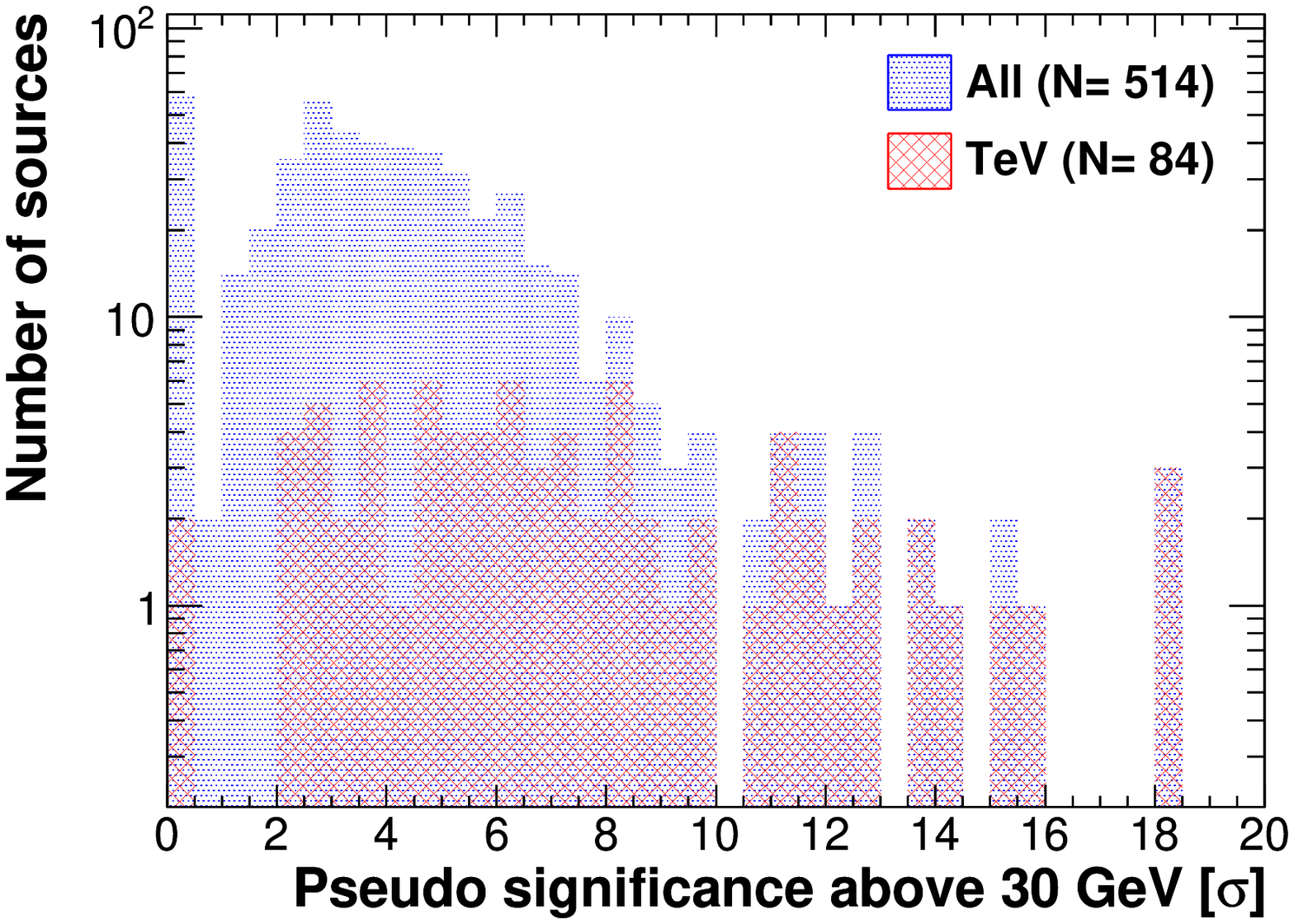}
\end{center}
\caption{\label{DistributionIndexAndSigma30GeV} 
Distribution of the power-law index resulting from the spectral fits above 10~GeV (left) and the pseudo significance of the detection above  30~GeV (right; see text for definition).
The blue histograms depict all the 1FHL sources, while the red filled histograms show the 1FHL sources that have already been detected at VHE.}
\end{figure}

We adopted the following criteria to select 1FHL sources that have not been detected at VHE but have properties similar to those that have associations with known VHE sources:  (a) $Sig_{30} >3$; (b)  $\Gamma_{10} < 3$; and (c) $F_{50} > 10^{-11}$ ph cm$^{-2}$ s$^{-1}$.

As one can infer from Figures \ref{DistributionFlux10And50GeV} and \ref{DistributionIndexAndSigma30GeV}, the cut on $F_{50}$ is the most restrictive, although the three cuts are strongly correlated. For instance, when applying the cuts in the order listed above, from the 84 TeV 1FHL sources, 
we reject 11 with the $Sig_{30}$ cut, then 4 additional sources with the $\Gamma_{10}$ cut and zero sources when applying the $F_{50}$ cut.  Therefore, VHE sources with low $F_{50}$ also have low $Sig_{30}$ and/or a low $\Gamma_{10}$.
From the 15 TeV 1FHL sources that were rejected, we find that most of them (10 out of 15) are pulsars (6 HPSR and 4 PSR) that are associated with an extended PWN TeV source. Even though positional associations exist, the sources of the GeV radiation are not the sources of the TeV radiation. Among these 15 TeV 1FHL sources removed by the selection cuts are also the core of the Cen~A  radio galaxy, the FSRQ 3C~279, and VER J2016+372, a possible PWN. Cen~A is an extremely weak TeV source whose detection required more than 120 hours of observation with H.E.S.S., 
and 3C~279 was detected by MAGIC only during two large outbursts in 2006 and 2007,  
but has not been detected during the \FermiLAT\ era.
As for VER J2016+372, it is positionally coincident (angular separation is $0\fdg068$) with the source 1FHL~J2015.8+3710 (2FGL~J2015.6+3709), which is associated with the FSRQ  MG2 J201534+3710 ($z$=0.859). However, the TeV source is probably associated with the PWN CTB~87, and not with the distant FSRQ \citep[see][]{VERJ2016}. The 2FGL source, which is mostly dominated by photons below 10 GeV, shows high variability and strong curvature in the spectrum, which is typical of bright, distant FSRQs. Above 10 GeV, the spectrum from 1FHL~J2015.8+3710 seems to be somewhat harder ($\Gamma$ = 2.3$\pm$0.4), which might suggest the presence of an additional component. But the spectral difference is not significant due to the low photon statistics and hence we cannot exclude a statistical fluctuation in the number of detected high-energy events.

Therefore, we do not consider the 13 LAT-detected sources discussed above (i.e., 10 pulsars plus Cen~A, 3C~279, and MG2 J201534+3710/VER J2016+372) to be good candidates for detection with IACTs. 
The selection criteria remove only two ``good TeV candidate'' sources: the blazar 1RXS J101015.9$-$311909, and the unidentified source HESS J1507$-$622.
We conclude that the above-mentioned selection cuts are very conservative and that they keep most of the 1FHL sources that have already been detected at VHE.

\subsection{Results from the Selection of TeV Candidates}

From the 430 sources in the 1FHL catalog without VHE associations, we reject 175 with the cut on $Sig_{30}$, an additional 14 with the cut on $\Gamma_{10}$ and finally 28 more with the cut on $F_{50}$. That is, the conservative selection criteria specified above remove about half of the 1FHL sources that have not yet been detected at VHE, and retain 212, among which are 128 with BL Lac associations, 12 with FSRQ associations, 32 AGUs, 2 PSRs, 3 SPP, 1 SNR (SNR G260.4$-$03.4), 1 LVB star (Eta Carinae), 1 radio galaxy (PKS 0625$-$35), 1 Galaxy (LMC), and 31 unassociated sources.
These sources are denoted with the designator ``C'' in the column TEVCAT\_FLAG in Table \ref{MainCatalogTable} and their locations are depicted in Figure~\ref{SkymapTeVSrcCandidates}. \\

\begin{figure}[th]
\begin{center}
\includegraphics[width=18.0cm]{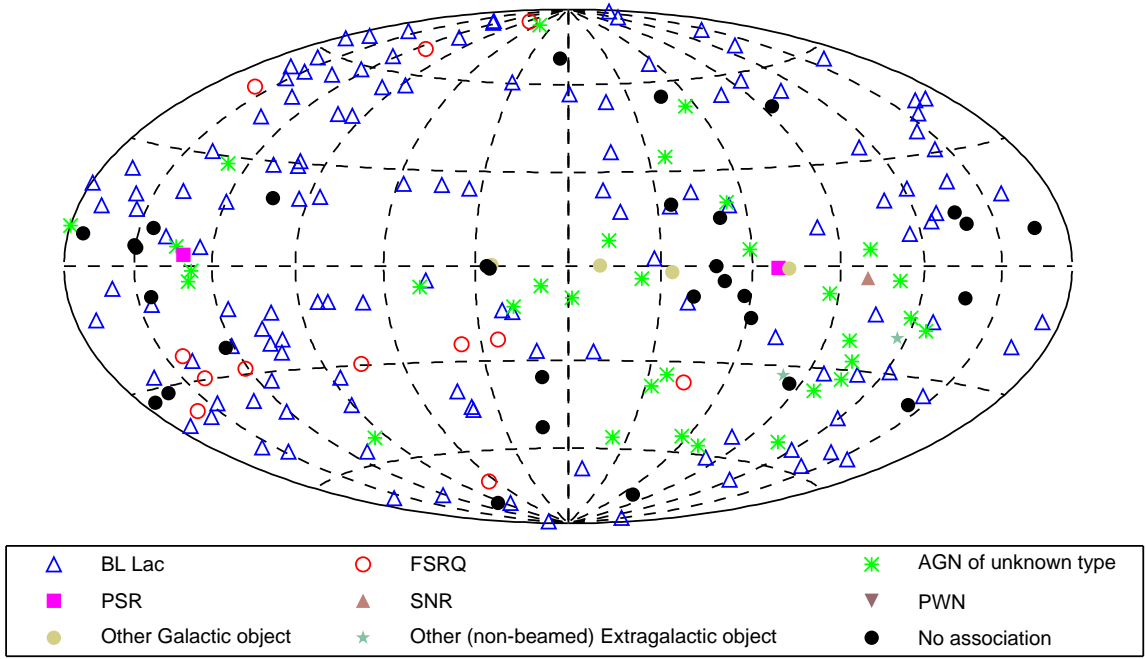}
\end{center}
\caption{\label{SkymapTeVSrcCandidates} 
Sky map showing the 1FHL sources that we identify as good candidates for VHE detection.
The markers represent  the source classes in Table \ref{TableStatisticsSrcClasses}.
{  The projection is Hammer-Aitoff in Galactic coordinates.}
}
\end{figure}

Many of these 1FHL sources should be detectable with the current generation of ground-based $\gamma$-ray instruments. 
As this manuscript was being finalized, two of the sources in the TeV candidates list, associated with MS1221.8+2452 and H1722+119, were detected in the VHE band \citep{2013ATel5080,2013ATel5038}. These were naturally not included in
the initial list of 84 1FHL sources that were detected at VHE, and so we have treated them as part of the 212-source VHE candidate list.  
The LAT detections above 10 GeV are already helping to substantially increase the number of VHE sources.  For example, 1ES~1215+303 was detected by MAGIC in observations initiated based on an early version of this catalog \citep{MAGIC1215}. 
The new generation of ground-based $\gamma$-ray observatories, namely HAWC, LHASSO and especially CTA, with lower energy thresholds and improved sensitivities,  would have an even greater chance to detect a large fraction of the TeV candidates reported here.

\cleardoublepage


\section{Population Studies}
\label{PopStudies}

\def\Breg{B_{reg}}
\def\Bran{B_{ran}}
\def\Bperp{B_{perp}}
\def\Btot{B_{tot}}
\def\B{{\bf B\ }} 
\def\microG{$\mu G$\ }
\def\Rscale{\Delta R}
\def\zscale{\Delta z}
\def\Lgamma{L_\gamma}
\def\Lgammamin{L_{\gamma,min}}
\def\Lgammamax{L_{\gamma,max}}
\def\Sgamma{S_\gamma}
\def\Sgammamin{S_{\gamma,min}}
\def\Sgammamax{S_{\gamma,max}}
\def\Lunits{ph~s$^{-1}$\ }
\def\perLunits{/ph~s$^{-1}$}
\def\Lergunits{erg\ ~s$^{-1}$\ }
\def\Sunits{ph~cm$^{-2}$~s$^{-1}$\ }
\def\kpc3{kpc$^{3}$}
\def\perkpc3{kpc$^{-3}$}
\def\gray{$\gamma$-ray\ }
\def\grays{$\gamma$-rays\ }
\def\emiss{\epsilon_\gamma}
\def\emissunits{ph~cm$^{-3}$~sr$^{-1}$~s$^{-1}$\ }

In the previous sections we reported results for \gray
sources that are significantly detected with {\it Fermi} LAT at energies above 10~GeV in
three years of accumulated data. 
The probability for a  \gray source to be detected at these high energies depends primarily on
its \gray flux and its location with respect to the Galactic plane.
As reported in \S~\ref{LatInstrumentIntro}, 
the PSF of the LAT is essentially independent of energy above 10~GeV, and the diffuse backgrounds are relatively dim (especially at
high Galactic latitudes), and so the detection efficiency does not
depend substantially on the spectral shapes of the sources. 
With the aid of Monte Carlo simulations, we can evaluate the detection
efficiency of the instrument and from source population models infer the true numbers of sources
above a given \gray flux below the detection limit and can infer the contribution of the resolved and
unresolved sources to the diffuse backgrounds. 

Given the substantial differences in the sensitivity of \FermiLAT for
sources located at high/low Galactic latitudes (see \S~\ref{LATExposure}), as 
well as the different natures of extragalactic and
Galactic sources, and of  the extragalactic (isotropic) and the Galactic
(non-isotropic) diffuse backgrounds, we address this problem
separately for extragalactic and Galactic sources in the following subsections.

\subsection{Evaluation of the Extragalactic Source Count Distribution Above 10 GeV}

In this subsection we determine the source-count distribution
(also known as $N(S)$ or log $N$--log $S$) of the $>$10\,GeV extragalactic sky.
Accurate knowledge of $N(S)$ allows us to understand
the contribution of sources to the isotropic $\gamma$-ray background 
\citep[IGRB,][]{lat_edb}, constrain the evolutionary properties of 
blazars \citep{fsrq12}, and predict the number of sources detectable
by future \gray\ instruments.  

We relate the observed flux distribution of sources to the intrinsic properties of the source population,
such as $N(S)$, by accounting for all of the observational
biases that led to the detection of that particular source sample.
Using the approach of \citet{pop_pap}, we performed detailed Monte Carlo simulations
in order to quantify these biases and correct for them.
In short, we performed 
five end-to-end simulations of the LAT sky resembling as closely
as possible the real observations. Each simulation was based on the real
pointing history of the {\it Fermi} satellite during the time spanned
by this analysis and comprises the Galactic and isotropic diffuse emissions
and an isotropic source population.

The isotropic source population is modeled on the basis of properties
of blazars determined in past {\it Fermi} observations.
In particular, each source was modeled with a power-law spectrum
in the 100\,MeV--500\,GeV band with flux and photon index randomly extracted
from the distribution of $N(S)$ and the power-law index distribution found by \citet{pop_pap}.
Each sky realization comprises $>$250,000 sources randomly distributed in the sky. 

Photons of the {\tt P7CLEAN} class in the whole 100\,MeV--500\,GeV band
were generated {  using the } {\tt P7CLEAN\_V6} {  IRFs} and the resulting simulated data
were treated exactly as the real data. This means that only photons
with measured energies $>$10\,GeV and zenith angles $<$105$\degr$ that were detected
during times when the spacecraft rocking angle was less than 52$\degr$
were  retained.  The source detection procedure
was performed as for the real data (see \S~\ref{SeedsAndLocalization}) for
all sources located at $|b|\geq$15$\degr$. We chose 
15$\degr$ Galactic latitude as a good compromise between maximizing source statistics
and minimizing systematic errors in the reconstructed source flux due to the strong
Galactic background \citep{pop_pap}.
In each simulation $\sim$500 sources were detected above 10\,GeV with 
$TS \geq 25$.

Figure~\ref{fig:logn_sim} compares the reconstructed
source fluxes ($Flux^{\rm OUT}$) with the simulated ones ($Flux_{\rm MC}$).
At very low fluxes, 
the fluxes of the few detected sources in the simulation tend to be systematically overestimated.
Due to the relatively low intensity of the diffuse background above 10\,GeV, sources with 
fluxes of  10$^{-10}$\,ph cm$^{-2}$ s$^{-1}$  are significantly detected with $\sim$10 photons. This number reduces to $\sim$4 for the weakest detected sources.
The large number of simulated sources (below the threshold) make it possible
for a number of them to fluctuate above the threshold and be detected.
This effect is often referred to as Eddington bias \citep{Eddington1913,Eddington1940}.  
The faintest source in the 1FHL catalog has a flux of $4.2 \times
10^{-11}$\,ph cm$^{-2}$ s$^{-1}$, for which the bias is about 1.5.
In any case, the efficiencies for source detection are evaluated as a
function of measured (i.e., Flux$^{\rm OUT}$) fluxes, hence automatically accounting for any bias. 

\begin{figure*}[ht!]
  \begin{center}
  \begin{tabular}{c}
    \includegraphics[scale=0.73]{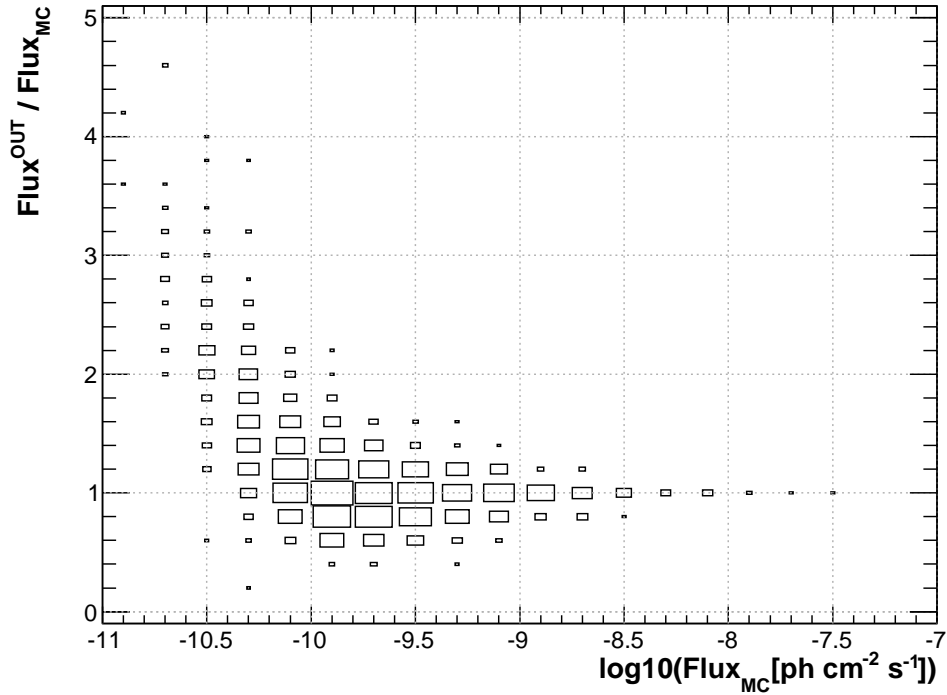} 
\end{tabular}
  \end{center}
  \caption{Ratio of measured to simulated flux versus simulated flux (all above 10~GeV)
for all sources with $TS \geq 25$ and $|b$$|\geq$15$\degr$. 
For each cell the area of the box is proportional to the number of sources contained.
\label{fig:logn_sim}}
\end{figure*}

Figure~\ref{fig:det_eff} shows the efficiency for detecting
(simulated) sources as function of the reconstructed source flux.
The detection efficiency is rather flat and about 100\% above a flux
of 10$^{-10}$\,ph cm$^{-2}$s$^{-1}$; 
{  a constant fit to the data points yields a detection efficiency of (98$\pm$4)\,\%. }
Below this flux, the detection efficiency decreases quickly, and at
a flux of $\sim$4$\times10^{-11}$\,ph cm$^{-2}$ s$^{-1}$, only 3 out of 100
(simulated) sources are detected. 


\begin{figure*}[ht!]
  \begin{center}
  \begin{tabular}{c}
    \includegraphics[scale=0.73]{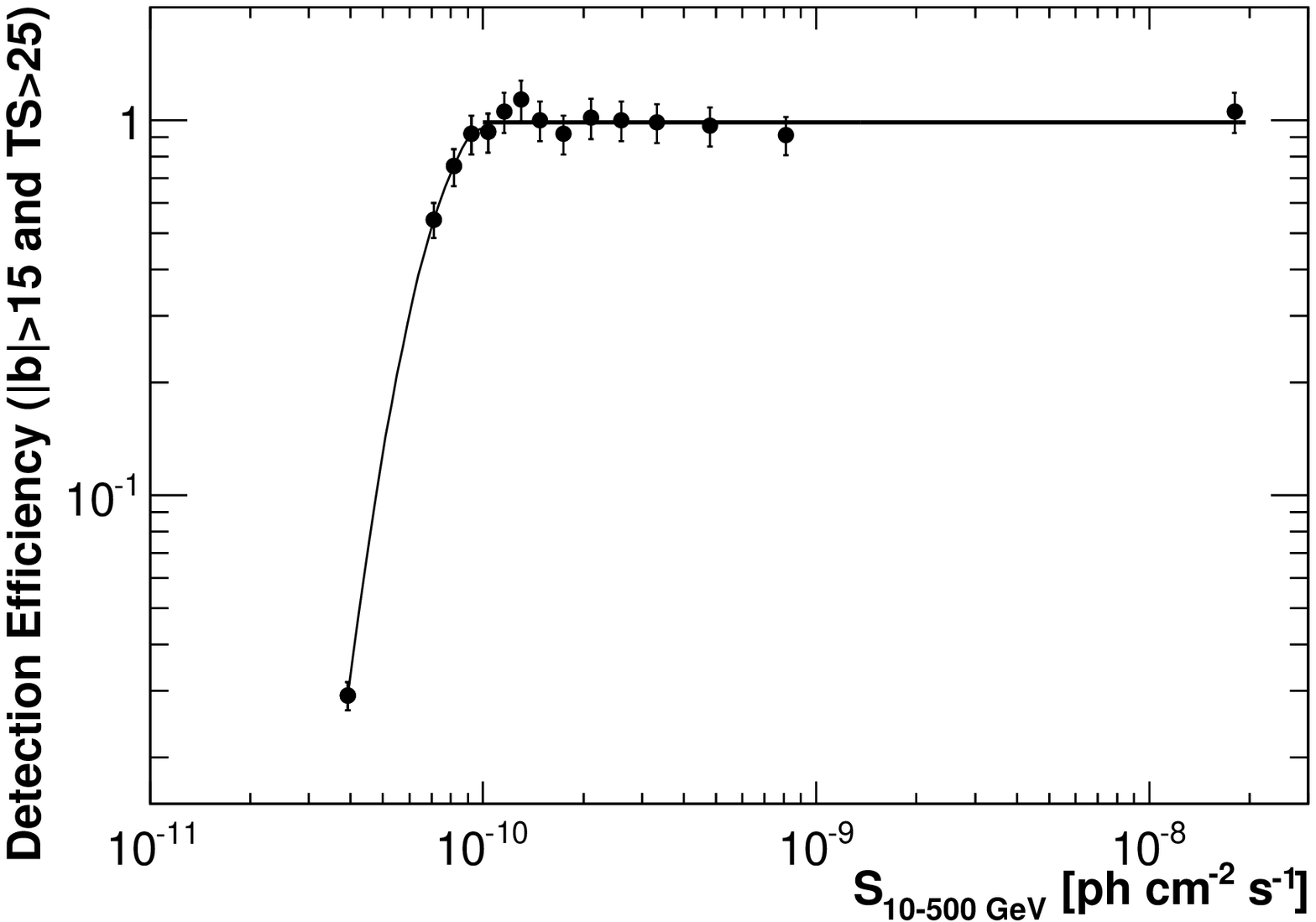} 
\end{tabular}
  \end{center}
  \caption{Detection efficiency as a function of measured flux for
$|b|\geq15\degr$ and  $TS \geq 25$.
The error bars represent uncertainties from the 
counting statistic of our Monte Carlo simulations.  {  The line above the
flux 10$^{-10}$\,ph cm$^{-2}$ s$^{-1}$ results from a fit with a
constant yielding 0.98$\pm$0.04, while below this flux the curve smoothly connects
the data points to guide the eye.}
\label{fig:det_eff}}
\end{figure*}

The source count distribution can be derived as:
\begin{equation}
\frac{dN}{dS} = \frac{1}{\Delta\ S}\
\sum_{i=1}^{N_{\Delta S}} \frac{1}{\Omega_i}
\end{equation}
where $N_{\Delta S}$ is the total number of detected sources with fluxes
in the $\Delta$S interval, and $\Omega_i$ 
 is the solid angle associated
with the flux of the $i_{th}$ source (i.e.,
 the detection efficiency multiplied by the survey solid angle).
For the $|b|\geq15\degr$ sample  the geometric solid angle
of the survey is 9.32~sr.   

In order to parametrize the source count distribution
we perform a maximum likelihood fit to the unbinned differential source counts using 
a simple power-law model: $dN/dS=A (S/10^{-7})^{-\beta}$.
The best-fit parameters are $A$=20.6$^{+6.7}_{-7.0}$
and $\beta$=2.19$^{+0.06}_{-0.04}$ where the errors 
were computed via a bootstrap procedure \citep[see ][]{pop_pap}.

Figure~\ref{fig:logn} shows the differential distribution, with
the power-law fit from the maximum-likelihood analysis (left),
and the cumulative distribution (right).
The cumulative distribution is also compared
to the source counts derived by \citet{pop_pap} for the 10--100\,GeV
band, who used only 11\,months of data.  This comparison required
converting the 10--100\,GeV source counts to the 10--500\,GeV
band, which we did by adopting a power-law spectrum with a photon index
of 2.5 (corresponding to a 3\% increase of flux).
 It is apparent that the new $N(S)$
extends to a factor $\sim$1.8 lower fluxes due to the increased sensitivity.

\begin{figure*}[ht!]
  \begin{center}
  \begin{tabular}{cc}
    \includegraphics[scale=0.42]{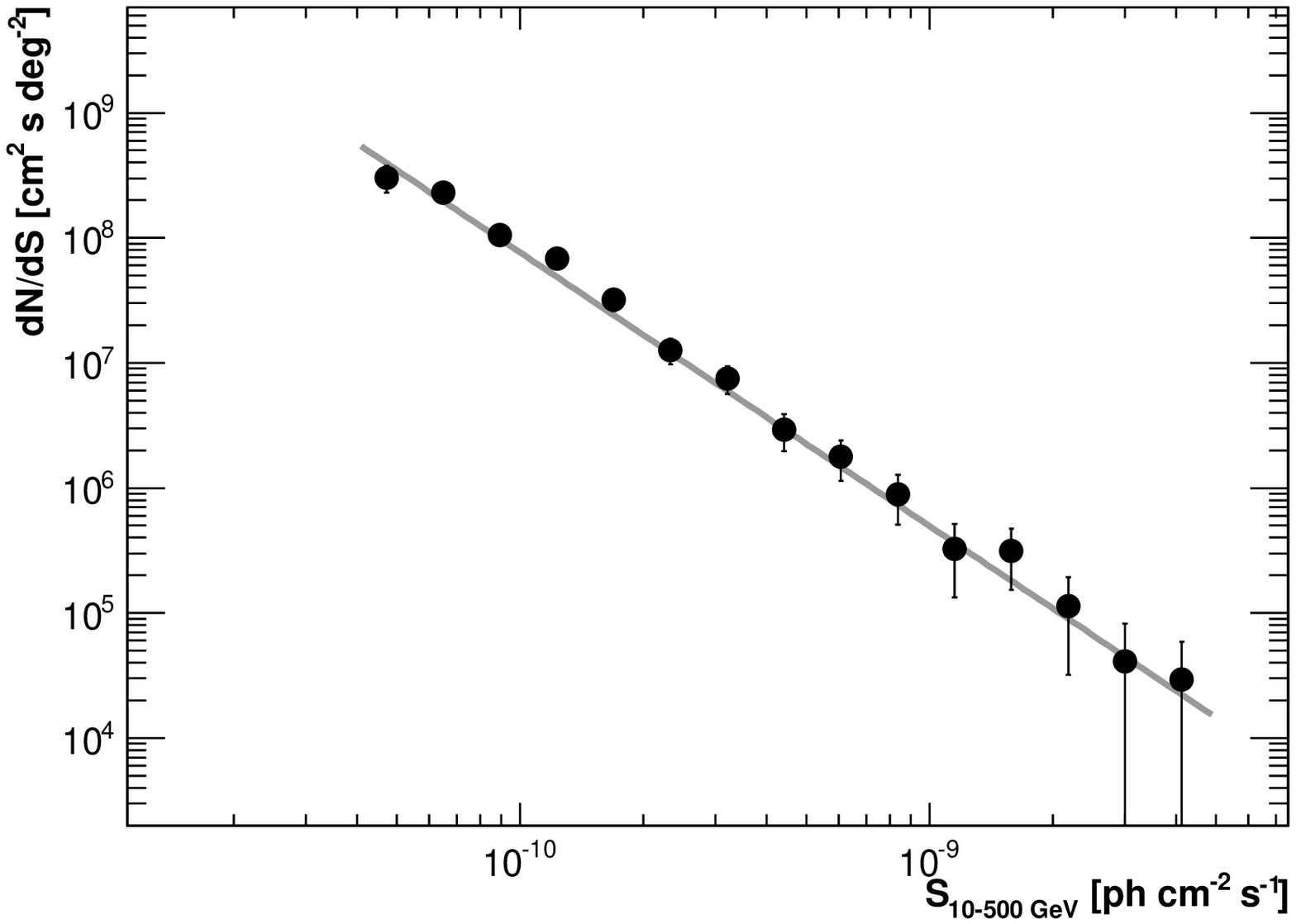} 
   \includegraphics[scale=0.42]{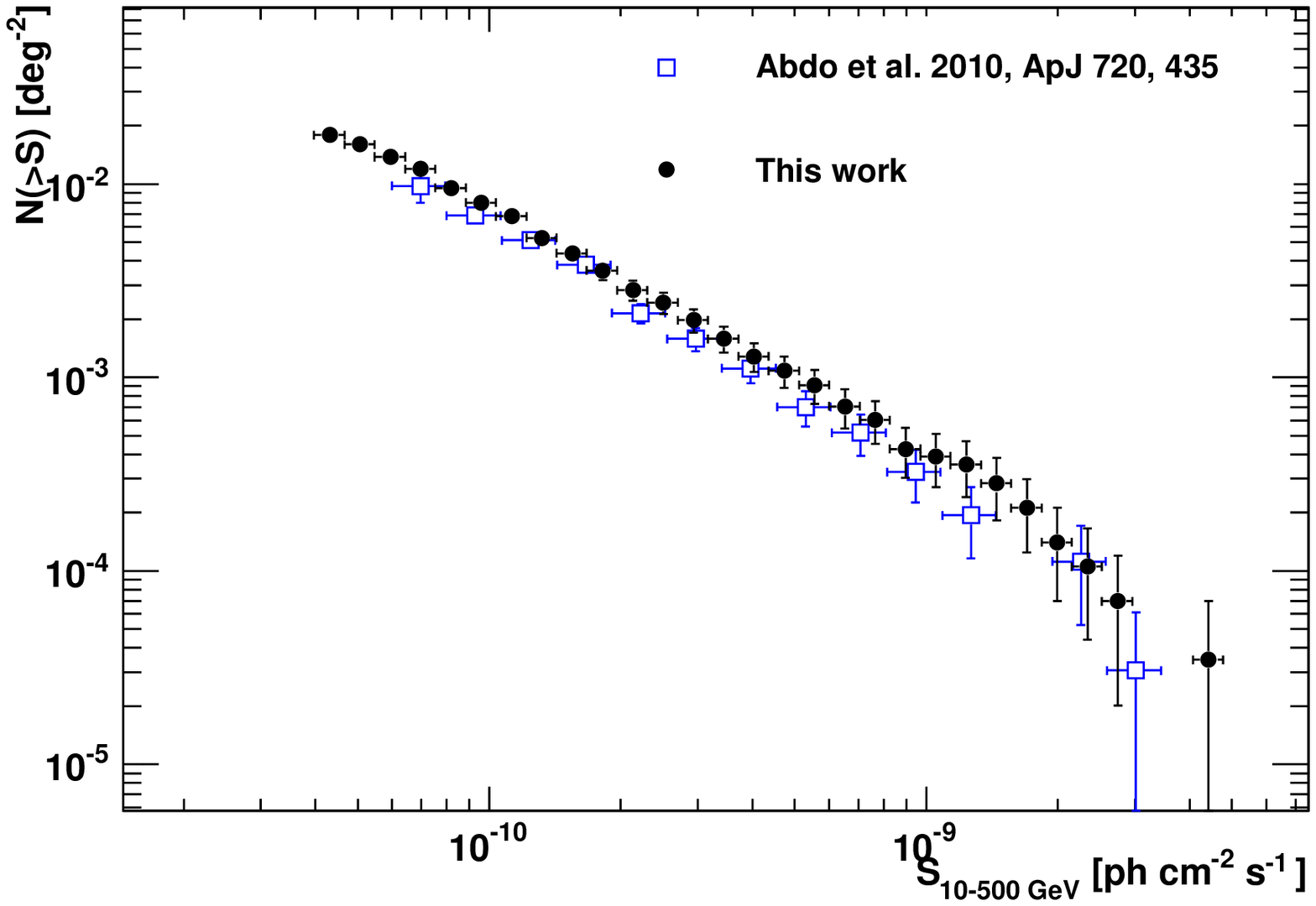} 
\end{tabular}
  \end{center}
  \caption{Left panel: differential $N(S)$ (data points)
and best-fit power-law model (grey line). Right panel: cumulative
$N(S)$ compared to the 10--100\,GeV $N(S)$
derived in \cite{pop_pap} using 11\,months of data (converted to the
10-500\,GeV band).
\label{fig:logn}}
\end{figure*}

As apparent from Figure~\ref{fig:logn}, the source count distribution
is compatible with a power law and does not show any significant flattening
down to the lowest measured fluxes. This is in contrast to the $N(S)$
of the full 100\,MeV--100\,GeV band \citep[see][]{pop_pap} and might have
important consequences for the generation of the IGRB at these high energies (see below).

Since the detection efficiency does not depend on the source spectrum
for energies $>$10\,GeV \citep[see][]{pop_pap}, the same efficiency curve
can be used to derive the source-count distribution of FSRQs and BL Lacs.
Figure~\ref{fig:logn_classes} shows the source counts for the FSRQ and BL Lac
source populations. In the
10--500\,GeV band and at the lowest fluxes measured by {\it Fermi} LAT, BL
Lacs are three times more numerous than FSRQs, reaching a density of
$\sim$0.01\,BL Lac deg$^{-2}$. Therefore, the ratio of the source
counts for BL Lacs and FSRQs (an estimate of the ``true'' relative numbers)
is similar to the ratio of the measured numbers of BL Lacs and FSRQs 
(see \S~\ref{SrcAssociations}). Since the spectral indices of FSRQs are
typically about one unit softer than those of BL Lacs (see
Fig.~\ref{DistributionOfIndexAt100MeVAnd10GeV}), 
this result confirms that, above 10 GeV, the detection
efficiency is not significantly affected by the different spectral
shapes of the sources, as indicated above. 
Moreover, we also note that the
$N(S)$ of BL Lacs does not flatten at the lowest measured fluxes while
that of FSRQs seems to flatten below $\sim$10$^{-10}$\,ph cm$^{-2}$ s$^{-1}$. 

\begin{figure*}[ht!]
  \begin{center}
  \begin{tabular}{cc}
    \includegraphics[scale=0.42]{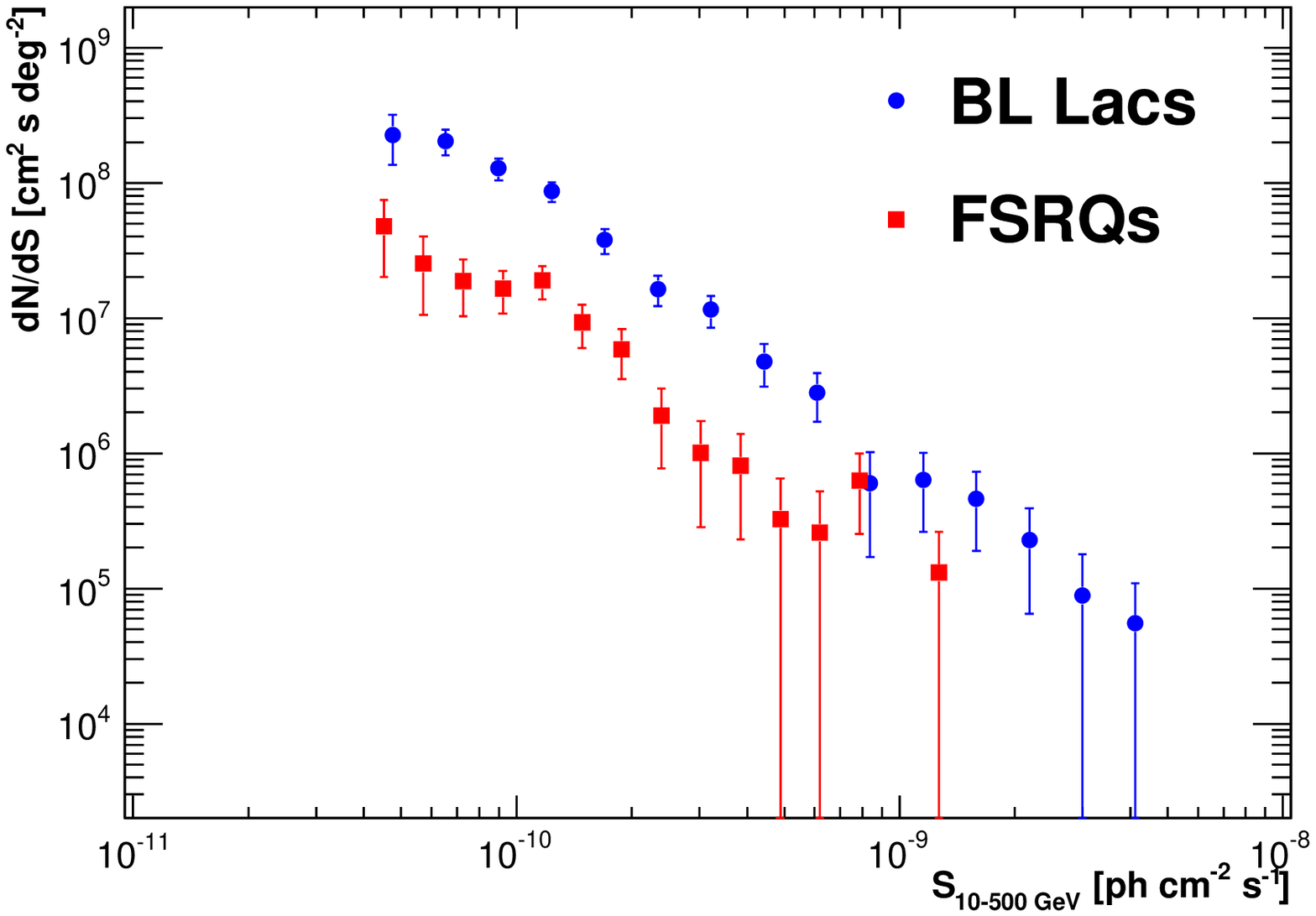} 
    \includegraphics[scale=0.42]{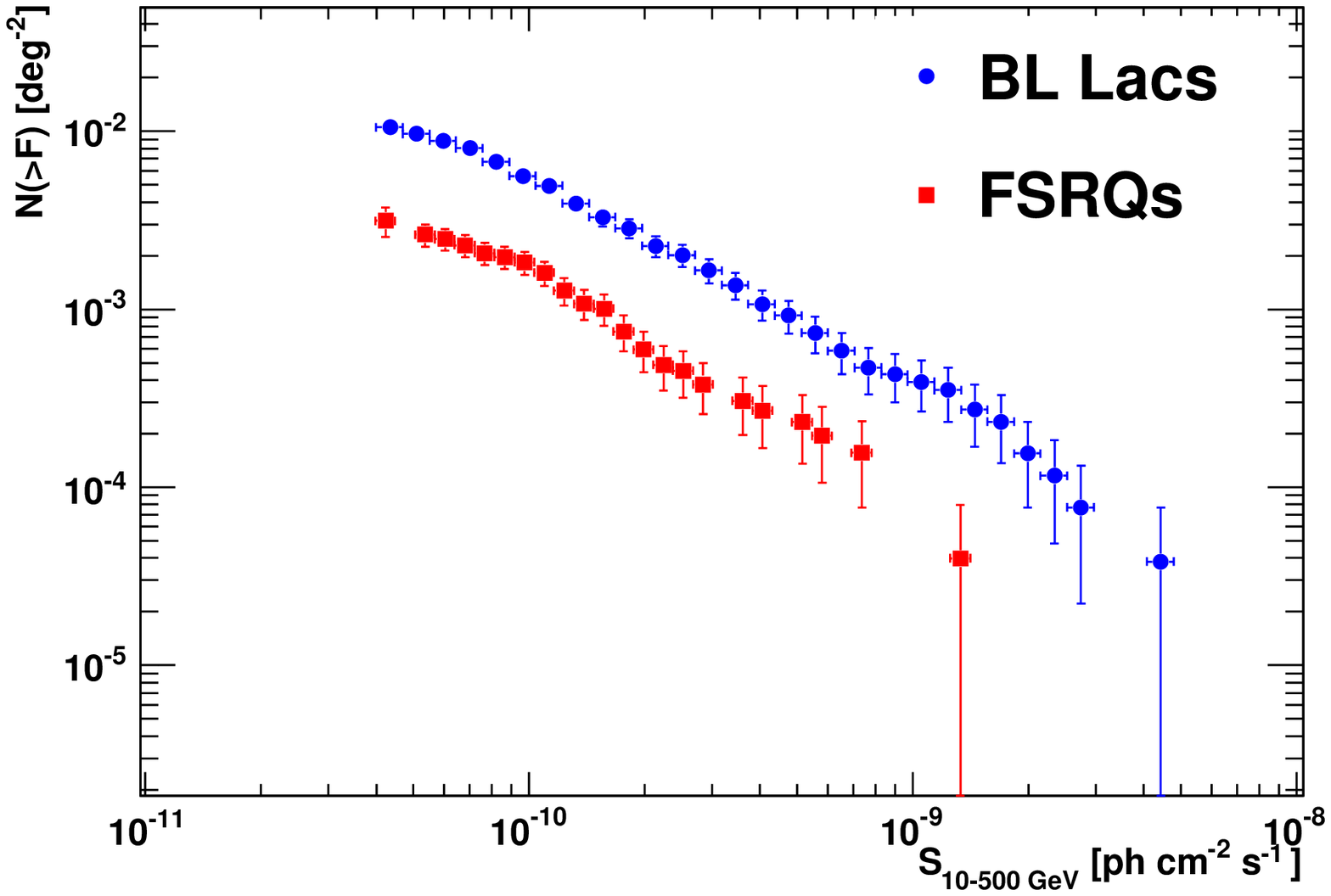} 
\end{tabular}
  \end{center}
\caption{Differential $N(S)$ (left panel) and cumulative $N(S)$
  (right panel) for BL Lacs  and FSRQs. 
\label{fig:logn_classes}}
\end{figure*}

From the source flux distribution we can determine how much of the intensity of the IGRB above
10\,GeV is due to 1FHL sources. The comparison between the $N(S)$
derived here and the IGRB measurement reported by  \citet{lat_edb} is not
straightforward. Indeed, the two works rely on sources detected 
on different timescales and above different thresholds.
Sources used by \citet{lat_edb} were detected with $TS \geq 25$ in the 
0.2--100\,GeV band using nine months of data while those used here
are detected with $TS \geq 25$ in the 10--500\,GeV band using three years of data.

The most straightforward comparison is between the {\it total} sky intensities, 
by which we mean the intensity of the IGRB plus the detected sources.
{  From fitting band-by-band intensities of the IGRB reported in Table 1 of \cite{lat_edb}
  with a power-law function above 10 GeV, and integrating the fitted function in the energy
  range 10--500~GeV, one can find that the intensity of the IGRB in the 10--500\,GeV band
is $(1.5 \pm 0.3) \times10^{-8}$\,ph cm$^{-2}$ s$^{-1}$ sr$^{-1}$.
The resolved sources account for a further
$(0.8 \pm 0.1) \times10^{-8}$\,ph cm$^{-2}$ s$^{-1}$ sr$^{-1}$, and the 
the {\it total} isotropic intensity in the 10--500\,GeV band is 
$(2.3 \pm 0.3) \times10^{-8}$\,ph cm$^{-2}$ s$^{-1}$ sr$^{-1}$.}

The diffuse flux produced by an {\it unresolved} source population can be obtained as:
\begin{equation}
S_{\rm diffuse} = \int^{S_{\rm max}}_{S_{\rm min}}dS  \frac{dN}{dS}
S \left ( 1-\frac{\Omega(S)} {\Omega_{\rm max}} \right )
\label{eq:diff}
\end{equation}
where $\Omega_{\rm max}$ is the geometrical sky area and the 
$\Omega(
S)/\Omega_{\rm max}$ term (which is the detection
efficiency reported in Figure~\ref{fig:det_eff}) takes into account 
the dependency of 
the LAT source detection efficiency on  the source
flux.

Setting $\Omega(S)/\Omega_{\rm max}=0$ allows us to evaluate the {\it total}
diffuse flux including resolved sources.
Integrating the $N(S)$ to the minimum observed flux of
4.2$\times10^{-11}$\,ph cm$^{-2}$ s$^{-1}$ we obtain $S_{total}=
(10.6 \pm 1.0)\times10^{-9}$\,ph cm$^{-2}$ s$^{-1}$ sr$^{-1}$, where
the error was computed through a bootstrap procedure, following \cite{pop_pap}.
This shows that 1FHL sources account for about half of the {\it total}
(IGRB plus sources) sky intensity in the energy band 10--500 GeV.

Most of the comparisons presented in the literature refer to the diffuse
emission arising from {\it unresolved} sources. Using 
Eq.~\ref{eq:diff}, the flux from the  {\it unresolved} sources can be computed to be 
$3.9^{+0.8}_{-0.6}\times10^{-9}$\,ph cm$^{-2}$ s$^{-1}$ sr$^{-1}$, where
the uncertainty is primarily due the statistical and systematic uncertainties
of the detection efficiency, and the contribution from the statistical uncertainty
of the bootstrap procedure is minor \citep{pop_pap}.

As a consistency check, one can compare the resolved source flux 
determined in different ways. The true flux of sources detected in this work,
which can be obtained  by averaging the fluxes of all the
$|b|\geq15\degr$ detected sources, amounts to 
$(8.2 \pm 0.1) \times10^{-9}$\,ph cm$^{-2}$ s$^{-1}$ sr$^{-1}$, 
and is compatible with the number derived from \cite{lat_edb} that was
reported above.
Additionally, one can also derive the source flux by 
subtracting from the total diffuse flux
($(10.6 \pm 1.0) \times 10^{-9}$\,ph cm$^{-2}$ s$^{-1}$
sr$^{-1}$) the unresolved source flux ($3.9 \pm 0.8\times10^{-9}$\,ph
cm$^{-2}$ s$^{-1}$ sr$^{-1}$), obtaining $(6.7 \pm 1.3)\times10^{-9}$\,ph
cm$^{-2}$ s$^{-1}$ sr$^{-1}$, which is comparable to the above-mentioned
estimates.

In conclusion, {\it unresolved} 1FHL sources with
$S\geq4\times10^{-11}$\,ph cm$^{-2}$ s$^{-1}$ sr$^{-1}$ account for 
$3.9^{+0.8}_{-0.6}\times10^{-9}$\,ph cm$^{-2}$ s$^{-1}$ sr$^{-1}$,
which is $27\pm 8$\,\% of the IGRB emission above 10~GeV reported in
\cite{lat_edb}.   We note that this
contribution to the IGRB is substantially larger than the 9\,\% lower limit reported by \cite{pop_pap} and is, in large part, due to the increased sensitivity 
(this $N(S)$ samples a factor $\sim$2 weaker fluxes), and also
to a better treatment of the {\it resolved} source flux. It is reasonable to expect that 1FHL sources produce an even
larger fraction of the diffuse emission than found in the earlier work since Eq.~\ref{eq:diff} has
been integrated only to the lowest flux observed, and the
$N(S)$ does not yet show any strong flattening.

\subsection {Galactic Sources} 


Here we analyze the population of Galactic sources to estimate the contribution of unresolved sources to the Galactic `diffuse' emission, following the method of \citet{Strong2007}.  We adopt a plausible reference model for the space density and luminosity function of Galactic sources and investigate the sensitivity of the results to the assumptions of the model.


The luminosity function at Galactocentric distance $R$ and distance from Galactic plane $z$
is the space density of sources per unit luminosity $\rho(\Lgamma,R,z)$.  
After \citet{Strong2007} we assume that the luminosity function depends on luminosity as $\Lgamma^{-\alpha}$ for  $\Lgammamin< \Lgamma< \Lgammamax$ and is zero outside these
limits.
The total space density of sources is  $\rho(R,z)=\int\rho(\Lgamma,R,z)\
d\Lgamma$, 
which we normalize to the value $\rho_\odot$ at $(R, z) = (R_\odot, {\bf 0})$. 
For a source of luminosity $\Lgamma$ at distance $d$ the flux is $\Sgamma=\Lgamma/4\pi d^2$.  
The differential source counts
are defined as $N(\Sgamma)$ sources per unit flux  
over the area of sky considered.   At lower $\Sgamma$, both the luminosity function and the spatial boundaries
influence $N(\Sgamma)$.
In practice the sources are binned in log($\Sgamma$) so that 
plotted distributions are proportional to $\Sgamma N(\Sgamma)$.

For $\Sgamma$ large enough that the spatial boundaries of the distribution have no influence on the detectability, the well-known relations $N(\Sgamma)\propto \Sgamma^{-5/2},\  \Sgamma^{-2}$ hold for 3-dimensional and 2-dimensional spatial source distributions, respectively,
independent of the shape of the luminosity function\footnote{Standard proof for uniform space density: 
for number $n(<R$) within  distance $R$ with luminosity $L$, $S\propto L/R^2$, $N(S)=dn/dS=dn/dR\times dR/dS = S^{-3/2} dn/dR$;
 3-dim. volume: $n\propto R^3,\ dn/dR\propto R^2\propto S^{-1}  \rightarrow N(S)\propto S^{-5/2}$;
 2-dim. disk:   $n\propto R^2,\ dn/dR\propto R\propto   S^{-1/2}\rightarrow N(S)\propto S^{-2}$.
 Integrating over a luminosity function $\rho(L)$  does not affect the dependence on $S$; hence the shape is independent of $\rho(L)$.
 This is valid when boundaries do not influence $N(S)$. 
 At lower $\Sgamma$, both the luminosity function and the spatial boundaries influence $N(\Sgamma)$.
 For a boundary at $R_{max}$, there will be a cutoff in $N(S)$ below $S=L_{min}/4\pi R_{max}^2$, where $L_{min}$ is the minimum luminosity contributing to  $\rho(L)$. }.
These apply to low luminosity /  high space density (quasi-isotropic) and high luminosity / low space density  (Galactic plane) populations respectively.

As in \citet{Strong2007} we use standard Monte Carlo techniques to 
sample  $\rho(\Lgamma,R,z)$ throughout  the  Galaxy.
using oversampling to reduce statistical fluctuations if necessary.
In these simulations we did not vary the source spectra because we did not consider spectral information in source detection, only the flux $>$~10 GeV.
We use the sources generated from such simulations to form simulated catalogs extending below the 1FHL flux limit and compare the flux distributions with the observations.


Our reference model for the luminosity function has  $\rho_\odot= 3$  kpc$^{-3}$, 
 and $\Lgamma^{-1.5}$ dependence on luminosity 
in the range $10^{34}$--$10^{37}$ \Lunits  above 10~GeV.  
The luminosity law is discussed in \citet{Strong2007}; the exact form is not critical and   will not be further addressed here.
The distribution {  in Galactocentric distance} is based on the model of  \citet{Lorimer2006} for the distribution of pulsars, taken as representative of Galactic sources. {  We adopt an exponential scale height
of 500~pc; the source count distribution $N(\Sgamma)$ depends only weakly on the scale height.}
This distribution peaks near  $R=4$ kpc and falls to zero at $R=0$;
it was chosen for illustration and has not been optimized for the 1FHL source counts.

Figures~\ref{PopulationGal_v3} and~\ref{PopulationGal_v4}  compare the simulated $N(\Sgamma)$ with the observed flux distributions of 1FHL sources at  low latitudes  ($|b|<10\degr$)
 and high latitudes  ($|b|>10\degr$),  respectively.
The unassociated sources at low latitudes are a mixture of Galactic and AGN sources, although the proportion is unknown.
The pure Galactic, and the combined Galactic and unassociated sample, can therefore be used to test the models.  
The reference model is consistent with the low-latitude source counts, having the observed dependence on flux above the source detection threshold; the slope
reflects the spatial distribution (independent of the shape of the luminosity function)
above $10^{-9}$ ph cm$^{-2}$ s$^{-1}$, while the distributions for both the model and observed source counts flatten at lower fluxes. 

Figure~\ref{PopulationGal_v3} shows that the distribution of simulated sources (in the reference model) continues down to fluxes $\sim100$ times below the detection threshold, the cutoff being due to the finite spatial 
extent of the Galaxy.
 The ratio of total flux below threshold to above threshold is 0.3, which gives an estimate of the contribution of the undetected sources 
to the `diffuse' emission (see below).

Figure~\ref{PopulationGal_v3}  also presents the source count distributions for  identified and associated Galactic sources only, indicating a reduction of low-flux sources relative to the counts distributions that also include unassociated sources. Pulsars are also shown separately; they account for about half of these sources, and this shows how their contribution compares with the unassociated ones.
The similarity of the observed $N(\Sgamma)$ for the total (non-blazar) and unassociated sources is consistent with their being similar populations.

In Figure~\ref{PopulationGal_v4} the reference model is seen to be consistent with the high-latitude $N(\Sgamma)$, since it lies below the observed source counts (which contain unidentified AGNs).
{
The identified high-latitude Galactic sources (all pulsars) are under-predicted by a factor 3 (but there are only 5 sources in the sample).
A higher density of Galactic sources would improve the agreement, and retain consistency with the low-latitude counts if the luminosities are correspondingly decreased, for example with $\rho_\odot$= 10  kpc$^{-3}$ and   $4\times 10^{33}$--$4 \times 10^{36}$ \Lunits.  This case is shown in  Figure~\ref{PopulationGal_v5} (upper row).
This model fits the Galactic sources at both low and high latitudes and is therefore another  possible combination of parameters consistent with the data.
Large deviations from these values  are excluded by the combination of low and high-latitude $N(\Sgamma)$.
We note that the full quoted luminosity range is required, the low end by high-latitude nearby low-luminosity sources,
 the high end by low-latitude distant high-luminosity sources.
 Therefore the  contribution to the unresolved emission from sources below threshold at low latitudes in  Figures~\ref{PopulationGal_v3} and \ref{PopulationGal_v5}
 is a necessary consequence of the observed $N(\Sgamma)$.

Although most high-latitude unassociated sources are probably AGNs,
a fraction may be pulsars or other objects, implying a greater density of Galactic sources.
To illustrate this, we increase the source density so that 30\% of the unassociated high-latitude sources are Galactic sources (Fig.~\ref{PopulationGal_v5}, lower row);
to satisfy the low-latitude $N(\Sgamma)$ the luminosity range has to be decreased to  $1.5 \times 10^{33}$--$1.5 \times 10^{36}$ ph s$^{-1}$.
In this case the  contribution to the unresolved emission from sources below threshold at low latitudes is larger (see below).



Using the reference model, we evaluate the contribution to the observed $\gamma$-ray intensity ($>$10~GeV) at low latitudes ($|b|<10\degr$, all longitudes).
Here we adopt a detection threshold of  $5 \times 10^{-10}$~\Sunits (\S~\ref{LATExposure}).
For the reference model shown in Figures~\ref{PopulationGal_v3} and \ref{PopulationGal_v4},
20\% of the emission is contributed by sources below the threshold. 
The total flux is  $7 \times 10^{-8}$ and $2 \times 10^{-8}$  \Sunits from above and below this threshold, respectively.
The total `diffuse' flux observed by {\it Fermi}-LAT from this region is  $\sim$$8 \times 10^{-7}$  \Sunits \citep{LATDiffuse2012}.
Hence about 2.5\% of the Galactic `diffuse' emission is from undetected sources.
For the `higher density' model shown in Figure~\ref{PopulationGal_v5} (upper row),
30\% of the emission is contributed by sources below the threshold, 
increasing to $\sim$4\% the fraction of `diffuse' emission from undetected sources.
For the `maximum density' model shown in Figure~\ref{PopulationGal_v5} (lower row),
50\% of the emission is contributed by sources below the threshold, and the contribution of unresolved sources to the overall Galactic `diffuse' emission is $\sim$8\%.
These results are similar to previous estimates at lower energies \citep{Strong2007}, but this is the first time a value for  $>$10~GeV has been derived.

A similar approach to using source counts to constrain the pulsar contribution to the inner Galaxy emission has been given by \cite{Hooper2013},
concluding that pulsars cannot account for the GeV excess.
A study of the MSP contribution to the Galactic emission, for energies above 100~MeV, has been given by \cite{Gregoire2011,Gregoire2013};
the contribution is at the few percent level.

Finally we consider the global picture.
The total luminosity of the source population $>$10~GeV based on the reference  model is
 $2.6 \times 10^{38}$ \Lunits or about $4 \times 10^{36}$ \Lergunits 
compared to the total luminosity of the Galaxy from interstellar emission in this range:
$3 \times 10^{39}$ \Lunits or  $5 \times 10^{37}$ \Lergunits  \citep{Strong2010}.
Point sources, resolved or not, therefore contribute at the several percent level to the total luminosity of the Galaxy, with a correspondingly larger contribution for the higher-density models.

\begin{figure}[th]
\begin{center}
\includegraphics[width=6.5cm]{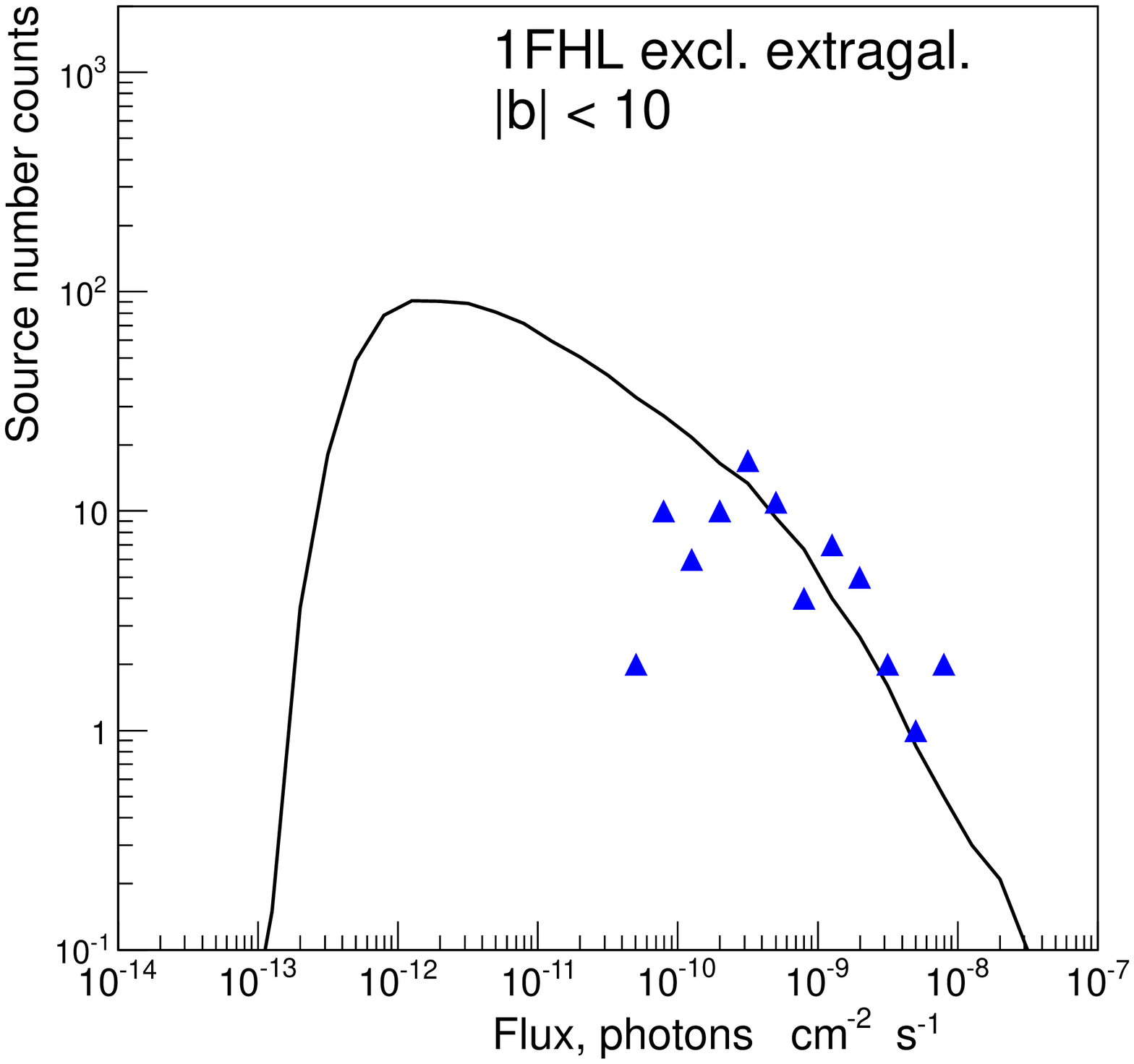} 
\includegraphics[width=6.5cm]{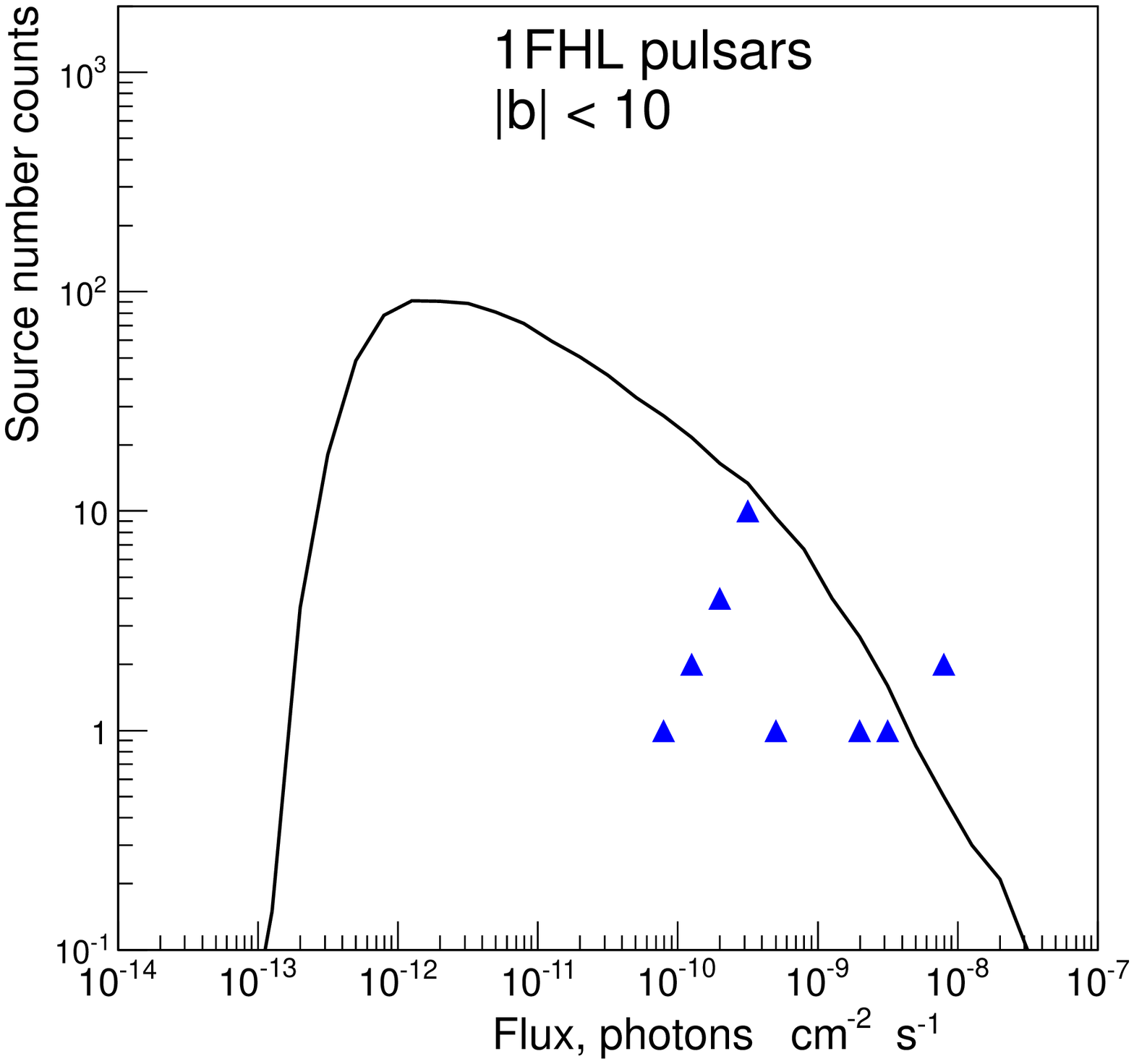} 
\includegraphics[width=6.5cm]{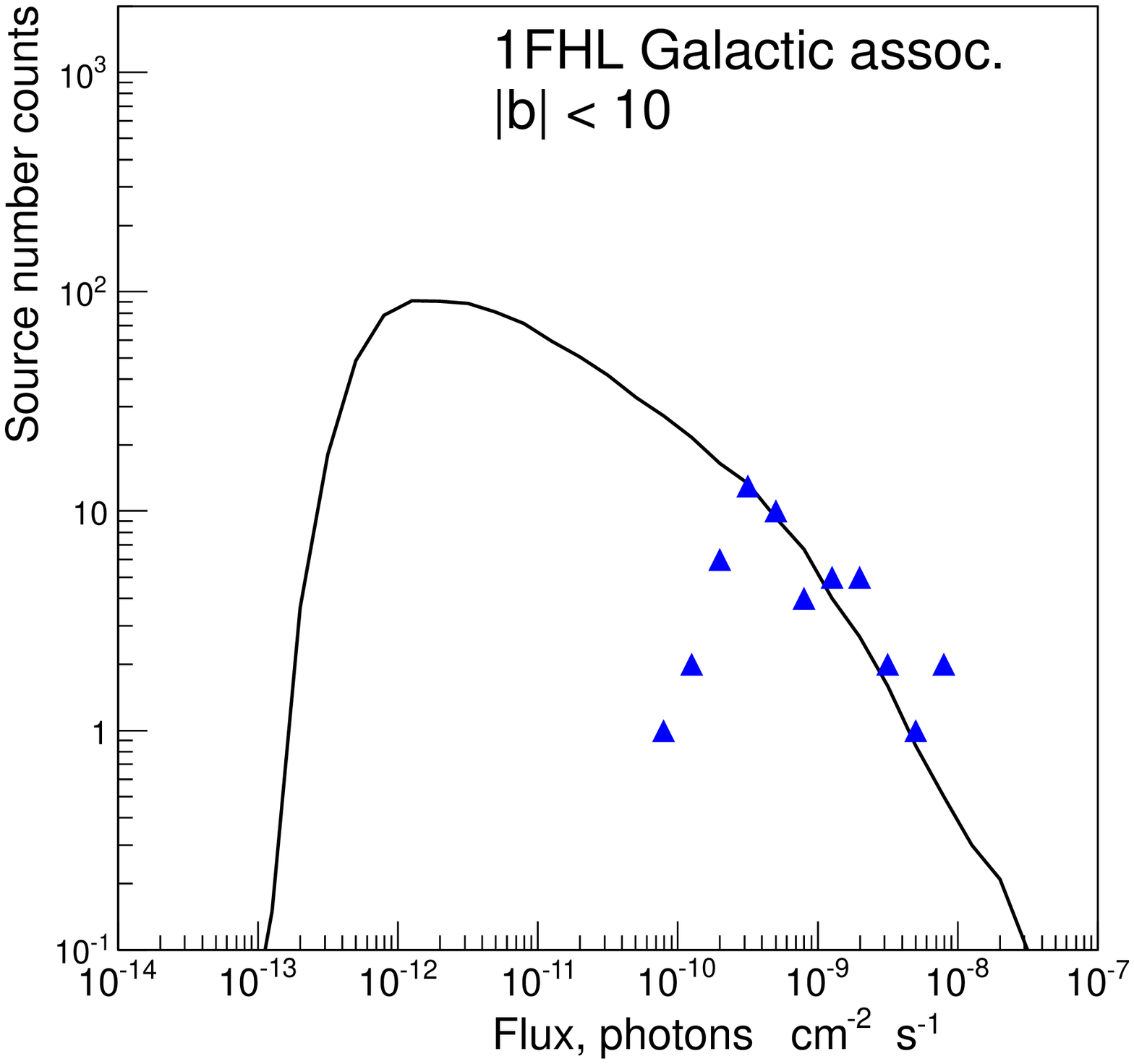} 
\includegraphics[width=6.5cm]{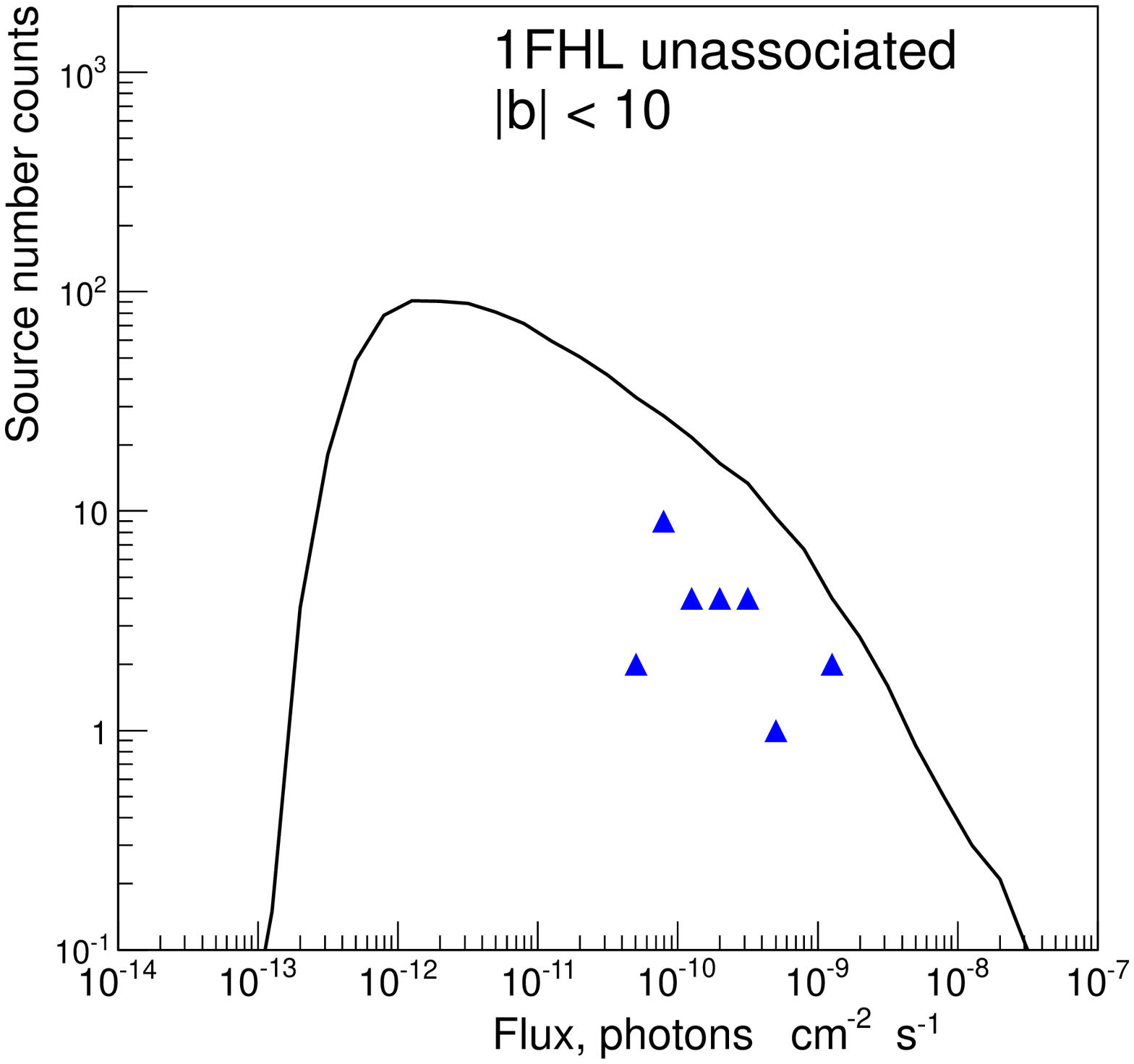} 
\end{center}
\caption{\label{PopulationGal_v3} 
{  Low-latitude ($|b|<10\degr$) source number counts for Galactic and unassociated sources above 10 GeV compared with the reference model described in the text.}
The blue triangles are source counts from the 1FHL catalog; 
Top left: all 1FHL sources, excluding those associated with extragalactic sources,
right:    1FHL pulsars,
Bottom left :   {  all 1FHL sources with Galactic associations (including pulsars),}
right:  unassociated sources.
}
\end{figure}

\begin{figure}[th]
\begin{center}
\includegraphics[width=6.5cm]{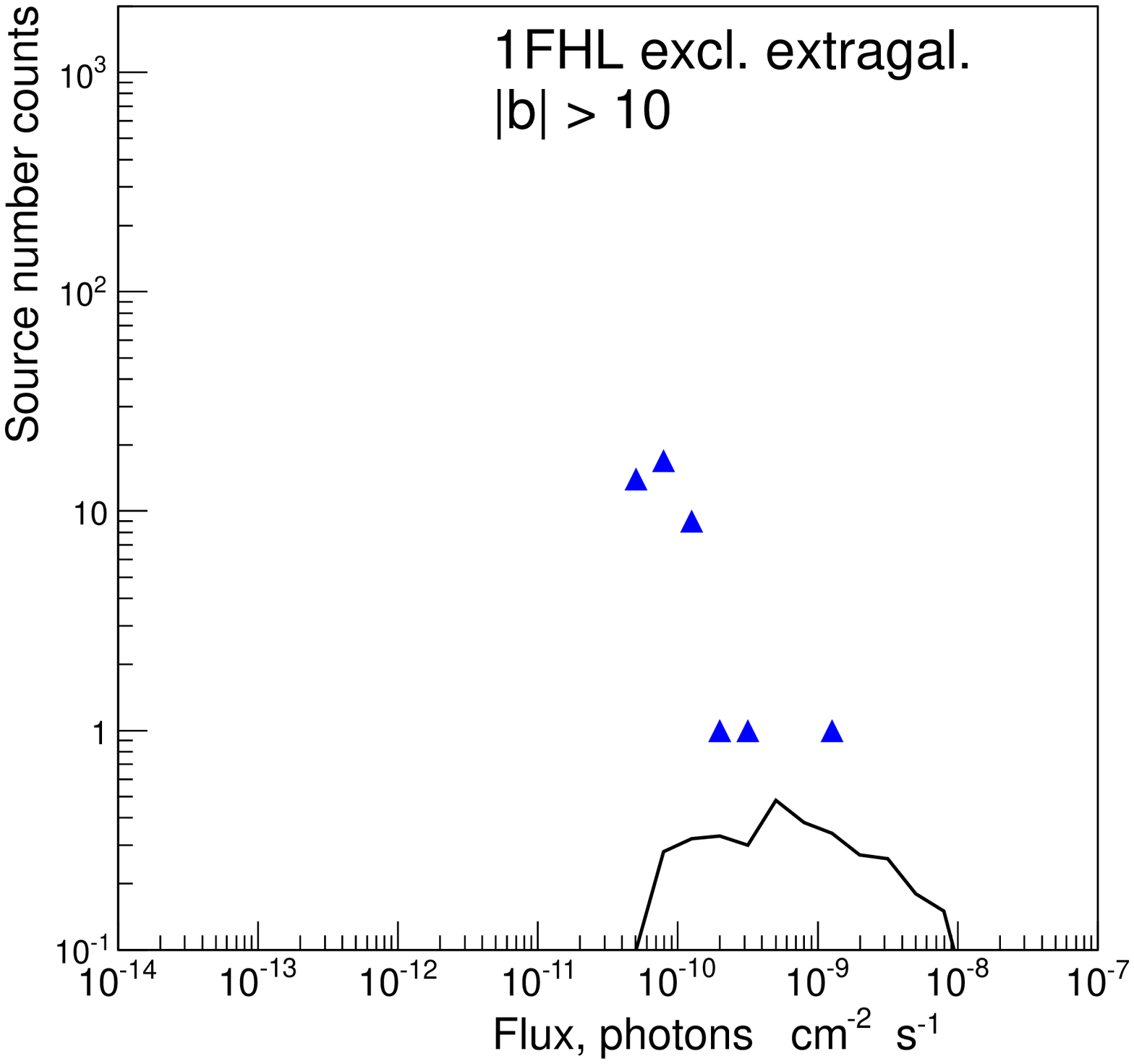} 
\includegraphics[width=6.5cm]{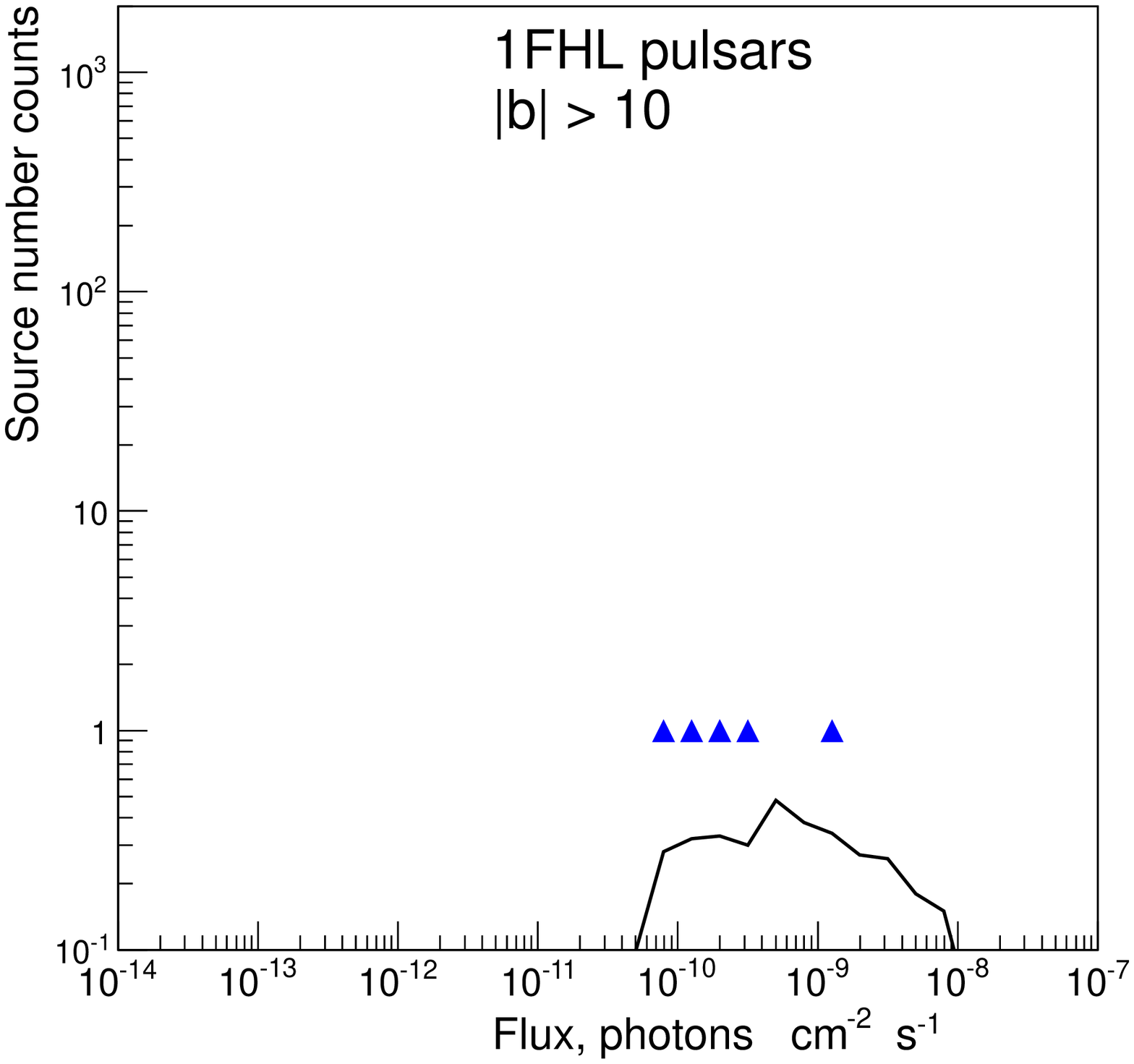} 
\includegraphics[width=6.5cm]{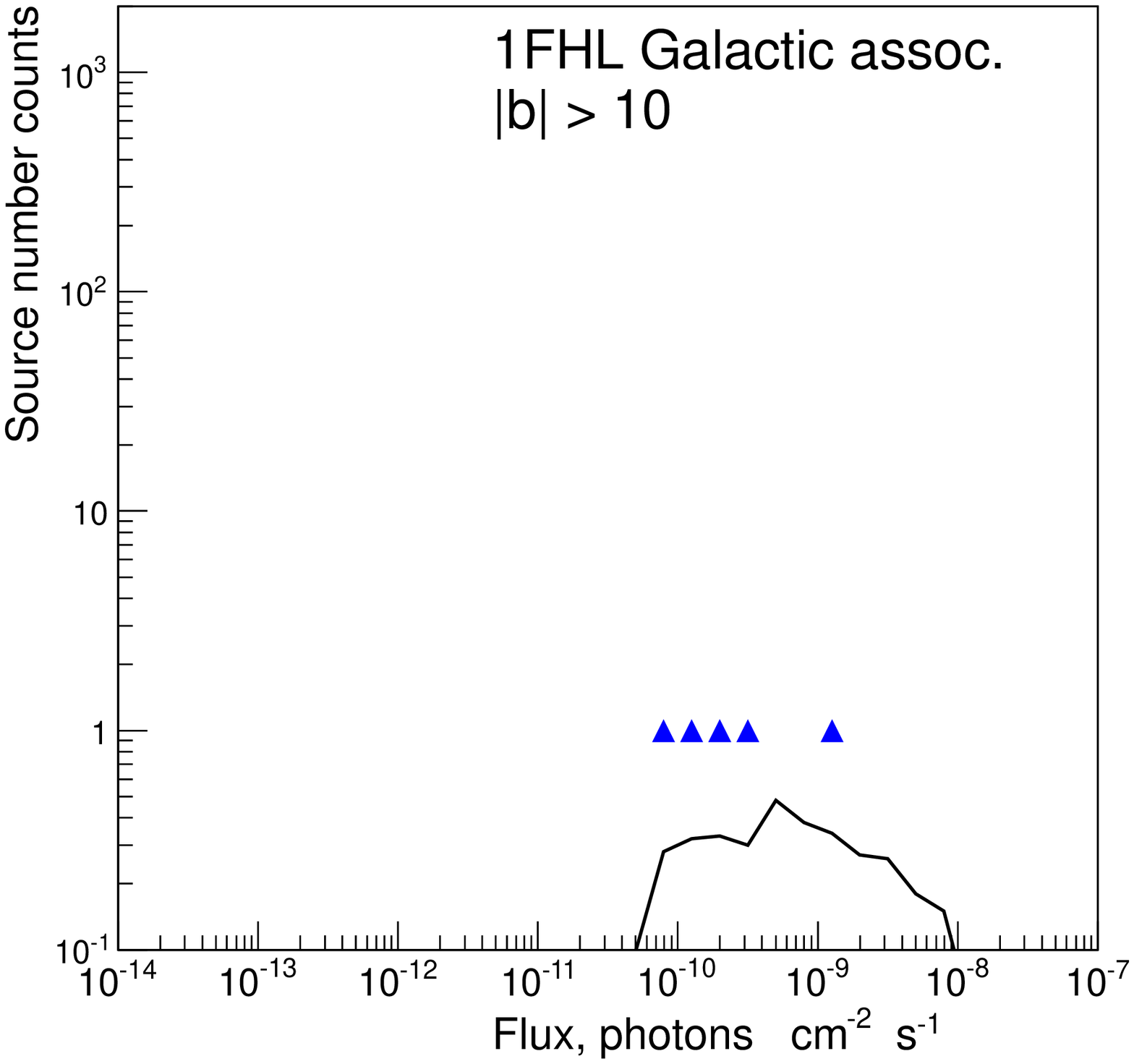} 
\includegraphics[width=6.5cm]{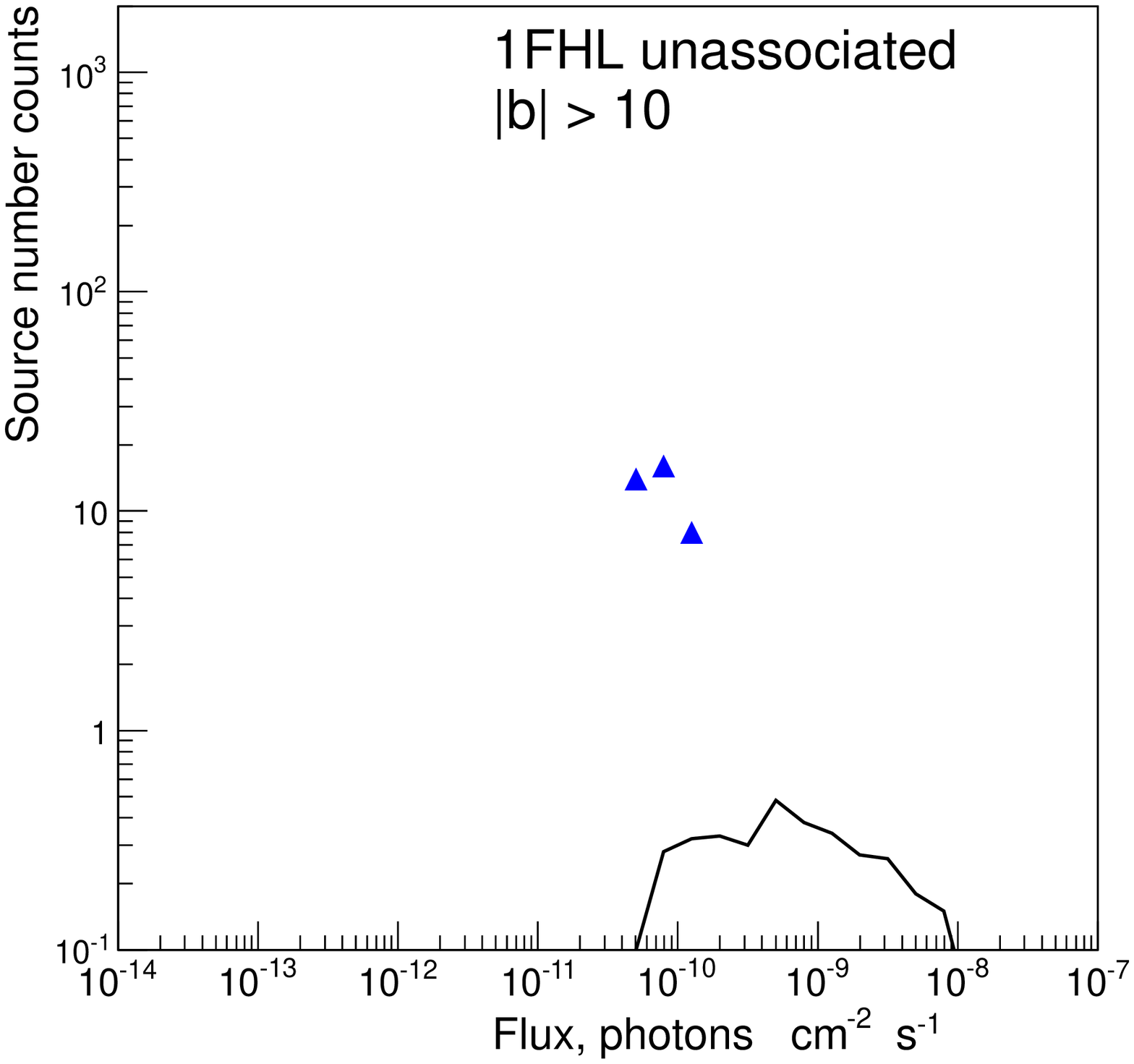} 
\end{center}
\caption{\label{PopulationGal_v4} 
{  High-latitude ($|b|>10\degr$) source number counts for Galactic and unassociated sources above 10 GeV compared with the reference model described in the text.}
The blue triangles are source counts from the 1FHL sources; 
Top left: all 1FHL sources, excluding those associated with extragalactic sources,
right:    1FHL pulsars,
Bottom left :  {  all 1FHL sources with Galactic associations (including pulsars),}
right:  unassociated sources.
}
\end{figure}

\begin{figure}[th]
\begin{center}
\includegraphics[width=6.5cm]{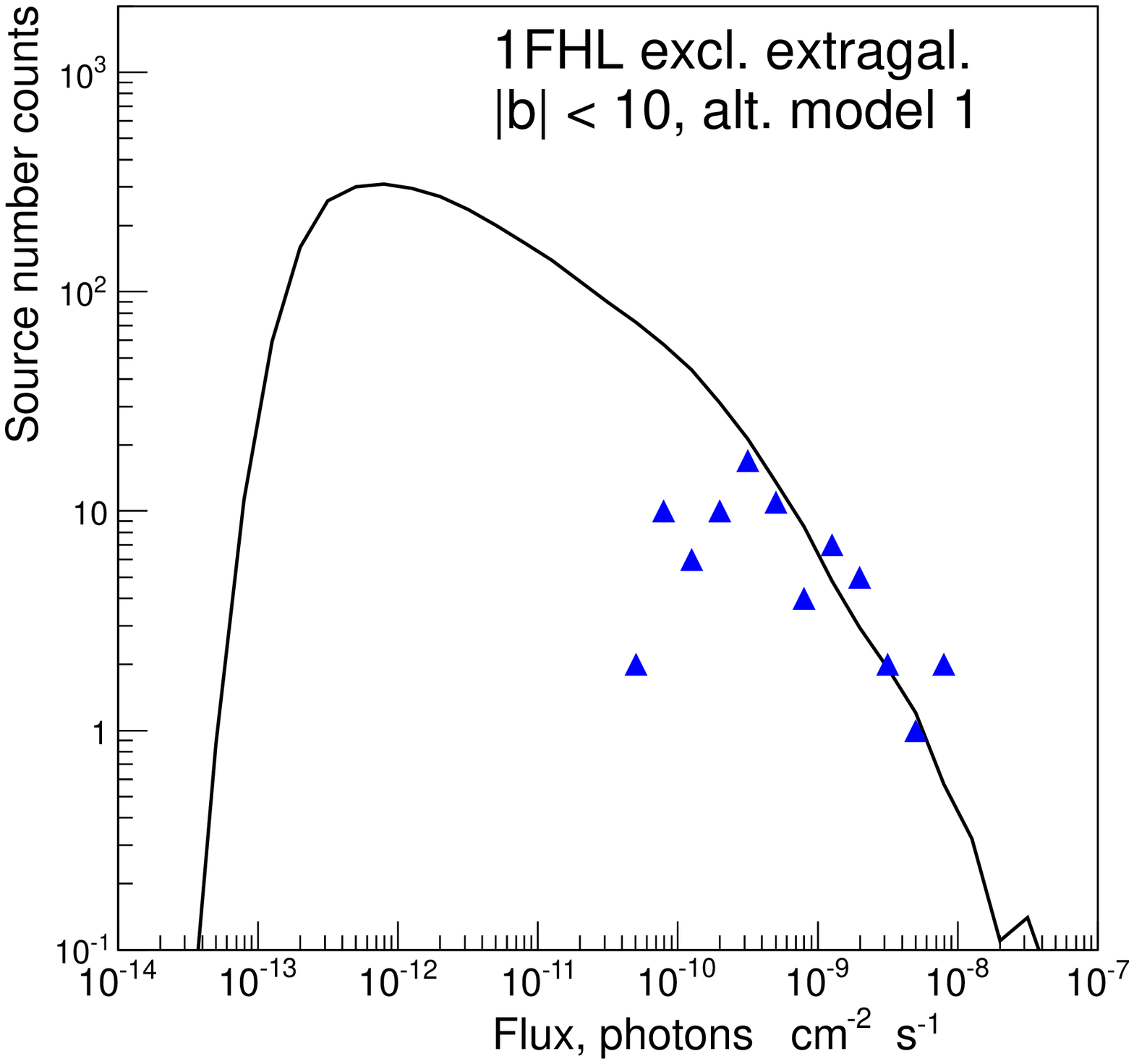} 
\includegraphics[width=6.5cm]{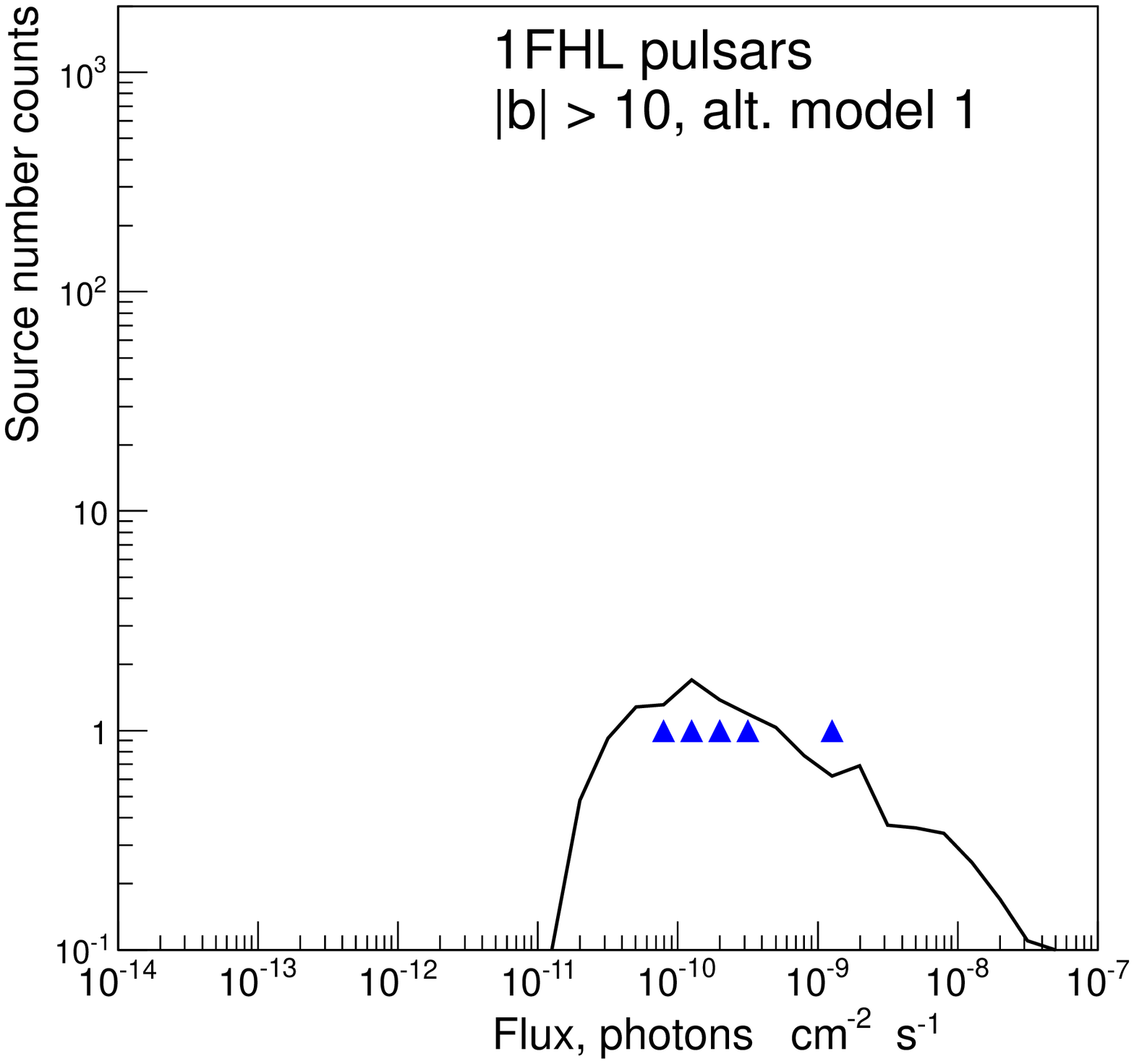} 
\includegraphics[width=6.5cm]{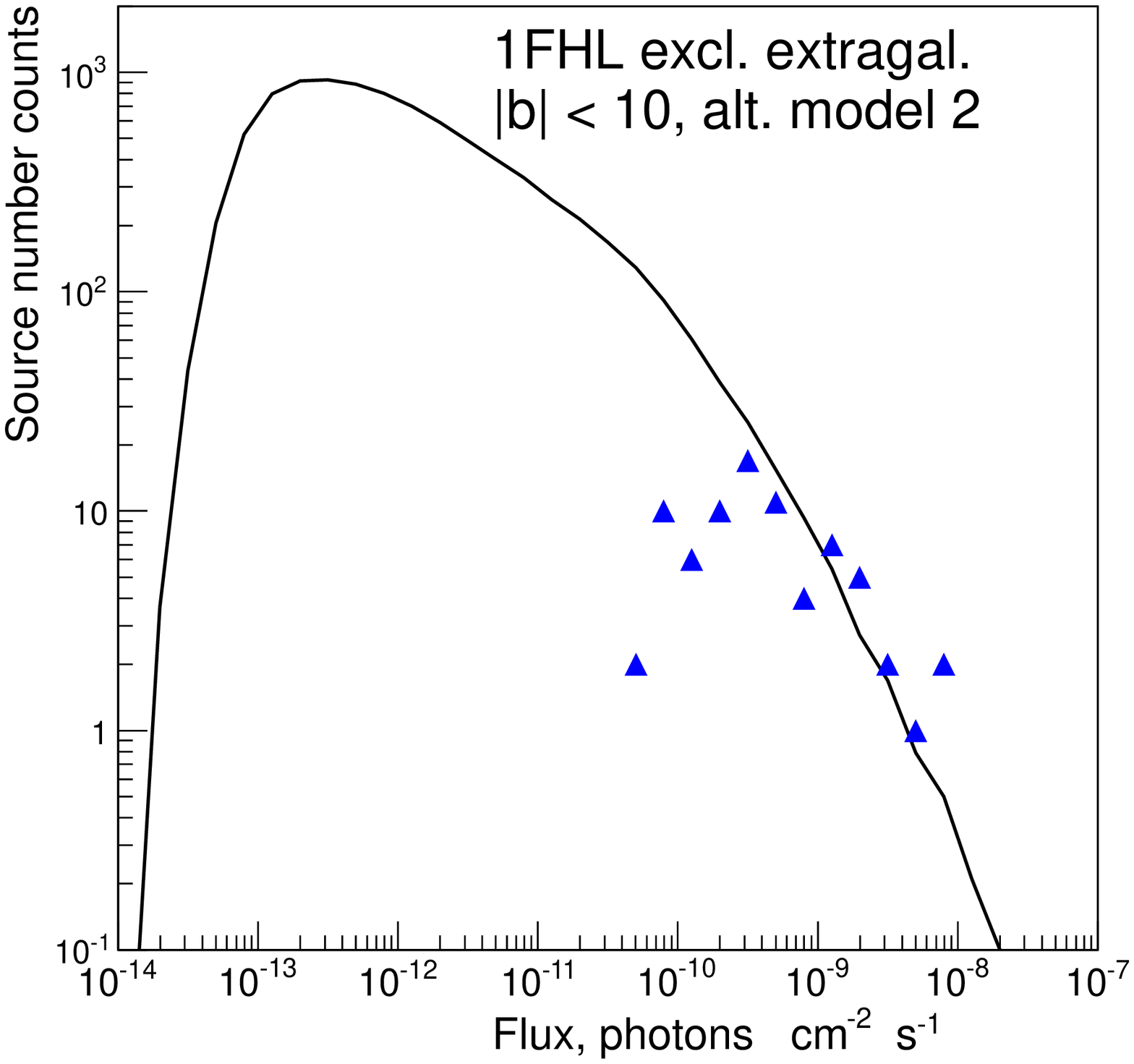} 
\includegraphics[width=6.5cm]{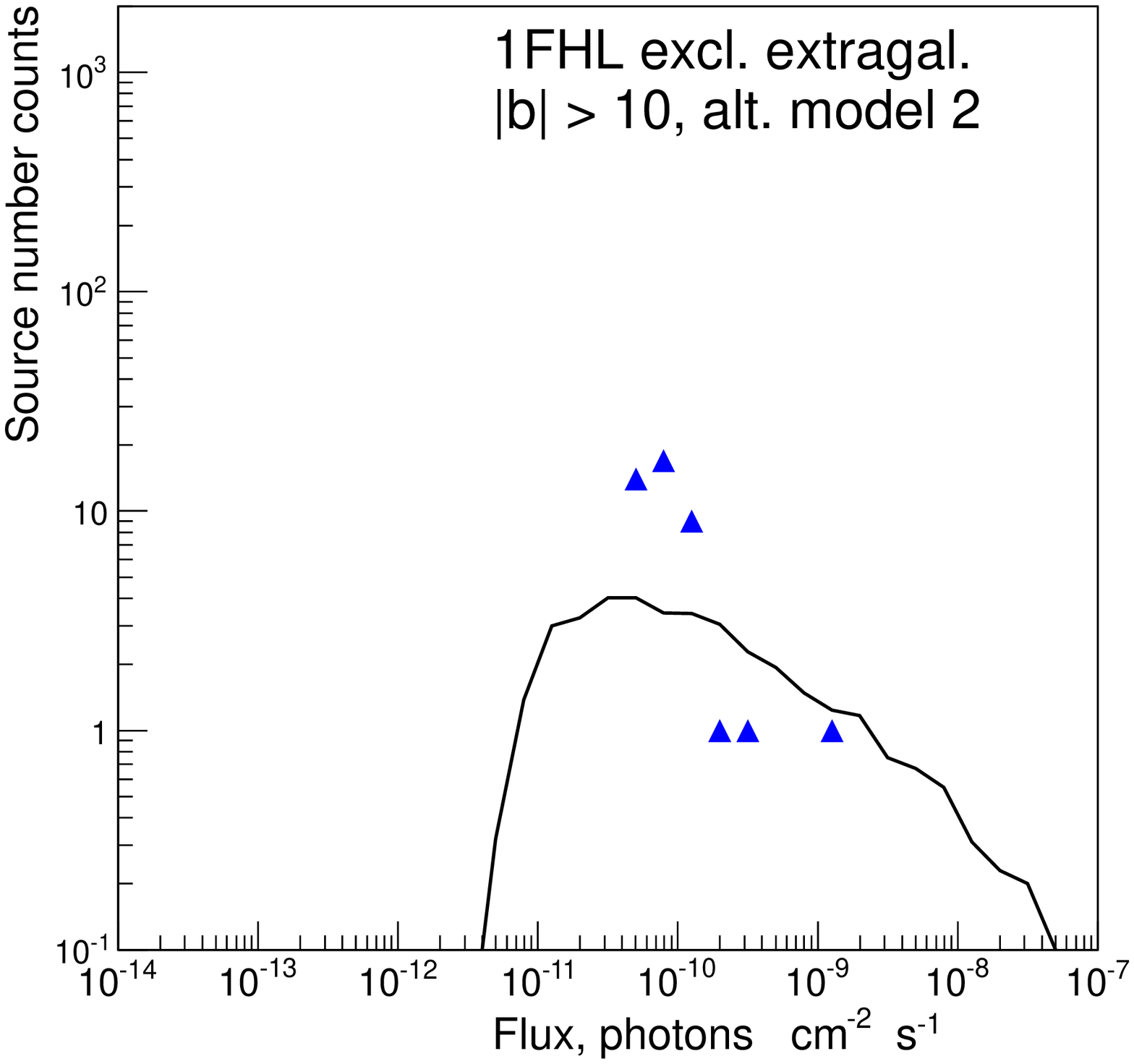} 
\end{center}
\caption{\label{PopulationGal_v5} 
Source number counts for  sources above 10 GeV at low and high latitudes, compared with modified models for the luminosity function.  For the upper row the local source density has been increased to $\rho_\odot$= 10 kpc$^{-3}$, and
$\gamma$-ray luminosity range decreased to   $4 \times 10^{33}$--$4 \times 10^{36}$ \Lunits above 10 GeV (labeled as alternative model 1).  For the lower row $\rho_\odot$= 30 kpc$^{-3}$ and the luminosity range is $1.5 \times 10^{33}$--$1.5 \times 10^{36}$ \Lunits,  for the same $\gamma$-ray luminosity law $\Lgamma^{-1.5}$ and spatial distribution as the reference model (labeled as alternative model 2). 
The blue triangles are derived from the 1FHL data; 
Upper left:  $|b|<10\degr$, all 1FHL sources, excluding those associated with extragalactic sources,
right:  $|b|>10\degr$,   1FHL pulsars,
Lower left:  $|b|<10\degr$, all 1FHL sources, excluding those associated with extragalactic sources,
right:  $|b|>10\degr$,  all 1FHL sources, excluding those associated with extragalactic sources. 
}
\end{figure}

\cleardoublepage

\section{\label{sec:conclusion}Discussion and Conclusions}
\label{Conclusion}

The first \FermiLAT\ catalog of sources above 10 GeV is a catalog of the highest-energy LAT sources.  With its focus on high-energy data, the 1FHL explores how the \gray\ Universe evolves between the 2FGL catalog (which is dominated by emission between 100~MeV and 10~GeV) and the VHE sources detected with ground-based \gray\ instruments (which are dominated by emission from 100~GeV and up).

The 1FHL catalog contains 514 sources.  Because of the steepness of the source count distribution $N(S)$, and the relatively low intensity of diffuse backgrounds (Galactic, extragalactic isotropic, and residual cosmic rays) at $>$10~GeV, which make source detection (TS$>$25) possible with only a few $\gamma$-rays, a
large number of the $>$10~GeV sources are detected close to the threshold, and the median number of $\gamma$ rays per source is 13. 
This very low photon count limits the possibilities for detailed spectral and variability analyses.  We have provided only power-law spectral fits, and applied the Bayesian Block \citep{BayesianBlocks1998} algorithm to study variability
without pre-defined temporal bins.  Our analysis treated 22 sources as
spatially extended, because they have been resolved in previous LAT analyses, typically at lower energies.  For these sources, we adopted their extents as
measured in the previous works.

We studied potential associations between 1FHL sources and counterparts at other wavelengths.  Approximately 75\% have likely associations with AGNs.  Galactic sources (pulsars, PWNs, SNRs, high-mass binaries, and star-forming regions) collectively represent 10\% of the sources.  The fraction of unassociated sources is only 13\%.  Among the 27 associations with known pulsars, we find 20 with significant pulsations above 10 GeV, and 12 with pulsations above 25~GeV, suggesting that the Crab pulsar will not remain the only pulsar to be detected by current and future IACTs.

We detected variability for 43 1FHL sources, all belonging to the blazar class. We found that the most variable of these belong to the SED class LSP, which in some cases have very bright (\gapp 10$\times$) and very short ($\sim$1 day) flaring episodes. 
This result is remarkable because HSP sources (rather than LSP) typically have the largest numbers of detected \grays\ above 10~GeV.  The implication is that the falling segment of the high-energy (presumably inverse-Compton) bump is more variable than the rising segment. This result is consistent with the trend reported at lower energies ($>$100~MeV) by \cite{LAT_2LAC}.

Based on the 84 associations between 1FHL sources and known VHE sources, we developed criteria to select other sources that are likely to be detectable with ground-based \gray\ instruments.  Of the 1FHL sources not already detected in the VHE range, we flagged 212 as good candidates based on their average properties for the 3-year time range of the analysis.

Using the source counts for blazars we estimate that $27 \pm 8$\% of the IGRB for energies $>$10 GeV can be attributed to blazars. 
This contribution to the IGRB in the range $>$10~GeV is well above the lower limit of \cite{pop_pap}; the measurement was enabled by the greater sensitivity here;
the 3-year $N(S)$ samples a factor $\sim$2 weaker fluxes than the 11-month $N(S)$.
Since the $N(S)$ does not show any flattening at the lowest measured fluxes, 
the contribution from blazars may be even larger.
\Fermi\ might survey the sky for 10\,years or more, potentially 
providing a further improvement in the $>$10~GeV sensitivity of the same
magnitude (a factor $\sim$2) as that provided in this work with respect to
the 11\,months of survey data analyzed by \cite{pop_pap}.
\Fermi\ LAT ultimately could be able to
directly resolve $>$40\,\% of the IGRB intensity above 10~GeV.

The source count distributions for sources in the Milky Way (i.e., those with associations with Galactic source classes) and more generally for sources without
extragalactic associations, can be well modeled with a power-law luminosity function for sources with characteristic luminosities in the range $10^{34}$--$10^{37}$ ph s$^{-1}$ above 10 GeV {  and a distribution in Galactocentric distance based on the pulsar distribution of \citet{Lorimer2006} and a scale height of 500~pc}.  From the models, we estimate that $\sim$5\% of the luminosity of the Milky Way above 10~GeV can be attributed to unresolved $\gamma$-ray point sources.


\acknowledgments
The \textit{Fermi} LAT Collaboration acknowledges generous ongoing support from a number of agencies and institutes that have supported both the development and the operation of the LAT as well as scientific data analysis. These include the National Aeronautics and Space Administration and the
Department of Energy in the United States, the Commissariat \`a l'Energie Atomique and the Centre National de la Recherche Scientifique / Institut National de Physique Nucl\'eaire et de Physique des Particules in France, the Agenzia Spaziale Italiana and the Istituto Nazionale di Fisica Nucleare in Italy, the Ministry of Education, Culture, Sports, Science and Technology (MEXT), High Energy Accelerator Research Organization (KEK) and Japan Aerospace Exploration Agency (JAXA) in Japan, and the K.~A.~Wallenberg Foundation, the Swedish Research Council and the Swedish National Space Board in Sweden.

Additional support for science analysis during the operations phase is gratefully acknowledged from the Istituto Nazionale di Astrofisica in Italy and the Centre National d'\'Etudes Spatiales in France.

M. Ajello acknowledges support from NASA grant NNH09ZDA001N for the study of the origin of the Isotropic Gamma-ray Background.  D. Paneque acknowledges support from NASA grant NNX10AP21G for the study of the highest-energy LAT sources.

{\it Facilities:} \facility{Fermi LAT}.




\bibliography{Fermi_Bibtex_LATHardSrc_v1}

\end{document}